\pdfoutput=1

\documentclass[12pt,ngerman,english,a4paper,twoside%
]{report}

\usepackage[ngerman, english]{babel}	

\usepackage[latin1]{inputenc}		  
\usepackage[T1]{fontenc}		  
\usepackage{graphicx,amsfonts,amsbsy} 
\usepackage{array,color,colortbl,rotating,supertabular,amssymb,amsmath}
\usepackage[numbers,sort&compress]{natbib}
\usepackage{mathptmx,courier}
\usepackage[scaled=.9]{helvet}

\newcommand{\dcauthorpre}{Herrn Dipl.-Phys.} 
\newcommand{\dcauthorsurname}{Guazzini} 
\newcommand{\dcauthorname}{Damiano} 
\newcommand{\dcauthoradd}{geboren am 15.11.1980 in Grosseto, Italien}

\newcommand{\dctitle}{Heavy-light mesons in lattice HQET and QCD} 
\newcommand{\dcsubtitle}{~}  

\newcommand{\dcapprovala}{Dr.~Rainer Sommer} 
\newcommand{\dcapprovalb}{Prof.~Dr.~Ulrich Wolff} 
\newcommand{\dcapprovalc}{Dr.~Mike Peardon} 

\newcommand{\dcdegree}{doctor rerum naturalium\\(Dr. rer. nat.)} 
\newcommand{\dcsubject}{Physik} 
\newcommand{\dcfaculty}{Mathematisch-Naturwissenschaftlichen Fakultät I}
\newcommand{\dcuniversity}{Humboldt-Universität zu Berlin}
\newcommand{\dcdean}{Prof. Dr. Christian Limberg}
\newcommand{\dcpresident}{Prof. Dr. Christoph Markschies}

\newcommand{\dcdatesubmitted}{29. März 2007} 
\newcommand{\dcdateexam}{28. Juni 2007} 

\newcommand{\dckeydea}{Gitter-QCD}
\newcommand{\dckeydeb}{HQET}
\newcommand{\dckeydec}{Hadronspektrum}
\newcommand{\dckeyded}{Chromo-magnetische Wechselwirkung}

\newcommand{\dckeywordsde}{\vfill \raggedright {\textbf{Schlagwörter:}}\\ \dckeydea, \dckeydeb, \dckeydec, \dckeyded \\}

\newcommand{\dckeyena}{Lattice QCD}
\newcommand{\dckeyenb}{HQET}
\newcommand{\dckeyenc}{Hadron spectrum}
\newcommand{\dckeyend}{Chromo-magnetic interaction}

\newcommand{\dckeywordsen}{\vfill \raggedright {\textbf{Keywords:}}\\ \dckeyena, \dckeyenb, \dckeyenc, \dckeyend \\}

\newcommand{\dcpdfsubject}{Dissertation}							

\newcommand{\heavy}{\psi_\mathrm{\phantom{\bar{h}^\dagger}\hspace{-0.28cm}h}}
\newcommand{\heavyb}{{\overline{\psi}}_\mathrm{\phantom{\bar{h}^\dagger}\hspace{-0.28cm} h}}
\newcommand{\aheavy}{\psi_{\mathrm{\phantom{\bar{h}^\dagger}\hspace{-0.28cm} \bar{h}}}}
\newcommand{\aheavyb}{\overline{\psi}_{\phantom{\bar{h}^\dagger}\hspace{-0.28cm}\bar{\mathrm{h}}}}

\newcommand{\lightb}{\overline{\psi}_\mathrm{l}}

\newcommand{\rholprime}{\rho_\mathrm{l}\kern1pt'}
\newcommand{\rholbprime}{\overline{\rho}_\mathrm{l}\kern1pt'}
\newcommand{\rhoh}{\rho_\mathrm{\phantom{\bar{h}^\dagger}\hspace{-0.28cm}h}}
\newcommand{\rhohbprime}{\bar{\rho}_\mathrm{\phantom{\bar{h}^\dagger}\hspace{-0.28cm} h}\kern1pt\!\!\!'\,}

\newcommand{\zetalprime}{\zeta_\mathrm{l}\kern1pt'}
\newcommand{\zetaiprime}{\zeta_{i}\kern1pt'}
\newcommand{\zetajprime}{\zeta_{j}\kern1pt'}
\newcommand{\zetalbprime}{\overline{\zeta}_\mathrm{l}\kern1pt'}
\newcommand{\zetaibprime}{\overline{\zeta}_{i}\kern1pt'}
\newcommand{\zetajbprime}{\overline{\zeta}_{j}\kern1pt'}
\newcommand{\zetah}{{\zeta_\mathrm{\phantom{\bar{h}^\dagger}\hspace{-0.28cm} h}}}
\newcommand{\zetahb}{\bar{\zeta}_\mathrm{\phantom{\bar{h}^\dagger}\hspace{-0.28cm} h}}
\newcommand{\zetahprime}{\zeta_\mathrm{h}'}
\newcommand{\zetahbprime}{\bar{\zeta}_\mathrm{\phantom{\bar{h}^\dagger}\hspace{-0.28cm} h}\kern1pt\!\!'}
\newcommand{\Astat}{A^\mathrm{stat}_0}

\newcommand{\Aren}{(A^\mathrm{stat}_\mathrm{R})_0}

\newcommand{\zetalbprimeone}{\overline{\zeta}_1\kern1pt'}
\newcommand{\zetalprimetwo}{\zeta_2\kern1pt'}

\newcommand{\gamSF}{\gamma^\mathrm{SF}}
\newcommand{\gamMSbar}{\gamma^{\overline{\mathrm{MS}}}}

\newcommand{\fastat}{f_\mathrm{A}^\mathrm{stat}}
\newcommand{\fastatimpr}{f_\mathrm{A}^\mathrm{stat, I}}

\newcommand{\fonestat}{f_{1}^\mathrm{stat}}

\newcommand{\fonehh}{f_{1}^\mathrm{hh}}

\newcommand{\zastat}{Z_\mathrm{A}^\mathrm{stat}}

\newcommand{\castat}{c_\mathrm{A}^\mathrm{stat}}

\newcommand{\bastat}{b_\mathrm{A}^\mathrm{stat}}

\newcommand{\ola}{\overleftarrow}

\newcommand{\vectn}{\mathit{\mathbf{0}}}
\newcommand{\vecs}{\mathit{\mathbf{s}}}

\newcommand{\Fm}{\mathrm{fm}}

\newcommand{\Cps}{C_\mathrm{PS}}

\newcommand{\mrm}[1]{\mathrm{#1}}

\renewcommand{\Im}{\mathrm{Im}\,}
\renewcommand{\Re}{\mathrm{Re}\,}

\newcommand{\proof}{\noindent{\normalfont\slshape Proof:}\kern0.6em}

\newcommand{\dual}{\mathstrut^*\kern-0.1em}

\newcommand{\lvec}[1]{\setbox0=\hbox{$#1$}
    \setbox1=\hbox{$\scriptstyle\gets$}
    #1\kern-\wd0\smash{
    \raise\ht0\hbox{$\raise1pt\hbox{$\scriptstyle\gets$}$}}
    \kern-\wd1\kern\wd0}
\newcommand{\rvec}[1]{\setbox0=\hbox{$#1$}
    \setbox1=\hbox{$\scriptstyle\to$}
    #1\kern-\wd0\smash{
    \raise\ht0\hbox{$\raise1pt\hbox{$\scriptstyle\to$}$}}
    \kern-\wd1\kern\wd0}

\newcommand{\nab}[1]{\nabla_{#1}}
\newcommand{\nabstar}[1]{\nabla\kern-0.5pt\smash{\raise 4.5pt\hbox{$\scriptstyle\ast$}}
               \kern-4.5pt_{#1}}
\newcommand{\drv}[1]{{\partial_{#1}}}
\newcommand{\drvstar}[1]{\partial\kern-0.5pt\smash{\raise 4.5pt\hbox{$\scriptstyle\ast$}}
               \kern-5.0pt_{#1}}
\newcommand{\drvsym}[1]{\widetilde{\drv{#1}}}

\newcommand{\momp}[2]{
    \setbox0=\hbox{${#1}$}\setbox1=\hbox{${#1}_{#2}$}
    {#1}_{#2}\kern-\wd1\kern\wd0
    \smash{\raise4.5pt\hbox{$\scriptscriptstyle +$}}}
\newcommand{\momm}[2]{
    \setbox0=\hbox{${#1}$}\setbox1=\hbox{${#1}_{#2}$}
    {#1}_{#2}\kern-\wd1\kern\wd0
    \smash{\raise4.5pt\hbox{$\scriptscriptstyle -$}}}
\newcommand{\mompm}[2]{
    \setbox0=\hbox{${#1}$}\setbox1=\hbox{${#1}_{#2}$}
    {#1}_{#2}\kern-\wd1\kern\wd0
    \smash{\raise4.5pt\hbox{$\scriptscriptstyle \pm$}}}
\newcommand{\smomp}[2]{
    \setbox0=\hbox{${#1}$}\setbox1=\hbox{${#1}_{#2}$}
    {#1}_{#2}\kern-\wd1\kern\wd0
    \smash{\raise3pt\hbox{$\scriptscriptstyle +$}}}
\newcommand{\smomm}[2]{
    \setbox0=\hbox{${#1}$}\setbox1=\hbox{${#1}_{#2}$}
    {#1}_{#2}\kern-\wd1\kern\wd0
    \smash{\raise3pt\hbox{$\scriptscriptstyle -$}}}
\newcommand{\smompm}[2]{
    \setbox0=\hbox{${#1}$}\setbox1=\hbox{${#1}_{#2}$}
    {#1}_{#2}\kern-\wd1\kern\wd0
    \smash{\raise3pt\hbox{$\scriptscriptstyle \pm$}}}

\newcommand{\MeV}{\textrm{MeV}}
\newcommand{\GeV}{\textrm{GeV}}

\newcommand{\psibar}{\overline{\psi}}

\newcommand{\rhoprime}{\rho\kern1pt'\!}
\newcommand{\rhobar}{\bar{\rho}}
\newcommand{\rhobarprime}{\rhobar\kern1pt'\!}
\newcommand{\rhobartilde}{\kern2pt\tilde{\kern-2pt\rhobar}}
\newcommand{\rhobartildeprime}{\kern2pt\tilde{\kern-2pt\rhobar}\kern1pt'}

\newcommand{\zetabar}{\overline{\zeta}}
\newcommand{\zzetaprime}{\zeta\kern1pt'}
\newcommand{\zetabarprime}{\zetabar\kern1pt'}
\newcommand{\zetar}{\zeta_{\raise-1pt\hbox{\mathrm{R}}}}
\newcommand{\zetabarr}{\zetabar_{\raise-1pt\hbox{\mathrm{R}}}}

\newcommand{\phiimpr}{\phi_{\kern0.5pt\mathrm{I}}}

\newcommand{\diracstar}[2]{
    \setbox0=\hbox{$\gamma$}\setbox1=\hbox{$\gamma_{#1}$}
    \gamma_{#1}\kern-\wd1\kern\wd0
    \smash{\raise4.5pt\hbox{$\scriptstyle#2$}}}

\newcommand{\csw}{c_\mathrm{sw}}
\newcommand{\ca}{c_\mathrm{A}}

\newcommand{\cs}{c_\mathrm{s}}
\newcommand{\ct}{c_\mathrm{t}}
\newcommand{\cttil}{\tilde{c}_\mathrm{t}}
\newcommand{\cstil}{\tilde{c}_\mathrm{s}}

\newcommand{\ba}{b_\mathrm{A}}

\newcommand{\Nf}{N_\mathrm{f}}

\newcommand{\SUthree}{\mathrm{SU(3)}}

\newcommand{\tr}{\,\hbox{tr}\,}

\newcommand{\op}[1]{{\mathcal O}_\mathrm{#1}}

\newcommand{\opprime}[1]{\setbox0=\hbox{${\mathcal{O}}$}\setbox1=\hbox{${\mathcal{O}}_\mathrm{#1}$}
    {\mathcal{O}}_\mathrm{#1}\kern-\wd1\kern\wd0
    \smash{\raise4.5pt\hbox{\kern1pt$\scriptstyle\prime$}}\kern1pt}
\newcommand{\ophat}[1]{\widehat{\mathcal{O}}_\mathrm{#1}}

\newcommand{\ophatprime}[1]{\setbox0=\hbox{$\widehat{\mathcal{O}}$}
    \setbox1=\hbox{$\widehat{\mathcal{O}}_\mathrm{#1}$}
    \widehat{\mathcal{O}}_\mathrm{#1}\kern-\wd1\kern\wd0
    \smash{\raise4.5pt\hbox{\kern1pt$\scriptstyle\prime$}}\kern1pt}

\newcommand{\bopprime}[1]{\setbox0=\hbox{${\mathcal{O}}$}\setbox1=\hbox{${\mathcal{O}}_\mathrm{#1}$}
    \mathcal{L}_\mathrm{#1}\kern-\wd1\kern\wd0
    \smash{\raise4.5pt\hbox{\kern1pt$\scriptstyle\prime$}}\kern1pt}

\newcommand{\blagprime}[1]{\setbox0=\hbox{$\mathcal{B}$}\setbox1=\hbox{$\mathcal{B}_{#1}$}
    \mathcal{B}_{#1}\kern-\wd1\kern\wd0
    \smash{\raise5.2pt\hbox{\kern1pt$\scriptstyle\prime$}}\kern1pt}

\newcommand{\gbar}{\bar{g}}
\newcommand{\gbsq}{\bar{g}^2}

\newcommand{\gbarMSbar}{\gbar_{\,\msbar}}
\newcommand{\gbarSF}{\gbar_\mathrm{SF\phantom{\MSbar}}\hspace{-0.48cm}}
\newcommand{\gbsqSF}{\gbar_\mathrm{SF\phantom{\MSbar}}^{2}\hspace{-0.48cm}}
\newcommand{\gbfourSF}{\gbar_\mathrm{SF\phantom{\MSbar}}^{4}\hspace{-0.48cm}}

\newcommand{\glat}{g_{\lat}}

\newcommand{\mbare}{m_0}
\newcommand{\mq}{m_\mathrm{q}}
\newcommand{\mqtilde}{\widetilde{m}_\mathrm{q}}

\newcommand{\mc}{m_\mathrm{c}}

\newcommand{\mbar}{\kern1pt\overline{\kern-1pt m\kern-1pt}\kern1pt}

\newcommand{\za}{Z_\mathrm{A}}

\newcommand{\zp}{Z_\mathrm{P}}

\newcommand{\gtilde}{\tilde{g}_0}

\newcommand{\msbar}{\mathrm{\overline{MS\kern-0.05em}\kern0.05em}}
\newcommand{\lat}{\mathrm{lat}}

\newcommand{\deltaoneprime}{\Delta\kern-1.0pt
    \smash{\raise 4.5pt\hbox{$\scriptstyle\prime$}}
    \kern-1.5pt_{1}}

\newcommand{\Tr}{\mathrm{Tr}}

\newcommand{\vecp}{\mathbf{p}}
\newcommand{\vecphat}{\mathbf{\hat{p}}}
\newcommand{\vecq}{\mathbf{q}}
\newcommand{\vecpprime}{\mathbf{p'}}

\newcommand{\delstar}[1]{\Delta\kern-1.0pt\smash{\raise 4.5pt\hbox{$\scriptstyle\ast$}}
               \kern-4.0pt_{#1}}
\renewcommand{\nab}[1]{{\nabla_{#1}}}
\renewcommand{\nabstar}[1]{\nabla\kern-0.5pt\smash{\raise 4.5pt\hbox{$\scriptstyle\ast$}}
               \kern-4.5pt_{#1}}
\newcommand{\lnab}[1]{{\overleftarrow{\nabla}_{#1}}}
\newcommand{\lnabstar}[1]{\overleftarrow{\nabla}\kern-0.5pt\smash
             {\raise 4.5pt\hbox{$\scriptstyle\ast$}}\kern-4.5pt_{#1}}
\newcommand{\cdev}[1]{D\kern-0.2pt\smash{\raise 4.2pt
            \hbox{$\scriptstyle\phantom{\ast}$}}
            \kern-4.8pt_{#1}}
\newcommand{\cdevstar}[1]{D\kern-0.2pt\smash{\raise 4.2pt
                \hbox{$\scriptstyle\ast$}}
                \kern-4.8pt_{#1}}

\renewcommand{\angle}{\phi^{\vphantom{\prime}}}
\newcommand{\angleprime}{\phi'}

\newcommand{\Dsl}{D \kern-.65em/}

\newcommand{\be}{\begin{equation}}
\newcommand{\ee}{\end{equation}}
\newcommand{\bi}{\begin{itemize}}
\newcommand{\ei}{\end{itemize}}
\newcommand{\bes}{\begin{align}}
\newcommand{\ees}{\end{align}}

\newcommand{\eq}[1]{eq.~(\ref{#1})}

\newcommand{\eqs}[1]{eqs.~(\ref{#1})}

\newcommand{\fig}[1]{Fig.~\ref{#1}}
\newcommand{\Fig}[1]{Fig.~\ref{#1}}
\newcommand{\sect}[1]{Sect.~\ref{#1}}

\newcommand{\app}[1]{App.~\ref{#1}}

\newcommand{\tab}[1]{Table~\ref{#1}}
\newcommand{\Tab}[1]{Table~\ref{#1}}
\newcommand{\Ref}[1]{Ref.~\cite{#1}}

\newcommand{\MSbar}{\overline{\mathrm{MS}}}

\newcommand{\Oa}{\mathrm{O}(a)}
\newcommand{\Oasq}{\mathrm{O}(a^2)}

\newcommand{\vecx}{\mathit{\mathbf{x}}}
\newcommand{\vecy}{\mathit{\mathbf{y}}}
\newcommand{\vecz}{\mathit{\mathbf{z}}}

\newcommand{\meffstat}{\Gamma_\mathrm{stat}}

\newcommand{\SF}{{Schr\"odinger functional }}
\newcommand{\Eume}{{Euclidean metric}}

\newcommand\mycaption[1]{\caption{\footnotesize \normalfont\sffamily #1}}

\newcommand{\fBs}{f_\mathrm{B_\mathrm{s}}}
\newcommand{\MBs}{M_\mathrm{B_\mathrm{s}}}
\newcommand{\ZRGI}{Z_\mathrm{RGI}}
\newcommand{\YRGI}{Y_\mathrm{RGI}}

\newcommand{\psif}{\psi^f}
\newcommand{\psib}{\psi^b}
\newcommand{\vecpplus}{\mathit{\mathbf{p^+}}}
\newcommand{\vcal}{\mathcal{V}}

\usepackage{fancyhdr}

\usepackage{calc}

\fancypagestyle{plain}{%
  \fancyhf{}
       \fancyhead{} 
}
\usepackage[left=25mm,textwidth=145mm,headheight=27.2pt]{geometry}
\usepackage{ifpdf}

\ifpdf
\usepackage[%
	pdftitle={\dctitle},
	pdfauthor={\dcauthorsurname\ \dcauthorname},
	pdfsubject={\dcpdfsubject}, 
	pdfkeywords={\dckeydea, \dckeydeb, \dckeydec, \dckeyded},
	pdfpagemode=UseOutlines,
  colorlinks=true,					
	linkcolor=black,					
	filecolor=black,					
	urlcolor=black,						
	citecolor=black,					
	pdftex=true,              
	plainpages=false,         
	hypertexnames=false,      
	pdfpagelabels=true,       
	hyperindex=true]{hyperref}
\else
	
\fi

\begin{document}

\pagenumbering{roman}

%
%
%
%

\author{von \\ \dcauthorpre\ \dcauthorname\ \dcauthorsurname\ \\ \dcauthoradd}

\title{\vspace{-5cm}\dctitle \\ 
\vspace{0.5cm}
\large{\dcsubtitle} \\ 
\vspace{0.5cm} {\Large{DISSERTATION}}\\ 
\vspace{0.5cm} \large{zur Erlangung des akademischen Grades \\ 
\dcdegree\\ im Fach \dcsubject \\ 
\vspace{0.5cm} eingereicht an der \\ 
\dcfaculty \\ 
\dcuniversity \\}}

\date{\vspace{0.5cm}
\raggedright{
Präsident der Humboldt-Universität zu Berlin:\\
\dcpresident \vspace{-0.3cm}
}\vspace{0.5cm}\\
\raggedright{
Dekan der \dcfaculty:\\
\dcdean \vspace{-0.3cm}
}\vspace{0.5cm}\\
\raggedright{
Gutachter:
\begin{enumerate} 
\item{\dcapprovala} \vspace{-0.3cm}
\item{\dcapprovalb} \vspace{-0.3cm}
\item{\dcapprovalc} \vspace{-0.3cm}
\end{enumerate}} \vspace{0.5cm}
\raggedright{
\begin{tabular}{lll}
eingereicht am: &  &\dcdatesubmitted\\ 
Tag der mündlichen Prüfung: & & \dcdateexam\\
&&\\[5ex]
&&\hspace{-1.8cm}\textrm{DESY-THESIS-2007-034} 
\end{tabular}}\\ 
}
\maketitle



\selectlanguage{english}

\begin{abstract}
\setcounter{page}{2} 
We present a study of a combination of HQET and relativistic
QCD to extract the b-quark mass and the $\mathrm{B_s}$-meson decay constant from 
lattice quenched simulations. We start from a small volume, where 
one can directly simulate the b-quark, and compute the connection to a 
large volume, where finite size effects are negligible, through a finite 
size technique. The latter consists of steps extrapolated to the 
continuum limit, where the b-region is reached through interpolations 
guided by the effective theory. 

With the lattice spacing given in terms of the Sommer's scale $r_0$ and the 
experimental $\mathrm{B_s}$ and K masses, we get the final results for the 
renormalization group invariant mass $M_\mathrm{b} = 6.88(10)$ GeV, translating into
$\overline{m}_{\mathrm{b}}(\overline{m}_{\mathrm{b}}) = 4.42(6)$ GeV in the $\MSbar$ scheme, 
and $f_\mathrm{B_s} = 191(6)$ MeV for the decay constant.

A renormalization condition for the chromo-magnetic operator, responsible, 
at leading order in the heavy quark mass expansion of HQET, for the  
mass splitting between the pseudoscalar and the vector channel in 
mesonic heavy-light bound states, is provided in terms of lattice correlations
functions which well suits a non-perturbative computation involving
a large range of renormalization scales and no valence quarks. 

The two-loop expression of the corresponding anomalous dimension in 
the Schrödinger functional (SF) scheme is computed starting from results in the 
literature; it requires a one-loop calculation in the SF scheme with
a non-vanishing background field. The cutoff effects affecting the scale 
evolution of the renormalization factors are studied at one-loop order, and 
confirmed by non-perturbative quenched computations to be negligible for the 
numerical precision achievable at present.
\dckeywordsen				
\end{abstract}


\selectlanguage{ngerman}

\begin{abstract}
\setcounter{page}{3} 
  Wir stellen eine Untersuchung einer Kombination zwischen HQET und
  relativistischer QCD vor, die das Ziel hat, die b-Quark Masse und 
  die Zerfallskonstante des $\mathrm{B_s}$-Mesons aus Gitter-Simulationen, 
  unter Nichtbeachtung virtueller Fermionenschleifen, zu gewinnen. 
  Wir beginnen mit einem kleinen Volumen, in dem man das b-Quark 
  direkt simulieren kann, und stellen die numerische Verbindung mit 
  einem großen Volumen, wo ``finite-size'' Effekte vernachlässigbar 
  sind, mit Hilfe einer ``finite-size'' Methode her.  Diese besteht aus 
  zum Kontinuum extrapolierten Schritten, wobei 
  der Massenpunkt, der der physikalischen 
  b-Quark Masse entspricht, durch eine Interpolation
  erreicht wird. In diese Interpolation fliessen die in der
  HQET erzielten Resultate ein.

  Mit dem durch die Sommersche Skale $r_0$ bestimmten Gitterabstand
  und den experimentalen Werten für die $\mathrm{B_s}$- und K-Massen
  erhalten wir die Endergebnisse für die renormierungsgruppeninvariante 
  Masse $M_\mathrm{b} = 6.88(10)$ GeV, äquivalent zu 
  $\overline{m}_{\mathrm{b}}(\overline{m}_{\mathrm{b}}) = 4.42(6)$ 
  GeV in dem $\MSbar$-Schema und $f_\mathrm{B_s} = 191(6)$ MeV für die 
  Zerfallskonstante.

  \hspace{-0.45pt}Eine Renormierungsbedingung für den Chromo-magne­ti­schen Ope­ra­tor,
  der in füh­ren­der Ord­nung der Entwicklung in der schweren Quarkmasse 
  in HQET für die Massenaufspaltung zwischen dem pseudoskalaren und 
  dem vektoriellen Kanal mesonischer schwer-leicht gebundener Zustände 
  verantwortlich ist, wird auf der Basis von Gitter-Korrelationsfunktionen 
  bereitgestellt.  Dies eignet sich gut für eine nicht-~\hspace{-0.2cm}störungstheoretische 
  Rechnung, welche einen großen Bereich der Renormierungsskala umfasst 
  und keine Valenz-Quarks beinhaltet.

  Die Zwei-Schleifen Ordnung der entsprechenden anomalen Dimension im \linebreak 
  Schrödinger-Funktional-Schema wird mit Hilfe von veröffentlichten 
  Ergebnissen berechnet; dies erforderte eine neue Ein-Schleifen Rechnung 
  im SF-Schema mit einem nicht verschwindenden 
  Hintergrundfeld. Die Gitterartefakte bezüglich der Skalenentwicklung 
  des Renormierungsfaktors werden zur Ein-Schleifen Ordnung untersucht, 
  und es wird von nicht-störungstheoretischen Simulationen, unter 
  Nichtbeachtung virtueller Fermionenschleifen, bestätigt, dass sie für 
  die gegenwärtige verfügbare numerische Präzision vernachlässigbar  
  sind.
\dckeywordsde
\end{abstract}

\selectlanguage{english}				
\setcounter{page}{4}  					




\tableofcontents

\listoffigures

\listoftables
\newpage

\pagenumbering{arabic}
\setcounter{page}{1}


\selectlanguage{english}
\chapter{Introduction}\label{chap:intro}
\pagenumbering{arabic}
\setcounter{page}{1}

In the seventeenth century the scientific method was born.
Its father, Galileo Galilei, proposed it as the correct way to 
understand nature, focusing the attention on the concepts of reproducibility
of a physical phenomenon and predictivity of its theoretical description.
Along this method, present-day particle physics research represents the
most ambitious and most organized effort to understand 
world's underlying structure. 

Besides the gravitational one, modern physicists have counted three
fundamental interactions: strong, weak and electromagnetic. 
To date,  almost all experimental tests of the three forces have well
agreed with the predictions of the Standard Model. This is not really surprising, 
since the latter has been built from the experimental observations. 
Its mathematical framework is represented by quantum field theory,
where each type of particle is described in terms of a field, obeying
to a local gauge principle, with the gauge group
\begin{displaymath}
\mathrm{SU(3)_c\times SU(2)_L\times U(1)_Y}\,.
\end{displaymath}
Quantum chromodynamics (QCD) is associated with the first 
of these groups, and is the theory of the strong interactions. 
The matter constituents are fermions; six quarks, organized in three generations
\begin{displaymath}
\left(
\begin{array}{c}
\mathrm{u} \\
\mathrm{d} \\
\end{array}
\right)\,,\quad
\left(
\begin{array}{c}
\mathrm{s} \\
\mathrm{c} \\
\end{array}
\right)\,,\quad
\left(
\begin{array}{c}
\mathrm{b} \\
\mathrm{t} \\
\end{array}
\right)\,,
\end{displaymath}
whose color interaction is mediated by vector 
bosons, the gluons. The latter have a color charge too, and,
like all other non-Abelian gauge theories (and unlike quantum electrodynamics), they interact with one another 
by the same force that affects the quarks.

QCD enjoys two peculiar properties:
asymptotic freedom and confinement. The former is relevant at high energies, or short distances,
where quarks and gluons are found to be weakly coupled. As the energy scales become arbitrarily large,
one is left with a non-interacting theory. This pattern has been verified in several 
experimental observations, especially in deep inelastic lepton-hadron scatterings. 
The theoretical explanation of the phenomenon came in 1973 thanks to D.~Gross, D.~Politzer 
and F.~Wilczek \cite{Politzer:1973fx,Gross:1973id}, in the framework of the renormalization group.
For sufficiently short distances or large momentum transfer, an asymptotically free theory, 
amenable to perturbation theory calculations using Feynman diagrams, emerges. 
Such situations are therefore more theoretically tractable, under the analytical point of view, than the long-distance, 
strong-coupling behavior, where confinement dominates. Although an analytical proof is still missing, confinement 
is widely accepted, and explains the consistent failure of free quark searches, as well as many other
accurate experimental data, like hadron masses.
In the low-energy regime, the coupling constant 
assumes values, which are too large to have a reliable perturbative expansion. It follows
that the theoretical predictions need other techniques. The answer came in 1974, when
K.~Wilson proposed \cite{Wilson} a way to quantize a gauge field theory on a discrete 
lattice in Euclidean space-time, preserving exact gauge invariance. The lattice gauge theory 
he developed has a computable strong-coupling limit, where it is possible to show the
color confinement, and thus the absence of free quarks.
Lattice QCD shares with the continuum QCD formulation the property of relying
on very few parameters, and allows first-principle predictions without 
any additional assumption. The lattice provides a regularization of the theory,
which becomes ultraviolet finite thanks to the presence of a cutoff proportional to
the inverse of the lattice spacing $a$,
while infrared divergences can be avoided 
by choosing either non-vanishing quark masses or particular boundary conditions. 
Such a theory suits well the implementation on a computer, or even on a cluster of them,
and is not restricted to the low-energy regime. Perturbative lattice computations 
are possible and useful; however, they are generally more complicated than in the continuum.\\
\indent Once a lattice setup has been chosen, the desired observable 
can be computed by evaluating the corresponding Feynman path integrals
via Monte Carlo techniques. Apart from statistical errors, 
such an evaluation is, however, only an approximation 
of the wished result, because of the finiteness of the lattice cutoff. As for any 
other regularization method, the regulator has to be removed. This amounts to take 
the continuum limit. The lattice discretization is made finer and finer, and, at 
the same time, one is asked to tune 
the bare parameters of the theory according
to the renormalization group equations, in order to 
keep fixed the corresponding physical quantities. 
The fineness of the lattice is strongly limited by the available computer facilities
as well as by the details of the Monte Carlo samplings. 
As a set of lattices have been simulated, one in general expects to find discretization 
errors vanishing linearly with the lattice spacing. This can be trusted if $a$ is
small enough compared to the other involved scales.
Furthermore, it often constrains one to vary $a$
over a large range before having a controlled estimate of the size of the 
lattice artifacts. The latter can also be quite large. The issue can be solved
by ``decorating'' the action and the correlation functions with irrelevant terms, 
i.e.~vanishing in the continuum limit, which, properly tuned, cancels the $\mathrm{O}(a)$-artifacts.

The Feynman path integrals involve the fermion determinant that is 
non-local and very expensive to evaluate numerically. In order to bring
the computational effort to an acceptable level, lattice physicists have introduced
the quenched approximation. The latter amounts to disregard the 
fermionic contributions in the generation of the gauge configurations. 
In the language of Feynman diagrams, it can be interpreted as
the removal of all virtual quark lines. The drawback is that one is confronted with 
an uncontrolled approximation. Deviations from full, or unquenched, QCD results are almost unpredictable,
as long as the latter are unknown. Still, quenched simulations have led to results
in many cases in very good agreement with experiments, and represent a powerful tool to test
techniques intended for later use in full QCD.

In spite of the richness of physical phenomena associated with the strong interactions, 
there are many phenomenologically interesting subjects which are not directly related with QCD alone.
Among others, it is the case of the weak interactions. Lattice QCD allows to study observables,
encoding important informations upon matrix elements of the effective weak Hamiltonian  
between QCD bound states. Outstanding examples are form factors and decay constants.  

However, a scan through a review of particle properties gives 
a rough idea of the abundance of open questions. 
As one intends to face all of them armed with lattice QCD, one discovers that 
a brute force computational approach is clearly doomed.

A way to solve the problem is to find alternative
discretizations of the theory, as well as more efficient 
computational algorithms. This is helpful, but in many cases not sufficient to 
obtain predictions precise enough to be competitive with most recent
experimental determinations.
 
A powerful tool is represented by the approximated 
symmetries of the theory. 
The light quarks $\mathrm{u}$ and $\mathrm{d}$ have masses which are much
smaller than the typical QCD scale $\Lambda_\mathrm{QCD}$, which
is of the order of a few hundreds of MeV. Consequently, for many QCD
processes it is a good approximation to take the
limit $m_\mathrm{u,d}\to 0$.
In this limit the theory shows a symmetry, associated with the
group $\mathrm{SU(2)_L\times SU(2)_R}$,
called chiral symmetry, and many low-energy properties of QCD are
related to a few matrix elements.
One may also approximate the strange quark to be massless, thus
having an enlarged symmetry $\mathrm{SU(3)_L\times SU(3)_R}$, but also
expect finite quark mass corrections to be not small.

Analogously, one observes that the remaining three quarks, namely
the charm, the bottom and the top quarks, have masses much bigger
than the QCD scale. Lattice gauge theory is then found to face a 
multi-scale problem. Let us consider the case of a heavy-light meson, 
i.e.~a hadron composed\footnote{Of course these considerations hold also for a meson
made of a heavy antiquark and a light quark, and even more generally
for heavy-light hadronic bound states.} of a heavy quark (e.g.~the bottom), and a light 
antiquark (e.g.~the antistrange). 
Due to the small Compton wavelength associated with the former, one would require
a cutoff $a^{-1}$ much bigger than the quark mass in order to have a good control over the 
aforementioned discretization errors. At the same time, the light antiquark wave function
introduces a widely spread object, demanding a large volume in physical units in order to 
have negligible finite volume effects.
It is not possible to fulfill both requirements at the same time, because one would need a
lattice resolution which is not affordable even by the most recent supercomputers.\\ 
\indent Several ways of dealing with this problem have 
been proposed and investigated. Instead of mentioning all of them,
we focus our attention on the conceptually simplest but all the while very
powerful approach, the Heavy Quark Effective Theory (HQET). When the Compton wavelength $1/m$ of the heavy quark is much smaller
than the size of the meson, the low energy properties of the latter are scantily sensitive
to the heavy quark mass. The non-perturbative dynamics of the meson can thus be approximated \cite{stat:eichten}
by considering the heavy quark to be static, i.e.~propagating only in time in the rest frame of the hadron.
While this approximation describes the correct 
asymptotic behavior of the bound state in the limit $m\to\infty$, and is 
interesting by itself, finite 
quark mass corrections can
be studied in a systematic way through an expansion in powers of
$1/m$, and allow to make closer contact with phenomenology.
In the static approximation the heavy quark is seen just as a static source of color
by the light degrees of freedom surrounding it. They are blind to the flavor and the spin 
of the static quark.
It means that one has at one's disposal further symmetries in addition to the ones QCD already comprises, and thus 
promising tools to extract interesting properties of heavy-light bound states by using 
a lattice formulation of HQET. How big the finite quark mass corrections are, it cannot be said a priori.
They depend on the involved masses as well as on the observable itself.\\
\indent Another interesting approach has roots in HQET, but exploits the QCD Lagrangian. 
Proposed by the Tor Vergata group \cite{Guagnelli:2002jd}, it is based on a finite size technique,
called Step Scaling Method,
where the heavy-light system is simulated, by using the QCD Lagrangian,
on a volume small enough, in physical units, to allow an accurate treatment of the heavy quark dynamics,
and to extrapolate to the continuum limit with confidence.
The light degrees of freedom are clearly squeezed in such a box. Through several steps, one then
performs the connection to an ideally infinite volume. In practice, the latter is a box large enough
to let consider the finite size effects negligible in comparison with the numerical precision on the 
computed observables. The steps consist in doubling the volume and halving the heavy quark mass in physical
units, in order to keep the discretization errors under control. Each step 
is performed for several heavy quark masses and extrapolated to the continuum limit. The main assumption
of the method is that the step scaling functions have a mild dependence on the heavy quark mass. One can thus join 
the heavy quark region of interest through an extrapolation, while the light quark mass is held
fixed to its phenomenological value during the whole evolution.\\ 
\indent The Step Scaling Method represents a very attractive approach to study the properties of 
heavy-light bound states; still, it entails extrapolations which 
cannot be considered reliable a priori, especially if the simulated 
heavy quark masses are much lighter than the extrapolated ones, and if a very high precision
is desired.\\
\indent By combining the static limit of HQET and the Step Scaling Method one can turn the extrapolations
of the latter into interpolations in all volumes. The upshot is a method which promises to allow to study 
heavy-light meson properties with astonishing confidence and precision also in the b-quark energy scale.\\
\indent The interplay of HQET and the QCD Step Scaling Method can be exploited also to extract important
informations concerning the magnitude of the finite heavy quark mass corrections to the static limit.
These can be directly studied in HQET too. At order $1/m$ the terms in the Lagrangian breaking
the spin-flavor symmetry of the static theory are two \cite{stat:eichhill2,Falk:1990pz}. One is 
the kinetic term, responsible
for the spatial motion of the heavy quark in the rest frame of the hadron, and the second is 
the chromo-magnetic term, describing the interaction of the heavy quark spin with the gluon field.
Restricting our attention to the heavy-light mesons, we stress that
the chromo-magnetic interaction is responsible, among other effects, for the mass splitting between the ground state
pseudoscalar and vector mesons. Since the splitting is known from experiments with 
very high accuracy, it represents an excellent testing ground for the effective theory, both 
in the bottom channel and in the charm one.\\ 
\indent In this context the renormalization of the chromo-magnetic operator plays a crucial role.
The renormalization procedure is simplified by the absence of operator mixing \cite{Falk:1990pz}. 
Furthermore, one would
like to relate the mass splitting to a scale and scheme independent matrix element. 
This is possible by exploiting the general strategy of the ALPHA collaboration 
\cite{alpha:sigma,alpha:letter,mbar:pap1}, which addresses the question how the perturbative
regime is related to the observed hadronic properties. The key idea is the introduction
of a renormalization scheme, the \SF, in which the scale evolution of the renormalization factor of the 
chromo-magnetic operator can be computed non-perturbatively from low to very high energies. 
In the high-energy regime
one can continue the scale evolution using perturbation theory, which allows to determine the 
renormalization group invariant operator. 
Working in a mass independent renormalization scheme,
the scale evolution of the renormalization factor is ruled by the renormalization
group equations, where the renormalized coupling, the corresponding $\beta$-function, and 
the anomalous dimension of the operator appear. A precise perturbative knowledge 
of the latter is an important brick to bridge 
the high-energy regime with an ideally infinite energy scale, where the RGI renormalization factor 
can be determined.\\   
\indent The rest of the thesis is organized as follows. In chapter \ref{chap:hqet_pheno}
an introduction to HQET with special care to the phenomenological aspects is given.
The static approximation and its symmetries are discussed, and the importance of a non-perturbative
renormalization is stressed. The final section concerns the matching of the effective theory with QCD.
The basic ideas are explained in order to make easier the reading of the following chapters.\\ 
\indent Chapter \ref{chap:SF} is dedicated to the \SF scheme. After its definition in the continuum,
the lattice discretization is introduced and followed by a short description of the Monte Carlo
methods, employed for the non-perturbative computations of this work. The basic concepts
of the $\Oa$-improvement of both relativistic QCD and HQET are explained, and are of central importance
throughout the following. The last section is strongly connected with chapter \ref{chap:SSM}.
In the former, explicit expressions for correlation functions and observables, stemming
from relativistic quark and static-light bilinears, are derived. In addition, their connection to
physical quantities like decay constants and hadron masses is set up. The improvement
of the correlation function is discussed, and the last subsection deals with their renormalization.\\ 
\indent With chapter \ref{chap:SSM} one enters the main part of the thesis. 
The first section explains how the interplay between HQET and the QCD Step Scaling Method 
works. Before presenting the results, the choice of the simulation parameters
is motivated, and all details necessary to compute them are provided. The definition of 
renormalization group invariant quark mass is expounded too. The rest of the chapter is split into
two parts, containing the main results of this work on the combination
of the effective theory and relativistic QCD. 
The first part discusses the non-perturbative computation of the b-quark mass, by starting 
simply from the $\mathrm{B_s}$-meson and Kaon masses. The second part
exploits the same experimental inputs, and achieves a non-perturbative computation of the $\mathrm{B_s}$-meson 
decay constant. All simulations are performed in the quenched approximation.\\
\indent Chapter \ref{chap:Zspin} provides an expression of the mass spitting between the pseudoscalar and 
the vector heavy-light mesons in terms of HQET. At order $1/m$, it can be expressed as a matrix element
of the chromo-magnetic operator, for which a proposal for the non-perturbative renormalization is furnished.
The main point is that it involves the computation of only pure gauge observables.
In order to get the total renormalization factor relating the bare operator to the renormalization group invariant one,
perturbation theory plays a fundamental role. After explaining the basics of the perturbative expansion
in the \SF scheme, and of the connection between the latter and the $\MSbar$ scheme, a formalism for 
the computation and the improvement of Wilson loops at one-loop order 
is provided, with special care dedicated to its implementation on a computer. The remaining
part of the chapter presents the main results on the subject: the two-loop expression of the anomalous dimension   
of the chromo-magnetic operator in the \SF scheme, and a study of the cutoff effects affecting the corresponding
renormalization factor. The latter are also compared with non-perturbative simulations at weak coupling
and in the quenched approximation.

In chapter \ref{chap:PTcode} the code used for the perturbative computations
of chapter \ref{chap:Zspin} is presented. It enables the user to compute up to and including
the one-loop order of any Wilson loop in the \SF scheme, as well as the improvement counterterms,
with or without non-vanishing background field. The first section is thought as a basic documentation,
which lets the reader understand the structure of the code and how to use it without entering
in the details of the underlying subroutines. The following sections are devoted to an explanation
of the modules constituting the pillars of the program. Each of them is an important part of the program,
but can also be used separately from the rest. Finally, two of the tests performed to check the correctness
of the program are discussed.

Part of the appendices are simply a report of the details of the numerical results, while
the remaining appendices present the details of the analysis procedure, and are thought as tools not restricted
to this work only.

Finally, the last chapter summarizes all results and discusses an outlook.


\chapter{Phenomenology of the heavy quarks}\label{chap:hqet_pheno}
\section{Strong interactions}\label{sect:strong_int}

A deep insight in the hadronic structure was given in the late 1960s
by the SLAC-MIT deep inelastic scattering experiments. These experiments
were a decisive test for models, in which hadrons behave like a complex
cloud of softly bounded constituents. In scatterings between an electron beam 
and a hydrogen target, these models predicted very low scattering rates.
Instead, the SLAC-MIT experiments showed a substantial rate
for hard scatterings of electrons from protons, where only in rare cases
did a single proton emerge from the process.

The explanation of these phenomena was given by the simple model
advanced by Bjorken and Feynman: the parton model. The latter predicts
that a hadron is a loosely bound assemblage of a small number of pointlike
constituents (quarks and antiquarks), obeying to the Fermi statistics, and 
carrying electric charge, and possibly other kinds of interaction,
responsible for their binding.

Historically Quantum Chromodynamics was born as a development of this model,
and it exploits the power of quantum field theory to explain the strong
interactions, which bind quarks and antiquarks together in the observed 
hadrons.

Quarks are identified by a flavor quantum number and are triplets
of the SU(3) gauge symmetry, the color. 
Under a gauge transformation $g(x)$, which is a matrix in the gauge
group representation, the quark field $\psi$ transforms according to
\be\label{eq:psi_under_gauge}
\psi(x)\to\psi^g(x)=g(x)\psi(x)\,.
\ee
Quarks interact through the corresponding gauge quanta, the gluons,
described by the gauge potential $A_\mu(x)$. The latter is a vector field
belonging to the $su(3)$ Lie algebra, and transforming as
\be\label{eq:A_under_gauge}
A_\mu(x)\to A_\mu^g(x)=g(x)A_\mu(x)g^{-1}(x)+g(x)\partial_\mu g^{-1}(x)\,.
\ee 
The covariant derivative $D_\mu$,
\be\label{eq:cov_dev_QCD}
D_\mu=\partial_\mu+A_\mu\,,
\ee
transforms as follows
\begin{align}
D_\mu(x)\to D_\mu^g(x)&= g(x)D_\mu (x)g^{-1}(x)\nonumber\\
&= \partial_\mu + g(x)A_\mu(x)g^{-1}(x)+g(x)\partial_\mu g^{-1}(x)\,.\label{eq:D_under_gauge}
\end{align}
The curvature tensor
\be\label{eq:F_munu_QCD}
F_{\mu\nu}=[D_\mu,D_\nu]=\partial_\mu A_\nu -\partial_\nu A_\mu + [A_\mu,A_\nu]\,.
\ee
is an element of the $su(3)$ Lie algebra as well, and transforms as
\be\label{eq:F_under_gauge}
F_{\mu\nu}(x)\to F_{\mu\nu}^g(x)=g(x)F_{\mu\nu}(x)g^{-1}(x)\,.
\ee 
With all these ingredients we can write down the QCD Lagrangian 
\be\label{eq:QCD_Lagrangian}
\mathcal{L}_\mathrm{QCD}=-\frac{ 1}{ 2g_0^2}
\tr\{F_{\mu\nu}F_{\mu\nu}\}+\sum_\mathrm{f=1}^{\Nf}\psibar_\mathrm{f}(\Dsl +m_\mathrm{f})
\psi_\mathrm{f}\,,
\ee 
where $g_0$ is the bare gauge coupling and the color indices are omitted.
The last term in the r.h.s. of \eq{eq:QCD_Lagrangian} is the Dirac Lagrangian
for $\Nf$ flavors, with the covariant derivative $\Dsl=D_\mu\gamma_\mu$,
and $m_\mathrm{f}$ is the bare quark mass of the flavor f. We have chosen
the \Eume, and the the definition of the Euclidean $\gamma$-matrices 
is given in \app{app:Not_Conv}.

Given an operator ${\mathcal{O}}$, a gauge invariant product of fields,
one computes its expectation value in QCD by evaluating
\be\label{eq:O_in_QCD}
\langle{\mathcal{O}} \rangle_\mathrm{QCD}=\frac{ 1}{\mathcal{Z}_\mathrm{QCD}^{\phantom{a}}}
\int_\mathrm{all\,fields}{\mathcal{O}}\,\mathrm{e}^{-\int\!\mathrm{d}^4 x\, \mathcal{L}_\mathrm{QCD}}\,,
\ee
where $\mathcal{Z}_\mathrm{QCD}$ is the QCD partition function 
\be\label{eq:Z_QCD}
\mathcal{Z}_\mathrm{QCD}=\int_\mathrm{all\,fields}\mathrm{e}^{-\int\!\mathrm{d}^4 x\, \mathcal{L}_\mathrm{QCD}}\,,
\ee
normalized in such a way that $\langle 1 \rangle_\mathrm{QCD}=1$. A more rigorous
definition of the partition function and of the evaluation of the integral in \eq{eq:O_in_QCD}
is provided in the next chapter within the lattice regularization.

As shown in \eq{eq:A_under_gauge}, the gauge field $A_\mu$ non trivially
transforms under the gauge group. As a consequence, the curvature tensor $F_{\mu\nu}$
is not gauge invariant, and thus not directly a physical observable. 

At high energies, as in the SLAC-MIT deep inelastic scattering experiments,
quarks behave almost like free particles, as indicated
by the parton model. This property has been theoretically
clarified in the early 1970s by Politzer, Gross and Wilczek \cite{Politzer:1973fx,Gross:1973id},
and is known under the name of asymptotic freedom. Quarks are still bound inside the hadrons,
and their binding force asymptotically vanishes as they get closer and closer.
Corrections to the free quark theory can be computed by performing
a perturbative expansion in powers of the coupling. As the coupling
itself, these corrections vanish only in the infinite energy limit.

Increasing the distance between the quarks, the perturbative
computations become less and less accurate. The coupling becomes
stronger and stronger, compromising the accuracy of the perturbative expansion.
As a consequence, no free quarks have ever been observed at large distance.

A complete study of the hadron properties requires one to be able to deal with both
energy regimes, and this can be achieved only through non-perturbative methods, like 
lattice QCD.

\section{Spin-flavor symmetry}\label{sect:spin-flav_symm}

 
Considering a meson composed by a heavy quark (e.g. the bottom quark)
and a light antiquark, one is found to face a non-perturbative two-scale problem,
which is difficult to be solved, with the Lagrangian in 
\eq{eq:QCD_Lagrangian}, by brute force computations.

By taking the limit of infinite quark mass, important symmetries
appear, which can be exploited to compute interesting meson properties.
In this limit the heavy quark is static, i.e.\hspace{4pt}it propagates only in time in the
rest frame of the meson. 
The quark is surrounded by a complicated cloud of gluons and quark-antiquark pairs,
and the light valence antiquark. They form the light degrees of freedom.
They can see neither the flavor nor the spin of the static quark, because the latter is infinitely
heavy; they just see it as a static source of color. 

Since the flavor and the spin of the static quark are irrelevant, the symmetry group is $\mathrm{SU}(2)$.
In general one can consider the case of $N_\mathrm{h}$ heavy flavors, 
thus getting an $\mathrm{SU}(2N_\mathrm{h})$ spin-flavor symmetry.

A simple physical picture of the system can be got by remembering that the heavy quark
and the light antiquark are particles with spin $1/2$. The total
spin operator and its component along the $\hat{z}$ axis are given respectively by 
\begin{align}
\vec{S}&=\vec{S_\mathrm{h}}+\vec{S_\mathrm{l}}\,,\label{eq:total_spi_op}\\
\vec{S^z}&=\vec{S_\mathrm{h}^z}+\vec{S_\mathrm{l}^z}\,,\label{eq:z_spi_op}
\end{align}
with eigenvalues $s(s+1)$ for $S^2=\vec{S}\cdot \vec{S}$ and $s_z$ for $\vec{S^z}$.
It is then simple to 
construct a singlet with spin and parity $s^{\pi}=0^-$,
and a triplet with $s^{\pi}=1^-$:
\begin{align}
s^{\pi}&=1^-\Leftrightarrow\left\{\begin{array}{l}
|1,+1\rangle = |\hspace{-0.1cm}\uparrow\uparrow\rangle\,, \\
\\
|1,0\rangle = \frac{1}{\sqrt{2}}(|\hspace{-0.1cm}\uparrow\downarrow\rangle 
+ |\hspace{-0.1cm}\downarrow\uparrow\rangle)\,,\label{eq:spin1}\\  
\\
|1,-1\rangle = |\hspace{-0.1cm}\downarrow\downarrow\rangle\,, 
\end{array}\right.\\
& \nonumber\\
s^{\pi}&= 0^-\Leftrightarrow|0,0\rangle=\frac{1}{\sqrt{2}}(|\hspace{-0.1cm}\uparrow\downarrow\rangle 
- |\hspace{-0.1cm}\downarrow\uparrow\rangle)\,,\label{eq:spin0}
\end{align}
where in the kets the numeric labels are the values of $s$ and $s_z$, and the first arrow refers to
the spin of the heavy quark along the $\hat{z}$ axis, while the second arrow refers to that
of the light antiquark.

In the infinite quark mass limit the spin $\vec{S_\mathrm{h}}$ of the heavy quark 
and the spin of the light antiquark $\vec{S_\mathrm{l}}$ are separately conserved
by the strong interactions, and the four states in eqs.~(\ref{eq:spin1},\,\ref{eq:spin0}) are degenerate in mass.\\
Nevertheless the system is still not trivial, because the interaction
between the static quark and the light degrees of freedom is
strong, and must be solved in a non-per\-tur\-ba\-ti\-ve way.

The physical picture that we have just discussed corresponds
to the leading order of the Heavy Quark Effective Theory (HQET) \cite{Isgur:1989vq}.

In nature the heavy quark has a finite mass, and the finite quark mass corrections
can be systematically computed in HQET. The starting
point for such computations is the Lagrangian derived in the next section.

\section{The effective Lagrangian}\label{sect:classic_lagrangian}

In this section we derive the continuum effective Lagrangian of HQET at the classical 
level, following the idea of \cite{Korner:1991kf,Sommer:Nara}.

The goal is to describe the dynamics of a hadron containing one heavy quark (e.g.\hspace{4pt}the b-quark)
and a light antiquark, in the frame where the hadron is at rest.
As in \sect{sect:strong_int} we keep the \Eume.
We start from the QCD Lagrangian appearing in \eq{eq:QCD_Lagrangian},
and focus our attention on the heavy flavor with mass $m$, dropping
the corresponding flavor index to simplify the notation.
The fermion part of the Lagrangian reads
\begin{align}
\mathcal{L}&=\psibar (\Dsl +m)\psi={\widetilde \psi}\mathcal{D}\psi\,,\label{eq:Lagr_hQCD}\\
\psibar&={\widetilde \psi}\gamma_0\,,\hspace{1.0cm}\mathcal{D}=m\gamma_0 + D_0 + \gamma_0D_k\gamma_k\,,\label{eq:Dcal}
\end{align}
and, in analogy to the derivation of the corrections to the nonrelativistic
approximation of QCD, we perform a Foldy-Wouthuysen transformation to derive
the Lagrangian of a static quark, with the symmetry properties described in \sect{sect:spin-flav_symm},
and take into account the finite quark mass corrections through an expansion
in powers of $1/m$.
The transformation 
\be\label{eq:Foldy_transf}
\psi\to \eta=\mathrm{e}^{iS}\psi\,,\hspace{0.5cm}{\widetilde \psi}\to 
{\tilde \eta}={\widetilde \psi}\mathrm{e}^{-iS}\,,\hspace{0.5cm}
\mathcal{D}\to \mathcal{D'}=\mathrm{e}^{iS}\mathcal{D}\mathrm{e}^{-iS}\,,
\ee
with $S=-\tfrac{i}{2m}D_k\gamma_k$,
leaves the Lagrangian invariant
\be\label{eq:Lag_after_Foldy}
\mathcal{L}\to {\tilde \eta}\mathcal{D'}\eta=\mathcal{L}\,,
\ee
and has the property
\be\label{eq:prop_of_S}
S=S^\dagger\,.
\ee
The hermicity in \eq{eq:prop_of_S} simply follows from the fact that in the \Eume~
$D_k^{\dagger}=(\partial_k^\dagger+A_k^\dagger)=-D_k$
and $\gamma_k^{\dagger}=\gamma_k$ as shown in \app{app:Not_Conv}.
Further, the assumptions 
\be\label{eq:order_D0_and_S}
D_0\psi=\mathrm{O}(m)\psi\,,\hspace{1.0cm}S\psi=\mathrm{O}(1/m)\psi\,,
\ee
are justified at the end of this section.

The new operator $\mathcal{D'}$ can be computed by making use of the general identity
\be\label{eq:gen_id_operators}
\mathrm{e}^AB\mathrm{e}^{-A}=\sum_{n=0}^{\infty}\tfrac{1}{n!}\underbrace{[A,[A,[\ldots [A}_{n\,\mbox{\scriptsize times}},B]]\ldots]]\,,
\ee
obtained by a formal Taylor expansion of $\mathrm{e}^{\epsilon A}B\mathrm{e}^{-\epsilon A}$ at $\epsilon=1$,
giving
\begin{align}
\mathcal{D'}&=\mathcal{D}+[iS,\mathcal{D}]+\tfrac{1}{2}[iS,[iS,\mathcal{D}]]+\ldots\nonumber\\
         &=\mathcal{D}+\tfrac{1}{2m}[D_k\gamma_k,\mathcal{D}]+\tfrac{1}{8m^2}[D_l\gamma_l,[D_k\gamma_k,\mathcal{D}]]
            +\mathrm{O}(\tfrac{1}{m^2})\,.\label{eq:Dprime_eq2}
\end{align}
In particular
\begin{align}
[D_k\gamma_k,\mathcal{D}]&=mD_k[\gamma_k,\gamma_0]+[D_k,D_0]\gamma_k+[D_k\gamma_k,\gamma_0D_n\gamma_n]\nonumber\\
                      &=-2m\gamma_0D_k\gamma_k+F_{k0}\gamma_k+[D_k\gamma_k,\gamma_0D_n\gamma_n]\,,\label{eq:D_kg_kD}
\end{align}
where $F_{\mu\nu}$ is defined in \eq{eq:F_munu_QCD}, and
\be\label{eq:Dg_gDg}
[D_k\gamma_k,\gamma_0D_n\gamma_n]=\gamma_0[D_k,D_n]\gamma_n\gamma_k-2\gamma_0D_kD_k\,.
\ee
Inserting \eq{eq:D_kg_kD}
into the r.h.s.~of \eq{eq:Dprime_eq2}, and using \eq{eq:Dg_gDg} in the second step, gives
\begin{align}
\hspace{-3.0cm}\mathcal{D'} &= \mathcal{D}+\tfrac{1}{2m}[D_k\gamma_k,\mathcal{D}]
               -\tfrac{1}{4m}[D_l\gamma_l,\gamma_0 D_k \gamma_k]
               +\mathrm{O}(\tfrac{1}{m^2})\label{eq:Dprime_eq2_new}\\
                         &= \mathcal{D}+\tfrac{1}{2m}\left(-2m\gamma_0D_k\gamma_k
                         +F_{k0}\gamma_k+D_\mathrm{spin}-\gamma_0{\boldsymbol{D}}^2\right)+\mathrm{O}(\tfrac{1}{m^2})\,,\nonumber
\end{align}
with 
\begin{align}
& D_\mathrm{spin}=-\tfrac{1}{2}\gamma_0[D_k,D_l]\gamma_k\gamma_l=-\tfrac{1}{2i}\gamma_0[D_k,D_l]\sigma_{kl}\,,\label{eq:D_spin}\\
& \sigma_{kl}=\tfrac{i}{2}[\gamma_k,\gamma_l]\,,\hspace{0.5cm}{\boldsymbol{D}}^2=D_kD_k\,.\label{eq:D_perp}
\end{align}
We are now ready to write \eq{eq:Dprime_eq2_new} in the form
\be
\mathcal{D'}=\gamma_0\Bigg\{\gamma_0 D_0+m+\tfrac{1}{2m}\left(-{\boldsymbol{D}}^2-\tfrac{1}{2i}F_{kl}\sigma_{kl}
            +F_{k0}\gamma_0\gamma_k\right)\Bigg\}+\mathrm{O}(\tfrac{1}{m^2})\,.\label{eq:Dprime_eq3}
\ee
Through the projectors $P_{\pm}=(1\pm \gamma_0)/2$ we split the fields $\psi$ and $\psibar$
in their upper and lower components
\begin{align}
\heavy=P_+\psi\, & \heavyb=\psibar P_+\,,\label{eq:big_and_small_psi_q}\\
\aheavy=P_-\psi\, & \aheavyb=\psibar P_-\,,\label{eq:big_and_small_psi_aq}
\end{align}
letting us write the Lagrangian as
\begin{align}
\mathcal{L}&=\mathcal{L}_\mathrm{h}^\mathrm{stat}+\mathcal{L}_\mathrm{\bar{h}}^\mathrm{stat}+\left(\mathcal{L}^{(1)}_\mathrm{h}+\mathcal{L}^{(1)}_\mathrm{\bar{h}}
+\mathcal{L}^{(1)}_\mathrm{h\bar{h}}\right)+\mathrm{O}(\tfrac{1}{m^2})\,,\\
\mathcal{L}_\mathrm{h}^\mathrm{stat}&=\heavyb\left(D_0+m\right)\heavy\,,\label{eq:Lstat_h}\\
\mathcal{L}_\mathrm{\bar{h}}^\mathrm{stat}&=\aheavyb\left(-D_0+m\right)\aheavy\,,\label{eq:Lstat_hb}\\
\mathcal{L}^{(1)}_\mathrm{h}&=\tfrac{1}{2m}\heavyb\left(-{\boldsymbol{D}}^2-\tfrac{1}{2i}F_{kl}\sigma_{kl}\right)\heavy\,,\label{eq:L1_h}\\
\mathcal{L}^{(1)}_\mathrm{\bar{h}}&=\tfrac{1}{2m}\aheavyb\left(-{\boldsymbol{D}}^2-\tfrac{1}{2i}F_{kl}\sigma_{kl}\right)\aheavy\,,\label{eq:L1_hb}\\
\mathcal{L}^{(1)}_\mathrm{h\bar{h}}&=\tfrac{1}{2m}\left\{\heavyb F_{k0}\gamma_k\aheavy
                                  +\aheavyb F_{k0}\gamma_k\heavy\right\}\,.
\end{align}
For our purposes we can drop the term $\mathcal{L}^{(1)}_\mathrm{h\bar{h}}$, because it does not
contribute in the computation of correlation functions involving only a heavy quark (or antiquark).
Only double insertions of this term would contribute, and they are of order $1/m^2$.

The fields $\heavy$, $\heavyb$, $\aheavy$ and $\aheavyb$ actually have only two Dirac components.
Nevertheless we keep the four components notation, with two vanishing components for each field.

Inspection of the Dirac equation, obtained 
from the static Lagrangians $\mathcal{L}_\mathrm{h}^\mathrm{stat}$ and $\mathcal{L}_\mathrm{\bar{h}}^\mathrm{stat}$,
justifies the kinematics $D_0\psi=\mathrm{O}(m)\psi$ given in (\ref{eq:order_D0_and_S}).
To prove the second equation in (\ref{eq:order_D0_and_S}), we start from the observation 
that a very heavy quark inside a hadron moves with essentially the hadron's velocity
and is almost on-shell. We can write its momentum as $P=m+k$, with $k$ ``measuring'' its off-shellness 
and $k\ll m$. This quantity is often referred to as residual momentum. 
Interactions of the heavy quark with the light degrees of freedom change it by an amount of 
order $\Delta k\sim\Lambda_\mathrm{QCD}$, and the lowest order, in
the $1/m$ expansion of the Lagrangian, where it is taken into account is in $\mathcal{L}^{(1)}_\mathrm{h}$,
through the operator $D_k$.
It is thus correct to state that $S\psi=\mathrm{O}(\Lambda_\mathrm{QCD}/m)\psi$, but in this section we 
are discussing the classical theory, where $\Lambda_\mathrm{QCD}$ makes no sense, and it would then be 
more appropriate to write  $S\psi=\mathrm{O}(\epsilon)\psi$. In a somewhat sloppy notation, we write
$\mathrm{O}(1/m)$ instead of $\mathrm{O}(\epsilon)$.
  
Finally we would like to get rid of the mass term appearing 
in $\mathcal{L}_\mathrm{h}^\mathrm{stat}$ and $\mathcal{L}_\mathrm{\bar{h}}^\mathrm{stat}$.
This can be easily achieved by observing that the operators appearing 
in the $\mathrm{O}(1/m)$ correction terms, namely ${\boldsymbol{D}}^2$ and $F_{kl}\sigma_{kl}$ 
in \eq{eq:L1_h} and \eq{eq:L1_hb} respectively, commute with $\gamma_0$,
and we redefine the fields according to
\begin{align}
\heavy\to \mathrm{e}^{-m\gamma_0x_0}\heavy\,,\hspace{0.3cm}& \hspace{0.3cm}\heavyb\to \heavyb \mathrm{e}^{m\gamma_0x_0}\,,\label{eq:redef_heavy}\\
\aheavy\to \mathrm{e}^{m\gamma_0x_0}\aheavy\,,\hspace{0.53cm}& \hspace{0.3cm}\aheavyb\to \aheavyb \mathrm{e}^{-m\gamma_0x_0}\,,\label{eq:redef_heavyb}
\end{align}
where $x_0$ is the usual time coordinate.
This has exactly the effect of cancelling the mass term in $\mathcal{L}_\mathrm{h}^\mathrm{stat}$ and $\mathcal{L}_\mathrm{\bar{h}}^\mathrm{stat}$,
which now read 
\begin{align}
\mathcal{L}_\mathrm{h}^\mathrm{stat}&=\heavyb\left(D_0+\epsilon^+\right)\heavy\,,\label{eq:Lstat_h_nom}\\
\mathcal{L}_\mathrm{\bar{h}}^\mathrm{stat}&=\aheavyb\left(-D_0+\epsilon^+\right)\heavyb\,,\label{eq:Lstat_hb_nom}
\end{align}
where the infinitesimal, positive $\epsilon^+$ prescription plays the role of selecting the appropriate time propagation.

Notice that dropping the mass term in $\mathcal{L}_\mathrm{h}^\mathrm{stat}$ and $\mathcal{L}_\mathrm{\bar{h}}^\mathrm{stat}$,
corresponds exactly to an energy shift by an amount of $m$ in all single heavy quark (or antiquark) states.

\section{Symmetries of the effective Lagrangian}\label{sect:symm_eff_lagrangian}

We focus our attention on the heavy quark static Lagrangian of \eq{eq:Lstat_h_nom} 
(the antiquark case is analogous),
and we now show that it is invariant under the symmetries described in \sect{sect:spin-flav_symm}, which
are absent in finite mass QCD.

The spin symmetry can be seen as invariance under the rotations \cite{Georgi:1990um,Isgur:1989vq,Isgur:1989ed}
\be\label{eq:spin_inv}
\heavy\to V\heavy\,,\hspace{0.5cm}\heavyb\to\heavyb V^{-1}\,,\hspace{0.5cm}V=\mathrm{exp}\left\{-i\phi_i \epsilon_{ijk}\sigma_{jk}\right\}\,, 
\ee
with $\phi_i$ an arbitrary parameter.
This simply follows from the fact that $\sigma_{jk}$ commutes with $\gamma_0$, as noticed in \sect{sect:classic_lagrangian}.

Another symmetry, corresponding
to the local conservation of the quark number, is the invariance under 
phase transformations \cite{Kurth:2000ki}
\be\label{eq:xphase_inv}
\heavy\to \mathrm{e}^{i\eta({\mathbf{x}})}\heavy\,,\hspace{0.5cm}\heavyb\to\heavyb \mathrm{e}^{-i\eta(\mathbf{x})}\,,
\ee
where it is important to underline that the parameter $\eta(\mathbf{x})$ is time independent;
it depends only on the spatial coordinates. The proof is trivial, because in the static Lagrangian
only the time derivative appears.
 
The symmetry under rotations in flavor space mentioned in \sect{sect:spin-flav_symm} 
is irrelevant in our case, because we have only
one heavy flavor.

These symmetries are broken by the $\mathrm{O(1/m)}$ correction terms.
We adopt a more compact notation for the operators in \eq{eq:L1_h} 
\begin{align}
\mathcal{L}_\mathrm{h}^{(1)}&=-\tfrac{1}{2m}({\mathcal{O}}_\mathrm{kin}+{\mathcal{O}}_\mathrm{spin})\,,\label{eq:L1_h_ops}\\
{\mathcal{O}}_\mathrm{kin}&=\heavyb D_kD_k \heavy=\heavyb {\boldsymbol{D}}^2 \heavy\,,\label{eq:O_kin}\\
{\mathcal{O}}_\mathrm{spin}&=\heavyb\tfrac{1}{2i}F_{kl}\sigma_{kl}\heavy=
\heavyb {\boldsymbol{\sigma}}\!\cdot\!{\boldsymbol{B}}\heavy\,.\label{eq:O_spin}
\end{align}
The kinetic operator ${\mathcal{O}}_\mathrm{kin}$ does not break the spin symmetry (\eq{eq:spin_inv}). It 
describes, at leading order in the $1/m$ expansion, the spatial motion
of the heavy quark.

The spin symmetry is broken by the ``hyperfine'' chromo-magnetic term ${\mathcal{O}}_\mathrm{spin}$,
which is obviously invariant under the phase transformations described in (\ref{eq:xphase_inv}).

\section{Renormalization of the effective theory}\label{sect:ren_HQET}

The leading order Lagrangian, \eq{eq:Lstat_h_nom}, contains local fields
of mass dimension $d\leq 4$, and it is thus power counting renormalizable. At this
order only a finite number of counterterms are needed. The latter must be compatible
with the symmetries described in the previous section, and the only possible counterterm
with dimension $d\leq 4$, involving the fields $\heavyb$ and $\heavy$, and invariant under
the symmetries of the static action, is $\heavyb\heavy$.

Indicating with $\delta m$ its coefficient, one is thus left with the formal static Lagrangian
\be\label{eq:stat_Lagr_formal}
\mathcal{L}_\mathrm{h}^\mathrm{stat}=\heavyb\left(D_0+\delta m\right)\heavy\,.
\ee
As noticed in \sect{sect:classic_lagrangian}, such a term amounts only to an energy shift 
of exactly $\delta m$ for all single heavy quark states.

If we now allow finite quark mass corrections, we can 
use the HQET Lagrangian 
\be\label{eq:HQET_Lagr}
\mathcal{L}^\mathrm{HQET}_\mathrm{h}=\mathcal{L}_\mathrm{h}^\mathrm{stat}+\mathcal{L}^{(1)}_\mathrm{h}+\mathrm{O}(\tfrac{1}{m^2})\,.
\ee
The renormalization of the operators appearing in $\mathcal{L}^{(1)}_\mathrm{h}$ may introduce
new terms compatible with the symmetries of the theory and dimension $d\leq 5$.
Under these constraints no new terms are needed, and the coefficients of the operators ${\mathcal{O}}_\mathrm{kin}$ and
${\mathcal{O}}_\mathrm{spin}$ are only parameters depending on the bare coupling
and on the mass $m$. The $\mathrm{O}(1/m)$ Lagrangian has thus the form
\begin{align}
\mathcal{L}_\mathrm{h}^{(1)}&=-(\omega_\mathrm{kin}{\mathcal{O}}_\mathrm{kin}+\omega_\mathrm{spin}{\mathcal{O}}_\mathrm{spin})\,,\label{eq:L1_h_omegas}\\
      \omega_\mathrm{kin}&=\mathrm{O}(\tfrac{1}{m})\,,\qquad\omega_\mathrm{spin}=\mathrm{O}(\tfrac{1}{m})\,.\label{eq:omegas_orders}         
\end{align}
Since $\mathcal{L}^{(1)}_\mathrm{h}$ contains dimension five operators, one may conclude that,
for dimensional reasons the theory is not renormalizable.
This happens to be true for NRQCD, but not for HQET \cite{hqet:pap1}.

We consider an operator ${\mathcal{O}}(x)$ containing light and heavy degrees
of freedom, and we expand it at tree-level in a power series in $1/m$,
\be\label{eq:O_exp_tl}
{\mathcal{O}}(x)=\sum_{\nu=0}^{n}{\mathcal{O}}^{(\nu)}(x)\,,\qquad {\mathcal{O}}^{(\nu)}(x)=\sum_i \alpha_i^{(\nu)}{\mathcal{O}}^{(\nu)}_i(x)\,,
\ee
with ${\mathcal{O}}^{(\nu)}_i$ of dimension $3+\nu$ and $\alpha_i^{(\nu)}=\mathrm{O}(1/m^{\nu})$.
The terms of the expansion can be obtained by applying a Foldy-Wouthuysen transformation
as in \sect{sect:classic_lagrangian}.
A typical example is the $\mu=0$ component of heavy-light axial vector current $A_\mu(x)$,
\be\label{eq:Amu}
A_\mu(x)=\lightb(x) \gamma_\mu \gamma_5 \psi(x)\,,
\ee
giving
\begin{align}
A^{(0)}(x)&=\alpha_0^{(0)}A_0^\mathrm{stat}(x)\,,\qquad A_0^\mathrm{stat}(x)=\lightb(x) \gamma_0 \gamma_5 \heavy(x)\,,\label{A0_exp}\\
A^{(1)}(x)&=\alpha_1^{(1)}A_1^{(1)}(x)\,,\qquad\hspace{0.235cm}A_1^{(1)}(x)=\lightb(x) \gamma_j \gamma_5 \overleftarrow{D}_{\!j}\heavy(x)\,,\label{A1_exp}
\end{align}
with $\alpha_0^{(0)}=1$ and $\alpha_1^{(1)}=1/m$.

To evaluate the expectation value (e.g. between two heavy-light meson states) of the operator in HQET,
we introduce the partition function 
\begin{align}
\mathcal{Z}_\mathrm{HQET}&=\int_{\heavyb,\heavy,U_\mathrm{l}}
                      \mathrm{e}^{-\int_x (\mathcal{L}^\mathrm{HQET}_\mathrm{h}
                     [\,\heavyb(x),\heavy(x),
                      U_\mathrm{l}(x)\,]+\mathcal{L}_\mathrm{light}[\,U_\mathrm{l}(x)\,])}\nonumber\\
                   &=\int_{\heavyb,\heavy,U_\mathrm{l}}
                      \mathrm{e}^{-\int_x (\mathcal{L}^\mathrm{stat}_\mathrm{h}
                     [\,\heavyb(x),\heavy(x),
                       U_\mathrm{l}(x)\,]+\mathcal{L}_\mathrm{light}[\,U_\mathrm{l}(x)\,])}\nonumber\\        
                   &\quad\times\left\{1-\int_x\mathcal{L}^{(1)}_\mathrm{h}[\,\heavyb(x),\heavy(x),U_\mathrm{l}(x)\,]
                       +\mathrm{O}(\tfrac{1}{m^2})\right\}\,,\label{eq:path_int_HQET}
\end{align}
where $U_\mathrm{l}$ is a shorthand for the light degrees of freedom, and $\mathcal{L}_\mathrm{light}$ 
the corresponding Lagrangian. A more rigorous
definition of the partition function is given in the next chapter.
From \eq{eq:O_exp_tl} it follows that the expectation value
of ${\mathcal{O}}(x)$ is defined as 
\begin{align}
\langle {\mathcal{O}}(x)\rangle_\mathrm{HQET}&\equiv\langle {\mathcal{O}}^{(0)}(x)\rangle_\mathrm{stat}+
                                         \langle {\mathcal{O}}^{(1)}(x)\rangle_\mathrm{stat}\nonumber\\
                                    &\quad +\omega_\mathrm{kin}\langle {\mathcal{O}}^{(0)}(x)\int_y 
                                         {\mathcal{O}}_\mathrm{kin}(y)\rangle_\mathrm{stat}\nonumber\\       
                                    &\quad +\omega_\mathrm{spin}\langle {\mathcal{O}}^{(0)}(x)\int_y 
                                         {\mathcal{O}}_\mathrm{spin}(y)\rangle_\mathrm{stat}
                                         +\mathrm{O}(\tfrac{1}{m^2})\,,\label{eq:expt_O}       
\end{align}
where
\be\label{eq:expt_stat}
\langle {\mathcal{O}}\rangle_\mathrm{stat}=\frac{1}{\mathcal{Z}_\mathrm{stat}}\int_{\heavyb,\psi_\mathrm{h\phantom{\bar{j}}}\hspace{-0.13cm},U_\mathrm{l}}
                                   {\mathcal{O}}\,\mathrm{e}^{-\int_x \left(\mathcal{L}^\mathrm{stat}_\mathrm{h}
                     [\,\heavyb(x),\psi_\mathrm{h\phantom{\bar{f}}}\hspace{-0.14cm}(x),U_\mathrm{l}(x)\,]+\mathcal{L}_\mathrm{light}[\,U_\mathrm{l}(x)\,]\right)}\,,
\ee
and the path integral is normalized by $\mathcal{Z}_\mathrm{stat}$ such that $\langle 1\rangle_\mathrm{stat}=1$. 
It is clear from \eq{eq:expt_O} and \eq{eq:expt_stat} that the expectation values are evaluated
with respect to the leading order of the HQET action, which is power counting renormalizable.

The operator ${\mathcal{O}}$ has still to be renormalized, and \eq{eq:expt_O} tells us that this
is just a problem of renormalizing correlation functions of local composite operators in 
the static effective theory. One has to include all local operators with dimension not
exceeding $n$, the one of the highest-dimensional operator, and compatible with the symmetries
of the theory. Being interested in the continuum theory up to $\mathrm{O}(1/m^2)$, we have $n=5$.
The coefficients are then computed by requiring the cancellation of the divergences
and the respect of the imposed renormalization conditions of all expectation values \cite{Collins:1984xc}.

The sums over all space-time coordinates in \eq{eq:expt_O} may lead to additional
singularities due to contact terms. However the terms needed to remove these singularities
must respect the dimensional and symmetry properties described above, and they
are already included in the expansion in \eq{eq:expt_O}.

As a regularization
of the theory breaks scale invariance, there exist power divergences in the $\mathrm{O}(1/m)$ corrections
to the HQET matrix elements \cite{stat:MMS}. These divergences must be subtracted non-perturbatively.
This fact is simply due to the mixing of operators of different dimensions in HQET Lagrangian.
To explain it in a simple way, let us consider the Lagrangian in \eq{eq:HQET_Lagr}
with the $\mathrm{O}(1/m)$ term given in \eq{eq:L1_h_omegas}. To regularize the theory
we can take a lattice discretization with lattice spacing $a$. Because of the mixing between
the operator ${\boldsymbol{D}}^2$ appearing in ${\mathcal{O}}_\mathrm{kin}$ and $D_0\gamma_0$ from the static Lagrangian,
a perturbative estimate of $\omega_\mathrm{kin}$ at order $g_0^{2l}$ would leave a perturbative remainder
\be\label{eq:pert_remainder}
\Delta \omega_\mathrm{kin}\sim g_0^{2(l+1)}a^{-1}\sim a^{-1}[\ln(a\Lambda_\mathrm{QCD})]^{-(l+1)}
\,\stackrel{a\to 0}{\longrightarrow}\,\infty\,.
\ee
This means that the theory is still affected by ultraviolet divergences and is not renormalizable in perturbation
theory.
If one considers higher orders in the $1/m$ expansion, further UV-divergences can appear because
of operator mixings, but they are not discussed here.

\section{Matching of the effective theory with QCD}\label{sect:matching_HQET}

The effective Lagrangian and the effective matrix elements derived in the previous sections
correctly reproduce the long-distance physics of the full theory.
However the light degrees of freedom interact with the heavy quark exchanging momenta,
which can be soft or hard. Soft momenta correspond to long wavelengths, and they are 
well identified by using the partition function in \eq{eq:path_int_HQET}. Hard momenta
can resolve the short-distance properties of the heavy quark, which are truncated in the expansion
of \eq{eq:expt_O}. It means that short-distance corrections are needed in order to produce
a predictive description of the heavy-light mesons. They introduce
a logarithmic dependence on the heavy quark mass $m$ through the strong 
coupling constant $\gbar(m)$. Given the smallness of the latter, 
they can be safely computed in perturbation theory, by performing the matching
to QCD of the matrix elements computed in the effective theory.

Here the quark mass $m$ is renormalized and defined at the scale $\mu=m$ itself.
The exact definition of the renormalization procedure is irrelevant
for the present discussion. The dependence on the space-time coordinates is omitted throughout
the following in order to lighten the notation.

We consider a matrix element $\langle {\mathcal{O}}(\mu,m)_\mathrm{R}\rangle_\mathrm{QCD}$ computed in QCD,
and renormalized at the scale $\mu$. The matching means that 
we require it to be equal to the matrix element of the same operator computed 
in the effective theory up to $\mathrm{O}(1/m)$. The equivalence is achieved through
the Wilson coefficient $C_{{\mathcal{O}},\mathrm{match}}(\mu,m)$ 
\be\label{eq:match1}
\langle {\mathcal{O}}_\mathrm{R}(\mu,m)\rangle_\mathrm{QCD}=C_{{\mathcal{O}},\mathrm{match}}(\mu,m)
\langle {\mathcal{O}}_\mathrm{R}(\mu)\rangle_\mathrm{stat}+\mathrm{O}(\tfrac{1}{m})\,.
\ee  
For our purposes it is convenient to renormalize the operator in the 
effective theory in the $\msbar$ scheme \cite{ms:thooft,msbar:gen}, thus having 
an expansion
\be\label{eq:Cmatch}
C_{{\mathcal{O}},\mathrm{match}}(\mu,m)=1+c_{1}^{{\mathcal{O}}}(\mu,m)\gbarMSbar^2(\mu)+\mathrm{O}(\gbarMSbar^4(\mu))\,,
\ee
with $\gbarMSbar$ being the renormalized coupling in the $\msbar$ scheme.
The relation between the renormalized operator appearing on the r.h.s. of \eq{eq:match1}
and the bare one can be written in the form
\be\label{eq:ren_O}
{\mathcal{O}}_{\MSbar}(\mu)=Z^{{\mathcal{O}},\MSbar}(\mu)\,{\mathcal{O}}\,.
\ee
The scale dependence of the renormalization factor
is governed by the renormalization group equation
\be\label{eq:ren_g_O}
\gamma^{{\mathcal{O}},\msbar}(\gbarMSbar(\mu))=
\mu\frac{\partial \ln Z^{{\mathcal{O}},\msbar}(\mu)}{\partial \mu\phantom{T}}\,,
\ee
and the renormalization group function $\gamma$, the anomalous dimension, has a perturbative
expansion ($\gbar=\gbarMSbar(\mu)$),
\be\label{eq:an_dim_O}
\gamma^{{\mathcal{O}},\msbar}(\gbar)\stackrel{\gbar\to 0}{\sim}-\gbar^2\left\{\gamma^{{\mathcal{O}},\msbar}_0
+\gamma^{{\mathcal{O}},\msbar}_1\,\gbar^{2}+\gamma^{{\mathcal{O}},\msbar}_2\,\gbar^{4}+\dots\right\}\,.
\ee
The same holds for the renormalized coupling
\be\label{eq:running_gbar}
\mu\frac{\partial \gbar}{\partial \mu}=
\beta_{\msbar}(\gbar)\stackrel{\gbar\to 0}{\sim}-\gbar^3\left\{b_0+b_1\gbsq+b_2^{\msbar}\gbar^4+\ldots\right\}\,.
\ee
The coefficients $b_0$ and $b_1$ are universal, whereas the higher order coefficients, here $b_2^{\msbar},b_3^{\msbar},\ldots$, depend 
on the chosen renormalization scheme.

The scale dependence of $\langle {\mathcal{O}}_\mathrm{R}(\mu)\rangle_\mathrm{stat}=\langle {\mathcal{O}}_\msbar(\mu)\rangle_\mathrm{stat}$
is removed by introducing the renormalization group invariant (RGI) 
matrix element $\langle {\mathcal{O}}_\mathrm{RGI}\rangle_\mathrm{stat}$, defined as
\begin{align}
\langle {\mathcal{O}}_\mathrm{RGI}\rangle_\mathrm{stat}&=\langle {\mathcal{O}}_{\msbar}(\mu)\rangle_\mathrm{stat}\,[2b_0\gbsq(\mu)]^{-\gamma^{{\mathcal{O}},\msbar}_0/(2b_0)}\nonumber\\
&\quad \times\, \mathrm{exp}\left\{-\int_0^{\gbar(\mu)}\mathrm{d}g\,
\left[\frac{\gamma^{{\mathcal{O}},\msbar}(g)}{\beta_{\msbar}(g)}
-\frac{\gamma^{{\mathcal{O}},\msbar}_0}{b_0g}\right]\right\}\,.\label{eq:RGI_matrix_element_O}
\end{align}
With the same technique one is able to compute the RGI matrix element in QCD, 
and by
setting the renormalization scale to the heavy quark mass, the matching condition now reads
\be\label{eq:match2}
\langle {\mathcal{O}}_\mathrm{RGI}(m)\rangle_\mathrm{QCD}=C_{\mathcal{O}}(m)\langle {\mathcal{O}}_\mathrm{RGI}\rangle_\mathrm{stat}+\mathrm{O}(\tfrac{1}{m})\,,
\ee
where the matching coefficient $C_{\mathcal{O}}(m)$ is defined through eqs.~(\ref{eq:match1},\,\ref{eq:match2})
\be\label{eq:def_C_O}
C_{\mathcal{O}}(m)= C_{{\mathcal{O}},\mathrm{match}}(m,m)\,
\frac{\langle {\mathcal{O}}_{\msbar}(m)\rangle_\mathrm{stat}}{\langle {\mathcal{O}}_\mathrm{RGI}\rangle_\mathrm{stat}}\,
\frac{\langle {\mathcal{O}}_\mathrm{RGI}(m)\rangle_\mathrm{QCD}}{\langle {\mathcal{O}}_\mathrm{R}(m,m)\rangle_\mathrm{QCD}}\,.
\ee
This equation allows to define the anomalous dimension $\gamma_\mathrm{match}$ 
in the matching scheme through
\begin{align}
C_{\mathcal{O}}(m)&= [2b_0\gbsq(m)]^{\gamma^{\mathcal{O}}_0/(2b_0)}\nonumber\\
&\quad \times\,\mathrm{exp}\left\{\int_0^{\gbar(m)}\mathrm{d}g\,
\left[\frac{\gamma^{\mathcal{O}}_\mathrm{match}(g)}{\beta_{\msbar}(g)}
-\frac{\gamma^{\mathcal{O}}_0}{b_0g}\right]\right\}\,,\label{eq:expr_C_O}\\
\,\nonumber\\
\gamma^{\mathcal{O}}_\mathrm{match}(\gbar)&\stackrel{\gbar\to 0}{\sim}-\gbar^2\left\{\gamma^{\mathcal{O}}_0
+\gamma^{\mathcal{O}}_1\,\gbar^{2}+\gamma^{\mathcal{O}}_2\,\gbar^{4}+\dots\right\}\,\quad \gbar=\gbarMSbar\,.\label{eq:def_gamma_match}
\end{align}
In the case where the operator ${\mathcal{O}}$ is the axial vector current defined 
in \eq{eq:Amu}, the corresponding renormalization is
fixed by the chiral Ward identities, which
imply the independence on a renormalization scale. This simplifies the r.h.s of \eq{eq:def_C_O},
because the second ratio from the left becomes one. 

From eqs.~(\ref{eq:Cmatch},\,\ref{eq:expr_C_O},\,\ref{eq:def_gamma_match})
it is straightforward to collect the several contributions entering in $\gamma^{\mathcal{O}}_\mathrm{match}$, namely
\be
\gamma^{\mathcal{O}}_0=\gamma^{{\mathcal{O}},\msbar}_0\,,\quad 
\gamma^{\mathcal{O}}_1=\gamma^{{\mathcal{O}},\msbar}_1+2b_0c_{1}^{{\mathcal{O}}}(m,m)\,,\quad\ldots\,.
\ee
Finally we would like to get rid of the scale dependence of the heavy quark mass.
The scale evolution of the latter is given by the renormalization group equation
\begin{align}
\mu\frac{\partial m}{\partial \mu}&=\tau(\gbar)m\,,\label{eq:RGE_mass}\\
\tau(\gbar)&\stackrel{\gbar\to 0}{\sim}-\gbar^2\left\{d_0+d_1\gbsq+d_2\gbar^4+\ldots\right\}\,,\label{eq:RGE_tau}
\end{align}
where only the coefficient $d_0$ is scheme-independent.
We thus introduce the RGI and scheme-independent quark mass $M$, as the asymptotic behavior of any mass $m(\mu)$
\be\label{eq:def_M}
M=\lim_{\mu\to\infty}m(\mu)\,[2b_0\gbsq(\mu)]^{-d_0/2b_0}\,.
\ee

\chapter{The \SF}\label{chap:SF}
\section{Definition}\label{sect:SF_def}

At the classical level the transition from some initial field 
configuration at time $x_0=0$ to a final configuration
at $x_0=T$ is given in terms of only one path,
the classical path, which satisfies the principle of least action.
In quantum field theory the transition can take place in more than one way,
and the total amplitude is the coherent sum of the amplitudes
for each way.

Following Feynman, this sum can be written as a path integral,
where, in the functional integral formalism, all field configurations 
are integrated over. 
The Schr\"odinger functional (SF) is defined as the integral kernel of the whole
transition amplitude.

The following description of the Schr\"odinger functional in SU(3) gauge theory
is carried out with the \Eume~following \cite{SF:LNWW} for the gauge part,
and \cite{SF:stefan1} for the fermionic part.
The transition amplitude that we would like to compute is
\be\label{eq:transition_amplitude}
\mathcal{Z}[\phi_\mathrm{fin},\phi_\mathrm{in}]=\langle \phi_\mathrm{fin}|\mathrm{e}^{-{\mathbb H}\,T}| \phi_\mathrm{in}\rangle\,,
\ee
where ${\mathbb H}$ is the Hamilton operator. The starting point is now the definition
of the quantum mechanical states in the Schr\"odinger representation.

In order to facilitate the passage to the lattice formulation we start with QCD
at a fixed time slice, say $x_0=0$, and on a finite spatial volume, a box
of size $L\times L\times L$ and periodic boundary conditions.
Gauge fields are represented by the vector potentials $A_\mu({\mathbf x})$, and we
work in the temporal gauge, i.e.\hspace{4pt}with $A_0({\mathbf x})=0$.
The fields $A_k({\mathbf x})$ are periodic in space and transform under a gauge
transformation according to
\be\label{eq:A_gauge_slice}
A_k^\Lambda({\mathbf x})=\Lambda({\mathbf x})A_k({\mathbf x})\Lambda^{-1}({\mathbf x})+
\Lambda({\mathbf x})\partial_k\Lambda^{-1}({\mathbf x})\,.
\ee
To preserve the periodicity
we admit only periodic gauge functions $\Lambda({\mathbf x})$
\be\label{eq:A_periodicity}
A_k({\mathbf x}+L\hat{k})=A_k({\mathbf x})\,,\qquad \Lambda({\mathbf x}+L\hat{k})=\Lambda({\mathbf x})\,.
\ee
The Schr\"odinger representation of a quantum mechanical state
is the wave functional $\psi[A]$, and a scalar product
of two functionals $\psi$ and $\chi$ is defined by
\be\label{eq:functional_scalar_prod}
\langle \psi|\chi\rangle=\int \mathrm{D}[A]\psi[A]^\ast\chi[A]\,,\qquad \mathrm{D}[A]=\prod_{{\mathbf x},k}\mathrm{d} A_k({\mathbf x})\,.
\ee 
Here, and in the following of this section, color indices have been dropped to lighten the notation. 
It is understood that $\mathrm{d} A_k({\mathbf x})=\prod_a \mathrm{d} A_k^a({\mathbf x})$.

In \eq{eq:functional_scalar_prod} only gauge invariant states with $\psi[A^\Lambda]=\psi[A]$ are physical.
For any functional $\psi[A]$ the projection onto the physical subspace
is achieved by applying the projector ${\mathbb P}$ according to
\be\label{eq:proj_P}
{\mathbb P}\psi[A]=\int \mathrm{D}[\Lambda]\psi[A^\Lambda]\,,\qquad \mathrm{D}[\Lambda]=\prod_{{\mathbf x}}\mathrm{d} \Lambda({\mathbf x})\,,
\ee
where $\mathrm{d} \Lambda({\mathbf x})$ denotes the SU(3) Haar measure.\\
The Hamilton operator appearing in \eq{eq:transition_amplitude} is defined through
the curvature tensor (\ref{eq:F_munu_QCD}), giving
\be\label{eq:Hamilton}
{\mathbb H}=\int_0^L \mathrm{d}^3\mathbf{x}\,\left\{\frac{g_0^2}{ 2}\tr\{F_{0k}({\mathbf x})F_{0k}(\vecx)\}
+\frac{1}{4g_0^4}\tr\{F_{kl}({\mathbf x})F_{kl}(\vecx)\}\right\}\,.
\ee
Each classical gauge field $C_k(\vecx)$ defines a state $|C\rangle$ and the functional
\be\label{eq:functional_C}
\langle C|\psi \rangle=\psi[C]\,.
\ee
In general the state $|C\rangle$ is not gauge invariant, but it can be projected
onto the corresponding physical subspace through the projector ${\mathbb P}$.

Given two classical gauge fields $C$ and $C'$,
the first living on the time slice at $x_0=0$, the second at $x_0=T$,
we define the Euclidean \SF as
\be\label{eq:SF_CC}
\mathcal{Z}[C',C]=\langle C'|e^{-{\mathbb H}\,T}{\mathbb P}|C\rangle\,.
\ee
Its gauge invariance 
\be\label{eq:SF_gauge_inv}
\mathcal{Z}[C'^{\Lambda'},C^\Lambda]=\mathcal{Z}[C',C]\,,
\ee
is guaranteed by the integral over $\Lambda$ in \eq{eq:proj_P}.

We now have all ingredients to express a matrix element of the time
evolution operator $\mathrm{e}^{-{\mathbb H}\,T}$ in terms of the gauge fields 
$A_\mu(x)$ and the corresponding gauge action in four dimensions, 
where the time extent is $0\leq x_0\leq T$. We impose Dirichlet
boundary conditions in time
\begin{align}
A_k(x)=\left\{\begin{array}{l}
C^\Lambda_k(\vecx)\quad\mbox{at}\,\,x_0=0\,,\\
\label{eq:A_Dirichlet_boundaries}\\
C'_k(\vecx)\quad\mbox{at}\,\,x_0=T\,,
\end{array}\right.
\end{align}
and we finally write down the functional integral as
\be\label{eq:SF_with_action}
\mathcal{Z}[C',C]=\int \mathrm{D}[\Lambda]\int \mathrm{D}[A]\,\mathrm{e}^{-S_\mathrm{gauge}[A]}\,,
\ee
where the gauge action can be obtained from \eq{eq:QCD_Lagrangian}
\be\label{eq:QCD_gauge_action}
S_\mathrm{gauge}[A]=-\frac{1}{2g_0^2}\int \mathrm{d}^4 x\,\tr\{F_{\mu\nu}F_{\mu\nu}\}\,,
\ee
and the integration measure is now
\be\label{eq:A_int_meas4}
\mathrm{D}[A]=\prod_{x,\mu}\mathrm{d}A_\mu(x)\,.
\ee
The integral (\ref{eq:SF_with_action}) and the 
boundary conditions (\ref{eq:A_Dirichlet_boundaries}) are invariant
under the gauge transformation
\begin{align}
A_\mu(x)&\to A_\mu^\Omega(x)=
\Omega(x)A_\mu(x)\Omega^{-1}(x)+\Omega(x)\partial_\mu\Omega^{-1}(x)\,,\label{eq:A_gauge transf_4d}\\
\Lambda(\vecx)&\to \Lambda^\Omega(\vecx)=\Omega(x)|_{x_0=0}\Lambda(\vecx)\,,\label{eq:L_gauge transf_4d}
\end{align}
with the necessary condition
\be\label{eq:O_gauge_transf}
\Omega(x)|_{x_0=T}=1\,.
\ee
For QCD, the inclusion of the quarks as dynamical variables in the path integral
can be carried out in a similar way. Nevertheless, imposing Dirichlet
boundary conditions on the quark fields requires some attention. This is
due to the fact that the Dirac equation is a first order differential
equation. In the \SF the dynamical degrees of freedom of the quark fields
are their components at $0<x_0<T$. In order to have
a unique solution of the Dirac equation,
at the boundaries only half of the Dirac
components are defined, and are fixed to be
\begin{align}
P_+\psi(x)|_{x_0=0}&= \rho(\vecx)\,,\qquad\hspace{0.16cm} P_-\psi(x)|_{x_0=T}=\rhoprime(\vecx)\,,\label{eq:quark_boundaries1}\\
\psibar(x)P_-|_{x_0=0}&= \rhobar(\vecx)\,,\qquad\psibar(x)P_+|_{x_0=T}=\rhobarprime(\vecx)\label{eq:quark_boundaries2}\,,
\end{align}
where the projectors $P_\pm=(1\pm\gamma_0)/2$ are the same as in \sect{sect:classic_lagrangian}.
It is then evident that 
\be\label{eq:quark_boundaries3}
P_-\rho(\vecx)=P_+\rhoprime(\vecx)=\rhobar(\vecx)P_+=\rhobarprime(\vecx)P_-=0\,.
\ee 
In the space direction $\hat{k}$ the quark fields are periodic 
up to a phase factor $\theta$
\be\label{eq:quark_boundaries4}
\psi(x+L\hat{k})=\mathrm{e}^{i\theta}\psi(x)\,,\qquad \psibar(x+L\hat{k})=\psibar(x)\,\mathrm{e}^{-i\theta}\,.
\ee
With these boundaries one finds that the fermionic part of the continuum action (for one flavor) is
\begin{align}
S_\mathrm{quark}[A,\psibar,\psi]&=\int_0^T\!\mathrm{d} x_0\int_0^L\!\mathrm{d}^3\mathbf{x}\,
                                \psibar(x)\left\{D_\mu\gamma_\mu+m\right\}\psi(x)\label{eq:SF_quark_action}\\
                            &\quad -\!\int_0^L\!\mathrm{d}^3\mathbf{x}\,[\,\psibar(x)P_-\psi(x)\,]_{x_0=0}
                              -\!\int_0^L\!\mathrm{d}^3\mathbf{x}\,[\,\psibar(x)P_+\psi(x)\,]_{x_0=T}\,.\nonumber
\end{align}
As pointed out in \cite{SF:stefan1}, the presence of the boundary terms is due to the requirement for the classical
action to be a parity-invariant functional acting on the space of analytical functions,
which satisfy the boundary conditions in eqs.~(\ref{eq:quark_boundaries1},\,\ref{eq:quark_boundaries2}).
The extension of the pure gauge \SF (\ref{eq:SF_with_action}) to QCD with one quark flavor
has the path integral representation
\be\label{eq:SF_with_quarks}
\mathcal{Z}[C',\rhobarprime,\rhoprime,C,\rhobar,\rho]=\int \mathrm{D}[\Lambda]\mathrm{D}[A]\mathrm{D}[\psibar]\mathrm{D}[\psi]\,
\mathrm{e}^{-S_\mathrm{gauge}[A]-S_\mathrm{quark}[A,\psibar,\psi]}\,.
\ee
We are left with the discussion of the static quarks. They propagate only forward in time,
and they cannot thus cross the spatial borders. However, for later 
convenience, we formally impose
periodic boundary conditions
\be\label{eq:static_quark_boundaries1}
\heavy(x+L\hat{k})=\heavy(x)\,,\qquad \heavyb(x+L\hat{k})=\heavyb(x)\,.
\ee
Here there is no need to introduce a phase like in \eqs{eq:quark_boundaries4},
because in the considered correlation functions, thanks to the U(1) 
symmetry under transformations like in (\ref{eq:xphase_inv}) such a term is irrelevant.

At the time boundaries one has Dirichlet boundary conditions as for QCD,
but the heavy quark fields have only two non-vanishing Dirac components, and the projectors
are unnecessary
\be\label{eq:static_quark_boundaries2}
\heavy(x)|_{x_0=0}=\rho_\mathrm{\phantom{\bar{h}^\dagger}\hspace{-0.28cm} h}(\vecx)\,,\qquad \heavyb(x)|_{x_0=T}=
\rhobar_\mathrm{\phantom{\bar{h}^\dagger}\hspace{-0.28cm} h}(\vecx)\,.
\ee
In analogy to \eq{eq:SF_quark_action}, the static quark action reads
\begin{align}
S_\mathrm{heavy}[A,\heavyb,\heavy]&= \int_0^T\!\mathrm{d} x_0\int_0^L\!\mathrm{d}^3\mathbf{x}\, 
                                \heavyb(x)\left\{D_0\gamma_0+\delta m\right\}\heavy(x)\label{eq:SF_static_quark_action}\\
                           &\quad -\!\int_0^L\!\mathrm{d}^3\mathbf{x}\,[\,\heavyb(x)\heavy(x)\,]_{x_0=T}\,,\nonumber
\end{align}
and the path integral representation of the \SF is
\be\label{eq:SF_with_heavy}
\mathcal{Z}[C',\rhobarprime_\mathrm{h},C,\rhoh]=\int \mathrm{D}[\Lambda]\mathrm{D}[A]\mathrm{D}[\heavyb]\mathrm{D}[\heavy]\,
\mathrm{e}^{-S_\mathrm{gauge}[A]-S_\mathrm{heavy}[A,\heavyb,\psi_\mathrm{h\phantom{\bar{\bar{\bar{\bar{p}}}}}}\hspace{-0.16cm}]}\,.
\ee
For completeness we give the heavy antiquark action too
\begin{align}
S_\mathrm{\overline{heavy}}[A,\aheavyb,\aheavy]&= \int_0^T\!\mathrm{d} x_0\int_0^L\!\mathrm{d}^3\mathbf{x}\, 
                                \aheavyb(x)\left\{-D_0\gamma_0+\delta m\right\}\aheavy(x)\nonumber\\
                           &\quad -\!\int_0^L\!\mathrm{d}^3\mathbf{x}\,[\,\aheavyb(x)\aheavy(x)\,]_{x_0=0}\,,\label{eq:SF_static_aquark_action}
\end{align}
with boundary conditions
\begin{align}
\aheavy(x+L\hat{k})&= \aheavy(x)\,,\qquad \aheavyb(x+L\hat{k})=\aheavyb(x)\,,\label{eq:static_aquark_boundaries1}\\
\aheavy(x)|_{x_0=T}&= \rhoprime_\mathrm{\phantom{\bar{h}^\dagger}\hspace{-0.33cm} h}(\vecx)\,,\qquad 
\aheavyb(x)|_{x_0=0}=\rhobar_\mathrm{\phantom{\bar{h}^\dagger}\hspace{-0.28cm} h}(\vecx)\,.\label{eq:static_aquark_boundaries2}
\end{align}

\section{Lattice formulation}\label{sect:SF_QCD_lat}

To be able to compute interesting QCD and HQET matrix elements
through the \SF we introduce the lattice regularization. The latter
is not unique, and a suitable choice, for our purposes, is explained in this section.

The space-time manifold is discretized in a regular hypercubic lattice. This
hypercube preserves the volume $L\times L\times L\times T$ of the continuum,
where $L$ and $T$ are now integers multiple of the lattice spacing $a$.
A lattice point $x$, or site, is identified by the space-time coordinates $(x_0,x_1,x_2,x_3)$,
where
\be\label{eq:SF_coordinates}
 x_\mu/a \in {\mathbb Z}^4\,,\qquad 0\leq x_0\leq T\,,\qquad 0\leq x_k <L\,.
\ee
We define the SU(3) gauge fields through the assignment of a link variable
$U_\mu (x)\in \SUthree$ to each pair of lattice points $(x,x+a\hat{\mu})$. The temporal
link variables $U_0(x)$ are defined only for $0\leq x_0<T$.

Under a gauge transformation $\Omega(x)$, where $\Omega(x)\in \SUthree$ and $\Omega(x+L\hat{k})=\Omega(x)$,
the gauge links transform as
\be\label{eq:U_under_gauge}
U_\mu(x)\to U_\mu^\Omega(x)=\Omega(x)U_\mu(x)\Omega^{-1}(x+a\hat{\mu})\,.
\ee
To be consistent with the continuum \SF, we require the gauge links to be periodic in space
\be\label{eq:U_space_periodicity}
U_\mu(x+L\hat{k})=U_\mu(x)\,,
\ee 
and to respect Dirichlet boundary conditions in time
\be\label{eq:U_time_periodicity}
U_k(x)|_{x_0=0}=W_k(\vecx)\,,\qquad U_k(x)|_{x_0=T}=W'_k(\vecx)\,.
\ee
Instead, the temporal gauge links $U_0(x)|_{x_0=0}$ remain unconstrained. 
The fields $W$ and $W'$ are related to the continuum fields $C$ and $C'$ through
\be\label{eq:rel_W_C}
W_k(\vecx)=\mathcal{P}\mathrm{exp}\left\{a\int_0^1\!\mathrm{d}t \,C_k(\vecx+a\hat{k}-ta\hat{k})\right\}\,,
\ee
and analogously for $W'$ and $C'$. 
In \eq{eq:rel_W_C} the symbol $\mathcal{P}$ denotes
the time-decreasing path ordering.
We follow \cite{Wilson} and introduce the gauge action
\be\label{eq:lattice_gauge_action}
S_\mathrm{G}[U]=\frac{1}{g_0^2}\sum_p \omega(p)\tr\{1-U(p)\}\,,
\ee
where the sum is extended to all oriented plaquettes $p$, and $U(p)$
is the parallel transporter around $p$:
\be\label{eq:Up}
U(p)=U_\mu(x)U_\nu(x+a\hat{\mu})U_\mu^{-1}(x+a\hat{\nu})U^{-1}_\nu(x)\,,\quad\mu\neq\nu\,.
\ee
In the tree-level improved theory the weight $\omega(p)$ is one except for the spatial plaquettes
at the temporal boundaries, where $\omega(p)=1/2$.
It is then straightforward to write down the lattice \SF in the pure gauge case
\be\label{eq:lSF_pure_gauge}
\mathcal{Z}[C',C]=\int \mathrm{D}[U]\mathrm{e}^{-S_\mathrm{G}[U]}\,,\qquad \mathrm{D}[U]=\prod_{x,\mu}\mathrm{d}U_\mu(x)\,.
\ee
We now introduce the QCD Dirac-Wilson action for the quark fields living on the lattice sites $x$.
To be able to write it in an elegant way, we define the discretized forward and backward
covariant derivatives
\begin{align}
\nab{\mu}\psi(x)&= \tfrac{1}{a}\left[U_\mu(x)\psi(x+a\hat{\mu})-\psi(x)\right]\,,\label{eq:lat_nab}\\
\nabstar{\mu}\psi(x)&= \tfrac{1}{a}\left[\psi(x)-U^{-1}_\mu(x-a\hat{\mu})\psi(x-a\hat{\mu})\right]\,.\label{eq:lat_nabstar}\
\end{align}
We extend the fields to all times $x_0$ by ``padding'' with zeros, i.e.~by setting
\be\label{eq:}
\psi(x)=\psibar(x)=0\,,\qquad \mbox{for}\,\, x_0<0\,\, \mbox{and}\,\, x_0>T\,,
\ee
and by supplying the boundary conditions of eqs.~(\ref{eq:quark_boundaries1},\,\ref{eq:quark_boundaries2}) with
\begin{align}
P_-\psi(x)|_{x_0=0}&= 0\,,\qquad P_+\psi(x)|_{x_0=T}=0\,,\label{eq:quark_boundaries1_pad}\\
\psibar(x)P_+|_{x_0=0}&= 0\,,\qquad\psibar(x)P_-|_{x_0=T}=0\label{eq:quark_boundaries2_pad}\,.
\end{align}
We can now write the Dirac-Wilson action
\be\label{eq:Wilson_action}
S_\mathrm{F}[U,\psibar,\psi]=a^4\sum_x \psibar(x)(D+m)\psi(x)
\ee
with the Dirac-Wilson operator
\be\label{eq:Wilson_operator}
D=\frac{1}{2}\left(\gamma_\mu(\nabstar{\mu}+\nab{\mu})-a\nabstar{\mu}\nab{\mu}\right)\,.
\ee
Before dealing with the heavy quark action, we would like to write \eq{eq:Wilson_action} in a different
way, which will be useful later on. By using the explicit expression of the Dirac-Wilson
operator we get
\begin{align}
S_\mathrm{F}&= a^4\sum_{x,\mu}\psibar(x)
\left\{\frac{1}{2a}\gamma_\mu\left[U_\mu(x)\psi(x+a\hat{\mu})-U_\mu^{-1}(x-a\hat{\mu})\psi(x-a\hat{\mu})\right]\right.\nonumber\\
&\quad -\left.\frac{1}{2a}\left[U_\mu(x)\psi(x+a\hat{\mu})+U_\mu^{-1}(x-a\hat{\mu})\psi(x-a\hat{\mu})
-(2+\frac{am}{2})\psi(x)\right]\right\}\nonumber\\
&= a^4\sum_{x,\mu}\psibar(x)\left\{\frac{1}{2a}\left(\gamma_\mu-1\right)U_\mu(x)\psi(x+a\hat{\mu})\right.\nonumber\\
&\quad  \left.-\frac{1}{2a}\left(\gamma_\mu+1\right)U_\mu^{-1}(x-a\hat{\mu})\psi(x-a\hat{\mu})+(2+\frac{am}{2})
\psi(x)\right\}\,.\label{eq:Wilson_action_exp1}
\end{align}
Instead of the bare quark mass $m$, it is convenient to work with the hopping
parameter $\kappa=(8+2am)^{-1}$, and, by rescaling the fields $\psi\to (2\kappa)^{1/2}\psi$
and $\psibar\to (2\kappa)^{1/2}\psibar$, we get the simple expression
\begin{align}
S_\mathrm{F}&= a^3\sum_x\Big\{\psibar(x)\psi(x)-\kappa\sum_\mu\left[\,\psibar(x)
(1-\gamma_\mu)U_\mu(x)\psi(x+a\hat{\mu})\right.\nonumber\\
&\quad  \left.+\psibar(x)(1+\gamma_\mu)U_\mu^{-1}(x-a\hat{\mu})\psi(x-a\hat{\mu})\right]\Big\}\,,\label{eq:Wilson_action_exp2}
\end{align}
which is usually written in the compact form
\begin{align}
S_\mathrm{F}&= a^3\sum_{y,x} \psibar(y)M(y,x)\psi(x)\,,\quad\mbox{with}\label{eq:def_M_q_action1}\\
   M(y,x)&= \delta_{y,x}-\kappa\sum_\mu\left(\delta_{y,x+a\hat{\mu}}(1+\gamma_\mu)U_\mu^{-1}(x)\right.\nonumber\\
         &  \hspace{1.87cm}\quad \left. +\,\delta_{y,x-a\hat{\mu}}(1-\gamma_\mu)U_\mu(x-a\hat{\mu}) \right)\,.\label{eq:def_M_q_action2}
\end{align} 
For static quarks the simplest discretized version of the action in \eq{eq:stat_Lagr_formal}
has been proposed by Eichten and Hill \cite{stat:eichhill1,stat:eichhill_za}:
\be\label{eq:EH_action}
S_\mathrm{h}[U,\heavyb,\heavy]=\frac{a^4}{1+a\delta m}\sum_x \heavyb(x)(\nabstar{0}+\delta m)\heavy(x)\,.
\ee
One can also consider a more general action
\be\label{eq:general_heavy_action}
S_\mathrm{h}^\mathrm{W}[U,\heavyb,\heavy]=\frac{a^4}{1+a\delta m_\mathrm{W}}\sum_x \heavyb(x)(D_0^\mathrm{W}+\delta m_\mathrm{W})\heavy(x)\,,
\ee
where 
\be\label{eq:D0W}
D_0^\mathrm{W}\heavy=\frac{1}{a}\left[\heavy(x)-W^\dagger_0(x-a\hat{0})\heavy(x-a\hat{0})\right]\,,
\ee
and $W_\mu$ is a general parallel transporter, differing from $U_\mu$ by $\Oasq$-terms.
Analogously for the heavy antiquarks
\begin{align}
S_{\bar{\mathrm{h}}}^\mathrm{W}[U,\aheavyb,\aheavy]&= \frac{a^4}{1+a\delta m_\mathrm{W}}\sum_x 
\aheavyb(x)(\bar{D}_0^\mathrm{W}+\delta m_\mathrm{W})\aheavy(x)\,,\label{eq:general_aheavy_action}\\
\bar{D}_0^\mathrm{W}\aheavy&= \frac{1}{a}\left[W_0(x)\aheavy(x+a\hat{0})-\aheavy(x)\right]\,.\label{eq:aD0W}
\end{align}
Under a gauge transformation $W_\mu$ transforms as $U_\mu$, and, together with gauge invariance,
parity and cubic symmetry, the actions $S_\mathrm{h}$, $S_\mathrm{h}^\mathrm{W}$ and $S_{\bar{\mathrm{h}}}^\mathrm{W}$ 
satisfy the symmetries 
of the static theory expounded in \sect{sect:symm_eff_lagrangian}. These alternative discretizations
are introduced to reduce, in a Monte Carlo simulation, the noise-to-signal ratio $(R_\mathrm{NS})$ of the observables
of interest, while remaining with roughly the same discretization errors.
This is sufficient to ensure to stay within the same universality class as well as the same 
$\Oa$-improvement of the Eichten-Hill action. The authors of \cite{stat:letter,stat:actpaper}
have proposed several possible discretizations. Here we present only the so-called HYP actions.

Following \cite{HYP} the parallel transporter $W_\mu(x)$ is constructed in three con\-ca\-te\-na\-ted steps.
They are\\
{\flushleft \fbox{Step 1}}
\begin{align}
W_\mu(x)&= \mathcal{P}_{\SUthree}\left[(1-\alpha_1)U_\mu(x)
+\frac{\alpha_1}{6}\sum_{\pm\nu\neq\mu}\!\tilde{K}_{\nu\mu}(x)\right]\label{eq:HYP_step1}\\
\tilde{K}_{\nu\mu}(x)&= \tilde{V}_{\nu;\mu}(x)
\tilde{V}_{\mu;\nu}(x+a\hat{\nu})\tilde{V}_{\nu;\mu}^\dagger(x+a\hat{\mu})\label{eq:HYP_step1_staple}
\end{align}
{\flushleft \fbox{Step 2}}
\begin{align}
\tilde{V}_{\nu;\mu}(x)&= \mathcal{P}_{\SUthree}\left[(1-\alpha_2)U_\mu(x)
+\frac{\alpha_2}{4}\sum_{\pm\rho\neq\nu,\mu}\!\bar{K}_{\rho\nu\mu}(x)\right]\label{eq:HYP_step2}\\
\bar{K}_{\rho\nu\mu}(x)&= \bar{V}_{\rho;\nu\mu}(x)
\bar{V}_{\mu;\rho\nu}(x+a\hat{\rho})\bar{V}_{\rho;\nu\mu}^\dagger(x+a\hat{\mu})\label{eq:HYP_step2_staple}
\end{align}
{\flushleft \fbox{Step 3}}
\begin{align}
\bar{V}_{\mu;\nu\rho}(x)&= \mathcal{P}_{\SUthree}\left[(1-\alpha_3)U_\mu(x)
+\frac{\alpha_3}{2}\sum_{\pm\eta\neq\rho,\nu,\mu}\!\!K_{\eta\mu}(x)\right]\label{eq:HYP_step3}\\
K_{\eta\mu}(x)&= U_\eta(x)U_\mu(x+a\hat{\eta})U^\dagger_\eta(x+a\hat{\mu})\label{eq:HYP_step3_staple}
\end{align}
$ $\\
Here the greek indices can assume also negative values and $U_{-\mu}(x)=U^\dagger_\mu(x-a\hat{\mu})$. The symbol
$\mathcal{P}_{\SUthree}$ indicates the projection onto SU(3), which is here defined by the rescaling
\be
W\to W/\sqrt{\tr{(WW^\dagger)}/3}\,,
\ee
followed by four iterations of
\be
W\to X\left(1-\frac{i}{3}\Im\!\left(\det (X)\right)\right)\,,\quad\mbox{where}\quad 
X=W\left(\frac{3}{2}-\frac{1}{2}W^\dagger W\right)\,.
\ee
The parameters $\alpha_1$, $\alpha_2$ and  $\alpha_3$ can be tuned by requiring an
approximate minimization of the $R_\mathrm{NS}$ of some matrix elements.

\section{Expectation values and Monte Carlo integration}\label{sect:MC_path_integral}

As it is already clear from \eqs{eq:lSF_pure_gauge}, the evaluation of the lattice SF requires
an integration over an enormous number of variables. If we consider a volume $L^3\times T$,
the number of real parameters to be integrated is approximately
\be\label{eq:number_real_variables}
[8]_\SUthree \times [4]_{\hat{\mu}} \times [(L/a)^3]_\mathrm{space}\times [(T/a)]_\mathrm{time}\,.
\ee
With $L/a=T/a=10$ we have 320000 real variables. In absence of
symmetries and/or approximations capable to drastically reduce the number (\ref{eq:number_real_variables}),
the recourse to statistical methods is necessary.

We start by considering a product ${\mathcal{O}}$ of fields on the lattice, and we want 
to evaluate its expectation value $\langle {\mathcal{O}}\rangle_\mathrm{QCD}$ by using the SF scheme in QCD.
We thus compute the integral
\be\label{eq:O_in_lQCD}
\langle {\mathcal{O}}\rangle_\mathrm{QCD}=\left\{\frac{1}{\mathcal{Z}}\int \mathrm{D}[U]\mathrm{D}[\psibar]\mathrm{D}[\psi]\,{\mathcal{O}}\,
\mathrm{e}^{-S[U,\psibar,\psi]}\right\}_{\rhobarprime=\rhoprime=\rhobar=\rho=0}\,,
\ee
where
\be\label{eq:Z_in_lQCD}
\mathcal{Z}=\int \mathrm{D}[U]\mathrm{D}[\psibar]\mathrm{D}[\psi]\,
\mathrm{e}^{-S[U,\psibar,\psi]}\,,\quad S[U,\psibar,\psi]=S_\mathrm{G}[U]+S_\mathrm{F}[U,\psibar,\psi]\,.
\ee
Apart from the dynamical variables $U$, $\psibar$ and $\psi$ integrated over, the product ${\mathcal{O}}$ 
may involve the boundary fields
\begin{align}
\zeta(\vecx)&= \frac{\delta}{\delta\rhobar(\vecx)}\,,
\quad\hspace{0.26cm} \zetabar(\vecx)=-\frac{\delta}{\delta\rho(\vecx)}\,,\nonumber\\
\zzetaprime(\vecx)&= \frac{\delta}{\delta\rhobarprime(\vecx)}\,,\quad 
\zetabarprime(\vecx)=-\frac{\delta}{\delta\rhoprime(\vecx)}\,,\label{eq:boundary_functionals}
\end{align}
having the form of functional derivatives acting on the Boltzmann factor in \eq{eq:O_in_lQCD}.

Since there is (until now) no efficient way of dealing with the Grassmann-valued quark fields
in a computer simulation, we exploit the property of the fermionic action $S_\mathrm{F}$
to be a bilinear in quark fields, and we analytically integrate out the fermionic
variables in \eq{eq:O_in_lQCD}. This defines the new observable 
\be\label{eq:O_F}
[{\mathcal{O}}]_\mathrm{F}[U]\equiv \left\{\frac{1}{\mathcal{Z_\mathrm{F}}}\int \mathrm{D}[\psibar]\mathrm{D}[\psi]\,{\mathcal{O}}\,
\mathrm{e}^{-S_\mathrm{F}[U,\psibar,\psi]}\right\}_{\rhobarprime=\rhoprime=\rhobar=\rho=0}\,,
\ee
depending only on the gauge fields, and
\be\label{eq:Z_F}
\mathcal{Z_\mathrm{F}}=\int\mathrm{D}[\psibar]\mathrm{D}[\psi]\,\mathrm{e}^{-S_\mathrm{F}[U,\psibar,\psi]}\,.
\ee
We can now rewrite the expectation value (\ref{eq:O_in_lQCD}) as
\be\label{eq:O_eff_in_lQCD}
\langle {\mathcal{O}}\rangle_\mathrm{QCD}=\frac{\int \mathrm{D}[U]\,[{\mathcal{O}}]_\mathrm{F}[U]\,\mathrm{e}^{-S_\mathrm{eff}[U]}}{ \int \mathrm{D}[U]\,\mathrm{e}^{-S_\mathrm{eff}[U]}}\,,
\ee
with the effective action given by
\be\label{eq:eff_lQCD_action}
S_\mathrm{eff}[U]=S_\mathrm{G}[U]-\ln(\det(M[U]))\,.
\ee
where $M$ has been defined in \eq{eq:def_M_q_action2}.
The computation of (\ref{eq:O_eff_in_lQCD}) is considerably simplified by replacing $\det(M[U])$ by $1$. This 
is widely referred to as quenched approximation. In perturbation
theory it can be interpreted with the omission of the Feynman diagrams consisting of fermion loops,
with an arbitrary number of gluon legs attached to it.

In the case where also static quarks are involved, we calculate in HQET the expectation value 
of the composite fields ${\mathcal{O}}$ according to the procedure developed in \sect{sect:ren_HQET}.
By using the static action (\ref{eq:general_heavy_action}), we have the path integral expression
\begin{align}
\hspace{-0.5cm}\langle {\mathcal{O}}\rangle_\mathrm{stat}&= \left\{\frac{1}{\mathcal{Z}}\int \mathrm{D}[U]\mathrm{D}[\psibar]
\mathrm{D}[\psi]\mathrm{D}[\heavyb]\mathrm{D}[\heavy]
\,{\mathcal{O}}\,
\mathrm{e}^{-S[U,\psibar,\psi,\overline{\psi}_\mathrm{h},\psi_\mathrm{h\phantom{\bar{f}}}\hspace{-0.16cm}
]}\right\}_{\{\rho\}=0}\,,\nonumber\\
\{\rho\}&= \{\rhobarprime,\rhoprime,\rhobar,\rho,\rhohbprime,\rhoh\}\,,\label{eq:O_in_lHQET}
\end{align}
where the action is now
\be\label{eq:action_in_lHQET}
S[U,\psibar,\psi,\heavyb,\heavy]=S_\mathrm{G}[U]+S_\mathrm{F}[U,\psibar,\psi]+S_\mathrm{h}^\mathrm{W}[U,\heavyb,\heavy]\,,
\ee
and the boundary fields
\be\label{eq:h_boundary_functionals}
\zetahprime(\vecx)=\frac{\delta}{\delta \rhohbprime(\vecx)}\,,\quad  
\zetahb(\vecx)=-\frac{\delta}{\delta \rhoh(\vecx)}\,,
\ee
may appear besides the ones in \eqs{eq:boundary_functionals}.
Of course an observable ${\mathcal{O}}$ made of only heavy quark and gauge fields is an interesting case too.
The analytical integration of the fermionic observables in \eq{eq:O_in_lHQET}
follows the scheme explained for QCD, without further complications. 

For later convenience we define also the boundary fields
\be\label{eq:ah_boundary_functionals}
\zetah(\vecx)=\frac{\delta}{\delta \rhobar_\mathrm{\phantom{\bar{h}^\dagger}\hspace{-0.28cm} h}(\vecx)}\,,\quad  
\zetahbprime(\vecx)=-\frac{\delta}{\delta \rhoprime_\mathrm{\phantom{\bar{h}^\dagger}\hspace{-0.38cm} h}(\vecx)}\,,
\ee
and through the action
\begin{align}
S[U,\psibar,\psi,\aheavyb,\aheavy,\heavyb,\heavy]&= S_\mathrm{G}[U]+S_\mathrm{F}[U,\psibar,\psi]\nonumber\\
&\quad +S_{\bar{\mathrm{h}}}^\mathrm{W}[U,\aheavyb,\aheavy]+S_\mathrm{h}^\mathrm{W}[U,\heavyb,\heavy]\,,\label{eq:action_lonlyH}
\end{align}
the path integral expression
\begin{align}
\hspace{-0.5cm}\langle {\mathcal{O}}\rangle_{\mathrm{h}\mathrm{h}}&= \left\{\frac{1}{\mathcal{Z}}\int \mathrm{D}[U]\mathrm{D}[\{\psi\}]
\,{\mathcal{O}}\,
\mathrm{e}^{-S[U,\psibar,\psi,\overline{\psi}_\mathrm{\bar{h}},\psi_\mathrm{\bar{h}},\overline{\psi}_\mathrm{h},\psi_\mathrm{h\phantom{\bar{f}}}\hspace{-0.16cm}]}\right\}_{\{\rho\}=0}\,,\nonumber\\
& \nonumber\\
\{\rho\}&= \{\rhobarprime,\rhoprime,\rhobar,\rho,\rhoprime_\mathrm{\phantom{\bar{h}^\dagger}\hspace{-0.38cm} h},\rhobar_\mathrm{\phantom{\bar{h}^\dagger}\hspace{-0.28cm} h},
\rhohbprime,\rhoh\}\,,\label{eq:O_onlyH}\\
&  \nonumber\\
\mathrm{D}[\{\psi\}]&= \mathrm{D}[\psibar]\mathrm{D}[\psi]\mathrm{D}[\aheavyb]\mathrm{D}[\aheavy]\mathrm{D}[\heavyb]\mathrm{D}[\heavy]
\,.\nonumber
\end{align}
On a computer, the expectation value $\langle {\mathcal{O}}\rangle$, in QCD or HQET, is calculated by a Monte Carlo
simulation, where an ensemble of $N$ gauge links configurations $U^{(i)}$, with $i=1,\ldots, N$,
is generated with probability 
\be\label{eq:prob}
P^{(i)}\propto \mathrm{D}[U^{(i)}]\,\mathrm{exp}\{-S_\mathrm{eff}[U^{(i)}]\}\,,
\ee
and $\langle {\mathcal{O}}\rangle$ is approximated by the average
\begin{align}
\langle {\mathcal{O}}\rangle&= \overline{{\mathcal{O}}}\pm\Delta\overline{{\mathcal{O}}}\,,\label{eq:MC_average}\\
\overline{{\mathcal{O}}}&= \frac{1}{N}\sum_{i=1}^N{\mathcal{O}}_i=\frac{1}{N}\sum_{i=1}^N[{\mathcal{O}}]_\mathrm{F}[U^{(i)}]\,,\quad \Delta\overline{{\mathcal{O}}}=\mathrm{O}(\frac{1}{\sqrt{N}})\,,\label{eq:MC_naive_errors}
\end{align}
where $\Delta\overline{{\mathcal{O}}}$ is the statistical error due to the finiteness of the statistical 
sample.

The non-perturbative computations reported in this thesis have been performed in the quenched
approximation, and we thus have $S_\mathrm{eff}=S_\mathrm{G}$ in \eq{eq:eff_lQCD_action}. 
Analogously, if the valence sector is just made of a static quark-antiquark pair,
in the action (\ref{eq:action_lonlyH}) one can eliminate $S_\mathrm{F}$ from the very beginning.
The ensemble of gauge configurations
is generated by means of the so-called local ``hybrid over-relaxation'' (HOR) algorithm (see e.g.~\cite{deDivitiis:1995yp}). The basic cycle consists
of 1 heathbath (HB) update sweep throughout the whole lattice, followed by $N_\mathrm{OR}$ over-relaxation (OR)
sweeps.  The cycle is iterated $N_\mathrm{UP}$ times, and with the obtained gauge configuration $U^{(i)}$ a
computation of $[{\mathcal{O}}]_\mathrm{F}[U^{(i)}]$ is performed. This last step is called measurement (MEAS).
The procedure can thus be summarized as
\be\label{MC_computation}
\Big(\mathrm{HB}\times(\mathrm{OR})^{N_\mathrm{OR}}\Big)^{N_\mathrm{UP}}\times\mathrm{MEAS}\,,
\ee
and the whole procedure is repeated $N$ times.
The algorithm used for the gauge update is local in the sense that one processes one time-slice after
the other, and each gauge link $U_\mu(x)$ is separately visited and updated.
The heathbath algorithm is a modification of Creutz's algorithm \cite{pot:creutz}
by Fabricius and Haan \cite{Fabricius:1984wp} and independently by Kennedy and Pendleton \cite{Kennedy:1985nu}. 
The over-relaxation sweeps are microcanonical and follow \cite{HOR1}. The whole update procedure
is executed in embedded SU(2) subgroups according to \cite{Cabibbo:1982zn}.
Optimized choices of $N_\mathrm{UP}$ and $N_\mathrm{OR}$, for which the statistical errors
on our observables are minimized at fixed simulation run time, 
are $N_\mathrm{UP}=5$ and $N_\mathrm{OR}=L/2a$. The latter is chosen to be a multiple of the correlation
length in lattice units in order to minimize the autocorrelation times \cite{alpha:SU3}.

The correct estimation of the statistical uncertainty $\Delta\overline{{\mathcal{O}}}$
plays a fundamental role for the reliability of the results. At this point
two issues must be discussed: thermalization and (auto-)correlation. The discussion
is presented in \app{MC_errors}.

When considering full QCD, i.e.~with an exact treatment of the quark determinant 
in \eq{eq:eff_lQCD_action}, the updating algorithm described above
is known to lead to unsatisfactory performances. Other techniques are employed,
and for a recent review one can consult \cite{Kennedy:2006ax,Giusti:2007hk}.

\section{Improvement}\label{sect:Oa_impr}

The lattice discretization introduces a cutoff in the Feynman 
integrals described in the previous section, making them ultraviolet
finite. Furthermore such a discretization is particularly suitable 
to be implemented on a computer.
As for other regularization methods, one then proceeds by applying
a renormalization program, consisting in defining renormalized 
Green functions, which approach a finite limit as the cutoff is removed.

The removal of the lattice structure amounts to studying the continuum limit.
This is a non-trivial task. Since the bare parameters of the theory depend on the 
cutoff, they have to be tuned as a function of the lattice spacing with respect to the imposed
renormalization conditions. The latter state that some physical quantities,
such as the coupling and the particle masses, are to be held fixed as the cutoff
is removed. This procedure defines in the parameter space the so-called Line
of Constant Physics (LCP).

In the approach to the continuum limit, the presence of effects linear in $a$
usually obliges to vary the size of the lattice spacing over a wide range
of physical values before having a reasonable control on the cutoff effects.
In addition, the latter can be quite large, compromising the reliability and precision
of the continuum extrapolation. It turns out that the effects linear in $a$ can be isolated
and cancelled by adding appropriate counterterms vanishing in the continuum. The resulting
theory is said to be $\Oa$-improved.
 
A successful and interesting way of implementing the improvement
has been introduced by
Symanzik \cite{Symanzik:1982,impr:Sym1,impr:Sym2}, who provided arguments
to cancel the $\Oa$-effects in on-shell quantities
by adding appropriate counterterms. On-shell quantities are for example particle
masses and correlation functions over physical space-time distances. 
Close to the continuum, the lattice theory can be described by a local effective theory
with action \cite{impr:pap1}
\be\label{eq:eff_Sym_action}
S_\mathrm{eff}=S_0+\sum_{k=1}^{\infty}a^k S_k\,,
\ee
where $S_0$ is the continuum action, and the consecutive terms are given by
\be\label{eq:eff_Sym_action_cons_terms}
S_k=\int \mathrm{d}^4x\,\mathcal{L}_k(x)\,.
\ee
The Lagrangians $\mathcal{L}_k(x)$ gather in a linear combination 
local and gauge-invariant composite fields respecting the 
symmetries of the theory, and having mass dimension $4+k$.
To give a precise meaning to all terms of the effective action, one can
regularize them by using a lattice spacing $a'\ll a$. 

Cutoff effects do not stem only from the action. The composite
fields, appearing in the observables that we want to compute, are possible
sources of discretization errors linear in the lattice spacing. Let us call such a 
field $\phi(x)$,
and examine the connected correlation function
\be\label{eq:conn_corr_f}
G_n(x_1,\ldots,x_n)=(Z_\phi)^n\langle\phi(x_1)\ldots\phi(x_n)\rangle_\mathrm{con}\,.
\ee 
The factor $Z_\phi$ accounts for the field renormalization and the points $x_1,\ldots,x_n$
are kept at non-zero physical distance. The mixing of $\phi$ with other fields 
under renormalization is excluded to lighten the discussion. 
We thus expect that, if the renormalization factor
is correctly chosen, the quantity $G_n$ has a well-defined continuum limit.

In the local effective theory an expansion akin to (\ref{eq:eff_Sym_action}) can be provided
for the renormalized lattice field $Z_\phi \phi(x)$, and reads
\be\label{eq:eff_Sym_field}
\phi_\mathrm{eff}(x)=\phi_0(x)+\sum_{k=1}^{\infty}a^k \phi_k(x)\,,
\ee
where the fields $\phi_k$ are linear combinations of composite and local fields
with the appropriate dimension and symmetries.
Let us now write down the expansion of the correlation function (\ref{eq:conn_corr_f}):
\begin{align}\label{eq:eff_conn_corr_f}
G_n(x_1,\ldots,x_n)&= \langle\phi_0(x_1)\ldots\phi_0(x_n)\rangle_\mathrm{con}\nonumber\\
                   &\quad  -a\int \mathrm{d}^4y\,\langle\phi_0(x_1)\ldots\phi_0(x_n)\mathcal{L}_1(y)\rangle_\mathrm{con}\\
                   &\quad  +a\sum_{k=1}^n\langle\phi_0(x_1)\ldots\phi_1(x_k)\ldots\phi_0(x_n)\rangle_\mathrm{con}+\mathrm{O}(a^2)\,.\nonumber
\end{align}
The reader may notice a strong similarity with the expression (\ref{eq:expt_stat}) given for HQET.
The expectation values on the r.h.s.~are to be taken by using the continuum Lagrangian $\mathcal{L}_0$,
which is power-counting renormalizable. 
The integral over $y$, in the second term
of the r.h.s.~of \eq{eq:eff_conn_corr_f}, can produce contact terms leading to divergences and/or to limitations in the use 
of the field equations. These terms amount to operator insertions, which 
are constrained by the dimensional analysis and the symmetries of the effective theory, and
can be compensated by a redefinition of the field $\phi_1$.
As hinted at the beginning of this section, 
the on-shell $\Oa$-improvement is achieved by modifying the action and the composite fields
through the addition of appropriate counterterms. The latter aim to let $\mathcal{L}_1$ and $\phi_1$
vanish up to term of higher order in $a$, and have the general form
\be\label{eq:Sym_counterterms}
\delta S=a\int \mathrm{d}^4x\,\sum_{l}c_l{\mathcal{O}}_l\,,\quad \delta \phi =a\sum_{l}c_l^\phi\mathcal{\phi}^{(l)}\,.
\ee 
In principle also higher orders of the expansion (\ref{eq:eff_conn_corr_f}) can be improved
using the same technique.
Nevertheless the number of counterterms to be fine-tuned rapidly increases,
and in practice the improvement is usually applied only to the terms linear in the lattice spacing. 

We start with the
\SF formulation of lattice QCD with two mass-degenerate quarks, and the action 
(\ref{eq:lattice_gauge_action}) for the gauge fields and the Dirac-Wilson action (\ref{eq:Wilson_action})
for the fermions. Later on we will discuss the heavy quark case.

We concentrate on the bulk of the lattice, and
observe that the Lagrangian $\mathcal{L}_1$ can be made to vanish by adding to the action a counterterm
of the form
\be\label{eq:Sym_L1}
a^5 \sum_{x_0=a}^{T-a}\sum_\vecx \sum_l c_l \widehat{\mathcal{O}}_l(x)\,,
\ee
where the operators $\widehat{\mathcal{O}}_l$ are a lattice representation of the basis
\begin{align}
{\mathcal{O}}_1&= \psibar\,\sigma_{\mu\nu}{F}_{\mu\nu}\psi\,,\nonumber\\
{\mathcal{O}}_2&= m\tr\{{F}_{\mu\nu}{F}_{\mu\nu}\}\,,\label{possible_basis}\\
{\mathcal{O}}_3&= m^2\psibar\psi\,.\nonumber
\end{align}
The discretized version $\widehat{F}_{\mu\nu}$ of the curvature tensor (\ref{eq:F_munu_QCD}) can be written as
\be\label{eq:F_munu_lQCD}
\widehat{F}_{\mu\nu}(x)=\frac{1}{8a^2}\{Q_{\mu\nu}(x)-Q_{\nu\mu}(x)\}\,,
\ee
with
\begin{align}
Q_{\mu\nu}(x)&= \Big\{U_\mu(x)U_\nu(x+a\hat{\mu})U_\mu^{-1}(x+a\hat{\nu})U_\nu^{-1}(x)\nonumber\\
             &  \nonumber\\
             &\quad  +U_\nu(x)U_\mu^{-1}(x+a\hat{\nu}-a\hat{\mu})U_\nu^{-1}(x-a\hat{\mu})U_\mu(x-a\hat{\mu})\nonumber\\
             &  \nonumber\\
             &\quad  +U_\mu^{-1}(x-a\hat{\mu})U_\nu^{-1}(x-a\hat{\mu}-a\hat{\nu})
                 U_\mu(x-a\hat{\mu}-a\hat{\nu})U_\nu(x-a\hat{\nu})\nonumber\\
             &  \nonumber\\
             &\quad  +U_\nu^{-1}(x-a\hat{\nu})U_\mu(x-a\hat{\nu})U_\nu(x+a\hat{\mu}-a\hat{\nu})U_\mu^{-1}(x)\Big\}\,.\label{eq:Q_munu_lQCD}
\end{align}
We notice that the operators $\ophat{2}$ and $\ophat{3}$ already appear 
in the action $S=S_\mathrm{G}+S_\mathrm{F}$, and they merely lead to a reparametrization of 
the bare coupling and mass, when the latter get renormalized.
We are thus left with the operator $\ophat{1}$, which can be added 
to the Dirac-Wilson operator
\be\label{eq:D_impr}
D_\mathrm{impr}=D+\csw\frac{ia}{4}\sigma_{\mu\nu}\widehat{F}_{\mu\nu}\,.
\ee
The coefficient $\csw$ depends on $g_0$ and must be appropriately tuned.
It has been introduced for the first time by Sheikholeslami and Wohlert \cite{impr:SW}. 
Another equivalent way of formulating this improvement is the addition to the action $S$
of the volume term 
\be\label{eq:impr_vol_term}
\delta S_\mathrm{v}=a^5\sum_{x_0=a}^{T-a}\sum_\vecx\csw\psibar \frac{i}{4}\sigma_{\mu\nu}\widehat{F}_{\mu\nu}\psi\,.
\ee
Finally we notice that no improvement terms are needed for the gauge action.

So far we have discussed only the bulk of the lattice. We still need to add to the action
the improvement terms which account for the effects at the temporal boundaries. 
The fully improved action can be written in the form
\be\label{eq:improved_action}
S_\mathrm{impr}[U,\psibar,\psi]=S[U,\psibar,\psi]+\delta S_\mathrm{v}[U,\psibar,\psi]
+\delta S_\mathrm{G,b}[U]+\delta S_\mathrm{F,b}[U,\psibar,\psi]\,.
\ee
The boundary counterterms for the gauge action read \cite{SF:LNWW}:
\begin{align}
\delta S_\mathrm{G,b}[U]&= \frac{1}{2g_0^2}(\cs-1)\sum_{p_\mathrm{s}}\tr\{1-U(p_\mathrm{s})\}\nonumber\\
                     &\quad  +\frac{1}{g_0^2}(\ct-1)\sum_{p_\mathrm{t}}\tr\{1-U(p_\mathrm{t})\}\,.\label{eq:improved_g_action}
\end{align}
The sums here are restricted to all oriented plaquettes attached to the 
temporal boundaries $x_0=0$ and $x_0=T$, being space-like $(p_\mathrm{s})$ or time-like $(p_\mathrm{t})$.

By exploiting the symmetries of the theory, the field equations, and by omitting the terms which can be reabsorbed
in the renormalization of the quark masses and of the quark and antiquark fields, a possible
choice for $\delta S_\mathrm{F,b}$ is \cite{impr:pap1}:
\begin{align}
\delta S_\mathrm{F,b}[U,\psibar,\psi]&= a^4\sum_\vecx\left\{(\cstil-1)[\ophat{s}(\vecx)
+\ophatprime{s}(\vecx)]\right.\nonumber\\
                                  &  \hspace{0.8cm}\quad+\left.
(\cttil-1)[\ophat{t}(\vecx)
+\ophatprime{t}(\vecx)]
\right\}\,,\label{eq:delta_SFb}
\end{align}
where
\begin{align}
\ophat{s}(\vecx)&= \tfrac{1}{2}\rhobar(\vecx)\gamma_k(\nabstar{k}+\nab{k})\rho(\vecx)\,,\label{eq:ophats}\\
& \nonumber\\
\ophatprime{s}(\vecx)&= \tfrac{1}{2}\rhobarprime(\vecx)\gamma_k(\nabstar{k}+\nab{k})\rhoprime(\vecx)\,,\label{eq:ophatsp}\\
& \nonumber\\
\ophat{t}(\vecx)&= \tfrac{1}{2}\{\,\psibar(y)P_+\nabstar{0}\psi(y)+\psibar(y)\lnabstar{0}P_-\psi(y)\}_{y=(a,\vecx)}\,,\label{eq:ophatt}\\
&  \nonumber\\
\ophatprime{t}(\vecx)&=\tfrac{1}{2}\{\,\psibar(y)P_-\nab{0}\psi(y)+\psibar(y)\lnab{0}P_+\psi(y)\}_{y=(T-a,\vecx)}\,.\label{eq:ophattp}
\end{align}
For our choices of the boundary conditions and observables, specified in the following chapters,
the terms proportional to $\cs$ and $\cstil$ do not contribute. For $N_\mathrm{f}=0$, the coefficient
$\csw$ has been non-perturbatively computed in \cite{impr:pap3}, where the parametrization
\be\label{eq:csw_parametrization}
\csw(g_0^2)=\frac{1-0.656g_0^2-0.152g_0^4-0.054g_0^6}{1-0.922g_0^2}\,,
\ee
has been proposed with a precision of 3\% in the range $0\leq g_0\leq 1$. The coefficients
$\ct$ and $\cttil$ are only perturbatively known
\begin{align}
\cttil(g_0^2)&= 1-0.01795(2)g_0^2+\mathrm{O}(g_0^4)\,,\label{eq:cttil_1loop}\\
\ct(g_0^2)&= 1-0.08900(5)g_0^2-0.0294(3)g_0^4+\mathrm{O}(g_0^6)\,,\quad N_\mathrm{f}=0\,,\label{eq:ct_2loop}
\end{align}
and have been computed in \cite{impr:pap5} and \cite{pert:2loop_fin} respectively.

For the heavy quarks the improvement procedure \cite{zastat:pap1} is much simpler than in
the relativistic case. 
We consider the action (\ref{eq:action_in_lHQET}), and besides the improvement terms for
$S_\mathrm{G}$ and $S_\mathrm{F}$ that have just been explained, we examine the possible additional
terms for $S_\mathrm{h}^\mathrm{W}$. On the bulk we can write down 
a possible basis of dimension five operators. Nevertheless only one of them will survive
after applying the field equations and the symmetries of the static theory. Their continuum expression
reads\footnote{The operator basis in the continuum can be used both for the Eichten-Hill 
action and a HYP action. They differ by $\mathrm{O}(a^2)$-terms.}
\begin{align}
\op{h,1}&= \heavyb\sigma_{0k}{F}_{0k}\heavy\,,\label{eq:Oh1}\\
\op{h,2}&= \heavyb\sigma_{jk}{F}_{jk}\heavy\,,\label{eq:Oh2}\\
\op{h,3}&= \heavyb\{ D_{0}D_{0}+\ola{D}_{0}\ola{D}_{0}\}\heavy\,,\label{eq:Oh3}\\
\op{h,4}&= m\heavyb\{D_0-\ola{D}_{0}\}\heavy\,,\label{eq:Oh4}\\
\op{h,5}&= \heavyb\{D_{k}D_{k}+\ola{D}_{k}\ola{D}_{k}\}\heavy\,,\label{eq:Oh5}\\
\op{h,6}&= m\heavyb\{\gamma_k D_k-\ola{D}_{k}\gamma_k\}\heavy\,,\label{eq:Oh6}\\
\op{h,7}&= m^2\heavyb\heavy\,,\label{eq:Oh7}
\end{align}
where $m$ is the mass of the relativistic quark. 
The operator $\op{h,1}$
vanishes because $P_+\sigma_{0k}P_+=0$, while $\op{h,2}$ is not invariant under
the spin rotations (\ref{eq:spin_inv}). The operators $\op{h,3}$ and $\op{h,4}$
vanish because of the formal equations of motion
\be\label{eq:formal_eq_of_motion}
D_0\heavy(x)=0\,,\quad\heavyb(x)\ola{D}_0=0\,.
\ee
The operators $\op{h,5}$ and $\op{h,6}$ break the U(1) symmetry, 
indeed they are not invariant under the phase
transformations (\ref{eq:xphase_inv}). We are thus left with $\op{h,7}$.
This is just a mass-dependent shift in $\delta m$, appearing in \eq{eq:stat_Lagr_formal},
and will be taken into account when dealing with the renormalization.

In complete analogy to the QCD case, we now analyze the possible boundary 
counterterms for the heavy quark action. In the continuum, a basis of local operators of dimension 4,
which are to be summed over the boundary lattice points at $x_0=0$ and $x_0=T$ and 
are compatible with the 
symmetries of the static theory is given by
\begin{align}
\op{h,8}&= \heavyb D_{0}\heavy\,,\label{eq:Oh8}\\
\op{h,9}&= \heavyb \ola{D}_{0}\heavy\,,\label{eq:Oh9}\\
\op{h,10}&= \heavyb\{ \gamma_k D_k-\ola{D}_k\gamma_k\}\heavy\,,\label{eq:Oh10}\\
\ophat{h,11}&= m\heavyb\heavy\,.\label{eq:Oh11}
\end{align}
As $\op{h,8}$ and $\op{h,9}$ vanish because of \eqs{eq:formal_eq_of_motion},
the operator $\op{h,10}$ is zero because $P_+\gamma_kP_+=0$. It remains only $\op{h,11}$,
which is of the same kind of $\op{h,7}$, and amounts to a shift in $\delta m$.

We have thus shown that, apart from a redefinition of $\delta m$, in the \SF static actions
of the form (\ref{eq:general_heavy_action})
are already $\Oa$-improved without additional counterterms.

So far we have not discussed the improvement of the effective fields (\ref{eq:eff_Sym_field}). This will
be performed for the correlation functions of interest after having appropriately defined them  
on the lattice.

\section{Correlation functions}\label{sect:corr_functs}

In this section we derive, within the \SF scheme, explicit 
expressions for correlation functions and observables,
originated from relativistic quark and static-light bilinear currents. 
We describe their behavior at large physical time separations by using
the transfer matrix formalism \cite{SF:stefan1,mbar:pap2}. Throughout
the first subsection we work it out in the unimproved theory,
where the presented relations hold exactly. In the second subsection
we determine the $\Oa$-improvement counterterms for the correlation functions of interest,
and in the third subsection we present their renormalized expressions.
Since universality implies that the renormalized correlation
functions of the improved theory and unimproved theory
agree in the continuum limit, and they are multiplicatively renormalized,
we expect that the results derived in the first subsection are valid also
in the improved theory, provided that the involved physical distances are large
compared to the lattice spacing.

\subsection{Definition}\label{subsect:cf_def}

Our starting point is the QCD partition function defined in \eqs{eq:Z_in_lQCD}
in the \SF scheme. In analogy to the pure gauge case it can be represented as the 
matrix element of the time evolution from an initial state $|\mathrm{i}\rangle$ to 
a final state $|\mathrm{f}\rangle$ through the Euclidean step evolution operator $\mathrm{e}^{-{\mathbb H}a}$
applied $T/a$ times, according to the expression
\be\label{eq:evo_i_to_f}
\mathcal{Z}[W',\rhobarprime,\rhoprime,W,\rhobar,\rho]=
\langle \mathrm{f}|\left(\mathrm{e}^{-{\mathbb H}a}\right)^{T/a}{\mathbb P}|\mathrm{i}\rangle=
\langle \mathrm{f}|\mathrm{e}^{-{\mathbb H}T}{\mathbb P}|\mathrm{i}\rangle\,.
\ee
The states $|\mathrm{i}\rangle$ and $|\mathrm{f}\rangle$ live on the Hilbert space
at the temporal boundaries, $x_0=0$ and $x_0=T$ respectively, defined by the product
space of the pure gauge theory Hilbert space and the fermionic Fock space. The states 
are given in terms of the boundary fields appearing as arguments of the partition
function in \eq{eq:evo_i_to_f}. Gauge invariance is guaranteed by the presence
of the projector ${\mathbb P}$. An equivalent way of
writing \eq{eq:evo_i_to_f} is by means of the transfer matrix ${\mathbb T}$ between 
two adjacent time-slices
\be\label{eq:Tmatrix}
\mathcal{Z}[W',\rhobarprime,\rhoprime,W,\rhobar,\rho]=
\langle \mathrm{f}|\left({\mathbb T}\right)^{T/a} |\mathrm{i}\rangle\,,
\ee
where the projector ${\mathbb P}$ is included in ${\mathbb T}$.
%
Let us introduce the boundary operators
\be\label{eq:O_operators}
{\mathcal{O}}=\frac{a^6}{L^3}\sum_{\vecy,\vecz}\zetabar_\mathrm{q_1}(\vecy)\gamma_5 \zeta_\mathrm{q_2}(\vecz)\,,\quad  
{\mathcal{O}}'=\frac{a^6}{L^3}\sum_{\vecy,\vecz}\zetabarprime_\mathrm{q_2}(\vecy)\gamma_5\zzetaprime_\mathrm{q_1}(\vecz)\,,
\ee
containing the boundary fields defined in \eqs{eq:boundary_functionals}. These operators are 
dimensionless and $\{\mathrm{q_1,q_2}\}$ are flavor indices. They define the correlation function 
\be\label{eq:f1}
f_1=-\frac{1}{2}\langle{\mathcal{O}}'{\mathcal{O}}\rangle\,,
\ee
whose quantum mechanical representation is
\be\label{eq:f1_qm}
f_1=-\frac{1}{2}\frac{\langle \mathrm{i_\pi}|\mathrm{e}^{-{\mathbb H}T}{\mathbb P}|\mathrm{i_\pi}\rangle}
{\langle \mathrm{i_0}|\mathrm{e}^{-{\mathbb H}T}{\mathbb P}|\mathrm{i_0}\rangle}\,.
\ee
Here $|\mathrm{i_\pi}\rangle$ indicates a state with the quantum numbers of a pseudoscalar meson 
with quark content $(q_1,\bar{q}_2)$ and zero momentum, while $|\mathrm{i_0}\rangle$ carries the quantum numbers
of the vacuum. The latter is obtained by setting to zero the spatial components of the 
gauge potentials and the fermion fields at the temporal boundaries.
Furthermore we associate to the composite operators\footnote{Of course other bilinears can be formed, but they
are not used in this work.}
\begin{align}
A_0(x)&= \psibar_\mathrm{q_2}(x)\gamma_0\gamma_5\psi_\mathrm{q_1}(x)\,,\label{eq:A0_operator}\\
P(x)&= \psibar_\mathrm{q_2}(x)\gamma_5\psi_\mathrm{q_1}(x)\,,\label{eq:P_operator}
\end{align}
the correlation functions
\be\label{eq:fX}
f_\mathrm{X}(x_0)=-\frac{L^3}{2}\langle X(x){\mathcal{O}}\rangle=\left\{
\begin{array}{ll}
f_\mathrm{A}& \textrm{if $X(x)=A_0(x)$}\\
f_\mathrm{P}&\textrm{if $X(x)=P(x)$}
\end{array}\right.\,,
\ee
with $a\leq x_0\leq T-a$ and the quantum mechanical representation
\be\label{eq:fX_qm}
f_\mathrm{X}(x_0)=-\frac{L^3}{2}\frac{\langle \mathrm{i_0}|\mathrm{e}^{-{\mathbb H}(T-x_0)}{\mathbb P} 
{\mathbb X}\mathrm{e}^{-{\mathbb H}x_0}{\mathbb P}
|\mathrm{i_\pi}\rangle}
{\langle \mathrm{i_0}|\mathrm{e}^{-{\mathbb H}T}{\mathbb P}|\mathrm{i_0}\rangle}\,.
\ee
The physical interpretation of the correlation functions that we have just defined is  
\bi
\item $f_1$ is the (normalized) amplitude of a meson state $|\mathrm{i_\pi}\rangle$
      travelling from the temporal 
      boundary at $x_0=0$ to the other boundary at $x_0=T$ of the four dimensional cylinder.\\[-4ex]
\item $f_\mathrm{X}$ is the (normalized) amplitude of a meson state $|\mathrm{i_\pi}\rangle$
      created at time $x_0=0$ and annihilated at time $a\leq x_0\leq T-a$ by the operator ${\mathbb X}$;
      the latter being the representation of the field $X$ in the Schr\"odinger picture.
\ei
We consider an orthonormal basis $|n,q\rangle$
of gauge invariant eigenstates of the Hamiltonian ${\mathbb H}$. Their properties
explicitly read
\begin{align}
& \nonumber\\[-2.0ex]
n=0,1,\ldots\,,&\quad {\mathbb H}|n,q\rangle=E^{(q)}_n|n,q\rangle\,,\nonumber\\
 & \label{eq:orthonormal_basis}\\
1=\sum\limits_{n,q}| n,q \rangle\langle n,q|\,,&\quad
\langle n',q' | n,q \rangle=\delta_{n',n}\,\delta_{q',q}\,,\nonumber
\end{align}
where $n$ is the energy level and $q$ is a shorthand for the full set of internal quantum numbers
of the corresponding state. We insert two times the complete set into the numerator of \eq{eq:fX_qm}, and obtain
\begin{align}
f_\mathrm{X}(x_0)=&- \tfrac{L^3}{2}\mathcal{Z}^{-1}\label{eq:fX_qm_ins}\\
&\times \sum_{n,q}\sum_{n',q'}
\langle \mathrm{i_0}|\mathrm{e}^{-{\mathbb H}(T-x_0)}| n,q \rangle\langle n,q| 
{\mathbb X}| n',q' \rangle\langle n',q'|\mathrm{e}^{-{\mathbb H}x_0}
|\mathrm{i_\pi}\rangle\,.\nonumber
\end{align}
Since we are interested in the asymptotic behavior of correlation functions
for large values of $x_0$ and $T-x_0$ and on the ground state, we keep only 
the $n=0$ terms, and get
\begin{align}
f_\mathrm{X}(x_0)&\approx -\tfrac{L^3}{2}\mathcal{Z}^{-1}\langle \mathrm{i_0}|\mathrm{e}^{-{\mathbb H}(T-x_0)}\,
G_{0}\,{\mathbb X}\,G_{0}\,\mathrm{e}^{-{\mathbb H}x_0}\,
|\mathrm{i_\pi}\rangle\,,\label{eq:fX_qm_ins01}\\
G_{0}&= |0,0\rangle\langle 0,0|+|0,\pi\rangle\langle 0,\pi|\,.\label{eq:G0}
\end{align}
By using the properties in \eqs{eq:orthonormal_basis} we obtain
\begin{align}
f_\mathrm{X}(x_0)\approx&- \tfrac{L^3}{2}\mathcal{Z}^{-1}
\langle \mathrm{i_0}|0,0\rangle\langle 0,0|{\mathbb X}
|0,\pi\rangle\langle 0,\pi|\mathrm{i_\pi}\rangle \nonumber\\
&\times \mathrm{exp}[-E_0^{(0)}(T-x_0)-E_0^{(\pi)}x_0]\,.
\label{eq:fX_qm_ins02}
\end{align}
The same procedure can be applied to $f_1$ and $\mathcal{Z}$
\begin{align}
f_1&\approx -\tfrac{1}{2}\mathcal{Z}^{-1}\langle \mathrm{i_\pi}|0,\pi\rangle\langle 0,\pi|
\mathrm{i_\pi}\rangle\mathrm{e}^{-E_0^{(\pi)}T}\,,\label{eq:f1_qm_ins}\\
\mathcal{Z}&\approx \langle \mathrm{i_0}|0,0\rangle\langle 0,0|\mathrm{i_0}
\rangle\mathrm{e}^{-E_0^{(0)}T}\,.\label{eq:Z_qm_ins}
\end{align}
We can now write down two compact expressions for the correlation functions $f_\mathrm{X}$ and $f_1$
under the form
\be\label{eq:f_X_and_f_1_largeT_0}
f_\mathrm{X}(x_0)\approx-\tfrac{L^3}{2}\rho\langle 0,0|{\mathbb X}|0,\pi\rangle\mathrm{e}^{-m_\pi x_0}\,,\quad
f_1\approx\tfrac{1}{2}\rho^2\mathrm{e}^{-m_\pi T}\,,
\ee
where $\rho=\langle 0,\pi|\mathrm{i_\pi}\rangle/\langle 0,0|\mathrm{i_0}\rangle$,
and $m_\pi=E_0^{(\pi)}-E_0^{(0)}$ is the mass of the ground state meson.
We notice that the correlation function $f_\mathrm{A}$ is proportional to the matrix element
$\langle 0,0|{\mathbb A_0}|0,\pi\rangle$, which is related to the decay constant $f_\pi$
through 
\be\label{eq:fpi}
Z_\mathrm{A}\langle 0,0|{\mathbb A_0}|0,\pi\rangle=f_\pi m_\pi(2m_\pi L^3)^{-1/2}\,,
\ee
where $Z_\mathrm{A}$ is the renormalization constant of the axial current, and  
the factor 
\be
(2m_\pi L^3)^{-1/2} 
\ee
accounts for the normalization of the one-particle states.
There is no unique normalization for the decay constant, and the convention
in \eq{eq:fpi} complies with an experimental value of the pion decay constant of 132 MeV.
The decay constant $f_\pi$ can be conveniently extracted from the ratio
\be\label{eq:fpi_extr}
Z_\mathrm{A}f_\mathrm{A}(T/2)/\sqrt{f_1}\stackrel{(\ref{eq:f_X_and_f_1_largeT_0})}{\approx} -\tfrac{1}{2}f_\pi \sqrt{m_\pi}L^{3/2}\,,
\ee
where the ratio $\rho$, which is a divergent quantity, cancels out.
It remains to compute $m_\pi$. 
The latter can be extracted in the pseudoscalar channel from the effective meson mass 
\be\label{eq:eff_mass}
m_\mathrm{eff}(x_0)=\frac{1}{2a}\ln\left(\frac{f_\mathrm{A}(x_0-a)}{f_\mathrm{A}(x_0+a)}\right)
\stackrel{(\ref{eq:f_X_and_f_1_largeT_0})}{\approx}m_{\pi}\,.
\ee
For large values of $x_0$ and $T-x_0$ the effective meson mass
is expected to exhibit a plateau when plotted versus $x_0$, where the ground
state dominates and no boundary effects are present. Hence we choose 
as best estimate of $m_\pi$ the value of $m_\mathrm{eff}$ at $x_0=T/2$. While corrections
are exponentially suppressed, in the limit of infinite
physical time extension the two masses coincide.

The definitions given so far in this subsection can be easily extended to HQET.
We introduce the boundary operators
\begin{align}
\op{hl}&= \frac{a^6}{L^3}\sum_{\vecy,\vecz}\zetahb(\vecy)\gamma_5\zeta^{\phantom{\dagger}}_\mathrm{l}(\vecz)\,,\label{eq:Ohl}\\
\opprime{hl}&= \frac{a^6}{L^3}\sum_{\vecy,\vecz}\zetabarprime_\mathrm{l}(\vecy)\gamma_5\zzetaprime_\mathrm{h}(\vecz)\,,\label{eq:Ohlp}
\end{align}
and the $\mu=0$ component of the static-light axial vector 
current\footnote{This operator has already appeared in \eqs{A0_exp}. Here we reintroduce it in order
to preserve the self-consistency of the discussion.}
\be\label{eq:A0stat}
\Astat(x)=\lightb(x)\gamma_0\gamma_5\heavy(x)\,.
\ee
It is not necessary to define an equivalent of $P(x)$ in the effective theory,
because $\lightb(x)\gamma_0\gamma_5\heavy(x)=-\lightb(x)\gamma_5\heavy(x)$, due to $P_+\heavy=\heavy$.
We can now define the correlation functions 
\begin{align}
\fastat(x_0)&=-\frac{1}{2}\langle\Astat(x)\op{hl}\rangle_\mathrm{stat}\,,\quad a\leq x_0\leq T-a\,,\\
\fonestat&=-\frac{1}{2}\langle\opprime{hl}\,\,\op{hl}\rangle_\mathrm{stat}\,,\\
\fonehh(x_3)&=-\frac{a^8}{2L^2}\!\sum_{x_1,x_2,\vecy,\vecz}\!\langle \zetahbprime(\vecx)\gamma_5\zetahprime(\vectn)
\zetahb(\vecy)\gamma_5\zetah(\vecz)   \rangle_{\mathrm{h}\mathrm{h}}\,.\label{eq:fonehh}
\end{align}
It is clear that $\fastat$ and $\fonestat$ have a physical meaning similar to $f_\mathrm{A}$ and $f_1$ respectively.
The meson state is now made of a static quark and a light antiquark. 
The correlation function $\fonehh$ in turn involves a static quark and a static antiquark 
in the valence sector. If we carry on the analysis of the correlation 
functions for large physical $x_0$ and $T-x_0$,
as we have done for the relativistic QCD correlators, we end up with 
\begin{align}
\fastat(x_0)&\approx -\frac{1}{2}\mathrm{h}\langle 0,0|{\mathbb A_0^\mathrm{stat}} |0,\mathrm{B}\rangle 
\mathrm{e}^{-E_\mathrm{stat}x_0}\,,\label{eq:fastat_approx}\\
\fonestat&\approx \frac{1}{2}\mathrm{h}^2\mathrm{e}^{-E_\mathrm{stat}T}\,,\label{eq:fonestat_approx}
\end{align}
with $\mathrm{h}=\langle 0,\mathrm{B}|\mathrm{i_{B}}\rangle/\langle 0,0|\mathrm{i_0}\rangle$, and the 
state $|\mathrm{i_{B}}\rangle$
has the quantum numbers of a pseudoscalar static-light meson with zero momentum. Furthermore $E_\mathrm{stat}$
is the binding energy of the static-light system, and the static-light pseudoscalar decay constant $\Phi^\mathrm{stat}$
can be defined as
\be\label{eq:Phi_stat}
\Phi^\mathrm{stat}\approx-\zastat\fastat(T/2)/\sqrt{\fonestat}\,.
\ee
One may be tempted to identify $\Phi^\mathrm{stat}$ with the (na\"ive) static limit of the r.h.s.~of \eq{eq:fpi_extr}.
However, it would be wrong, because one has to take care of the dependence on the renormalization
scale introduced by $\zastat$, and of the necessary matching coefficient as discussed in \sect{sect:matching_HQET}.
The relation between the two decay constants will be given in the next chapter, after the discussion on
the renormalization of the correlation functions in QCD and HQET.
As for $m_\pi$, the quantity $E_\mathrm{stat}$ can be extracted from the effective energy
\be\label{eq:E_eff}
E_\mathrm{eff}(x_0)=\frac{1}{2a}\ln\left(\frac{\fastat(x_0-a)}{\fastat(x_0+a)}\right)
\stackrel{(\ref{eq:fastat_approx})}{\approx}E_\mathrm{stat}\,.
\ee

\subsection{Improvement}\label{subsect:cf_impr}

In \sect{sect:Oa_impr} we pointed out that, in order to
achieve the $\Oa$-improvement of the desired correlation functions, the
supplement of the action with appropriate counterterms may not be sufficient.
One has also to use improved fields. As for the action, an improved field
is given in terms of the originally (unimproved) field $\phi(x)$ and 
an additional counterterm $a\delta \phi(x)$
\be\label{eq:impr_field}
\phi_\mathrm{I}^{\phantom{\dagger}\!}(x)=\phi(x)+a\delta \phi(x)\,.
\ee
The counterterm is a linear combination of basis fields with mass dimension $\mathrm{dim}(\phi)+1$,
which have not been included in the lattice action, and have the same lattice symmetries as $\phi(x)$.

We start with $A_0$, and, by using the field equations
to reduce the number of basis fields, we are left with the continuum basis
\begin{align}
\op{A,1}&= \psibar_\mathrm{q_2}\gamma_5\{D_0+\ola{D}_0\}\psi_\mathrm{q_1}\,,\label{eq:OhatA1}\\
\op{A,2}&= m\psibar_\mathrm{q_2}\gamma_5\gamma_0\psi_\mathrm{q_1}\,,\label{eq:OhatA2}
\end{align}
Since 
the counterterm associated with $\op{A,2}$ amounts to a renormalization of $A_0$, we
discuss it in the next subsection. We thus conclude that
\be\label{eq:deltaA0}
\delta A_0(x)=\ca\frac{1}{2}(\nab{0}+\nabstar{0})P(x)\,.
\ee 
In the static approximation the improvement of $\Astat$ is achieved
by starting from the continuum basis
\begin{align}
\op{A,h,1}&= \lightb\gamma_5D_0\heavy\,,\label{eq:OhatAh1}\\
\op{A,h,2}&= \lightb\ola{D}_0\gamma_5\heavy\,,\label{eq:OhatAh2}\\
\op{A,h,3}&= \lightb\gamma_5\gamma_jD_j\heavy\,,\label{eq:OhatAh3}\\
\op{A,h,4}&= \lightb\ola{D}_j\gamma_j\gamma_5\heavy\label{eq:OhatAh4}\\
\op{A,h,5}&= m\lightb\gamma_0\gamma_5\heavy\,.\label{eq:OhatAh5}
\end{align}
We observe that $\op{A,h,1}$ vanishes because of the first equation in 
(\ref{eq:formal_eq_of_motion}), and $\op{A,h,3}$ breaks the U(1) symmetry of the
static approximation, because it is not invariant 
under the transformations (\ref{eq:xphase_inv}). 
One of the remaining three operators can be omitted, because it can be expressed in 
terms of the other two by means of the light quark equation of motion.
We thus neglect $\op{A,h,2}$ and notice that $\op{A,h,5}$ is again a term
which amounts to a renormalization of $\Astat$. It is discussed in the next subsection.
The improvement counterterm of the time component of the static-light axial current is thus
\be\label{eq:deltaAstat}
\delta \Astat(x)=\castat\frac{1}{2}\lightb(\lnab{j}+\lnabstar{j})\gamma_j\gamma_5\heavy\,.
\ee
The form of this counterterm is the same for all static actions, which have been considered.
However, the improvement coefficient $\castat$ depends on the chosen action,
and must be determined consequently \cite{stat:actpaper}.

The boundary fields appearing in the correlation functions defined in the 
previous subsection need to be improved too. The counterterms can be reabsorbed
in the renormalization of the fields and are discussed in the next subsection.


\subsection{Renormalization}\label{subsect:cf_ren}

Since we want to relate the bare quantities computed on the
lattice to physical observables, we need to define a renormalization
scheme. Furthermore we want to reach the continuum limit, by
decreasing $a/L$ and keeping fixed the ratio $T/L$ and the physical length of $L$,
with a rate proportional to $a^2$. The bare parameters must thus be scaled
to satisfy these requirements, and it is technically advantageous
to employ a mass-independent renormalization scheme.
In the plane of the bare parameters we define the critical line
\be\label{eq:def_critic_line}
m_0=m_\mathrm{c}(g_0)\,,
\ee
where the physical quark mass vanishes\footnote{As it will be clear in the
next chapter, the critical line is not unambigously defined in a regularization
without chiral symmetry. It depends on the exact definition of the quark mass.}. It allows to define the subtracted mass
\be\label{eq:def_sub_qmass}
\mq=\mbare-\mc\,.
\ee
We then define a modified bare coupling and bare quark mass
\begin{align}
\gtilde^2&= g_0^2(1+b_\mathrm{g}a\mq)\,,\label{eq:gtilde}\\
\mqtilde&= \mq(1+b_\mathrm{m}a\mq)\,,\label{eq:mqtilde}
\end{align}
where the b-coefficients are functions of $g_0^2$, and let $\gtilde$ and $\mqtilde$
be $\Oa$-improved. Along the critical line the modified coupling coincides with
the ordinary one. It is then natural to define the corresponding renormalized quantities
as
\begin{align}
g_\mathrm{R}^2&= \gtilde^2Z_\mathrm{g}(\gtilde^2,a\mu)\,,\label{eq:gR}\\
m_\mathrm{R}&= \mqtilde Z_\mathrm{m}(\gtilde^2,a\mu)\,.\label{eq:mR}
\end{align}
Another important point is that, in the scaling of the bare
parameters required to reach the continuum limit, the modified coupling is tuned such that
it scales independently of the quark mass. Similarly for the modified
bare quark mass $\mqtilde$, which is scaled by a factor dependent on $a\mu$ only.

We extend this procedure to the renormalization 
of the local field $\phi(x)$. Besides the improvement
counterterms considered in the previous subsection, it is natural
to include a factor of the form $1+b(g_0^2)a\mq$, thus getting a renormalized
field with the expression
\be\label{eq:ren_phi}
\phi_\mathrm{R}^{\phantom{\dagger}\!}(x)=Z_\phi (\gtilde^2,a\mu)(1+b_\phi a\mq)\phi_\mathrm{I}(x)\,.
\ee
Since the renormalization condition is imposed at zero quark mass, the
coefficient $b_\phi$ does not depend on the details of $Z_\phi$.

Let us now apply this scheme to QCD by considering the 
composite fields (\ref{eq:A0_operator}, \ref{eq:P_operator}). The renormalized
and improved axial density and time component of the axial current are given by
\begin{align}
(P_\mathrm{R})(x)&= \zp(\gtilde^2,a\mu)(1+b_\mathrm{P}a\mq)P(x)\,,\label{eq:P_R}\\
&  \nonumber\\
(A_\mathrm{R})_0(x)&= \za(\gtilde^2)(1+b_\mathrm{A}a\mq)\{A_0(x)+a \delta \!A_0(x)\}\,,\label{eq:A_R}
\end{align}
while the renormalized boundary field $\zeta_\mathrm{R}$ has the form
\be\label{eq:zeta_R}
\zeta_\mathrm{R}^{\phantom{\dagger}\!}(\vecx)=Z_\zeta(\gtilde^2,a\mu)(1+b_\mathrm{\zeta}a\mq)\zeta(\vecx)\,,
\ee
and the other boundary fields are renormalized similarly. 
The renormalization factor $\za$ has no scale dependence. This is a
consequence of the fact that the renormalization condition of 
the axial current can be fixed by imposing a continuum chiral Ward identity, which
is locally valid.
It is now straightforward
to write down the renormalized correlation functions
\begin{align}
[f_\mathrm{A}(x_0)]_\mathrm{R}&= \za(1+b_\mathrm{A}a\mq)Z_\zeta^2(1+b_\mathrm{\zeta}a\mq)^2f_\mathrm{A}^\mathrm{I}(x_0)\,,\label{eq:fA_R}\\
                \mbox{with}\quad f_\mathrm{A}^\mathrm{I}(x_0)&= f_\mathrm{A}(x_0)+\ca\tfrac{1}{2}a(\drvstar{0}+\partial_0)f_\mathrm{P}(x_0)\,,\label{eq:fA_I}\\
& \nonumber\\
\left[f_\mathrm{P}(x_0)\right]_\mathrm{R}&=\zp(1+b_\mathrm{P}a\mq)Z_\zeta^2(1+b_\mathrm{\zeta}a\mq)^2 f_\mathrm{P}(x_0)\,,\label{eq:fP_R}\\
& \nonumber\\
\left[f_{1}\right]_\mathrm{R}&= Z_{\zeta}^4(1+b_\zeta a\mq)^4 f_{1}\,.\label{eq:f1_R}
\end{align}
Here we have used the lattice derivatives
\begin{align}
\partial_\mu f(x)&= \tfrac{1}{a}\{f(x+a\hat{\mu})-f(x)\}\,,\label{eq:lat_dev_simple}\\
\drvstar{\mu}f(x)&= \tfrac{1}{a}\{f(x)-f(x-a\hat{\mu})\}\,.\label{eq:lat_devstar_simple}
\end{align}
In a mass-independent renormalization scheme the renormalized
version of the static-light axial density $\Astat(x)$ can be written as
\be\label{eq:Astat_R}
\Aren(x)=\zastat(\gtilde^2,a\mu)(1+\bastat am_\mathrm{l})(\Astat(x)+a\delta\Astat(x))\,.
\ee
As for QCD, the renormalization factor $\zastat$ can be written as a function
of $g_0$ instead of $\gtilde$ in the quenched approximation. 
Nevertheless the dependence on the scale $\mu$
remains, because in the heavy quark sector of HQET chiral symmetry is not present at all.
Furthermore $\zastat$ depends on the used action. 
We can now write down the renormalized and improved correlation functions
\begin{align}
\hspace{-0.5cm}[\fastat(x_0)]_\mathrm{R}&= (1+a\delta m)^{-x_0/a}\zastat (1+\bastat am_\mathrm{l})Z_\zeta(1+b_\zeta am_\mathrm{l})\nonumber\\
                      & \quad \times Z_\mathrm{h}(1+b_\mathrm{h} am_\mathrm{l})\fastatimpr(x_0)|_{\delta m=0}\,,\label{eq:fastat_R}\\
\mbox{with}\quad\fastatimpr(x_0)&= \fastat(x_0)-\tfrac{a}{2}\langle \delta\Astat(x){\mathcal{O}}_\mathrm{hl}\rangle_\mathrm{stat}\,,\label{eq:fastat_I}\\
& \nonumber\\
\left[\fonestat\right]_\mathrm{R}&= (1+a\delta m)^{-T/a}Z_\zeta^2(1+b_\zeta am_\mathrm{l})^2\nonumber\\
& \quad \times Z_\mathrm{h}^2(1+b_\mathrm{h} am_\mathrm{l})^2\fonestat|_{\delta m=0}\,,\label{eq:fonestat_R}\\
& \nonumber\\
\left[\fonehh(x_3)\right]_\mathrm{R}&= (1+a\delta m)^{-2T/a}Z_\mathrm{h}^4(1+b_\mathrm{h} am_\mathrm{l})^4\fonehh(x_3)|_{\delta m=0}\,.\label{eq:fonehh_R}
\end{align}
We finally remark that, both in QCD and in HQET, the renormalization
factors and the improvement coefficients of the boundary fields do not need
to be computed. In the next chapter we introduce only observables, which
can be obtained from functions of the renormalized correlators, where these terms
cancel out. The same considerations apply to the factors which are powers of $(1+a\delta m)$. 
They will drop out in all computed observables, 
and we will implicitly set $\delta m=0$ from now on.
\chapter{Combining HQET and relativistic QCD}\label{chap:SSM}
\section{The Step Scaling Method}\label{sect:SSM}

The evaluation of the $\mrm{B_s}$ meson properties
represents a challenging problem for lattice QCD.
The difficulties arise from the presence of two largely
separated energy scales. One of them is the heavy quark
mass ($m_\mathrm{b}\sim 5$~GeV), while the other one is given
by the typical QCD scale $\Lambda_\mathrm{QCD}$, which is of the order
of a few hundreds of \MeV. The light antiquark mass is of the
order of the latter, and, by borrowing the language of HQET,
the light antiquark belongs to the light 
degrees of freedom\footnote{With this choice of valence quarks, the most common
in the literature about the subject, the corresponding ground state meson
is the $\mathrm{\overline{B}_s^{\hspace{0.03cm}0}=\overline{s}b}$.
However, thanks to the invariance of the action under charge conjugation,
we will keep for the meson the notation $\mathrm{B_s}$ throughout the following.}.

An accurate simultaneous treatment of these two scales 
would require the ability to simulate a very fine lattice
to resolve the propagation of the heavy quark, and, at the same time,
a large physical volume to correctly accommodate the light degrees of
freedom with negligible finite size effects. Such a computation,
even in the quenched approximation, is unfeasible with
the present computational facilities.

The recourse to an effective theory is a natural
approach to the problem. In this context HQET represents a promising
way to solve it.
Another kind of approach, which keeps the relativistic
QCD Lagrangian in the form (\ref{eq:improved_action}),
is the Step Scaling Method (SSM)
proposed by the authors of \cite{Guagnelli:2002jd}.
Here we show how to combine the two methods in such a way that
we can exploit their respective advantages. 

Let us explain the basic ideas of the SSM.
We consider an observable ${\mathcal{O}}(E_\mathrm{l},E_\mathrm{h},L)$ depending
on two different scales $E_\mathrm{l}\ll E_\mathrm{h}$, and computed
on a lattice with physical extension $L$. It is easy to identify
$E_\mathrm{h}$ with the heavy quark mass, and $E_\mathrm{l}$ with the QCD scale
mentioned above. Since the light scale is kept
fixed in physical units in all simulations, we can simplify
the notation, and just write ${\mathcal{O}}(m_\mathrm{h},L)$. The computation
of the observable using the SSM is based on the identity
\be\label{eq:id_SSM}
{\mathcal{O}}(m_\mathrm{h},L_\infty)={\mathcal{O}}(m_\mathrm{h},L_0)
\frac{{\mathcal{O}}(m_\mathrm{h},L_1)}{{\mathcal{O}}(m_\mathrm{h},L_0)}
\ldots
\frac{{\mathcal{O}}(m_\mathrm{h},L_N)}{{\mathcal{O}}(m_\mathrm{h},L_{N-1})}
\frac{{\mathcal{O}}(m_\mathrm{h},L_\infty)}{{\mathcal{O}}(m_\mathrm{h},L_N)}\,.
\ee
Here $L_0$ is a volume small enough to allow
to simulate with a lattice cutoff much larger
than $m_\mathrm{b}$, where the latter can match 
its phenomenological value. Clearly the motion of the light
antiquark is squeezed in such a box, and the following
steps perform the evolution to bigger and bigger volumes
until $L_\infty$ is reached. Ideally, the latter is an 
infinite volume; in practice, it is a box large enough
to let consider negligible the finite size effects   
affecting ${\mathcal{O}}$. 
One in fact reaches the situation where 
the last ratio
\be\label{eq:last_sigma_SSM}
\frac{{\mathcal{O}}(m_\mathrm{h},L_\infty)}{{\mathcal{O}}(m_\mathrm{h},L_N)}=1\pm\delta_\infty\,,
\ee
is consistent with unity within the numerical precision $\delta_\infty$.
In other words, one has reached a physical situation, where a measurement of ${\mathcal{O}}$ is no 
more sensitive to changes
of the volume.
The intermediate correcting factors are the step scaling functions
\be\label{eq:sigma_SSM}
\sigma_{\mathcal{O}}(m_\mathrm{h},L_i)=\frac{{\mathcal{O}}(m_\mathrm{h},L_i)}{{\mathcal{O}}(m_\mathrm{h},L_{i-1})}\,,
\ee
where for the sake of simplicity we choose a fixed ratio $s=L_i/L_{i-1}$ in all steps.
The number $N$ and the scale ratio $s$ of the steps are in principle dependent
on the considered observable and on the desired level of accuracy. It has been
numerically shown \cite{romeII:mb,romeII:fb} that $(N,s)=(2,2)$ is a suitable 
choice for the mass and the decay constant of the $\mrm{B_s}$ meson. Indeed,
it represents a good compromise between the computational effort and
the precision on the physical results.

The main assumption of the method is that the step scaling functions
have a mild dependence on the heavy quark mass. A total decoupling
would mean that the steps are insensitive to variations of $m_\mathrm{h}$
as long as $m_\mathrm{h}\gg E_\mathrm{l}$:
\be\label{eq:decoupling_SSM}
\sigma_{\mathcal{O}}(m_\mathrm{h},L_i)\simeq \sigma_{\mathcal{O}}(L_i)\,.
\ee
To understand it we can get help from HQET. If we imagine
a meson made of a light antiquark and a static quark, the latter
is just a pointlike source of color. The step scaling function $\sigma_{\mathcal{O}}$
measures the error made by computing ${\mathcal{O}}$ in a volume $L_{i-1}$
instead of $L_i$. Indeed this measurement cannot depend on the heavy
quark; only the light degrees of freedom are squeezed. It follows
that (\ref{eq:decoupling_SSM}) becomes an equation.
As the heavy quark mass is made finite, one can expect that
a total decoupling never takes place, but the step scaling 
functions can be expanded as
\be\label{eq:expansion_steps}
\sigma_{\mathcal{O}}(m_\mathrm{h},L_i)=\sigma_{\mathcal{O}}^{(0)}(L_i)+
\frac{\sigma_{\mathcal{O}}^{(1)}(L_i)}{L_i m_\mathrm{h}}+\mathrm{O}\left(\frac{1}{(L_i m_\mathrm{h})^2}\right)\,.
\ee 
Since we want to keep the discretization errors roughly at
the same level in all steps, and each of them is to be extrapolated
to the continuum, we must scale the simulated heavy quark masses according
to the physical size of the involved volumes.
To clarify this point we consider again the starting volume
$L_0$. Here one can simulate with heavy quark masses around
the b-quark mass with an acceptable confidence. If 
we move to $L_1=2L_0$ and want to keep the discretization errors,
which are of the order of $am_\mathrm{h}$, of the same magnitude
as in $L_0$ we have to halve the heavy mass. This simply means 
that we are not simulating a b-quark any more. Nevertheless
the SSM predicts that the b-region can be reached through
a mild extrapolation lead by the expansion (\ref{eq:expansion_steps}).
As one doubles the volume extension again, the heavy mass
is halved once more, and the extrapolation to the b-region 
becomes more difficult. One can then expect that, as the volume
becomes bigger and bigger, while the heavy quark mass smaller and smaller, 
one arrives to a point where the expansion (\ref{eq:expansion_steps})
is unreliable.\\
\indent On the other side, if one computes the coefficients $\sigma_{\mathcal{O}}^{(0)}(L_i)$,
for all volumes, by using the static theory developed in HQET, all extrapolations
can be turned into interpolations. This implies higher confidence in the approach
to the b-quark energy scale as well as increased precision in all steps.
Furthermore, one can exploit HQET to directly compute some coefficients  
$\sigma_{\mathcal{O}}^{(n)}(L_i),\,n\geq 1$ in order to have 
a better control on the heavy scale expansion. The combination
of the SSM with HQET thus represents a very appealing way to go about
studying the heavy-light meson properties \cite{Guazzini:2007ja}.

\vspace{-0.4cm}

\section{Simulation parameters}\label{sect:SSM_simu_para}

\subsection{Action}\label{sect:stat_action}

All simulations carried out to get the results of this
chapter have been performed within the quenched 
approximation (cf.~\sect{sect:MC_path_integral}).
The choice of the static action has been restricted to
the HYP2 action \cite{stat:actpaper} only, which is 
a HYP action, as described in \sect{sect:SF_QCD_lat}, 
with parameters $(\alpha_1,\alpha_2,\alpha_3)=(1.0,\ 1.0,\ 0.5)$.\\
\indent In an early stage of this work, simulations on a small 
volume ($L=0.4$ fm) for the static-light 
pseudoscalar decay constant have shown that other choices
of the parameters $\alpha_i$ as well as other discretizations
of the static action give results consistent with HYP2 in the continuum limit,
as we expect from universality. However, the precision is sensibly lower,
and studies on the subject \cite{stat:actpaper} have shown that this
pattern is preserved in bigger volumes too. The simultaneous
employment of several actions does not pay off under the point of view
of the statistical precision, because whenever all data share the same gauge
configurations, they are found to be strongly correlated. 

\vspace{-0.4cm}

\subsection{Scale setting}\label{sect:r0}

In order to let lattice QCD computations
be able to be physically predictive, one has
to spend as many experimental input as are the free
parameters of the theory. In pure gauge theory
the only free parameter is the bare coupling $g_0$,
and each dimensionful quantity is expressed in units
of the lattice spacing. It is therefore of
fundamental importance to have an accurate
knowledge of the lattice spacing in physical units.
The functional dependence of $a$ from $g_0$ is predicted
in the high energy regime by the asymptotic freedom once the 
scale $\Lambda$, appearing in the renormalization group equation 
of the coupling, is known. 
As one enters into the low energy
regime, a perturbative evaluation of the scale becomes unsatisfactory.

The setting of the scale is thus achieved by using suitable observables,
and a widely used one is the hadronic length $r_0$ \cite{pot:r0},
whose definition relies on the force $F(r)$ between two static color sources.
To be more precise, $r_0$ is defined as the distance where
\be\label{eq:def_of_r0}
r^2F(r)|_{r=r_0}=1.65\,.
\ee
The authors of \cite{pot:r0_SU3,pot:intermed} have performed a direct
computation of the ratio $a/r_0$ for several values of the bare coupling,
and they have provided a phenomenological representation of $\ln(a/r_0)$
as a polynomial in $\beta$. In fact, by choosing the ansatz
\be\label{eq:a_r0_ansatz}
\ln(a/r_0)=\sum_{n=0}^{p}a_n(\beta-6)^n
\ee
they have found that
\begin{align}
\ln(a/r_0)&= -1.6804-1.7331(\beta-6)+0.7849(\beta-6)^2-0.4428(\beta-6)^3\,,\nonumber\\
          &\quad  \mbox{for}\quad 5.7\leq\beta\leq 6.92\,,\label{eq:a_r0}
\end{align}
is an excellent approximation of the Monte Carlo results. The accuracy of $a/r_0$ in \eq{eq:a_r0}
has been estimated to decrease from about 0.5\% at low $\beta$ to 1\% at $\beta=6.92$.

For values of $\beta$ greater than 6.92 the validity of parametrization (\ref{eq:a_r0})
cannot be blindly trusted any more. However, the authors of \cite{Guagnelli:2002ia} have shown that one
can use the non-perturbative results coming from the computation of the \SF renormalized 
coupling \cite{alpha:SU3,mbar:pap1} and a renormalization group analysis,
to obtain a parametrization of $(a/r_0)$ as a function of the bare coupling for $\beta>6.92$.
The parametrization reads
\be\label{eq:scale_RII}
\ln(a/r_0)=-\ln(\lambda_\mathrm{L})-\frac{b_1}{2b_0^2}\ln(b_0g^2_0)-\frac{1}{2b_0g_0^2}-I(g_0)\,,
\ee
where 
\begin{align}
I(g_0)&= \int_0^{g_0}\mathrm{d}x\left[\frac{1}{\beta(x)}+\frac{1}{b_0x^3}-\frac{b_1}{b_0^2\,x}\right]\,,\label{eq:I_scale_RII}\\
      &  \nonumber\\
\beta(x)&= -x^3\{b_0+b_1x^2+b_2x^4+b_3x^6+\ldots\}\,.\label{eq:b_scale_RII}
\end{align}
The function in \eq{eq:b_scale_RII} has already appeared in \eq{eq:running_gbar},
and it is reported also here only for reading convenience. 
Restricting the analysis to the four-loop expression of $\beta(x)$, the coefficients
read
\begin{align}
\lambda_\mathrm{L}&= 0.0203\,\quad b_0=11/(4\pi)^2\,,\quad b_1=102/(4\pi)^4\,,\nonumber\\
b_2&= -0.0015998323314\,,\quad b_3=-0.0025\,.\nonumber
\end{align}
With these values the uncertainty on $a/r_0$ for $6.92<\beta<7.5$ can be estimated to be 2\%, worsening
to 3\% at $\beta=8.5$. One has to remark that the coefficient $b_3$ is the result of a fit,
and not of a direct computation.

The choice of the value $1.65$ in \eq{eq:def_of_r0} is dictated by phenomenological
potential models, which predict an approximate value for $r_0$ of 0.5 fm. The uncertainty
on the physical value of $r_0$ is around 10\%, and this is translated in large systematic
errors for the quantities converted in physical units through this scale. However, as long
as the lattice results are kept in units of $r_0$, the precise parametrizations of \eq{eq:a_r0} 
and \eq{eq:scale_RII} allow to compare determinations of the same physical quantity,
obtained from different choices of the lattice setup (action, lattice spacing, volume, \ldots).

\subsection{Quark masses}\label{sect:quark_masses_SSM}

Since quarks are confined inside hadrons, their masses are 
not physical observables.
They are free parameters of the theory, and, as it happens for 
the gauge coupling, they have to be determined by introducing an 
experimental input. In addition, there is no universal definition of
the quark mass; one is free to use the most suitable definition for
one's purposes. 
We start by defining the bare current quark mass through the $\Oa$-improved PCAC relation
\begin{align}
m_\mathrm{q_2q_1}&= \left.\frac{\drvsym{0}f_\mathrm{A}(x_0)+a\ca \drvstar{0}\drv{0}f_\mathrm{P}(x_0)} 
{2f_\mathrm{P}(x_0)}\right|_{x_0=T/2}\,,\quad\mbox{with}\label{eq:bare_current_qmass}\\
\drvsym{0}&= \frac{1}{2}(\drvstar{0}+\drv{0})\,.\label{eq:symm_deriv}
\end{align}
Quark masses defined in this way are not affected by the choice of the kinematical 
parameters $(L,T,\theta,x_0)$ up to corrections of $\Oasq$. 
This is a consequence of the improvement and of the very derivation of the PCAC relation, relying
on the symmetries of the continuum action. The spatial components do not appear on the
r.h.s.~of \eq{eq:bare_current_qmass}, because they vanish under periodic boundary conditions. 
The choice $x_0=T/2$ in \eq{eq:bare_current_qmass} simply follows from the intention
of computing the correlation functions $f_\mathrm{A}$ and $f_\mathrm{P}$ with the bulk operator 
inserted as far as possible from the temporal boundaries, and therefore to minimize the cutoff effects. 

The coefficient $\ca$ has been computed by the authors of \cite{impr:pap3},
who provided the parametrization
\be\label{eq:ca_determination}
\ca(g_0^2)=-0.00756 g_0^2 \times \frac{1-0.748 g_0^2}{1-0.977 g_0^2}\,,
\ee
valid in the range of bare couplings $0\leq g_0 \leq 1$.

In \sect{subsect:cf_ren} we have seen that the \SF allows to define
a finite volume renormalization scheme, where the correlation functions
$f_\mathrm{A}$ and $f_\mathrm{P}$ are multiplicatively renormalized. We exploit this property
to define from \eq{eq:bare_current_qmass} a renormalized and improved quark mass
\begin{align}
\overline{m}_\mathrm{q_2q_1}&= \left.\frac{\drvsym{0}[f_\mathrm{A}(x_0)]_\mathrm{R}}{[f_\mathrm{P}(x_0)]_\mathrm{R}}
                            \right|_{x_0=T/2}\label{eq:mbar1}\\
                    &= \frac{Z_\mathrm{A}[1+\tfrac{1}{2}b_\mathrm{A}(am_\mathrm{q_2}+am_\mathrm{q_1})]} 
                       {Z_\mathrm{P}[1+\tfrac{1}{2}b_\mathrm{P}(am_\mathrm{q_2}+am_\mathrm{q_1})]}m_\mathrm{q_2q_1}\label{eq:mbar2}\\
                    &= \frac{Z_\mathrm{A}}{Z_\mathrm{P}}\left[1+(b_\mathrm{A}-b_\mathrm{P})\frac{am_\mathrm{q_2}
                        +am_\mathrm{q_1}}{2}\right]m_\mathrm{q_2q_1}\label{eq:mbar3}\,,
\end{align}
where each equation is valid up to $\Oasq$ corrections.
At the end of \sect{sect:matching_HQET} we have introduced 
the RGI quark mass $M$, as the asymptotic behavior of any renormalized 
running mass. This mass has the advantage of being scale and scheme independent, 
while $\overline{m}_\mathrm{q_2q_1}$ misses this property because of the presence of $Z_\mathrm{P}$.
However, the two masses can be related by the renormalization group equation
\be\label{eq:M_from_RGE}
M_\mathrm{q_2q_1}=\overline{m}_\mathrm{q_2q_1}(2b_0\gbsq)^{-d_0/2b_0}\times \mathrm{exp}
               \left\{-\int_0^{\gbar} \mathrm{d}g \left[\frac{\tau(g)}{\beta(g)}-\frac{d_0}{b_0 g}\right]\right\}\,.
\ee 
Actually, there is no unique way of normalizing the mass $M$. 
One could multiply the r.h.s.~of \eq{eq:M_from_RGE}
by a factor 2, and obtain a valid mass definition. Here we 
choose to comply with the conventions
of Gasser and Leutwyler \cite{Gasser:1982ap,chir:GaLe1,Gasser:1984gg}.
We rewrite \eq{eq:M_from_RGE} by showing the explicit dependence 
on the renormalization scale 
\begin{align}
M_\mathrm{q_2q_1}&= \frac{M_\mathrm{q_2q_1}}{\overline{m}_\mathrm{q_2q_1}(\mu)}\frac{Z_\mathrm{A}(g_0)}{Z_\mathrm{P}(g_0,L)}
\left[1+(b_\mathrm{A}-b_\mathrm{P})\frac{am_\mathrm{q_2}+am_\mathrm{q_1}}{2}\right]m_\mathrm{q_2q_1}+\Oasq\,,\nonumber\\
&  \nonumber\\
&  \mbox{with}\quad\mu=1/L\,.\label{eq:M_RGI1}
\end{align}
We can now define the total renormalization factor
\be\label{eq:Z_M}
Z_\mathrm{M}(g_0)=\frac{M_\mathrm{q_2q_1}}{\overline{m}_\mathrm{q_2q_1}(\mu)}\frac{Z_\mathrm{A}(g_0)}{Z_\mathrm{P}(g_0,L)}\,,\quad \mu=1/L\,,
\ee
such that
\be\label{eq:Z_M2}
M_\mathrm{q_2q_1}=Z_\mathrm{M}(g_0)m_\mathrm{q_2q_1}(g_0)+\Oasq\,.
\ee
It relates (up to terms of $\Oasq$) the bare current quark mass with the RGI one, and consists
of a regularization independent (but scale dependent) part $M/\overline{m}$, and the 
ratio $Z_\mathrm{A}/Z_\mathrm{P}$, depending on the regularization details.

A single quark mass is obtained by choosing two mass degenerate flavors $\mathrm{q_1}$ and $\mathrm{q_2}$
\be\label{eq:M_deg}
M_\mathrm{q}=M_\mathrm{qq}=Z_\mathrm{M}m_\mathrm{qq}+\Oasq\,,
\ee
which we label as diagonal definition. Up to $\Oasq$ it can be replaced
by the off-diagonal definitions
\be\label{eq:M_offdiag}
M_\mathrm{q\{j\}}=2M_\mathrm{qj}-M_\mathrm{qq}\,.
\ee
The quantities $r_0M_\mathrm{q}$ and $r_0M_\mathrm{q\{j\}}$ have the same continuum limit,
even if the cutoff effects, still remaining of $\Oasq$, may strongly depend
on the choice of the j-flavor. It is reasonable to have a light j-flavor, 
in order to minimize the improvement term proportional to $(b_\mathrm{A}-b_\mathrm{P})am_\mathrm{j}$.

In addition we consider current quark masses with improved derivatives, by replacing
in \eq{eq:bare_current_qmass},
\be
\drvsym{0}\to\drvsym{0}\left(1-\tfrac{1}{6}a^2\drvstar{0}\drv{0}\right)\,,\quad
\drvstar{0}\drv{0}\to\drvstar{0}\drv{0}\left(1-\tfrac{1}{12}a^2\drvstar{0}\drv{0}\right)\,.
\ee
when acting on smooth functions these lattice derivatives have errors of $\mathrm{O}(a^4)$ only.

Another way of computing the RGI quark mass starts from the subtracted quark mass
$\mq$ defined in \eq{eq:def_sub_qmass}, and is completed by a multiplicative renormalization
\be\label{eq:M_deg2}
\hat{M}_\mathrm{q}=Z_\mathrm{M}(g_0)Z(g_0)[1+b_\mathrm{m}a\mq]\mq\,.
\ee  
The renormalization factor $Z_\mathrm{M}$ has been computed in \cite{mbar:pap1,lat06:damiano}
for the range of $\beta$-values $[6.0,7.6101]$, and can be parametrized in this range
by the expression
\be\label{eq:Z_M_parametrization}
Z_\mathrm{M}(\beta)=1.755+0.188(\beta-6.0)-0.024(\beta-6.0)^2\,.
\ee
The uncertainty relative to the regularization independent part $M/\overline{m}$ amounts to 0.9\% at $\mu=1/L_0=2.5~\Fm^{-1}$,
while for the ratio $Z_\mathrm{A}/Z_\mathrm{P}$ we have estimated the errors $\Delta (Z_\mathrm{A}/Z_\mathrm{P})$ reported in \tab{tab:err_ZAZP}.
\begin{table}
\centering
\begin{tabular}[h]{c|cccc}
\hline
\hline
        & & & & \\[-1ex]
$\beta$ & $[6.0,6.5]$& $6.7370,\,6.9630$   & $7.1510$ & $7.3000, 7.5480$\\ 
        & & & & \\[-1ex]
\hline
        & & & & \\[-1ex]
$\Delta (Z_\mathrm{A}/Z_\mathrm{P})$   & $1.1\%$& $1.0\%$   & $0.8\%$ & $0.6\%$\\[-1ex]
        & & & & \\
\hline
\hline
\end{tabular}
\mycaption{Uncertainties on the regularization dependent part of the total renormalization 
factor $Z_\mathrm{M}$.}\label{tab:err_ZAZP}
\end{table}
The value of $\Delta (Z_\mathrm{A}/Z_\mathrm{P})$ in the range $6.0\leq\beta\leq 6.5$ has been quoted 
in \cite{mbar:pap1} to be $1.1\%$, while the uncertainties for smaller values of the coupling 
have been estimated from the errors on $Z_\mathrm{A}$ and $Z_\mathrm{P}$ quoted in \cite{mbar:pap1} and \cite{impr:pap4}
respectively. They are reported in \tab{tab:err_ZAZP} for the $\beta$-values needed in our computations.
The factor $Z$ has been defined and computed in \cite{impr:babp}, where one can read off the parametrization
\be\label{eq:Z_parametrization}
Z(g_0^2)=(1+0.090514g_0^2)\times \frac{1-0.9678 g_0^2+0.04284g_0^4-0.04373g_0^6}{1-0.9678 g_0^2}\,,
\ee
with a relative precision better than 0.04\% in the range $0.8881\leq g_0^2\leq 1.0$. The authors
of \cite{HQET:pap2} have shown that the parametrization (\ref{eq:Z_parametrization}) can be 
used also for values of $g_0^2$ smaller than 0.8881, but we did not exploit it in our computations.
The reason relies in the improvement coefficient $b_\mathrm{m}$, as it explained in the following discussion.

The quantities $b_\mathrm{A}-b_\mathrm{P}$ and $b_\mathrm{m}$ have been computed in \cite{impr:babp},
whose authors provided the parametrizations
\begin{align}
(b_\mathrm{A}-b_\mathrm{P})(g_0^2)=-0.00093 g_0^2 &\times  \frac{1+23.3060g_0^2-27.3712g_0^4}{1-0.9833g_0^2}\,,\label{eq:babp_para}\\
b_\mathrm{m}(g_0^2)=(-0.5-0.09623g_0^2)&\times \frac{1-0.6905g_0^2+0.0584g_0^4}{1-0.6905g_0^2}\,.\label{eq:bm_para}
\end{align} 
In the range $0.8881\leq g_0^2\leq 1.0$ the parametrization (\ref{eq:babp_para}) represents the 
computed data with an absolute deviation smaller than 0.3\%, and the authors of \cite{HQET:pap2}
have shown that this can be believed also for smaller values of the gauge coupling, covering
the ones needed in our computations. The parametrization (\ref{eq:bm_para}) describes
the available non-perturbative data in the range $0.8881\leq g_0^2\leq 1.0$ with an absolute deviation smaller that 1.3\%,
but, as it has been demonstrated in \cite{HQET:pap2}, this parametrization cannot be trusted for smaller
values of the coupling. 
For our computations we thus decided to use the mass definition given in \eq{eq:M_deg2} only for $0.8881\leq g_0^2\leq 1.0$.

\begin{figure}[t]
\centering
\includegraphics[scale=0.5]{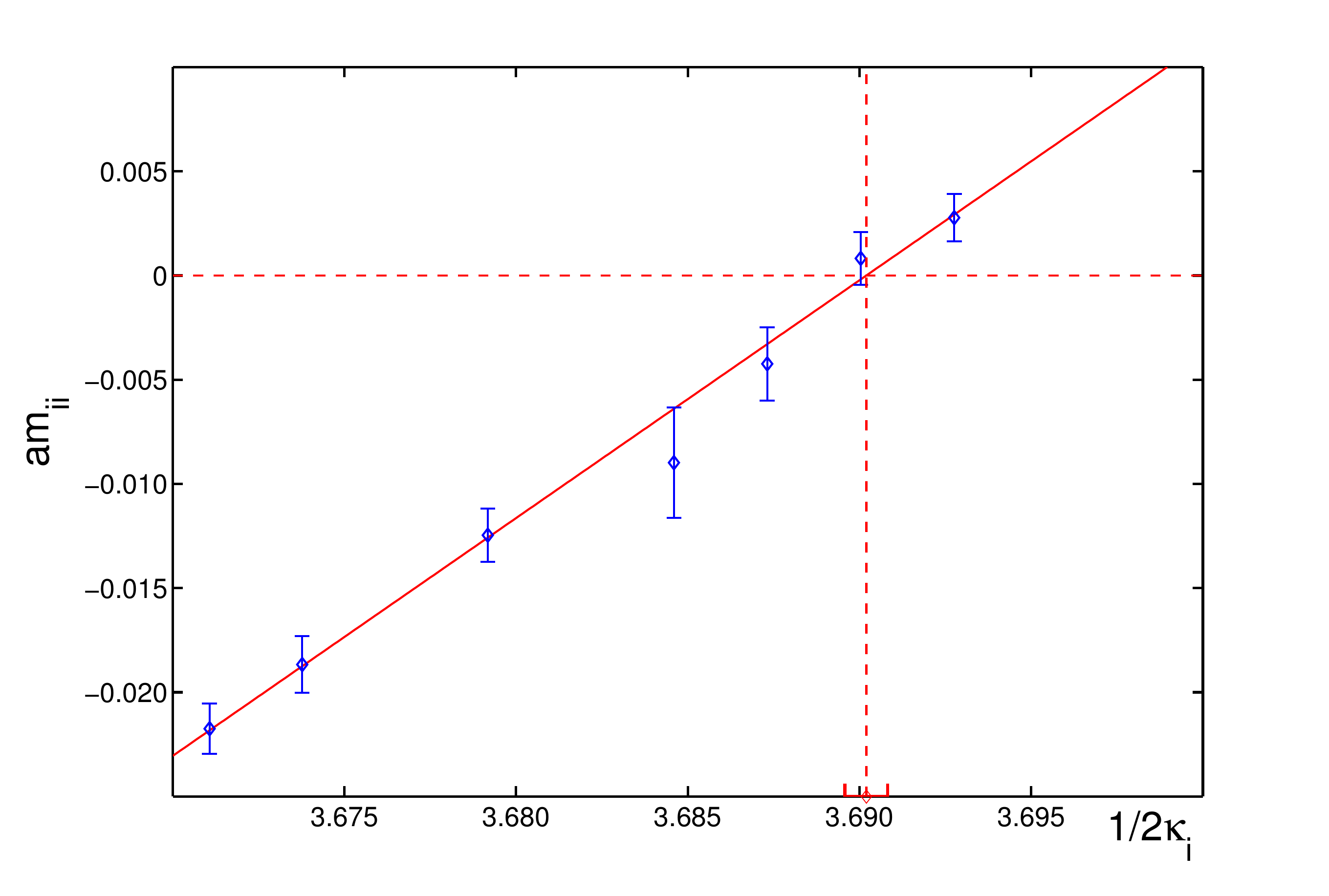}
\mycaption{Linear interpolation for $\kappa_\mathrm{crit}$. The data refer to
a lattice with $\beta=6.0914$, $L/a=10$, $T=2L$ and $\theta=0.0$. All points
are statistically independent.}\label{fig:kappac_L10}
\end{figure}

In terms of the hopping parameter $\kappa$ introduced in \sect{sect:SF_QCD_lat},
the subtracted bare mass (\ref{eq:def_sub_qmass}) reads
\be\label{eq:def_sub_qmass2}
a\mq=\frac{1}{2}\left(\frac{1}{\kappa}-\frac{1}{\kappa_\mathrm{crit}}\right)\,.
\ee
For $\kappa=\kappa_\mathrm{crit}$
the physical quark mass vanishes, as we require in our renormalization scheme.
The value of $\kappa_\mathrm{crit}$ is determined, for each choice of the lattice setup, 
by choosing a set of $\kappa$-values $\kappa_\mathrm{i}$ such that,
after a short Monte Carlo simulation, part of the corresponding (flavor-degenerate)
values of the bare current quark mass (\ref{eq:bare_current_qmass}) are positive,
and the remaining part is negative or consistent with zero. The data are then represented
in a plot $am_\mathrm{ii}$ vs. $1/2\kappa_\mathrm{i}$, and, for the values of $am_\mathrm{ii}$ which are closest
to zero, a linear interpolation is performed. The fitted function $am_\mathrm{ii}=a_1+a_2/2\kappa_\mathrm{i}$
usually shows $a_2\approx 1$, and the extracted $-a_2/2a_1$ is our estimate of $\kappa_\mathrm{crit}$.
The latter can be used as hopping parameter in a further simulation to check that the
computed current mass vanishes within statistical errors. An example is shown in \fig{fig:kappac_L10}.
For the simulations with the relativistic QCD action \cite{romeII:mb,romeII:fb}, 
the hopping parameters
corresponding to the strange quark have been determined in such a way that,
for each choice of the lattice setup, one can arrange three RGI quark masses,
according to the definition (\ref{eq:M_deg}), lying around the value
\be\label{eq:Mstrange}
M_\mathrm{s}=0.1346(55)\,\GeV\,,
\ee
computed in \cite{mbar:pap3}, where the chosen experimental input is the 
Kaon mass. The dependence of the computed observables on the light
quark is very mild, and they are interpolated as linear functions of the simulated
light quarks to match \eq{eq:Mstrange}.

The results published in \cite{romeII:mb,romeII:fb} have been used
to match the hopping parameters corresponding to the strange quark,
by performing a linear fit of the light quark masses $M_\mathrm{i}=a+b/\kappa_\mathrm{i}$,
and extracting the desired $\kappa_\mathrm{s}$-values as $\kappa_\mathrm{s}=b/(M_\mathrm{s}-a)$,
with $M_\mathrm{s}$ quoted in \eq{eq:Mstrange}. These values have been used as input
for the simulations with the heavy quark Lagrangian. However, the data appearing
in \cite{romeII:mb,romeII:fb}, do not cover all lattice setups used in the HQET computations.
The missing values of the hopping parameters have then been determined ex novo,
by requiring the corresponding RGI quark masses to be consistent with \eq{eq:Mstrange}.
A list of all simulation parameters is given in \app{app:CL_SSM}.

\subsection{Improvement coefficients and renormalization factors}\label{sect:further_impr_and_ren}

All improvement coefficients and renormalization factors needed
to determine the simulation parameters have been presented in the previous
subsections. However, they are not sufficient to achieve the $\Oa$-improvement
and the renormalization of all computed correlation functions. Here we report the missing
ingredients. 
The improvement coefficient $b_\mathrm{A}$ appearing in \eq{eq:A_R} has been computed
\cite{impr:pap5} at 1-loop, and reads
\be\label{eq:ba_pert}
b_\mathrm{A}(g_0^2)=1+0.15219(5)g_0^2+\mathrm{O}(g_0^4)\,.
\ee
The renormalization factor $Z_\mathrm{A}$ has been non-perturbatively computed in \cite{impr:pap4},
where the rational expression
\be\label{eq:ZA}
Z_\mathrm{A}(g_0^2)=\frac{1-0.8496 g_0^2+0.0610 g_0^4}{1-0.7332 g_0^2}\,,
\ee
has been proposed as good representation of the numerical results in the
range of bare couplings $0\leq g_0\leq 1$.
The uncertainty associated with this parametrization is reported
for the couplings of interest in the following table.\vspace{0.3cm}
\begin{center}
\begin{tabular}[h]{c|ccc}
\hline
\hline
        & & & \\[-1ex]
$\beta$ & $7.1510$ & $7.3000$ & $7.5480$\\ 
        & & &  \\[-1ex]
\hline
        & & &  \\[-1ex]
$\Delta (Z_\mathrm{A})$   & $0.8\%$ & $0.6\%$& $0.6\%$ \\[-1ex]
        & & &  \\
\hline
\hline
\end{tabular}
\end{center}\vspace{0.3cm}
For HQET the 
coefficients $\castat$ and $\bastat$ have been
computed \cite{stat:actpaper} for the static HYP2 action at 1-loop
\be\label{eq:castat_and_bastat}
\castat=0.220(14)g_0^2+\mathrm{O}(g_0^4)\,,\quad \bastat=1/2+0.259(13)g_0^2+\mathrm{O}(g_0^4)\,.
\ee
The renormalization factor $\zastat$ for the HYP2 action has been computed
by the authors of \cite{stat:actpaper}, but in a range of couplings, which does not cover all
our simulations. Its computation represents part of the work of this thesis. 

\subsection{Quark mass scheme conversion}\label{sect:RGItoMSmass}

Once the quark mass has been computed using one of the RGI definitions
given in \sect{sect:quark_masses_SSM}, it is useful to translate it to the mass
in the widely used $\MSbar$ scheme. This is achieved
through the equation
\be\label{eq:movermbar}
\frac{M}{\overline{m}(\overline{m})}  =  (2b_0\bar g^2(\overline{m}))^{-d_0/2b_0}  \exp\left\{-\int\limits_0^{\bar g{(\overline{m})}}
\mathrm{d}g\left[\frac{\tau^{\MSbar}(g)}{\beta^{\MSbar}(g)}-\frac{d_0}{b_0g}\right] \right\}\,,
\ee
where $\gbar=\gbarMSbar$,  $\overline{m}=\overline{m}_{\MSbar}$, and the expression on the r.h.s.~can be evaluated
with a 4-loop accuracy thanks to the coefficients reported in \tab{tab:betantau} 
and the result $\Lambda_{\MSbar}r_0=0.602(48)$ \cite{alpha:SU3,pot:intermed}. 

\begin{table}
\centering
\begin{tabular}{rcl|rcl}
\hline
\hline
    & & & & & \\[-1ex]
\multicolumn{3}{c|}{$\beta^{\overline{\mathrm{MS}}}$} &\multicolumn{3}{c}{$\tau^{\overline{\mathrm{MS}}}$} \\
    & & & & & \\[-1ex]
\hline
\vspace{-.4cm}&&&&\\
$b_0$&=&$\frac{11}{(4\pi)^2}$                    &$d_0$&=& $\frac{8}{(4\pi)^2}$\\[1.5ex]
$b_1$&=&$\frac{102}{(4\pi)^4}$                    &$d_1$&=& $\frac{404}{3 (4\pi)^4}$\\[1.5ex]
$b_2$&=&$\frac{2857}{ 2(4\pi)^6}$                 &$d_2$&=& $\frac{2498}{(4\pi)^6}$\\[1.5ex]
\cite{MS:4loop1}\,\, $b_3$&=&$\frac{29243-5033/18}{(4\pi)^8}$          &$d_3$&=& $\frac{50659}{(4\pi)^8}$\,\, \cite{MS:4loop3} \\[.5ex]
\hline
\hline
\end{tabular}
\mycaption{4-loop $\beta$-function and anomalous dimension of 
the mass in the $\overline{\mathrm{MS}}$ scheme. The references are only for the highest order computations; there
one finds also the references for the lower orders too.}\label{tab:betantau}
\end{table}

\section{Computation of the b-quark mass}\label{sect:b-mass}

\subsection{General strategy}\label{sect:b-mass_general}

Our lattice setup is given by the discretized Schr\"odinger functional, as in \sect{sect:SF_QCD_lat},
with topology $T\times L^3$ and parameters
\be\label{eq:lat_setup}
T=2L\,,\quad C=C'=0\,,\quad\mbox{and}\quad \theta=0.0\,,
\ee
for all correlation functions, unless stated otherwise.
We first consider the case of a finite volume pseudoscalar meson mass\footnote{The mass $m_\mathrm{h}$ 
and the lenght $L$
are given in physical units through the scale $r_0$ defined in \sect{sect:r0}. If more
rigor is desired, one may identify the observable ${\mathcal{O}}$ with the dimensionless 
quantity $LM_\mathrm{PS}(L m_\mathrm{h},L/r_0)$.}
\be\label{eq:O_eq_M}
{\mathcal{O}}(m_\mathrm{h},L)=M_\mathrm{PS}(m_\mathrm{h},L)\,,
\ee
where $m_\mathrm{h}$ stands generically for a heavy quark mass, whose precise 
definition is needed only later. The dependence on the light quark mass,
which is fixed in all volumes to the physical strange 
quark mass as described in \sect{sect:quark_masses_SSM},
has been dropped. The meson mass is defined in
analogy to the effective mass in \eq{eq:eff_mass}, as
\be\label{eq:def_Mps}
M_\mathrm{PS}(m_\mathrm{h},L)=\frac{1}{2a}\ln\left[\frac{f_\mathrm{A}^\mathrm{I}(m_\mathrm{h},L,x_0-a)}{f_\mathrm{A}^\mathrm{I}(m_\mathrm{h},L,x_0+a)}\right]_{x_0=L}\,.
\ee
The choice of taking $f_\mathrm{A}^\mathrm{I}$ around 
the middle of the temporal lattice
is a definition which well suits the Step Scaling Method, and aims 
also to reduce the cutoff effects in all steps.
It is straightforward to conclude that
\be\label{eq:M_PS_ev}
M_\mathrm{PS}(m_\mathrm{h},L)\stackrel{m_\mathrm{h}\to m_\mathrm{b}}{=}M_\mathrm{B_s}(L)
\stackrel{L\to\infty}{=}M_\mathrm{B_s}\,.
\ee
In the expansion (\ref{eq:expansion_steps}) we simply have $\sigma^{(0)}_{\mathcal{O}}=1$,
and the first non-trivial term $\sigma^{(1)}_{\mathcal{O}}$ is computable in the static approximation
of HQET. We further define
\be\label{eq:x_def}
x(m_\mathrm{h},L)\equiv\frac{1}{LM_\mathrm{PS}(m_\mathrm{h},L)}=\frac{1}{Lm_\mathrm{h}}+\mathrm{O}\left(\frac{1}{(Lm_\mathrm{h})^2}\right)\,,
\ee
as the natural non-perturbative dimensionless mass variable. It is given
in terms of physical observables, and there is no uncertainty deriving
from a choice of quark mass definition\footnote{This is certainly true 
once the continuum limit has been reached.}. The step scaling function for the meson
mass assumes then the simple form
\begin{align}
\sigma_\mathrm{m}(x,L_i)&\equiv \frac{M_\mathrm{PS}(m_\mathrm{h},L_i)}{M_\mathrm{PS}(m_\mathrm{h},L_{i-1})}=
1+\sigma^\mathrm{stat}_\mathrm{m}(L_i)\cdot x +\mathrm{O}(x^2)\,,\nonumber\\
& \label{eq:sigma_m}\\
x&= x(m_\mathrm{h},L_i)\,,\nonumber
\end{align}
and it is defined for all $x,L$. Our strategy for its precise numerical evaluation
is to compute $\sigma^\mathrm{stat}_\mathrm{m}$ explicitly in the static approximation, and fix
the small remainder by means of relativistic QCD data with (heavy) quarks
of masses of the physical charm and higher. In analogy to the mass (\ref{eq:M_PS_ev}) we define
\be\label{eq:Gamma_stat}
\meffstat(L)=\frac{1}{2a}\ln\left[\frac{\fastatimpr(L,x_0-a)}{\fastatimpr(L,x_0+a)}\right]_{x_0=L}\,,
\ee
whose large volume limit gives
\be\label{eq:Gamma_stat_ev}
\meffstat(L)\stackrel{L\to\infty}{=}E_\mathrm{stat}\,.
\ee
By considering
the case $L_i=2L_{i-1}=2L$ we can write a more explicit form for 
$\sigma_\mathrm{m}$ appearing in \eqs{eq:sigma_m}, that reads
\begin{align}
\frac{M_\mathrm{PS}(m_\mathrm{h},2L)}{M_\mathrm{PS}(m_\mathrm{h},L)}&= \frac{m_\mathrm{h}+\meffstat^\mathrm{R}(2L)+
\mathrm{O}\left(1/L m_\mathrm{h}\right)} 
{m_\mathrm{h}+\meffstat^\mathrm{R}(L)+\mathrm{O}\left(1/L m_\mathrm{h}\right)}\nonumber\\
&= 1+\frac{\meffstat(2L)-\meffstat(L)}{m_\mathrm{h}}+\mathrm{O}\left(1/L^2m_\mathrm{h}^2\right)\nonumber\\
&= 1+\frac{2L[\meffstat(2L)-\meffstat(L)]}{2Lm_\mathrm{h}}+\mathrm{O}\left(1/L^2m_\mathrm{h}^2\right)\nonumber\\
&= 1+\frac{2L[\meffstat(2L)-\meffstat(L)]}{2LM_\mathrm{PS}(m_\mathrm{h},2L)}
+\mathrm{O}\left(1/L^2M_\mathrm{PS}^2(m_\mathrm{h},2L)\right)\nonumber\\
&= 1+\sigma^\mathrm{stat}_\mathrm{m}(2L)\cdot x +\mathrm{O}(x^2)\,,\label{eq:sigma_m_explicit}
\end{align}
where we can identify the static step scaling function 
\be\label{eq:sigma_stat_m}
\sigma^\mathrm{stat}_\mathrm{m}(L_i)=L_i[\meffstat(L_i)-\meffstat(L_{i-1})]\,.
\ee
On the r.h.s.~of the first equation in 
the expansion (\ref{eq:sigma_m_explicit}), the renormalized
static effective mass $\meffstat^\mathrm{R}$ appears. It is related to $\meffstat$,
computed at $\delta m=0$, by
\be\label{eq:Gamma_stat_R}
\meffstat^\mathrm{R}=\meffstat+\frac{1}{a}\ln(1+a\delta m)\,.
\ee
However, this additive renormalization term cancels out
in the following equations, and we do not need to take care of it. 

It has been shown in \cite{romeII:mb}, that the choice $L_\infty\approx L_2=1.6$ fm guarantees
to have a box, where lattice computations of $M_\mathrm{PS}(m_\mathrm{h},L_\infty)$ are affected 
by negligible finite size effects. With the experimental value for the mass
of the $\mathrm{B_s}$ meson, $M_\mathrm{B_s}=5.3675(18)$ GeV we fix $x_2=1/L_2M_\mathrm{B_s}$,
and the physical points corresponding to the b-quark are then given by
\be\label{eq:x_recursion}
x_2=1/L_2M_\mathrm{B_s}\,,\quad x_{i-1}=2\sigma_\mathrm{m}(x_i,L_i)\cdot x_i\,.
\ee
The numerical results will have to be evaluated at these points. 
After performing the steps $L_2\to L_1=0.8$ fm and $L_1\to L_0=0.4$ fm,
we arrive at the small volume\footnote{Thanks to \eqs{eq:lat_setup} the volume can be identified
without ambiguities with the spatial extension. The adjective ``small'' is here
referred to the smallest box where our simulations took place.} $L_0$, which plays a special role.
Here we relate the meson mass to the renormalization 
group invariant quark mass $M_\mathrm{h}$ defining
\be\label{eq:rho_def}
\rho(x,L_0)=\frac{M_\mathrm{PS}(M_\mathrm{h},L_0)}{M_\mathrm{h}}
=\rho^{(0)}(L_0)+\rho^{(1)}(L_0)\cdot x+\mathrm{O}(x^2)\,.
\ee
We finally gather \eqs{eq:x_recursion} and \eqs{eq:rho_def}
to get the connection between the $\mathrm{B_s}$ meson mass and the RGI b-quark
mass $M_\mathrm{b}$ in the compact form
\be\label{eq:MBs_Mb}
M_\mathrm{b}=\frac{M_\mathrm{B_s}}{\rho(x_0,L_0)\cdot \sigma_\mathrm{m}(x_1,L_1)\cdot \sigma_\mathrm{m}(x_2,L_2)}\,.
\ee


\subsection{The results}\label{sect:b-mass_results}

The computation of $\sigma_\mathrm{m}(x,L_2)$ is performed
at finite heavy quark masses around the charm quark mass,
on lattices with $\beta=5.9598$,
$6.2110,\ 6.4200$, corresponding to resolutions $L_2/a=16,\ 24,\ 32$ respectively; the continuum limits
for the three heaviest quark masses are shown in \app{app:CL_SSM}.

For the static step scaling function we took the results for the bigger
volume from \cite{Estat:me}. There the spatial extension
$L_2\approx 1.5$ fm, and the ratio $T/L$, with $T>L$, is not fixed to
be the same for all values of the bare coupling. However the box
is big enough to assume $L_2\approx L_\infty$ for a $\mathrm{B_s}$ meson,
and we have
\be\label{eq:sigma_2_m}
\sigma_\mathrm{m}^\mathrm{stat}(L_2)=2L_1[E_\mathrm{stat}-\meffstat(L_1)]\,.
\ee
The energy $E_\mathrm{stat}$ is available from \cite{Estat:me} for four values of the coupling,
while $\meffstat(L_1)$ is computed with the setup in \eqs{eq:lat_setup} for the couplings in the third row
of the following table.\vspace{0.3cm}
\begin{center}
\begin{tabular}{c|ccccc}
\hline
\hline
        & & & & & \\[-1ex]
$\beta$ for $\sigma_\mathrm{m}(x,L_2)$ & $5.9598$ & - & $6.2110$ & $6.4200$ &-\\ 
        & & & & & \\[-1ex]
\hline
        & & & & & \\[-1ex]
$\beta$ for $E_\mathrm{stat}$ & $6.0291$ & $6.2885$ & $6.4500$ & $6.4956$ &- \\ 
        & & & & & \\[-1ex]
\hline
        & & & & & \\[-1ex]
$\beta$ for $\meffstat(L_1)$& $5.9598$& $6.0914$   & $6.2110$ & $6.4200$ &$6.7370$\\[-1ex]
        & & & & & \\
\hline
\hline
\end{tabular}
\end{center}\vspace{0.3cm}
After a quadratic fit of $a\meffstat(L_1)$ vs.~$\beta$, the computed coefficients
are used to interpolate $a\meffstat(L_1)$ at the same values
of $\beta$ used for the computation of $E_\mathrm{stat}$. 
This makes possible to take the continuum limit for the static step scaling function, and get
\be\label{eq:sigma_2_m_cl}
\sigma_\mathrm{m}^\mathrm{stat}(L_2)=1.561(53)\,.
\ee
Since the interpolated values of $a\meffstat(L_1)$ are all obtained from the same
set of data, it is clear that they are correlated. However, such correlation
affects the continuum limit (\ref{eq:sigma_2_m_cl}) by an amount, which is several orders
of magnitude smaller than the error quoted in (\ref{eq:sigma_2_m_cl}).
The static step scaling function is used 
to fix (within errors) the first non-trivial term in the expansion (\ref{eq:sigma_m}). In practice it consists of a
constraint on the slope of the fitting curve. For a quadratic fit, the $\chi^2$ to be minimized 
has the expression
\be\label{eq:chisq_constraint}
\chi^2=\sum_i \left(\frac{\sigma_\mathrm{m}(x_{(i)},L_2)-1-bx_{(i)}-cx_{(i)}^2}{\Delta\sigma_\mathrm{m}(x_{(i)},L_2)}\right)^2
+\left(\frac{b-\sigma_\mathrm{m}^\mathrm{stat}(L_2)}{\Delta\sigma_\mathrm{m}^\mathrm{stat}(L_2)}\right)^2\,.
\ee
A correct estimate of the uncertainties has to consider the concomitant
presence of the errors associated with $\sigma_\mathrm{m}$ and the ones with $x$.
First of all, a fit with only the uncertainties on $\sigma_\mathrm{m}$ is performed,
dealing with them as if they were uncorrelated. Then the correlation between the $\sigma_\mathrm{m}$'s
associated with different heavy quark masses is handled by means of the method
described in \app{app:fit_QCD_corr}. Finally the errors on $x$ are taken into account
according to the procedure described in \app{app:error_xy}. However, given the flatness
of the fitting curves, these corrections to the errors are found to be negligible compared to the
uncertainty on $\sigma_\mathrm{m}$ itself. For this reason also the correlation
between the $x$'s associated with different heavy quark masses, as well as
the correlation between $x$ and $\sigma_\mathrm{m}$ can be neglected.
The results of the 
interpolation are given in \tab{tab:sigma_m}. 

The step $L_1\to L_0=0.4$ fm is performed in complete analogy
with the previous step. 
The continuum limit
for the static step scaling function 
\be\label{eq:sigma_1_m_cl}
\sigma_\mathrm{m}^\mathrm{stat}(L_1)=0.233(36)\,,
\ee
is used as constraint on the fitting curves. The results of the interpolations are
given in \tab{tab:sigma_m}.

On the small volume ($L_0=0.4$ fm) only the relativistic QCD data
are needed to establish a finite volume relationship between
the pseudoscalar meson mass $M_\mathrm{PS}(L_0)$ and the RGI quark mass $M_\mathrm{h}$,
and consequently interpolate the bottom quark mass. 
First of all a linear fit $\rho(x,L_0)$ vs.~$x$ is performed considering
only the uncertainties on $\rho$, without correlation between the data. The latter
is taken into account in a successive step through the method described in \app{app:fit_QCD_corr}.
Finally, with the computed slope $\rho^{(1)}$,  one adds to $\rho$ the uncertainty on x and,
for each point, the correlation between the uncertainties on $x$ and $\rho$ according to
the procedure of \app{app:error_xy}:
\be\label{error_xy_SV}
(\Delta \rho)^2\to (\Delta \rho)^2+(\rho^{(1)}\Delta x)^2
-2\rho^{(1)}\frac{\partial \rho}{\partial M_\mathrm{PS}}\frac{\partial x}{\partial M_\mathrm{PS}}(\Delta M_\mathrm{PS})^2\,,
\ee
This last step amounts to a small correction to the errors, justifying the neglect 
of the correlation between the $x$'s associated with different heavy quark masses.

\begin{figure}[b]
\centering
\hspace{0.23cm}\includegraphics[width=11.9cm,height=8.3cm]{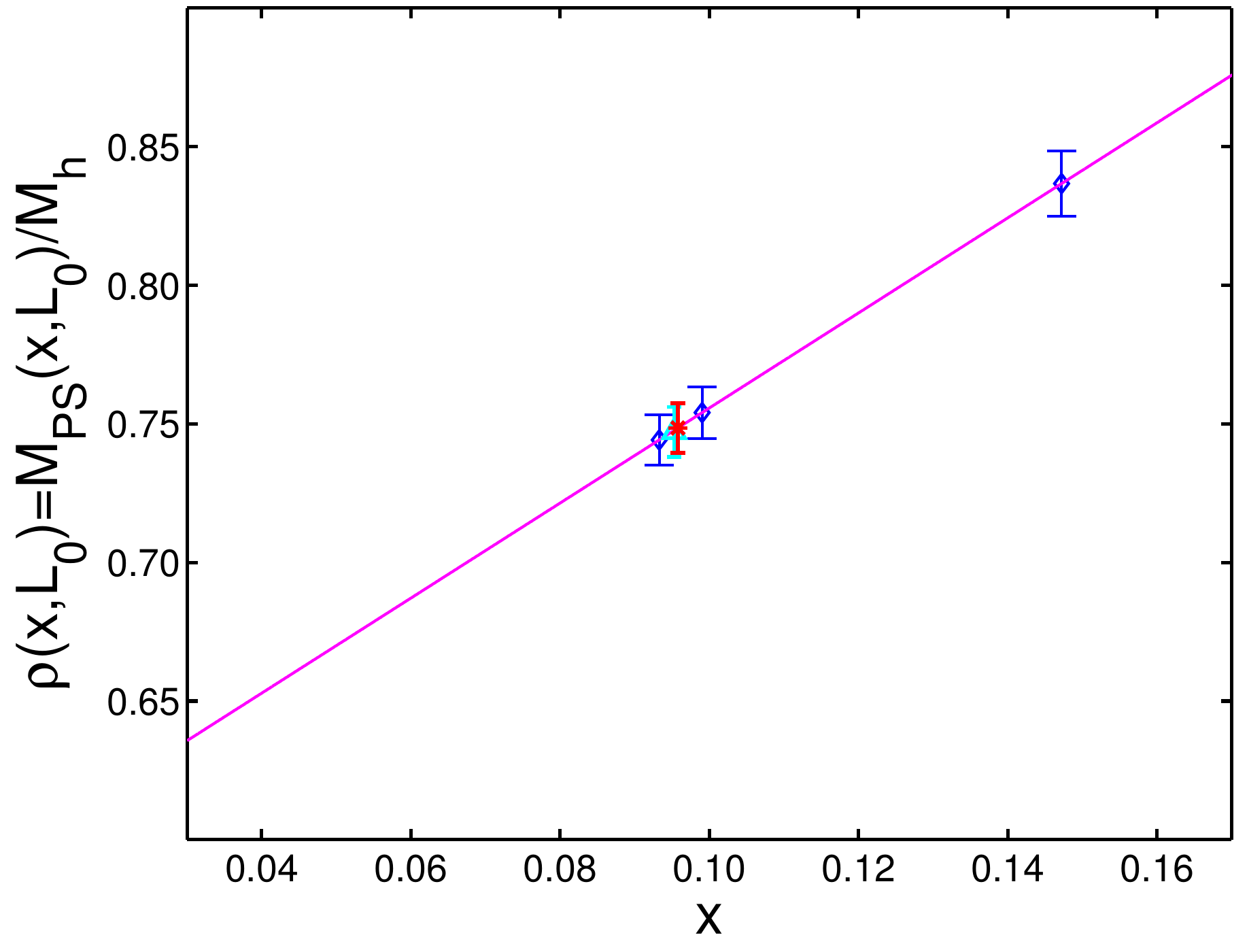}
\mycaption{Plot of $\rho(x,L_0)$ vs.~$x=x(m_\mathrm{h},L_0)$.}\label{fig:fig_Msv}
\end{figure}

The data are shown in \Fig{fig:fig_Msv}, where the red point (asterisk) corresponds
to the interpolation at the value of $x_0$ obtained from the step scaling
functions with the static constraint. The cyan point (triangle) corresponds to the interpolation
at the value of $x_0$ computed with only the data at finite heavy quark mass.

Using 
\eq{eq:MBs_Mb}, the interpolated value
\be\label{eq:rho_res}
\rho(x_0,L_0)=0.748(11)\,,
\ee
is combined with the above step scaling functions to find the scale and scheme independent number
\be\label{eq:M_b}
M_\mathrm{b}=6.879(104)\,\GeV\hspace{0.2cm}\Rightarrow\hspace{0.2cm}\overline{m}_{\mathrm{b},\MSbar}
(\overline{m}_{\mathrm{b},\MSbar})=4.416(60)\,\GeV.
\ee
By performing the whole analysis with only the data at finite heavy quark mass, we get
\be\label{eq:M_b_star}
M_\mathrm{b}^\star=6.953(108)\,\GeV\hspace{0.2cm}\Rightarrow\hspace{0.2cm}\overline{m}_{\mathrm{b},\MSbar}^\star
(\overline{m}_{\mathrm{b},\MSbar}^\star)=4.458(62)\,\GeV.
\ee 
The conversion of the mass to the $\MSbar$ scheme is perfomed according to the method 
outlined in \sect{sect:RGItoMSmass}, and the associated perturbative uncertainty can be safely neglected.
Starting from a precisely specified input, in our case the scale $r_0$ and 
the experimental $\mathrm{K}$ and $\mathrm{B_s}$ masses, the result for the b-quark mass is unambiguous
in the quenched approximation. 

In \cite{mb:nf0} the b-quark mass has been non-perturbatively determined in HQET, up to 
and including the $\mathrm{O}(1/m_\mathrm{h})$ terms, by starting from $r_0$ and the experimental 
values of $m_\mathrm{K}$ and of the spin-averaged mass $(M_\mathrm{B_s}+3M_\mathrm{B_s^\ast})/4$, 
obtaining $\overline{m}_{\mathrm{b},\MSbar}(\overline{m}_{\mathrm{b},\MSbar})=4.347(48)~\GeV$. This 
result can be compared with our determination (\ref{eq:M_b})
even if the experimental spin-averaged meson mass 
instead of the pseudoscalar one is used. 
These different choices affect the value of the b-quark mass by $\mathrm{O}(\Lambda_\mathrm{QCD}^3/m^2_\mathrm{b})$.
The agreement between the two determinations is evident.

Other quenched determinations of $\overline{m}_{\mathrm{b},\MSbar}(\overline{m}_{\mathrm{b},\MSbar})$
are $4.41(5)(10)~\GeV$ in HQET \cite{mbstat:MaSa}, extended to smaller values of 
the lattice spacing  in \cite{mb:Gimenez2} getting consistent results, and $4.34(7)~\GeV$  from 
NRQCD \cite{mb:NRQCD}.
However, if other inputs are used, one cannot perform a real comparison because $r_0$ is only 
approximatively known and the quenched approximation is not real QCD. 
See \cite{Onogi:2006km} for a recent review.

\clearpage

\begin{figure}\begin{minipage}[t]{\linewidth}
\centering
\includegraphics[width=12cm,height=9.0cm]{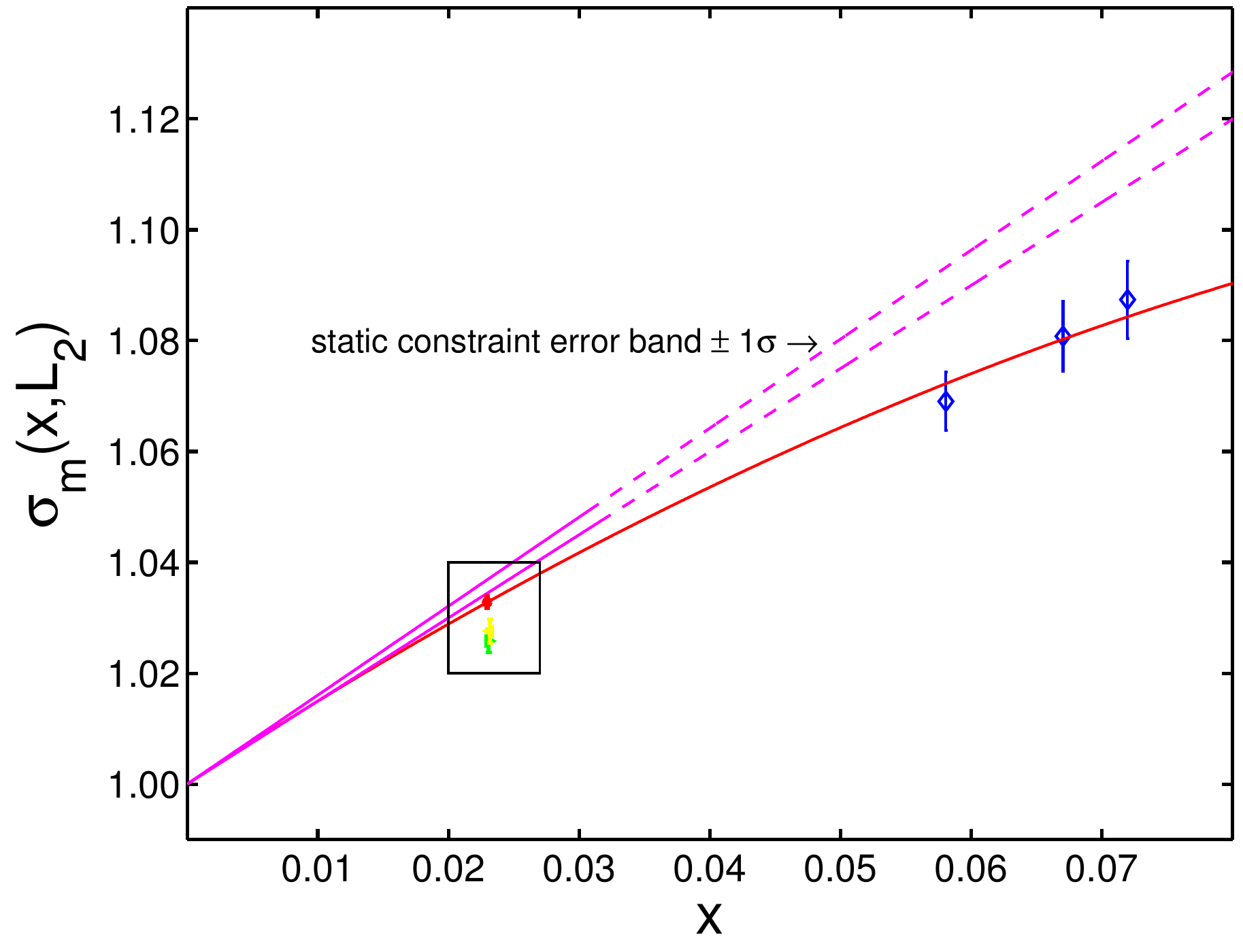}\\
\vspace{1.0cm}
\hspace{-0.4cm}\includegraphics[width=12.4cm,height=9.0cm]{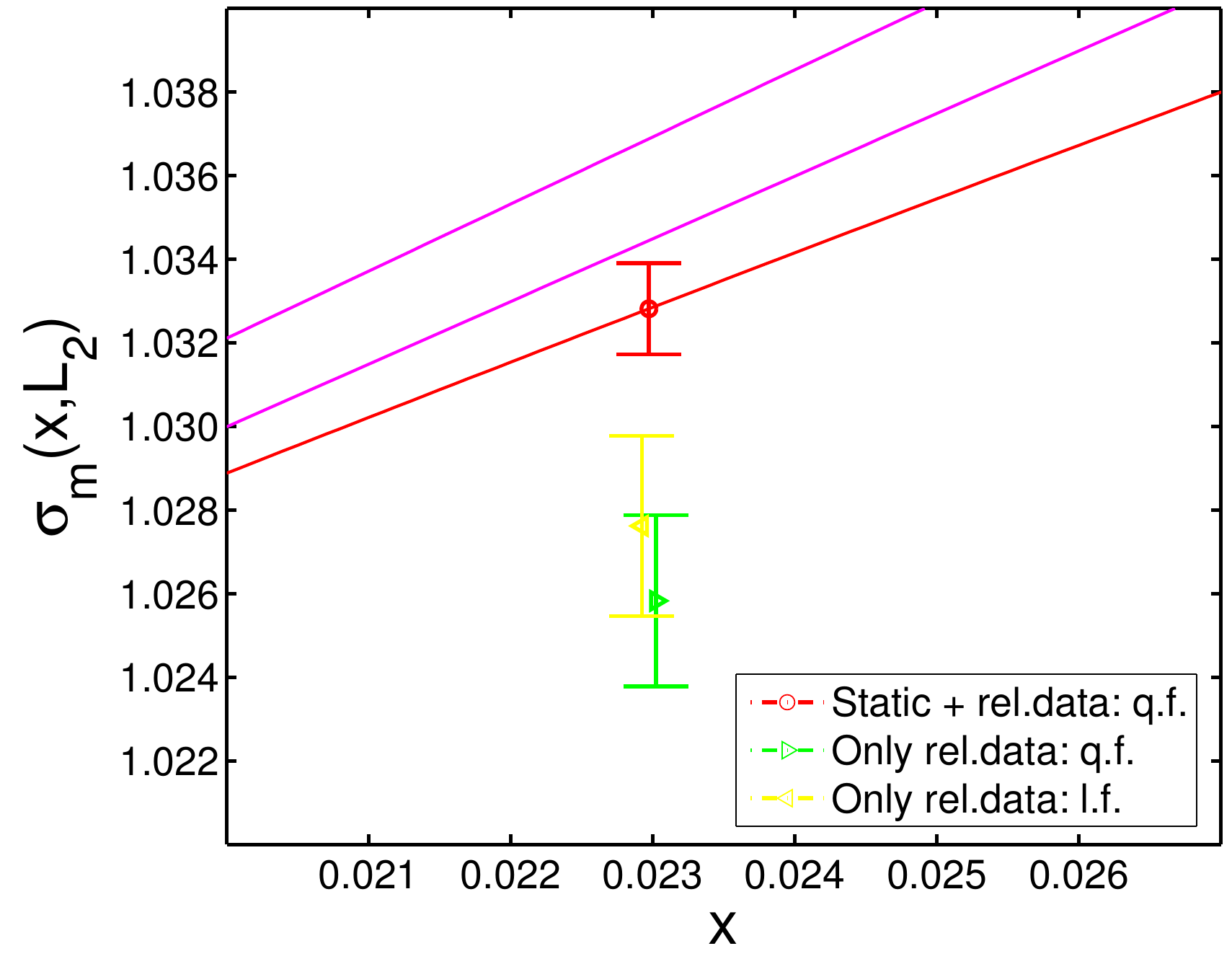}
\end{minipage}
\mycaption{Plot of the step scaling function $\sigma_\mathrm{m}(x,L_2)$ vs.~$x=x(m_\mathrm{h},L_2)$.}\label{fig:fig8}
\end{figure}

\clearpage

\begin{figure}[t]
\centering
\includegraphics[width=12cm,height=8.3cm]{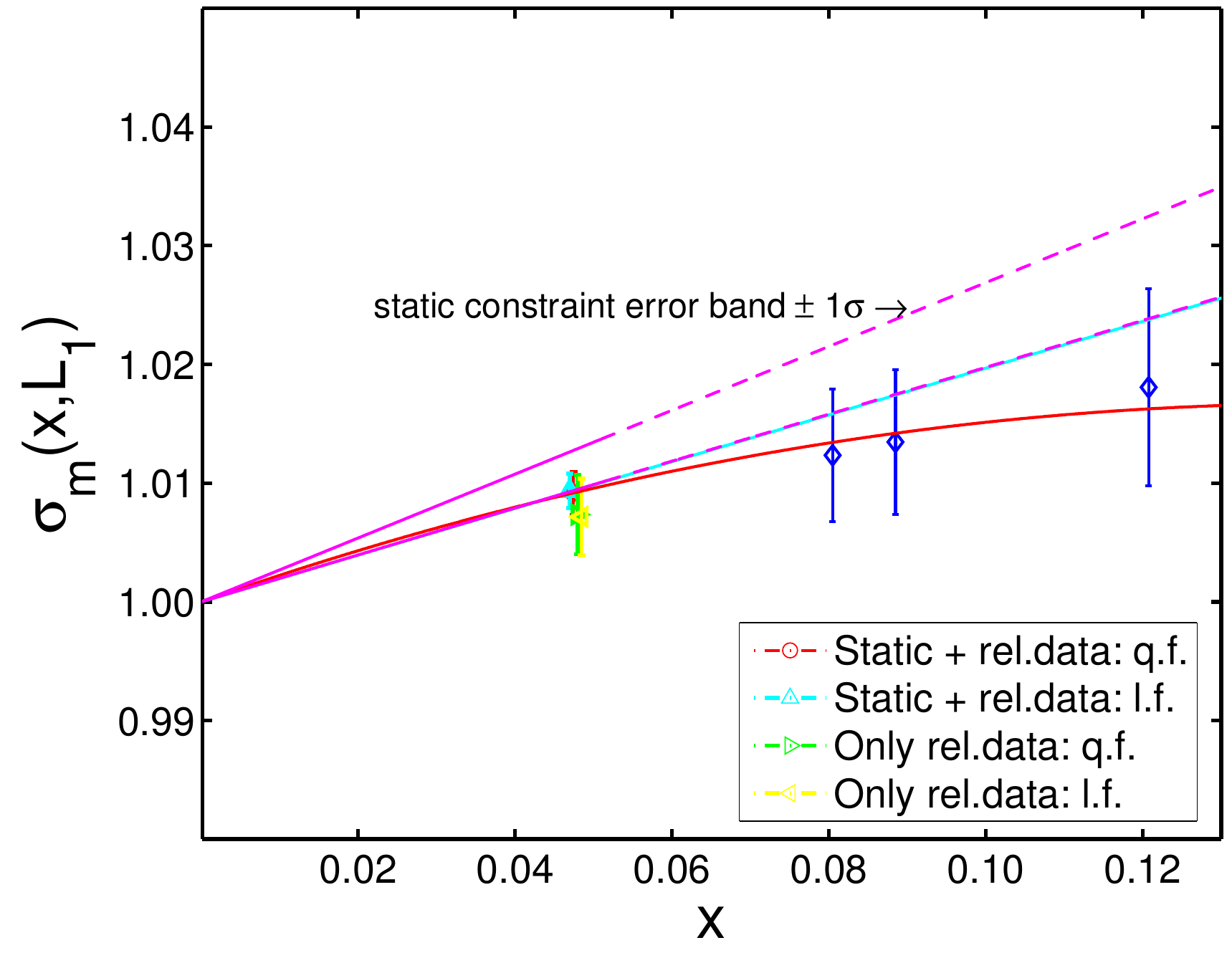}
\mycaption{Plot of the step scaling function $\sigma_\mathrm{m}(x,L_1)$ vs.~$x=x(m_\mathrm{h},L_1)$.}\label{fig:fig5}
\end{figure}

\clearpage

\section{Computation of the meson decay constant}\label{sect:Bs_dc}

\subsection{General strategy}\label{sect:Bs_dc_general}

With the lattice setup given in \eqs{eq:lat_setup}, we consider the case
where our observable is a finite volume meson decay constant times the square root of the
corresponding meson mass
\be\label{eq:O_eq_fPS}
{\mathcal{O}}(m_\mathrm{h},L)=f_\mathrm{PS}(m_\mathrm{h},L)\sqrt{M_\mathrm{PS}(m_\mathrm{h},L)}\,.
\ee
The decay constant is defined in analogy to (\ref{eq:fpi_extr}) by
\be\label{eq:fPS}
f_\mathrm{PS}(m_\mathrm{h},L)=\frac{-2}{\sqrt{L^3M_\mathrm{PS}(m_\mathrm{h},L)}}\left.\frac{[f_\mathrm{A}(m_\mathrm{h},L,x_0)]_\mathrm{R}} 
{\sqrt{[f_{1}(m_\mathrm{h},L)]_\mathrm{R}}}\right|_{x_0=L}\,.
\ee
The renormalized and improved correlators $[f_\mathrm{A}]_\mathrm{R}$ 
and $[f_{1}]_\mathrm{R}$ are defined in \eq{eq:fA_R} and \eq{eq:f1_R} respectively. It is
worth to notice that in the definition of the decay constant the renormalization factors
and the improvement terms of the boundary fields cancel out.
In accordance
with the large volume behaviour (\ref{eq:fpi_extr}) we write 
\be\label{eq:fPs_to_fBs}
f_\mathrm{PS}(m_\mathrm{h},L)\stackrel{m_\mathrm{h}\to m_\mathrm{b}}{=}f_\mathrm{B_s}(L)
\stackrel{L\to\infty}{=}f_\mathrm{B_s}\,.
\ee
The step scaling function
\be
\sigma_\mathrm{f}(x,L_i)\equiv\frac{f_\mathrm{PS}(m_\mathrm{h},L_i)\sqrt{M_\mathrm{PS}(m_\mathrm{h},L_i)}} 
{f_\mathrm{PS}(m_\mathrm{h},L_{i-1})\sqrt{M_\mathrm{PS}(m_\mathrm{h},L_{i-1})}}\,,\label{eq:sigma_f}                  
\ee
with expansion
\be
\sigma_\mathrm{f}(x,L_i)=\sigma_\mathrm{f}^\mathrm{stat}(L_i)+\sigma_\mathrm{f}^{(1)}(L_i)\cdot x + \mathrm{O}(x^2)\,,\label{eq:sigma_f_exp}
\ee
yields straightforwardly the connection between the finite volume decay constant
and the infinite volume one. Since the renormalization factor
$Z_\mathrm{A}$ of the axial current depends only on the bare
coupling, it cancels out in the ratio (\ref{eq:sigma_f}). The first order term in the expansion (\ref{eq:sigma_f_exp})
is computable in the static approximation of HQET. It is convenient to define
the renormalized ratio
\be\label{eq:YSF}
Y_\mathrm{SF}(L,\mu)=\zastat(1/\mu)\left.\frac{\fastatimpr(L,x_0)}{\sqrt{\fonestat(L)}}\right|_{x_0=L}=\zastat(1/\mu)X_\mathrm{SF}(L)\,,
\ee
where $\zastat$ is defined by the ``new'' renormalization 
scheme \cite{zastat:pap3}. The latter is based on the condition
\be\label{eq:zastat_def}
\zastat(1/\mu)\,\Xi(L')=\Xi^{(0)}(L')\,,\quad\mbox{at vanishing quark mass,}
\ee
with $\mu=1/L'$, and
\be\label{eq:Xi_def}
\Xi(L')=\left.\frac{\fastatimpr(L',x_0)}{[f_1(L')\fonehh(L',x_3)]^{1/4}}\right|_{x_0=x_3=L'/2}\,,
\ee
where $\Xi^{(0)}(L')$, computed in \cite{zastat:pap3}, is the tree-level value of $\Xi(L')$. The lattice setup
of the renormalization scheme slightly differs from \eqs{eq:lat_setup} and reads
\be\label{eq:lat_setup2}
T'=L'\,,\quad C=C'=0\,,\quad\mbox{and}\quad \theta=0.5\,.
\ee
This setup is used exclusively for the computation of $\zastat$. Furthermore,
the values of the critical hopping parameter needed to realize, up to $\Oasq$, the condition
of vanishing quark mass, are computed along the method discussed in \sect{sect:quark_masses_SSM}
with the setup in \eqs{eq:lat_setup}. 

We now proceed to relate the renormalized matrix element $Y_\mathrm{SF}(L,\mu)$
to the renormalization group invariant one defined by
\begin{align}
Y_\mathrm{RGI}(L)&=Y_\mathrm{SF}(L,\mu)\label{eq:Y_RGI}\\
&\quad \times [2b_0\gbsq(\mu)]^{-\gamma_0/2b_0}\mathrm{exp}\left\{
-\int_0^{\gbar(\mu)}\mathrm{d}g\left[\frac{\gamma(g)}{\beta(g)}-\frac{\gamma_0}{b_0g}\right]
\right\}\,,\nonumber
\end{align}
where the universal leading order coefficients of the $\beta$- and $\gamma$-functions
are given by $b_0=11/(4\pi)^2$ and $\gamma_0=-1/(4\pi)^2$ respectively. The 
ratio $\YRGI/Y_\mathrm{SF}$ depends only on $\mu$, 
\be\label{eq:Istat}
I^\mathrm{stat}(\mu)=\frac{\YRGI(L)}{Y_\mathrm{SF}(L,\mu)}\,,
\ee
and it
has been non-perturbatively evaluted in \cite{zastat:pap3}
for several orders of magnitude in $\mu$, covering all scales involved in our work.
The whole procedure can be summarized by introducing the renormalization factor $\ZRGI$
\begin{align}
Y_\mathrm{RGI}(L)&= \ZRGI\left.\frac{\fastatimpr(L,x_0)}{\sqrt{\fonestat(L)}}\right|_{x_0=L}\,,\label{eq:Y_RGI2}\\
\mbox{with}\quad \ZRGI&= I^\mathrm{stat}(\mu)\zastat(1/\mu)\,,\quad \mu=1/L'\,.\label{eq:Z_RGI}
\end{align}
The $L$-dependence of $Y_\mathrm{RGI}$ does not originate from the scale introduced by the
renormalization factor $\zastat$, but only from 
the fact that the correlation functions $\fastatimpr$
and $\fonestat$ have been determined on the volume $L$. Indeed, the 
renormalization factor $Z_\mathrm{RGI}$
depends only on the bare coupling. Furthermore, one is free to set $L'=L$. This choice
will be kept throughout the following. 
The RGI ratio $Y_\mathrm{RGI}$ is related to the QCD decay
constant $f_\mathrm{PS}$ via
\be\label{rel_YRGI_fPS}
\frac{f_\mathrm{PS}(m_\mathrm{h},L)\sqrt{L^3 M_\mathrm{PS}(m_\mathrm{h},L)}}{-2\Cps(M_\mathrm{h}/\Lambda_\msbar)}=
Y_\mathrm{PS}(x,L)=\YRGI(L)+\mathrm{O}(x)\,.
\ee
The function $\Cps(M_\mathrm{h}/\Lambda_\msbar)$, which
can be accurately evaluated in perturbation theory, 
allows to match the 
HQET matrix element with the QCD one; through the 3-loop anomalous dimension
computed in \cite{ChetGrozin}, the authors of \cite{hqet:pap3} 
provided the parametrization 
\begin{align}
\hspace{-0.7cm}\Cps(x)
& = 
\left\{
\begin{array}{ll}
t^{\gamma_0^\mathrm{PS}/(2b_0)}\left\{\,1-0.065\,t+0.048\,t^2\,\right\}
& \quad \mbox{2-loop $\gamma^\mathrm{PS}$}\\\vspace{0.5cm}
\label{eq:e_cpsfit}\\
t^{\gamma_0^\mathrm{PS}/(2b_0)}
\left\{\,1-0.068\,t-0.087\,t^2+0.079\,t^3\,\right\}
& \quad \mbox{3-loop $\gamma^\mathrm{PS}$}
\end{array}
\right. \,
\end{align}
where $t=1/\ln\left(M_\mathrm{h}/\Lambda_{\msbar}\right)$, 
the scale $\Lambda_\msbar=238(19)$ MeV from \cite{mbar:pap1}, and the coefficients $\gamma_0^\mathrm{PS}$
and $b_0$ are the same appearing in \eq{eq:Y_RGI}. This parametrization guarantees at least $0.2$\%
precision for $t\leq 0.6$. The function $\Cps$ is in practice needed only in the small volume. It
drops out in the ratio (\ref{eq:sigma_f}).
The step scaling function 
for the static decay constant, entering the expansion (\ref{eq:sigma_f_exp}), is given by
\be\label{eq:sigmaf_stat}
\sigma_\mathrm{f}^\mathrm{stat}(L_i)=\frac{1}{2^{3/2}}\frac{\YRGI(L_i)}{\YRGI(L_{i-1})}\,.
\ee
In this ratio the renormalization factor $\ZRGI$ cancels out. It is
needed only in the small volume. We have now all ingredients to combine 
the static data with the QCD ones, and interpolate at the scale of the b-quark.
The final result for $\fBs$ can be expressed in the compact form
\be\label{fBs_final}
\fBs=Y_\mathrm{PS}(x_0,L_0)\cdot \sigma_\mathrm{f}(x_1,L_1)\cdot \sigma_\mathrm{f}(x_2,L_2)
\cdot \frac{-2\Cps(M_\mathrm{b}/\Lambda_\msbar)}{\sqrt{L_0^3 \MBs}}\,,
\ee
where the interpolation points $x_i$ are computed through \eqs{eq:x_recursion}, and the b-quark
mass through \eq{eq:MBs_Mb}.

\subsection{The results}\label{sect:dc_results}

\begin{figure}[t]
\centering
\includegraphics[width=11.5cm,height=8.2cm]{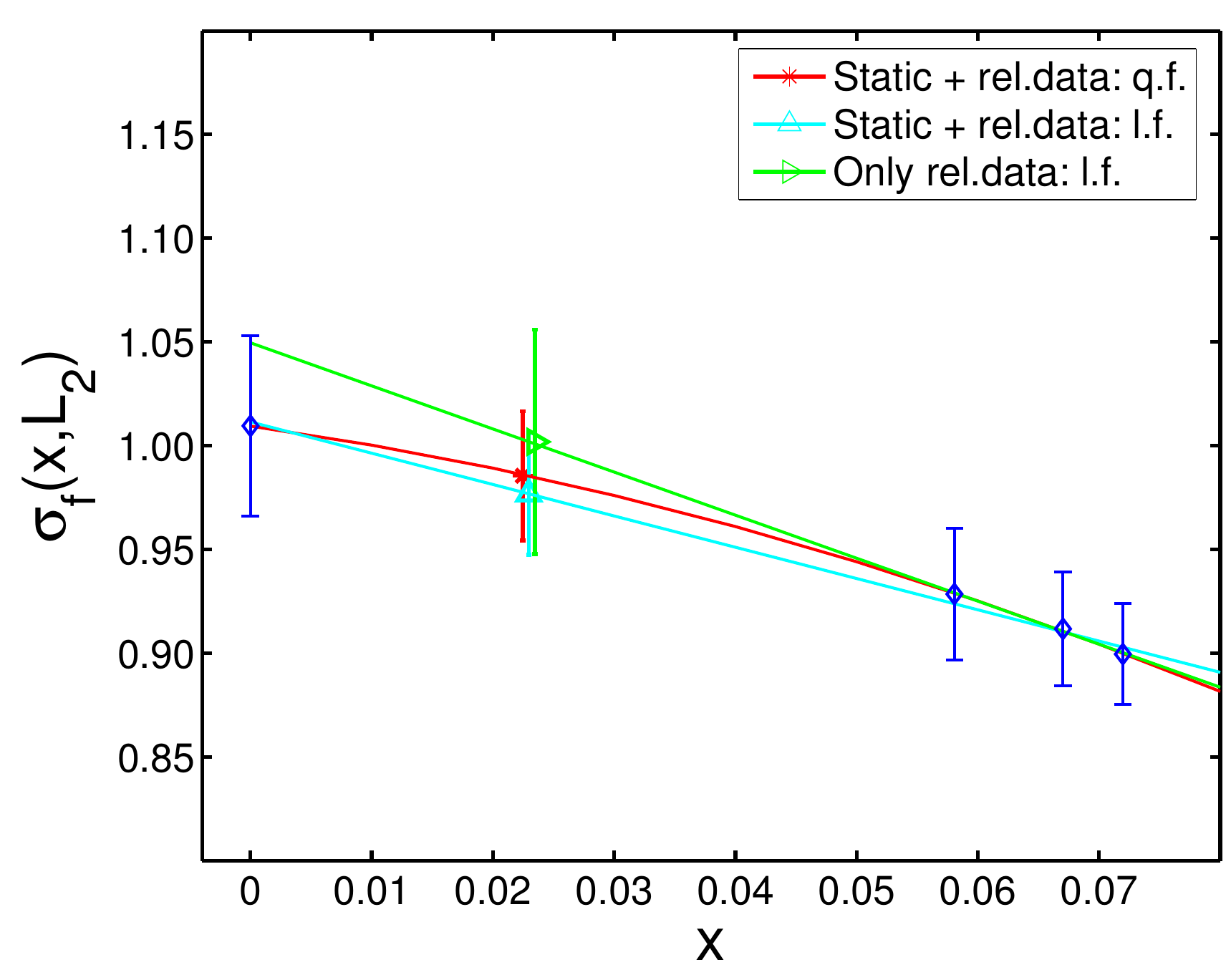}
\mycaption{Plot of the step scaling function $\sigma_\mathrm{f}(x,L_2)$ vs.~$x=x(m_\mathrm{h},L_2)$.}\label{fig:DC_S2}
\end{figure}

For the computation of $\sigma_\mathrm{f}(x,L_2)$ the relativistic data originate
from the same gauge configurations used earlier for the pseudoscalar meson mass.
In the static case the decay constant in the bigger volume
\be\label{eq:YRGI_L2}
\YRGI(L_2)=-4.65(19)\,,
\ee
has been computed and extrapolated to the continuum limit in \cite{Estat:me}. The continuum
extrapolation of the same quantity in the intermediate volume $L_1$ gives
\be\label{eq:YRGI_L1}
\YRGI(L_1)=-1.628(19)\,,
\ee
for which the renormalization factor $\zastat$ has been computed 
on the volume $L_1$. The results are presented in \app{app:CL_SSM}. The regularization
independent part of $\ZRGI$ can be extracted from the results of \cite{zastat:pap3},
and reads\footnote{Here the uncertainty does not contain 
the error coming from $I^\mathrm{stat}(\mu=(1.436r_0)^{-1})$, given in \cite{zastat:pap3},
because the latter factor cancels out in the step scaling function (\ref{eq:sf_stat_L2}). The aforementioned
error is included in the factor (\ref{eq:Istat_L0}) needed in the small volume.}
\be\label{eq:Istat_L1}
I^\mathrm{stat}(\mu)=0.9280(20)\,,\quad \mu=1/L_1\,.
\ee
These results allow to compute
\be\label{eq:sf_stat_L2}
\sigma_\mathrm{f}^\mathrm{stat}(L_2)=1.010(43)\,,
\ee 
which is used together with the relativistic data as shown in \fig{fig:DC_S2} to get
the results in the upper part of \tab{tab:sigma_f}. The interpolation procedure
as well as the estimate of the uncertainties follow the method employed
for the step scaling functions of the meson mass.

\begin{figure}\begin{minipage}[c]{\linewidth}
\centering

\includegraphics[width=12cm,height=9.0cm]{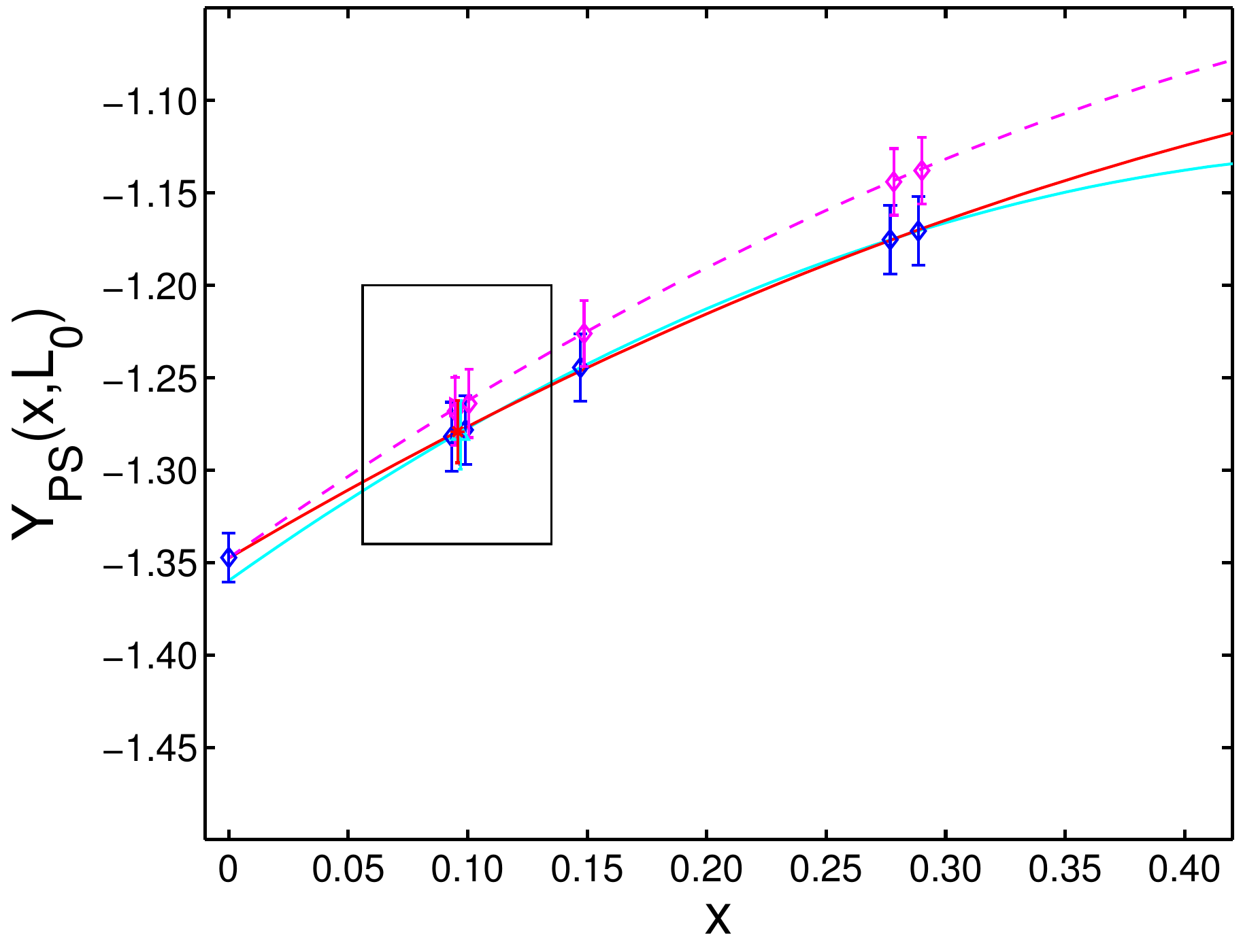}\\

\vspace{1.0cm}

\hspace{-0.2cm}\includegraphics[width=12.25cm,height=9.0cm]{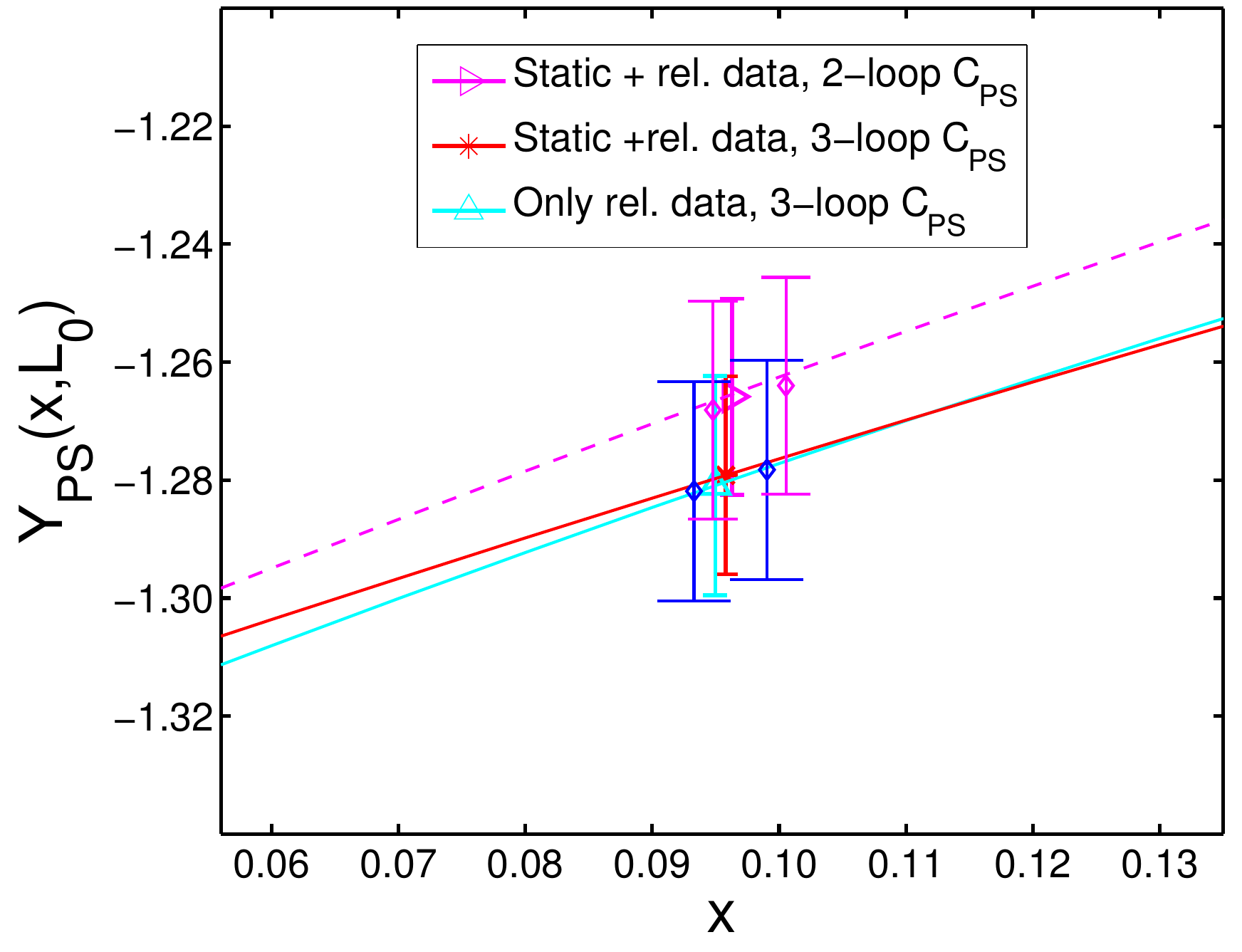}

\end{minipage}
\mycaption{Plot of the renormalized pseudoscalar decay constant 
on the small volume  $Y_\mathrm{PS}(x,L_0)$ vs.~$x=x(m_\mathrm{h},L_0)$.}\label{fig:DC_SV}
\end{figure}

Similarly, but by extrapolating the static step scaling function to the continuum limit rather
than $\YRGI(L_1)$ and $\YRGI(L_0)$ separately, we obtain
\be\label{eq:sf_stat_L1}
\sigma_\mathrm{f}^\mathrm{stat}(L_1)=0.4337(44)\,,
\ee 
which is combined with the relativistic data to give the results in the middle part
of \tab{tab:sigma_f}.
The small volume static result 
\be\label{eq:SV_DC_stat_results}
\YRGI(L_0)=-1.347(13)\,,
\ee
is obtained by exploiting the computations of \cite{zastat:pap3}, from which one extracts
\be\label{eq:Istat_L0}
I^\mathrm{stat}(\mu)=0.8462(62)\,,\quad \mu=1/L_0\,.
\ee
Together with the relativistic QCD data we obtain
\be\label{eq:SV_DC_result}
Y_\mathrm{PS}(x_0,L_0)=-1.279(17)\,,
\ee
as result of the quadratic fit, performed according
to the procedures of \app{app:error_xy} and \app{app:fit_QCD_corr}, of the 
data with the 3-loop expression of $\Cps$ shown in \fig{fig:DC_SV}. 
For other interpolated numbers we refer to \tab{tab:sigma_f}.
Also in this case the correction to the uncertainties originating from $x$
has been estimated to be negligible in comparison to the error on $Y_\mathrm{PS}$,
thus justifying the neglect of the correlation between the several $x$'s associated with different
heavy quark masses. 
We finally arrive through \eq{fBs_final} to
\be\label{eq:fBs_Joi_result} 
\fBs=191(6)\,\MeV\,.
\ee
By performing the whole analysis with only the relativistic QCD data, we obtain
\be\label{eq:fBs_RII}
\fBs^\star=195(11)\,\MeV\,.
\ee 
As last check, the whole analysis is repeated with the 2-loop expression
of $\Cps$, giving results which differ from (\ref{eq:fBs_Joi_result}) and
(\ref{eq:fBs_RII}) by an amount of $\mathrm{O(1\,eV)}$.

Our results are consistent with other quenched determinations using the same input
parameters. The authors of \cite{lat03:juri} did a non-perturbative calculation of the decay constant,
by combining the results \cite{stat:letter} in the static approximation of HQET 
with computations in relativistic QCD with heavy quark masses around the physical charm and slightly
heavier. They obtained $\fBs=206(10)$ MeV. An extension \cite{Estat:me} of this 
work produces a more precise and still consistent result. The computations are performed
in a large volume and extrapolated to the continuum limit. 
Another recent non-perturbative determination \cite{AliKhan:2007tm} is obtained by 
extrapolating to the b-region large volume computations in relativistic QCD.
The result is not extrapolated to the continuum, but computed at a fine lattice spacing
with $a^{-1}=4.97$ GeV, and reads $\fBs=206(7)(26)$ MeV, where the first error is statistical and 
the second systematic.

As for the b-quark mass, a real comparison with other quenched determinations can be made only
if the same input parameters are chosen. For completeness, we cite here 
the results of a few recent quenched computations
\begin{align}
\fBs/\MeV
& = 
\left\{
\begin{array}{ll}
&220(6)\binom{+23}{-28}\qquad\hspace{0.59cm}\mbox{\cite{Bowler:2000xw}}\,,\nonumber\\[1ex]
&220(2)(15)\binom{+8}{-0}\qquad\mbox{\cite{AliKhan:2000eg}}\,,\nonumber\\[1ex]
&199(5)\binom{+23}{-22}\qquad\hspace{0.59cm}\mbox{\cite{Bernard:2002pc}}\,.\nonumber
\end{array}\right.
\end{align}
For a recent review on the subject we suggest \cite{Onogi:2006km}.

\clearpage

\begin{figure}[b]
\centering
\includegraphics[width=11.5cm,height=8.2cm]{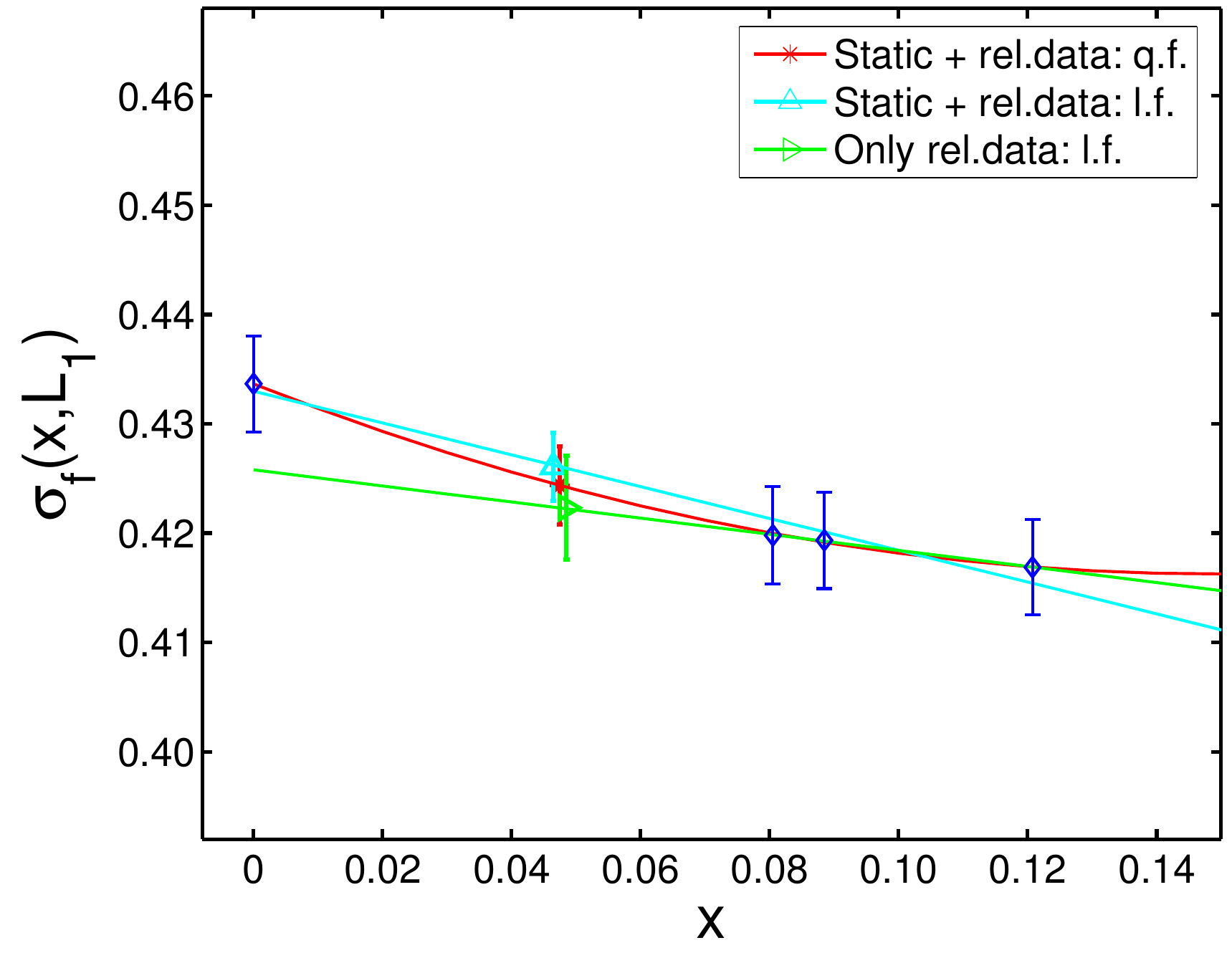}
\mycaption{Plot of the step scaling function $\sigma_\mathrm{f}(x,L_1)$ vs.~$x=x(m_\mathrm{h},L_1)$.}\label{fig:DC_S1}
\end{figure}

\clearpage

\clearpage

\begin{table}[ht]
\centering
\begin{tabular}[ht]{cccc}
\hline
\hline
        & & \\[-1ex]
Constraint on the slope & Fit & &$\sigma_\mathrm{m}(x_2,L_2)$ \\ 
        & & &\\[-1ex]
\hline
        & & &\\[-1ex]
Yes& quadratic& &$1.0330(11)$\\
Yes& linear& &$1.0319(11)$\\
No& quadratic& &$1.0258(21)$\\
No& linear& &$1.0276(22)$\\
        & & &\\[-1ex]
\hline
        & & &\\[-1ex]
Constraint on the slope & Fit & &$\sigma_\mathrm{m}(x_1,L_1)$ \\ 
        & & &\\[-1ex]
\hline
& & &\\[-1ex]
Yes& quadratic& &$1.0092(18)$\\
Yes& linear& &$1.0093(15)$\\
No& quadratic& &$1.0074(33)$\\
No& linear& &$1.0072(32)$\\
& & &\\[-1ex]
\hline
\hline
\end{tabular}
\vspace{-0.25cm}
\mycaption{Results of the interpolation of $\sigma_\mathrm{m}(x_2,L_2)$ and $\sigma_\mathrm{m}(x_1,L_1)$.
For the computation of the masses (\ref{eq:M_b}) and (\ref{eq:M_b_star}) the quadratic fit results
have been used.}\label{tab:sigma_m}
\end{table}

\begin{table}[ht]
\centering
\begin{tabular}[ht]{cccc}
\hline
\hline
        & & & \\[-1ex]
Inclusion of the static data & Fit & &$\sigma_\mathrm{f}(x_2,L_2)$ \\ 
        & & & \\[-1ex]
\hline
        & & & \\[-1ex]
Yes& quadratic&  &$0.985(31)$\\
Yes& linear&  &$0.977(29)$\\
 No& linear&  &$1.002(54)$\\
        & & & \\[-1ex]
\hline
        & & & \\[-1ex]
Inclusion of the static data & Fit &  &$\sigma_\mathrm{f}(x_1,L_1)$ \\ 
        & & & \\[-1ex]
\hline
& & & \\[-1ex]
Yes& quadratic&  &$0.4243(36)$\\
Yes& linear&  &$0.4260(31)$\\
 No& linear&  &$0.4223(48)$\\
& & & \\[-1ex]
\hline
        & & & \\[-1ex]
Inclusion of the static data & Fit & $\Cps$ &$Y_\mathrm{PS}(x_0,L_0)$ \\ 
        & & & \\[-1ex]
\hline
& & & \\[-1ex]
 Yes& quadratic&\hspace{0.1cm}3-loop  &$-1.279(17)$\\
 Yes& quadratic&\hspace{0.1cm}2-loop  &$-1.266(17)$\\
 No& quadratic&\hspace{0.1cm}3-loop &$-1.281(19)$\\
& & & \\[-1ex]
\hline
\hline
\end{tabular}
\mycaption{Results of the computation of $\sigma_\mathrm{f}(x_2,L_2)$, $\sigma_\mathrm{f}(x_1,L_1)$ and 
$Y_\mathrm{PS}(x_0,L_0)$. For the computation of (\ref{eq:fBs_Joi_result}) and (\ref{eq:fBs_RII}) 
the linearly interpolated step scaling functions have been used.}\label{tab:sigma_f}
\end{table}

\section{Summary}\label{sect:ssm_summary}

The combination of the Tor Vergata strategy to compute properties
of the heavy-light mesons \cite{romeII:mb,romeII:fb} with the expansion
of all quantities in HQET \cite{hqet:pap1}, changes extrapolations in the 
former computations into interpolations. The upshot is a very precise 
and controlled lattice determination, within the quenched approximation, of 
the b-quark mass and the $\mathrm{B_s}$-meson decay constant. The final result 
for the former is affected by an uncertainty of 1.5\%, mostly due to the 
mass renormalization factor.
The decay constant 
is computed with a precision slightly worse than 3\%, mostly relying 
on the large volume computations, which have been
recently improved in the HQET sector by the authors of \cite{Estat:me}. 

The numerical results of all steps show that the interpolations
are very well behaved. The step scaling functions for the pseudoscalar meson
mass show small deviations, roughly a few percents, from the static limit, and the HQET constraint
on the slope of the fitting curves allows to halve the uncertainty on the interpolated
points as well as to have a better control on the fitting parameters. All interpolated points are 
consistent with the HQET predictions within two standard deviations.  
However the error associated with the step scaling functions is much smaller than the one stemming from the 
small volume interpolation. There only the relativistic QCD data are needed,
and their uncertainty is dominated by the total renormalization factor $Z_\mathrm{M}$
of the RGI quark mass, whose error amounts to roughly 80\% of the final error on
the b-quark mass. 

The step scaling functions of the pseudoscalar decay constant
show an evident flatness, as expected from the Step Scaling Method. In the large
volume the presence of the static point plays a important role in 
the precision and confidence of the interpolation. 
This is not unexpected. The relativistic QCD data lie in a region
where the heavy quark mass is around the physical charm quark mass, and a linear extrapolation
to the b-region cannot be considered safe a priori.
The pattern is similar in the intermediate volume, where the flatness of the step scaling function
is even more evident. In all but one steps the static approximation alone gives
very accurate results. The one exception is the decay constant in the small volume,
where the $\mathrm{O}(1/m_\mathrm{b})$ corrections are around 5\%.

The elaborated error analysis, described 
in Apps.~\ref{app:CL_SSM}, \ref{app:error_xy}, \ref{app:fit_QCD_corr} and \ref{app:err_r0},
ensures that our results do not suffer from any systematic errors apart
from the use of the quenched approximation; the uncertainty of the 
regularization dependent quantities is taken into account before performing
the continuum limits. Small systematic errors quoted in \cite{romeII:mb,romeII:fb}
stemming from the extrapolations in the inverse of the heavy quark mass 
have been eliminated.


In order to significantly reduce the error on the b-quark mass
the total renormalization factor $Z_\mathrm{M}$ has to be computed even more precisely, while 
for an even more precise determination of decay constant,
the major issue is represented by the large volume computations, both in HQET and in QCD.

Concerning dynamical fermion computations, the challenge in this strategy is to simulate
in a large volume (such as $L_2$) with small enough lattice spacings, where quark masses 
of around $m_\mathrm{charm}$ and higher can be simulated with confidence.

\chapter{Renormalization of the chromo-magnetic operator in HQET}\label{chap:Zspin}

\section{Spin splitting}\label{sect:spin_splitting}

In chapter \ref{chap:hqet_pheno} we have discussed how 
heavy-light quark bound states can be described by an expansion in the inverse
of the heavy quark mass according to HQET. There we focused our attention on 
the phenomenological implications residing in the leading order 
of the expansion, the static limit. In this chapter we go
further on, and inspect the finite heavy quark mass corrections in 
a systematic way, by considering the mass splitting between vector
and pseudoscalar heavy light mesons. 

We follow \cite{Neubert:1993mb} and expand the mass of the ground 
state pseudoscalar ($\mathrm{PS}$) and vector ($\mathrm{V}$)
heavy-light mesons as
\be\label{eq:MM_expansion}
M_\mathrm{X}=m_\mathrm{h}+\bar{\mathrm{\Lambda}}+\frac{1}{2m_\mathrm{h}}\Delta M_\mathrm{X}^2+\mathrm{O}(\Lambda_\mathrm{QCD}^3/m_\mathrm{h}^2)\,,
\quad \mathrm{X}=\mathrm{PS,\!V}\,,
\ee
where 
\be\label{eq:Delta_MX}
\Delta M_\mathrm{X}^2=-\lambda_1-d_\mathrm{X}\lambda_2\,,
\ee
with $d_\mathrm{PS}=3$ and $d_\mathrm{V}=-1$. 
The details of the heavy quark mass definition are irrelevant for the present discussion.
One may think of a mass renormalized at the scale $m_\mathrm{h}$ itself.
The parameter $\lambda_1$ is associated with the kinetic operator (\ref{eq:O_kin}),
and $\lambda_2$ with the chromo-magnetic (bare) operator (\ref{eq:O_spin})
\be\label{eq:O_spin2}
{\mathcal{O}}_\mathrm{spin}(x)=\heavyb(x)\frac{1}{2i}F_{kl}\sigma_{kl}\heavy(x)= 
\heavyb(x){\boldsymbol{\sigma}}\!\cdot\!{\boldsymbol{B}}(x)\heavy(x)\,.
\ee
The latter is responsible
for the splitting
\be\label{eq:ssp1}
M_\mathrm{V}^2-M_\mathrm{PS}^2=4\lambda_2+\mathrm{O}(\Lambda_\mathrm{QCD}^3/m_\mathrm{h})\,.
\ee
The parameter $\lambda_2$ is a key quantity in HQET. It encodes, at order $1/m_\mathrm{h}$, 
the information on the deviations from the static limit,
where $M_\mathrm{V}=M_\mathrm{PS}$, stemming from the spin-dependent interactions
inside the heavy-light mesons.

So far we did neither exactly specify how $\lambda_2$ is defined and renormalized, nor discuss
the matching of the effective theory to QCD. 
The latter can be worked out by following
the procedure discussed in \sect{sect:matching_HQET}; the upshot is
\be\label{eq:ssp2}
M_\mathrm{V}^2-M_\mathrm{PS}^2=4C_\mathrm{mag}\lambda_2^\mathrm{RGI}+\mathrm{O}(\Lambda_\mathrm{QCD}^3/m_\mathrm{h})\,,
\ee
where the matrix element of the effective theory\footnote{Here the state $|\mathrm{B}\rangle$
is a shorthand for $|0,\mathrm{B}\rangle$ defined in \sect{subsect:cf_def}.}
\be\label{eq:lambda_2_RGI}
\lambda_2^\mathrm{RGI}=\frac{1}{3}\frac{\langle \mathrm{B}|{\mathcal{O}}_\mathrm{spin}^\mathrm{RGI}|\mathrm{B}\rangle}{\langle \mathrm{B}|\mathrm{B}\rangle}\,,
\ee
appears. The renormalization group invariant operator ${\mathcal{O}}_\mathrm{spin}^\mathrm{RGI}$,
is related
to the bare local operator (\ref{eq:O_spin2})
by a multiplicative, scale and scheme independent renormalization factor $Z_\mathrm{mag}^\mathrm{RGI}$,
which depends only on the bare coupling,
\be\label{eq:Ospin_RGI2}
{\mathcal{O}}^\mathrm{RGI}_\mathrm{spin}=Z_\mathrm{mag}^\mathrm{RGI}(g_0){\mathcal{O}}_\mathrm{spin}\,.
\ee
It follows that the parameter $\lambda_2^\mathrm{RGI}$
is independent of any scale and scheme. It is an unambiguous QCD observable, because the 
l.h.s.~of \eq{eq:ssp1} can be directly measured by experiments, and the corrections on the r.h.s.~vanish
in the limit\footnote{Or equivalently $M_\mathrm{X}\to\infty$.} $m_\mathrm{h}\to\infty$, where also
the matching factor $C_\mathrm{mag}$ is unambiguous and perturbatively computable.
The latter originates from the matching between the HQET matrix element and the QCD one.

In order to have a deeper insight into the subject, let us assume that the operator ${\mathcal{O}}_\mathrm{spin}$ has been renormalized 
\be\label{eq:Ospin_S}
{\mathcal{O}}_\mathrm{spin}^\mathrm{S}(\mu)=Z_\mathrm{mag}^\mathrm{S}(\mu){\mathcal{O}}_\mathrm{spin}\,,
\ee
in a scheme S, and that the matrix element
\be\label{eq:lambda_2_S}
\lambda_2^\mathrm{S}(\mu)=\frac{1}{3}\frac{\langle \mathrm{B}|{\mathcal{O}}_\mathrm{spin}^\mathrm{S}(\mu)|\mathrm{B}\rangle}
{\langle \mathrm{B}|\mathrm{B}\rangle}\,,
\ee
has been computed. The latter is related to the RGI one by
\be\label{eq:l2RGI_over_l2S}
\lambda_2^\mathrm{RGI}/\lambda_2^\mathrm{S}(\mu)= [2b_0\bar{g}^2_\mathrm{S}(\mu)]^{-\gamma_0/2b_0}
\exp\left\{-\int_0^{\bar{g}_\mathrm{S}(\mu)}\mathrm{d}g\left[\frac{\gamma^\mathrm{S}(g)}
{\beta^\mathrm{S}(g)}-\frac{\gamma_0}{b_0g}\right]\right\}\,,
\ee
where the anomalous dimension $\gamma^\mathrm{S}$ and the $\beta$-function 
in the $\mathrm{S}$ scheme with their leading order weak coupling expansion 
coefficients $\gamma_0$ and $b_0$ appear. The ratio (\ref{eq:l2RGI_over_l2S}) then allows
to relate the renormalization factors through
\be\label{eq:ZRGI_ZS}
Z_\mathrm{mag}^\mathrm{RGI}=Z_\mathrm{mag}^\mathrm{S}(\mu)\lambda_2^\mathrm{RGI}/\lambda_2^\mathrm{S}(\mu)\,.
\ee
The matrix element of the bare chromo-magnetic operator can be computed non-perturbatively
by lattice HQET simulations \cite{stat:bbstar,Gimenez:1996av,Aoki:2003jf}. As stated
in these references, the major source of uncertainty is the renormalization factor $Z_\mathrm{mag}$,
which is so far only perturbatively known. Our aim is to compute $Z_\mathrm{mag}^\mathrm{RGI}$ with high accuracy
for values of the bare coupling of interest for phenomenological applications, by starting from a low
energy regime (low $\mu$), to arrive to energy scales where the r.h.s.~of \eq{eq:l2RGI_over_l2S} can be safely evaluated
in perturbation theory. It is then clear that an accurate knowledge of the perturbative expression
of the anomalous dimension plays a fundamental role. This is the subject of the following sections.

\section{Definition of the renormalization scheme}\label{sect:SF_Zspin}

Our aim is to formulate a renormalization condition for ${\mathcal{O}}_\mathrm{spin}$ in a
finite volume, which suits the non-perturbative computation of the associated
renormalization factor $Z_\mathrm{mag}^\mathrm{RGI}$ along the general strategy of \cite{mbar:pap1}.
The lattice \SF fits well our requirements, and allows accurate non-perturbative computations 
in the low energy regime as well as the evolution to very high energies.
The latter can be achieved through a finite-size scaling technique,
which lets reach scales,
where perturbation
theory can be used to further continue the evolution,
and determine the renormalization group invariant renormalization factor with high accuracy.
In addition, perturbation theory allows to exactly match renormalization factors
computed in different schemes in a very easy way, as explained in the next section. 

For the static quarks we consider the Eichten-Hill action (\ref{eq:EH_action}), bearing in mind
that one could also employ other actions of the form (\ref{eq:general_heavy_action})
without compromising the validity of the following discussion. 
In the definition of ${\mathcal{O}}_\mathrm{spin}$, 
the product ${\boldsymbol{\sigma}}\!\cdot\!{\boldsymbol{B}}$ between the Pauli matrices $\sigma_k$
and the magnetic field $B_k$ is a shorthand for
\be\label{eq:sigma_dot_B}
{\boldsymbol{\sigma}}\!\cdot\!{\boldsymbol{B}}\otimes I_2=\sum_{k,j}\sigma_{kj}\frac{1}{2i}\widehat{F}_{kj}=
\left(
\begin{array}{cc}
\sigma_k& 0\\
0 & \sigma_k \\
\end{array}
\right)B_k\,, 
\ee
where $\sigma_{kj}=\tfrac{i}{2}[\gamma_k,\gamma_j]$, and the $B$-field 
is related to the lattice version $\widehat{F}_{\mu\nu}$, defined
in \eq{eq:F_munu_lQCD}, of the gauge field tensor by 
\be
B_k=i\epsilon_{qjk}\widehat{F}_{qj}\,,
\ee
with indices $q,j$ not summed over. For later convenience we report also the relation between the electric field
and the tensor $\widehat{F}_{\mu\nu}$:
\be\label{eq:E_k_def}
E_k=i\widehat{F}_{0k}\,.
\ee
The lattice definition of the magnetic field represents a further motivation 
for the choice of the Schr\"odinger functional.
With any kind of periodic boundary conditions, any correlation function with 
${\mathcal{O}}_\mathrm{spin}$ vanishes at tree-level. In order to avoid this, we introduce
boundary conditions inducing a non-trivial background field 
for which $\widehat{F}_{\mu\nu}$ does not vanish at tree-level. This ensures
a good signal in the MC simulations at weak coupling and means that a 1-loop computation
is sufficient to compute the renormalization factor up to and including $\mathrm{O}(g^2_0)$.

Inspection of the operator ${\mathcal{O}}_\mathrm{spin}$ reveals that it does not contain any
light fermion fields. This feature is exploited in the definition of the following 
correlation functions, where we completely avoid the introduction of light quarks.
The upshot is that for $N_\mathrm{f}=0$ we have a pure gauge quantity, without 
valence quarks. In perturbation theory, the diagrams owing their existence 
to $N_\mathrm{f}\neq 0$ are of order $g^{2n}_0$, with $n\geq 1$. 

Another important observation is that ${\mathcal{O}}_\mathrm{spin}$ does not mix with
other operators of the same or lower dimension under renormalization. This has been shown
in \cite{Falk:1990pz}, and can be easily understood by surveying that all but one operators with
dimension less or equal to five, and compatible with the symmetries of HQET, trivially transform
under spin rotations. The one exception is ${\mathcal{O}}_\mathrm{spin}$.

Our lattice setup is a volume of extension $L_0\times L_1 \times L_2 \times L_3$,
with $L_\mu=L=T$, and 
Dirichlet boundary conditions inducing an Abelian non-vanishing background field
as in \cite{SF:LNWW}, with the only difference that here they are imposed
in the 3-direction (instead of the temporal one),
\be\label{eq:U_x3}
  U_\mu(x)|_{x_3=0} = \exp(a C)\,, \quad
  U_\mu(x)|_{x_3=L} = \exp(a C')\,, \quad \mu=0,1,2\,,
\ee
while periodic boundary conditions are kept with respect to $x_0,x_1,x_2$.
The temporal coordinate is specified by $x_0$ with no misunderstandings.
It is distinguished from the very beginning in the static action,
which describes static quarks propagating only forward in time.
The parametrization of the matrices $C,C'$ can be found in \app{app:Not_Conv}.
A natural choice for the renormalization condition is then
\be\label{eq:zspindef}
Z_\mathrm{mag}^\mathrm{SF}(L)\frac{L^2\langle S_1(x+\tfrac{L}{2}\hat0){\mathcal{O}}_\mathrm{spin}(x)\rangle}  
{\langle S_1(x+\tfrac{L}{2}\hat0)S_1(x)\rangle}=
\left.\frac{L^2\langle S_1(x+\tfrac{L}{2}\hat0){\mathcal{O}}_\mathrm{spin}(x)\rangle}{\langle 
S_1(x+\tfrac{L}{2}\hat0)S_1(x)\rangle}\right|_{g_0=0}\,,
\ee
where $x_3=L/2$. The spin operator
\be\label{eq:Sk_current}
S_1(x)=\frac{1}{1+a\delta m}\heavyb(x)\sigma_1 U_0^\dagger(x-a\hat0)\heavy(x-a\hat0)\,,
\ee
is introduced in order to obtain a non-vanishing trace in spin space. It does not
need to be renormalized, because it is the (local) N\"other charge obtained
from the invariance of the action under the spin rotations (\ref{eq:spin_inv}).
There is no special reason for the choice of $S_1$ instead of $S_2$. Instead, if we had chosen
$B_3$, due to the symmetries of the background field, we would end up with a vanishing numerator
at tree-level.
If $N_\mathrm{f}$ light quarks are present in the action, we intend the condition (\ref{eq:zspindef})
with all of them having vanishing renormalized quark mass. 
The latter is provided in \eq{eq:M_deg}.
However, in \sect{sect:2loop_AD} we will
see that the details of the quark mass renormalization play no role in our computations.
Furthermore the angle $\theta$, appearing in the quark boundary conditions (\ref{eq:quark_boundaries4}),
is fixed to the value $-\pi/3$. This choice is motivated in \sect{sect:2loop_AD}, and represents a part
of the definition of our renormalization scheme for $N_\mathrm{f}\neq 0$.

By integrating out the static quark fields we can rewrite \eq{eq:zspindef}
in a simple form, which suits non-perturbative as well as perturbative computations.
An important ingredient is    
the explicit expression of the static quark propagator \cite{zastat:pap1}:
\be\label{eq:static_propagator}
G_\mathrm{h}(x,y)=\theta(x_0-y_0)\delta(\vecx-\vecy)(1+a\delta m)^{-(x_0-y_0)/a}W^\dagger(y,x)P_+\,,
\ee
with
\begin{align}
& \hspace{-1.2cm}W(x,x)=1\,,\label{eq:def_Wxx1}\\
& \hspace{-1.2cm}W(x,x+R\hat\mu)=U_\mu(x)U_\mu(x+a\hat\mu)\ldots U_\mu(x+(R-a)\hat\mu)\,,\quad R>0\,.\label{eq:def_Wxx2}
\end{align}
Our lattice $\delta$-functions are 
\be\label{eq:lattice_delta_function}
\delta(x_\mu)=a^{-1}\delta_{x_\mu,0}\,,\quad \delta(\vecx)=\prod_{k=1}^3\delta(x_k)\,,
\quad \delta(x)=\prod_{\mu=0}^3\delta(x_\mu)\,,
\ee
while for the $\theta$-function we have
\begin{align}
\left\{
\begin{array}{lcl}
\theta(x_\mu)&= 1\,,\quad\mbox{for $x_\mu\geq 0$}\,,\\
&  \label{eq:lat_theta_fct}\\
\theta(x_\mu)&= 0\,,\quad\mbox{otherwise}\,.
\end{array}
\right.
\end{align}
It is used in combination with the property
of the Pauli matrices
\be
\{\sigma_k,\sigma_j\}=2\delta_{kj}\,,
\ee 
to rewrite the ratio in \eq{eq:zspindef} as
\be\label{eq:zspindef_ratio}
\frac{\langle S_1(x+\tfrac{L}{2}\hat0){\mathcal{O}}_\mathrm{spin}(x)\rangle}  
{\langle S_1(x+\tfrac{L}{2}\hat0)S_1(x)\rangle}=\frac{\langle \Tr(\mathcal{P}_0(x)B_1(x)P_+)\rangle}
{\langle\Tr(\mathcal{P}_0(x)P_+)\rangle}\,,\quad B_1(x)=i\widehat{F}_{23}(x)\,,
\ee
where the Polyakov loop operator
\be\label{eq:Polyakov_loop_def}
\mathcal{P}_\mu(x)=U_\mu(x)U_\mu(x+a\hat\mu)\ldots U_\mu(x+(L-a)\hat\mu)
\ee 
enters. It follows that the renormalization condition (\ref{eq:zspindef})
is expressed in terms of observables where no valence quarks appear. At the denominator we have
the (traced) Polyakov loop $\mathcal{P}_0$, and at the numerator the (traced) Polyakov loop
$\mathcal{P}_0$ with the insertion of a $B$ field. 
We stress that the additive renormalization term $\delta m$, appearing both in 
the operator (\ref{eq:Sk_current}) and in the propagator (\ref{eq:static_propagator}), cancels
out in the ratios (\ref{eq:zspindef_ratio}), and we do not need to take care of it. In the following,
all correlation functions are intended to be computed with $\delta m=0$. The trace in Dirac space
concerns only the matrix $P_+$, giving an overall factor 2, which cancels out between 
numerator and denominator.
 
We come back to our renormalization condition, and note that it is natural to use
the equivalence of all coordinates in Euclidean space to switch to the usual \SF boundary conditions
\cite{SF:LNWW}, and, by using \eq{eq:E_k_def}, get
\be\label{eq:ren_cond_rot}
Z^\mathrm{SF}_\mathrm{mag}(L)\frac{L^2\langle \Tr(\mathcal{P}_3(x)E_1(x))\rangle}{\langle \Tr(\mathcal{P}_3(x))\rangle}=
\left.\frac{L^2\langle \Tr(\mathcal{P}_3(x)E_1(x))\rangle}{\langle \Tr(\mathcal{P}_3(x))\rangle}\right|_{g_0=0}\,,
\ee
where $x_0=L_0/2$, and $E_1(x)=i\widehat{F}_{01}(x)$. Here it is understood that one has
Dirichlet boundary conditions in time. Their exact definition is provided in \app{app:Not_Conv}.

\section{Connection between different schemes}\label{sect:scheme_connection}

In the following it is assumed that the theory has been regularized according
to the prescriptions of the previous section, with Dirichlet boundary 
conditions in time. Although some of the following results have
very general validity, we further assume to be working at an energy scale
where perturbation theory can be applied with confidence.

The scale evolution of the SF renormalized chromo-magnetic
operator is ruled by the renormalization group equation
\be\label{eq:RGE_Ospin}
\mu\frac{\partial}{\partial\mu}{\mathcal{O}}^\mathrm{SF}_\mathrm{spin}=\gamSF(\gbarSF){\mathcal{O}}^\mathrm{SF}_\mathrm{spin}\,,
\ee  
where the anomalous dimension $\gamSF$ appears. It can be expanded according to
\be
\gamma^\mathrm{SF}(\gbarSF)=-\gbsqSF(\gamma_0+\gamSF_1\gbsqSF+\ldots)\,,\label{eq:gamma_sf_expansion}
\ee
where \cite{stat:eichhill2,Falk:1990pz}: 
\be\label{eq:gamma_0_spin}
\gamma_0=3/(8\pi^2)\,,
\ee
is scheme-independent, and the two-loop anomalous dimension $\gamSF_1$ can be computed
from the one in the $\MSbar$ scheme
\be\label{eq:gamma1_MSbar}
\gamMSbar_1=(\tfrac{17}{2}-\tfrac{13}{12}N_\mathrm{f})/(32\pi^4)\,,
\ee
known from \cite{Amoros:1997rx,Czarnecki:1997dz}, by relating the two schemes according to the following method.

In our method an essential ingredient is represented by the lattice minimal subtraction scheme (``lat'').
The latter is defined by the requirement that, at each order of perturbation theory,
divergences are cancelled by introducing renormalization constants, which are polynomials
in $\ln(a\mu)$, without constant parts. At one-loop order,
the operator in the minimal subtraction scheme is related to the bare one by
\be\label{eq:Ospin_lat}
{\mathcal{O}}^\mathrm{lat}_\mathrm{spin}(\mu)=[1-g_\mathrm{lat}^2\gamma_0\ln(a\mu)]{\mathcal{O}}_\mathrm{spin}\,.
\ee
Any two mass independent renormalization schemes ``a'' and ``b'' can be related by a finite 
parameter renormalization of the form \cite{mbar:pert}:
\begin{align}
\bar{g}_\mathrm{a}&= \bar{g}_\mathrm{b}\sqrt{\chi_{\mathrm{g},(\mathrm{a,b})}(\bar{g}_\mathrm{b})}\,,\label{eq:rel_couplings}\\
{\mathcal{O}}_\mathrm{spin}^\mathrm{a}&= {\mathcal{O}}_\mathrm{spin}^\mathrm{b}\cdot\chi_\mathrm{a,b}(\bar{g}_\mathrm{b})\,,\label{eq:rel_ospins}
\end{align}
where, for simplicity, we assumed that the renormalization scale $\mu$ is the same
in both schemes. It follows that the operator (\ref{eq:Ospin_lat})
can be related to the ones in the SF and $\MSbar$ schemes by\footnote{A more rigorous notation
would require $\bar{g}_\mathrm{lat}$ instead of ${g}_\mathrm{lat}$. However, we keep the latter to
be consistent with the literature on the subject.} 
\begin{align}
{\mathcal{O}}_\mathrm{spin}^\mathrm{SF}(\mu)&= \chi_\mathrm{SF,lat}(\glat(\mu)){\mathcal{O}}^\mathrm{lat}_\mathrm{spin}(\mu)\,,\label{eq:Ospin_SF_lat}\\
{\mathcal{O}}_\mathrm{spin}^\mathrm{\MSbar}(\mu)&= \chi_\mathrm{\MSbar,lat}(\glat(\mu)){\mathcal{O}}^\mathrm{lat}_\mathrm{spin}(\mu)\,,\label{eq:Ospin_MS_lat}
\end{align}
with the expansion
\be\label{eq:chi_ab}
\chi_\mathrm{a,b}^{\phantom{(1)}}(g)=1+\chi_\mathrm{a,b}^{(1)}\,g^2+\ldots\,,
\ee
which lets us write the one-loop relation between the SF and $\MSbar$ schemes as
\be\label{eq:chi_SF_MSbar}
\chi_\mathrm{SF,\MSbar}^{(1)}=\chi_\mathrm{SF,lat\phantom{\MSbar}}^{(1)}\hspace{-0.5cm}-\chi_\mathrm{\MSbar,lat}^{(1)}\,.
\ee
Thanks to \eq{eq:rel_couplings}, we write down the relation between the couplings as 
\be\label{eq:gbar_SF_MSbar}
\gbsqSF=\chi_\mathrm{g\phantom{\MSbar}}\hspace{-0.4cm} \gbarMSbar^2\,,\quad \chi_\mathrm{g\phantom{\MSbar}}\hspace{-0.5cm}=\chi_{\mathrm{g},(\mathrm{SF},\MSbar)}=
1+\chi_\mathrm{g}^{(1)}\gbarMSbar^2+\ldots\,.
\ee
where for the precise definition of the renormalized coupling in the SF scheme we refer to \cite{alpha:SU3}.
The scale dependence is usually indicated by $\gbarSF(L)$ and $\gbarMSbar(\mu)$. In \eq{eq:gbar_SF_MSbar} it is implicitly
assumed that $\mu=1/L$.
The renormalization group function $\beta$, governing the scale evolution of the coupling, and the
anomalous dimension of the chromo-magnetic operator in the SF and $\MSbar$ schemes are related by
\begin{align}
& \beta^\mathrm{SF}(\gbarSF)=\Bigl\{\beta^{\MSbar}(\gbarMSbar)\frac{\partial\gbarSF}{\partial
\gbarMSbar}\Bigr\}_{\gbarMSbar=\gbarMSbar(\gbarSF\hspace{0.08cm})}\,,\label{eq:beta_SF}\\
& \gamSF(\gbarSF)=\Bigl\{\gamMSbar(\gbarMSbar)+\beta^{\MSbar}(\gbarMSbar)\frac{\partial}{\partial\gbarMSbar}
\ln \chi_{\mathrm{SF},\MSbar}(\gbarMSbar)\Bigr\}_{\gbarMSbar=\gbarMSbar(\gbarSF\hspace{0.08cm})}\,.\label{eq:gamma_SF}
\end{align}
We now work out \eq{eq:gamma_SF} by exploiting the perturbative expansion of the $\beta$- and $\gamma$-functions,
and in particular \eq{eq:running_gbar},
\begin{align}
\gbsqSF\left\{\gamma_0+\gamma_1^\mathrm{SF}\gbsqSF+\mathrm{O}(\gbfourSF)\right\}&=
\gbarMSbar^2\left\{\gamma_0+\gamma_1^{\MSbar}\gbarMSbar^2+\mathrm{O}(\gbarMSbar^4)\right\}\nonumber\\
& \hspace{-3.5cm}+\gbarMSbar^3(b_0+b_1\gbarMSbar^2)
\frac{\partial}{\partial \gbarMSbar}\ln\left\{1+\chi_\mathrm{SF,\MSbar}^{(1)}\gbarMSbar^2+
\mathrm{O}(\gbarMSbar^4)\right\}\nonumber\\
&=\gbarMSbar^2\left\{\gamma_0+\gamma_1^{\MSbar}\gbarMSbar^2+\mathrm{O}(\gbarMSbar^4)\right\}\nonumber\\
& \hspace{-3.5cm}+\gbarMSbar^3(b_0+b_1\gbarMSbar^2)
\cdot 2\frac{\chi_\mathrm{SF,\MSbar}^{(1)}\gbarMSbar^{\phantom{(2)}}+
\mathrm{O}(\gbarMSbar^3)}{ 1+\chi_\mathrm{SF,\MSbar}^{(1)}\gbarMSbar^2+
\mathrm{O}(\gbarMSbar^4)}\,.\nonumber
\end{align}
We divide both sides of the equation by $\gbsqSF$ and, through the Taylor expansion, obtain the
relation valid up to $\mathrm{O}(\gbarMSbar^4)$,
\begin{align}
\gamma_0+\gamma_1^\mathrm{SF}\gbsqSF&= (\gbarMSbar^2/\gbsqSF)\left\{\gamma_0+\gamma_1^{\MSbar}\gbarMSbar^2\right\}
+(\gbarMSbar^4/\gbsqSF)2b_0\chi_\mathrm{SF,\MSbar}^{(1)}\nonumber\\
&\hspace{-0.245cm}
\stackrel{(\ref{eq:gbar_SF_MSbar})}{=} (1-\chi_\mathrm{g}^{(1)}\gbarMSbar^2)\left\{\gamma_0+\gamma_1^{\MSbar}\gbarMSbar^2\right\}
+(\gbarMSbar^4/\gbsqSF)2b_0\chi_\mathrm{SF,\MSbar}^{(1)}\nonumber\\
&= \gamma_0+\left(\gamma_1^{\MSbar}-\gamma_0\chi_\mathrm{g}^{(1)}\right)
\gbarMSbar^2+(\gbarMSbar^4/\gbsqSF)2b_0\chi_\mathrm{SF,\MSbar}^{(1)}\,.\nonumber
\end{align}
We subtract $\gamma_0$ to both sides of the last equation, and then divide again by $\gbsqSF$, getting
\be\label{eq:gamma_1SF}
\gamma_1^\mathrm{SF}=\gamma_1^{\MSbar}-\gamma_0\chi_\mathrm{g}^{(1)}+2b_0\chi_\mathrm{SF,\MSbar}^{(1)}\,, 
\ee 
where we again used the relation (\ref{eq:gbar_SF_MSbar}) between the couplings.
Therefore, the two-loop anomalous dimension in the SF scheme can be extracted, 
through the exact relation (\ref{eq:gamma_1SF}),
from the one in the $\MSbar$ scheme, if the one-loop relation between the couplings, i.e.~$\chi_\mathrm{g\phantom{(}}^{(1)}\!$,
and the one-loop relation between the schemes, i.e.~$\chi_\mathrm{SF,\MSbar}^{(1)}$, are known.
The former has been computed in \cite{impr:pap2,pert:1loop}, and reads 
\be
\chi_\mathrm{g}^{(1)}=-\frac{1}{4\pi}(c_{1,0}+c_{1,1}N_\mathrm{f})\,,\quad c_{1,0}=1.25563(4)\,,\,\,c_{1,1}=0.039863(2)\,,
\ee
the coefficient $b_0=(11-\tfrac{2}{3}N_\mathrm{f})/(16\pi^2)$ is known from \cite{Politzer:1973fx,Gross:1973id}, while $\gamma_0$ appears 
in \eq{eq:gamma_0_spin}.
Since\footnote{We warmly thank Jonathan Flynn for clarifications on this result.} Flynn and Hill provided in \cite{Flynn:1991kw}:
\be\label{eq:chi_MSbar_lat_FH}
\chi_\mathrm{\MSbar,lat}^{(1)}=0.3824(3)\,,
\ee
inserting it in \eq{eq:chi_SF_MSbar}, we recognize that the only missing
ingredient is $\chi_\mathrm{SF,lat}^{(1)}$. To the computation of the latter
we dedicate the rest of the chapter.

\section{Perturbation theory in the SF scheme}\label{sect:SF_PT}

\subsection{The gauge fixed action}\label{sect:gauge_fix}

One of the pillars of perturbation theory consists in fixing
the gauge. This is necessary, because after having performed the saddle point
expansion around the induced background field, the resulting minimal action  
configuration is found to be unique only up to gauge transformations.
Indeed, once that a point on the gauge orbit of the minimal action 
has been chosen, the perturbative study amounts to parametrize
the infinitesimal fluctuations around this minimum. However, not
all of them are genuine physical field fluctuations. The gauge
fixing procedure separates the latter from the infinitesimal
gauge directions.

The following discussion is 
carried out in pure gauge theory, 
because the minima of the action are determined 
by the pure gauge theory alone, and the quark fields
only play a secondary role.  

The lattice \SF is invariant under all gauge transformations
$\Omega$ leaving the boundary fields $W$ and $W'$, defined 
in \eq{eq:rel_W_C}, intact.
It has been demonstrated by the authors of \cite{SF:LNWW}
that this condition is satisfied only by the gauge functions
that are constant and diagonal at the temporal boundaries, i.e.~
\begin{align}
\hspace{-0.7cm}\Omega(x)
& = 
\left\{
\begin{array}{ll}
z_m\,,\quad&\mbox{at}\,\,x_0=0\,,\\
\label{eq:omega_zm}\\
z_{m'}\,,\quad&\mbox{at}\,\,x_0=T\,,
\end{array}
\right. \,
\end{align}
with $z_m=\mathrm{exp}(2\pi im/3)$, and some integer numbers $m$
and $m'$. These gauge functions form a group, that we call $\hat{\mathcal{G}}$. 
Nevertheless, not the whole group is relevant for the gauge fixing.
The constant diagonal gauge functions $\Omega(x)$ form a subgroup of $\hat{\mathcal{G}}$,
which is isomorphic to the Cartan subgroup $\mathrm{C}_3$ of $\SUthree$, and trivially act 
on the background field. In addition, there are no further transformations
with this property \cite{SF:LNWW}. This group can be left out, and survives as a global
symmetry of the theory. The group which needs to be fixed is thus
\be\label{eq:def_G} 
\mathcal{G}=\hat{\mathcal{G}}/\mathrm{C}_3\,,
\ee
and it can be simply identified with the $m'=0$ component of $\hat{\mathcal{G}}$.

In an arbitrary but small neighborhood of the background field $V$, the gauge fields
$U$ may be parametrized by
\begin{align}
U_\mu(x)&= \mathrm{exp}(g_0 a q_\mu(x))V_\mu(x)\nonumber\\
&= \left\{1+g_0 a q_\mu(x)+\frac{1}{2}g_0^2 a^2 q^2_\mu(x)+\mathrm{O}(g_0^3)\right\}V_\mu(x)\,,\label{eq:U_parametrization}
\end{align}
where the lattice vector fields $q_\mu(x)$ form a linear space, which we name $\mathcal{H}$,
and whose spatial components respect the constraints
\be\label{eq:q_k_boundaries}
q_k(0,\vecx)=q_k(T,\vecx)=0\,,
\ee
while the temporal components are defined for $x_0\in [0,T)$ and we leave them unconstrained.
The condition (\ref{eq:q_k_boundaries}) guarantees that $U$ satisfies the boundary 
conditions (\ref{eq:U_x0}) once $V$ does.
The inner product of two vector fields $q$ and $r$
is given by
\be\label{eq:qr_product}
(q,r)=-2a^4\sum_{x,\mu}\tr\{q_\mu(x)r_\mu(x)\}\,.
\ee
The gauge fixing function that we are looking for is a linear mapping
from $\mathcal{H}$ to $\mathcal{L_{G}}$, where the latter is the Lie algebra of $\mathcal{G}$,
consisting of all fields $\omega(x)$, such that the infinitesimal transformation
\be\label{eq:omega_inf_transf}
\Omega(x)=1-g_0\omega(x)+\mathrm{O}(g_0^2)
\ee
belongs to $\mathcal{G}$. We thus introduce the operator
\be\label{operator_d}
d:\,\mathcal{L_{G}}\mapsto \mathcal{H}\,,\qquad (d\omega)_\mu(x)=D_\mu \omega(x)\,,
\ee
where the covariant derivative $\cdev{\mu}$ and its backward correspondent $\cdevstar{\mu}$ act according to
\begin{align}
\cdev{\mu}\,\omega(x)&= \frac{1}{a}\left[V_\mu(x)\omega(x+a\hat{\mu})V^{-1}_\mu(x)-\omega(x)\right]\,,\label{eq:D_mu}\\
\cdevstar{\mu}\,\omega(x)&= \frac{1}{a}\left[\omega(x)-V^{-1}_\mu(x-a\hat{\mu})\omega(x-a\hat{\mu})V_\mu(x-a\hat{\mu})\right]\,.\label{eq:D_mu_ast}
\end{align}
Analogously, $d^\ast$ maps any vector field $q\in \mathcal{H}$ onto an element of $\mathcal{L_G}$ such that
\be\label{d_ast} 
(d^\ast q,\omega)=-(q,d\omega)\,.
\ee
According to \cite{FSE:martin1}, the desired gauge fixing function has to vanish on the background field
and has to be non-vanishing on the gauge modes $d\omega$. These requirements are fulfilled by
the function
\be\label{eq:def_F}
F(U)=d^\ast q\,,
\ee
which lets us write the gauge fixing term for the action as
\be\label{eq:S_gf}
S_\mathrm{gf}[B,q]=\frac{\lambda_0}{2}(d^\ast q,d^\ast q)\,.
\ee
At this point a remark upon the boundary conditions on the fields $q$
is due. A careful analysis of the authors of \cite{SF:LNWW} shows that it is
useful to formally extend the time component $q_0(x)$ of the lattice field to all points
with $x_0=-a$ and $x_0=T$, thus allowing to write on the whole lattice
\be\label{eq:d_ast2}
d^\ast q(x)=\cdevstar{\mu}q_\mu(x)\,.
\ee
Besides \eqs{eq:q_k_boundaries}, one has to specify the boundary conditions
for the component $q_0$. It is worth to express them in momentum space;
we exploit the fact that all fields are periodic in space and perform the Fourier transform
\be\label{eq:FT_q0}
q_0(\vecp,x_0)=\sum_\vecx \mathrm{e}^{-i\vecp\cdot\vecx}q_0(x)\,,
\ee
and decompose it in a basis of the Lie Algebra of $\SUthree$
\be\label{eq:q0_SU3}
q_0(\vecp,x_0)=\tilde{q}_0^a(\vecp,x_0)I^a\,.
\ee
The basis $I^a$ used in all our computations can be found in \app{app:Not_Conv}.
With our choice of the background field
we have a mixture of Dirichlet and Neumann boundary conditions
\begin{align}
\tilde{q}_0^a(\vecp,-a)&= 0\,,\mbox{if}\,\,I^a\in \mathrm{C}_3\,\,\mbox{and}\,\,\vecp=\vectn\,,\nonumber\\[-2ex]
\hspace{-5.0cm}\fbox{lower boundary}\hspace{4.3cm}&  \nonumber\\[-2ex]
\drvstar{0}\,\tilde{q}_0^a(\vecp,x_0)|_{x_0=0}&= 0\,,\mbox{otherwise},\nonumber\\
& \label{eq:boundaries_q0}\\
\fbox{upper boundary}\hspace{1.5cm}\drvstar{0}\,\tilde{q}_0^a(\vecp,x_0)|_{x_0=T}&= 0\,.\nonumber
\end{align}
The associated Fadeev-Popov ghosts $c$ and $\bar{c}$ can be seen as infinitesimal
gauge transformations, except that they obey the Fermi statistics. The corresponding
action is given by
\be\label{eq:S_FP}
S_\mathrm{FP}[B,q,c,\bar{c}]=-(\bar{c},d^\ast \delta_c q)\,,
\ee
with $\delta_c q$ denoting the first order variation of $q$ under the gauge
transformation generated by $c$. We expand it to order $g_0^2$, and get
\begin{align}
\delta_c q_\mu&= \cdev{\mu} c +g_0 \mathrm{Ad}\,q_\mu\,c\nonumber\\
&\quad +\left[\frac{1}{2}g_0a\mathrm{Ad}\,q_\mu + \frac{1}{12}(g_0a\mathrm{Ad}\,q_\mu)^2+\ldots\right]\cdev{\mu} c\,,\label{eq:delta_c}
\end{align}
without summing over $\mu$. The boundary conditions for the ghost fields
can be obtained analogously to the gluon fields $q$. The analysis performed in \cite{pert:2loop_SU2}
reveals that, after a Fourier transform as in \eq{eq:FT_q0} and a group decomposition
as in \eq{eq:q0_SU3}, one is left 
\begin{align}
\drvstar{0}\,\tilde{c\kern+0.5pt}^a(\vecp,x_0)|_{x_0=0}&= 0\,,\mbox{if}\,\,
I^a\in \mathrm{C}_3\,\,\mbox{and}\,\,\vecp=\vectn\,,\nonumber\\[-2ex]
\hspace{-5.0cm}\fbox{lower boundary}\hspace{5.02cm}&  \nonumber\\[-2ex]
\tilde{c\kern+0.5pt}^a(\vecp,-a)=\tilde{c\kern+0.5pt}^a(\vecp,0)&= 0\,,\mbox{otherwise},\nonumber\\
& \label{eq:boundaries_c}\\
\fbox{upper boundary}\hspace{3.65cm}\tilde{c\kern+0.5pt}^a(\vecp,T)&= 0\,,\nonumber
\end{align}
and similarly for $\bar{c}$.

\subsection{The total action}\label{sect:total_action}

After having fixed the gauge, we are ready to write
down the gauge fixed functional integral in a form suitable
for our perturbative computations.
We gather the fully improved action (\ref{eq:improved_action}),
the gauge fixing term (\ref{eq:S_gf}) and the Fadeev-Popov action (\ref{eq:S_FP}) into
the total action 
\be\label{eq:total_action}
S_\mathrm{tot}[b,q,\bar{c},c,\psibar,\psi]=S_\mathrm{impr}[U,\psibar,\psi]+S_\mathrm{gf}[b,q]+S_\mathrm{FP}[b,q,c,\bar{c}]\,,
\ee
where it is understood that $U$ and $q$ are related by \eq{eq:U_parametrization}, and 
the $b$ is defined in  \eqs{deq:def_cal_B}.
We now perform
a change of integration variables in the functional integral; we relate the measure $\mathrm{D}[U]$
appearing in \eqs{eq:lSF_pure_gauge}, to the measure
\be\label{eq:Dq}
\mathrm{D}[q]=\prod_{x,\mu,a}\mathrm{d}\tilde{q}_\mu^a(x)\,,
\ee
by
\be\label{eq:DU_Dq}
\mathrm{D}[U]=\mathrm{D}[q]\mathrm{e}^{-S_\mathrm{m}[q]}=\mathrm{D}[q]\{1+\mathrm{O}(g_0^2)\}\,.
\ee
The additional contribution to the action stemming from this change of variables
is thus of $\mathrm{O}(g_0^2)$, and, since in our computations the action is needed
only up to $\mathrm{O}(g_0)$, we can neglect it from now on. The explicit form
of the \SF then reads
\be\label{eq:SF_total_action}
\mathcal{Z}=\int \mathrm{D}[q]\mathrm{D}[\bar{c}]\mathrm{D}[c]\mathrm{D}[\psibar]\mathrm{D}[\psi]\mathrm{e}^{-S_\mathrm{tot}}\,.
\ee
We consider now an observable ${\mathcal{O}}$, consisting of a product of gauge links. An eventual dependence
on the quark fields is not discussed here.
The situation looks like the one discussed in \sect{sect:MC_path_integral}; the difference is that 
here we want to find out a path integral expression, which suits a perturbative evaluation.
We keep the quark determinant resulting from an integration over the quark variables, and write down
the expectation value for ${\mathcal{O}}$ as
\be\label{eq:expect_value_O}
\langle {\mathcal{O}}\rangle_\mathrm{G}=\frac{1}{\mathcal{Z}}
\int \mathrm{D}[q]\mathrm{D}[\bar{c}]\mathrm{D}[c]\mathrm{D}[\psibar]\mathrm{D}[\psi]{\mathcal{O}}\,\mathrm{e}^{-S_\mathrm{tot}}\,.
\ee
We expand ${\mathcal{O}}$ in a series of $g_0$:
\be\label{eq:O_expansion}
\langle {\mathcal{O}}\rangle_\mathrm{G}={\mathcal{O}}^{(0)}+g_0\langle {\mathcal{O}}^{(1)}\rangle_\mathrm{G}
+g_0^2\langle{\mathcal{O}}^{(2)}\rangle_\mathrm{G}+\mathrm{O}(g_0^3)\,,
\ee
very simple to write down, thanks to \eq{eq:U_parametrization} and the fact that ${\mathcal{O}}$
is just a product of link variables. It is important to bear in mind 
that the term ${\mathcal{O}}^{(n)}$ contains the product of $n$ gluon fields.
In turn, the action can be expanded according to
\be\label{eq:S_expansion}
S_\mathrm{tot}=S_\mathrm{tot}^{(0)}+g_0S_\mathrm{tot}^{(1)}+g_0^2S_\mathrm{tot}^{(2)}+\mathrm{O}(g_0^3)\,.
\ee
Actually, there should also be a term proportional $S_\mathrm{tot}^{(-2)}$ coming from the gluon action.
However, such a term depends only on the background field and not on the integration variables
appearing in the functional integral, and it drops out in \eq{eq:expect_value_O}. It is then straightforward
to write down the expansion
\be\label{eq:eS_expansion}
\mathrm{e}^{-S_\mathrm{tot}}=\left[1-g_0S_\mathrm{tot}^{(1)}+g_0^2\left(\frac{1}{2}(S_\mathrm{tot}^{(1)})^2-
S_\mathrm{tot}^{(2)}\right)+\mathrm{O}(g_0^3)\right]\mathrm{e}^{-S_\mathrm{tot}^{(0)}}\,.
\ee
With this expression we can now write the expansion of the expectation value of ${\mathcal{O}}$ as
\be\label{eq:PT_expect_value}
\langle {\mathcal{O}}\rangle_\mathrm{G}={\mathcal{O}}^{(0)}+g_0^2\left[\langle {\mathcal{O}}^{(2)}\rangle_{0}-
\left\langle {\mathcal{O}}^{(1)}\left[S_\mathrm{tot}^{(1)}\right]_\mathrm{F}\right\rangle_{0}\right]+\mathrm{O}(g_0^4)\,,
\ee
where $\langle \rangle_0$ means that the expectation value is computed using the action
$S_\mathrm{tot}^{(0)}$ instead of the complete one. Furthermore there are no terms of order $g_0$ or $g_0^3$,
because they would involve an integral over an odd number of gluon fields, and vanish. 
The brackets $[\,]_\mathrm{F}$ indicate that the quark fields have been integrated out 
as in \sect{sect:MC_path_integral}. Finally
we mention that the term containing $S_\mathrm{tot}^{(1)}$ appears only because of the presence
of a non-vanishing background field. 

\section{Expectation values of Wilson loops at one-loop order}\label{sect:Wilson_loops}

\subsection{Parametrization of the observable}\label{sect:par_observable}

Our aim is to gather all necessary steps
for the computation of the expectation value of an arbitrary Wilson loop at one-loop order
in the \SF scheme, in a way which suits its numerical implementation. 
We bear in mind that in the end our observable is a Polyakov loop with, eventually, the insertion
of a clover operator. Due to the space-time locality of such an observable, it will
be advantageous to compute the gluon loops in $x$-space, while the tadpole contributions
are proportional to the zero-momentum gluon propagator. 

To achieve this, we parametrize the loop by a starting point $x^{(\mathrm{start})}$ 
and an ordered list $\vec\ell$ of
length $\ell$. The entries of the list are directions $\mu_i^{\vec\ell}$, $i=1,\dots,\ell$.
These directions take non-zero integer values between $-4$ and $+4$.  An electric plaquette
in the $(03)$ plane is thus parametrized by $\vec\ell=(3~4~-3~-4)$. Clearly the loop is closed
if and only if each integer appears as many times with the $+$ sign as it does with the $-$ sign,
modulo $L/a$ for the spatial directions.
The latter is a consequence of the spatial periodic boundary conditions. It means that
if we have a lattice setup with $L/a=4$ and the path is parametrized by $\vec\ell=(3~3~3~3)$,
the latter specifies a closed loop.
We normally drop the $\vec\ell$ in $\mu_i^{\vec\ell}$ since we will be dealing only with one path
at a time.

The sequence of points the loop goes through is obtained as follows:
\be
x^{(1)}=x^{(\mathrm{start})}\,, \qquad  x^{(i+1)}=x^{(i)} ~+~  a\hat{\mu_i},\quad i=1,\dots,\ell-1\,.
\ee
$\hat\mu= \mathrm{sign}(\mu) \widehat{|\mu|}$ are unit vectors pointing in the four $\pm$ directions of the lattice.
Here and throughout the following, the temporal direction is indicated by $|\mu|=4$. 
This choice is motivated 
by the necessity of indicating through the index $\mu$ the forward and backward direction 
also in time. For any unspecified notation we refer to \app{app:Not_Conv}.
  
At tree-level, the expectation value of the Wilson loop is
\be
W_{\vec\ell}[V] = \prod_{i=1}^\ell V_{\mu_i}(x^{(i)})\,.
\ee
In general, for any 4-vector we introduce negative-index components
\be
p_{-\mu} = - p_{\mu}\,.
\ee
Because of the way the path is parametrized,
for any link variable we introduce ne\-ga\-ti\-ve-index components 
by imposing\footnote{Compared to \eq{eq:q0_SU3} the twiddle on the color components
of the gluon field is dropped. We keep this notation also for the ghost and quark fields
throughout the following. The color index $a$ always appears as an index, unless specified otherwise.}
\be
U_\mu(x) = U^\dagger_{-\mu}(x+a\hat\mu),\qquad
q_{\mu}^a(x) =  - q_{-\mu}^{a}(x+a\hat\mu)\,.
\ee
The Fourier representation is defined for all $\mu$ as follows:
\be
q_\mu^a(x) = \frac{1}{L^3} \sum_\mathbf{p} e^{i\mathbf{p}\cdot \mathbf{x}} ~ e^{i\theta_a(\mathbf{p},x_0,\mu)} ~
                           q_\mu^a(\mathbf{p},x_0)\,.
\ee
where
\be
q_{\mu}^a(\mathbf{p},x_0)  = q_{|\mu|}^{a}(\mathbf{p},x_0-a\delta_{\mu+4,0})\,,
\ee
and
\begin{align}
e^{i\theta_a(\mathbf{p},x_0,\mu)} &=  \left\{ \begin{array}{l@{\qquad}l}
1 & \mathrm{if} ~ \mu =4 \\
e^{i(p_k+\phi_a(x_0))/2} & \mathrm{if} ~ \mu =k \\
-e^{i(-p_k+\phi_a(x_0))/2} & \mathrm{if} ~ \mu =-k \\
-1                          & \mathrm{if} ~ \mu =-4  \end{array}\right.  \nonumber\\
&  \nonumber\\
&=  \mathrm{sign}(\mu)\biggl(\delta_{|\mu|,4} + (1-\delta_{|\mu|,4}) e^{i( p_{\mu}+\phi_a(x_0))/2}\biggr)\,.
\end{align}
With these notations we have
\be
\langle q_\mu^a(\mathbf{p},x_0) q_\nu^b(\mathbf{p'},y_0) \rangle=
\delta_{b,\bar a} L^3 \delta_{\mathbf{p}+\mathbf{p'},\vectn}
D_{|\mu||\nu|}^{a}\left(\mathbf{p}; \tilde{x}_{0,\mu},\tilde{y}_{0,\nu}\right)\,,
\ee
where $\tilde{x}_{0,\mu}=x_0-a\delta_{\mu+4,0}$, and analogously for $\tilde{y}_{0,\nu}$.

\subsection{Single gluon radiative corrections}\label{sect:radiative corrections}

In this subsection we deal with the perturbative corrections
associated with the term ${\mathcal{O}}^{(2)}$ in \eq{eq:PT_expect_value}.
We consider separately two contributions
\be
{\mathcal{O}}^{(2)} = {\mathcal{O}}^{(\mathrm{2a})} + {\mathcal{O}}^{(\mathrm{2b})}\,.
\ee
We observe that our observable is nothing but
the product of link variables. To compute ${\mathcal{O}}^{(\mathrm{2a})}$ we expand each link $U$
according to the parametrization (\ref{eq:U_parametrization}),
and keep, for each link variable, the terms up to order $g_0$. The gluon fields $q_\mu$
are decomposed as in \eq{eq:q0_SU3} to stress the color structure.
We then contract the gluon fields and express the contractions in terms
of gluon propagators. Each contraction is multiplied by the trace
of the product of two SU(3) matrices $I^a$ and several $V$ matrices. The latter
would coincide with the identity matrix if there was no background field.
The order of the matrices is determined by the contraction
and by the path defining the observable.
In this way we obtain
\be
\hspace{0.5cm}\tr\{ W_{\vec\ell}^{(\mathrm{2a})} \} =
\sum_{j=1}^{\ell} \sum_{j'=j+1}^\ell
\tr\left\{q^{(j)} W_{\vec\ell}(j~\sigma_j| j'~\sigma_{j'})
     q^{(j')} W_{\vec\ell}(j'~\sigma_{j'}|j~\sigma_j) \right\}\,,
\label{eq:W2a1}
\ee
where we have used the cyclicity of the trace and the shorthand
\be
q^{(j)} \equiv q_{\mu_j}(x^{(j)})\,.
\ee
If we now consider for each gauge link only the terms quadratic in $g_0$ in the 
pa\-ra\-me\-tri\-za\-tion (\ref{eq:U_parametrization}), we obtain for ${\mathcal{O}}^{(\mathrm{2b})}$:
\be
\tr\{ W_{\vec\ell}^{(\mathrm{2b})} \} = \frac{1}{2} \sum_{j=1}^\ell
\tr\left\{ \left(q_{\mu_j}(x_j)\right)^2 W_{\vec\ell}[V] \right\}\,,
\ee
which is $1/2$ of the term $j=j'$ on the r.h.s.~of \eq{eq:W2a1}.
We also need the notation
\be
\sigma_j\equiv \frac{1-\mathrm{sign}(\mu_j)}{2}\,,
\ee
and for $n\geq 1$,
\be
\bar n=1+\mathrm{mod}(n-1,\ell)\,.
\ee
We can now formulate the definition
\begin{align}
W_{\vec\ell}(j~\sigma_j| j'~\sigma_{j'})
& = 
\left\{
\begin{array}{ll}
&W_{\vec\ell}[V]\,,\quad\mbox{if $\overline{j+\sigma_j}+\overline{j'+\sigma_{j'}}$ and $\sigma_j =0$}\,,\\
& \label{eq:W_vec}\\
&W_{\vec\ell}(\overline{j+\sigma_j}\to \overline{j'+\sigma_{j'}})\,,\quad\mbox{otherwise,}
\end{array}
\right. \,
\end{align}
that invokes the parallel transporter along the loop from $x^{(j)}$ to  $x^{(j')}$:
\be
W_{\vec\ell}(j\to j') = \left\{ \begin{array}{l@{\qquad} l}
1\,, & \mathrm{if} ~ j= j'\,, \\
\prod_{i=j}^{j'-1} V(x^{(i)},\mu_i)\,, & \mathrm{if} ~ j < j'\,, \\
\prod_{i=j}^{\ell} V(x^{(i)},\mu_i) \prod_{i=1}^{j'-1} V(x^{(i)},\mu_i)\,,& \mathrm{if} ~ j > j'\,.
\end{array}\right.
\ee
One then finds
\begin{align}
 \langle \tr\{ W_{\vec\ell}^{(\mathrm{2a})} \} \rangle_0 &=
 \frac{1}{L^3} \sum_{j=1}^\ell \sum_{j'=j+1}^\ell \sum_{a=1}^8 \sum_\mathbf{p}
 e^{i\mathbf{p}\cdot(\mathbf{x}^{(j)}-\mathbf{x}^{(j')})} ~
e^{i\theta_a(\mathbf{p},x_0^{(j)},\mu_j)} ~
e^{i\theta_{\bar a}(-\mathbf{p},x_0^{(j')},\mu_{j'})}\nonumber \\
& \nonumber\\[-1ex]
& \hspace{-1.7cm}{\scriptstyle \times}~ \tr\{\Omega^a_{\vec\ell}(\mu_j,\mu_{j'})\}
D^a_{|\mu_j||\mu_{j'}|}\left(\mathbf{p};x_0^{(j)}
-a\delta_{\mu_j+4,0} ; x_0^{(j')}-a\delta_{\mu_{j'}+4,0} \right)\,,\\
& \nonumber\\[-1ex]
\Omega^a_{\vec\ell}(\mu_j,\mu_{j'})
&=I^a W_{\vec\ell}(j~\sigma_j| j'~\sigma_{j'}) I^{\bar a}
W_{\vec\ell}(j'~\sigma_{j'}| j~\sigma_j)\,.
\end{align}
We introduce the propagator completely in $x$-space,
\begin{align}
\Delta^a_{\mu\nu}(\mathbf{x};x_0,y_0)&\equiv
\frac{1}{L^3} \sum_\mathbf{p}e^{i\mathbf{p}\cdot\mathbf{x}} ~
e^{i\theta_a(\mathbf{p},x_0,\mu)} ~
e^{i\theta_{\bar a}(-\mathbf{p},y_0,\nu)}\nonumber\\
&\quad \,\,{\scriptstyle \times}~D^a_{|\mu||\nu|}\left(\mathbf{p}; x_0-a\delta_{\mu+4} ; y_0-a\delta_{\nu+4} \right)\,,\label{eq:gluon_x}
\end{align}
which lets us now write
\begin{align}
\langle \tr\{ W_{\vec\ell}^{(\mathrm{2a})} \} \rangle_0 &= \sum_{j=1}^\ell \sum_{j'=j+1}^\ell \sum_{a=1}^8
\tr\{\Omega^a_{\vec\ell}(\mu_j,\mu_{j'})\}\nonumber\\
& \quad\,\,{\scriptstyle \times} ~\Delta^a_{\mu_j\mu_{j'}}(\mathbf{x}^{(j)} - \mathbf{x}^{(j')} ;x_0^{(j)},x_0^{(j')})\,,\label{eq:W2a}\\
&\nonumber\\
\langle \tr\{ W_{\vec\ell}^{(\mathrm{2b})} \} \rangle_0 &= \frac{1}{2} \sum_{a=1}^8
\tr\{ I^a I^{\bar a } W_{\vec\ell}[V] \} \sum_{j=1}^\ell
\Delta^a_{\mu_j\mu_j}(\mathbf{0} ;x_0^{(j)},x_0^{(j)})\,. \label{eq:W2b}
\end{align}

\subsection{Tadpole contributions}\label{sect:tadpole_contributions}

In this subsection we deal with the one-loop terms stemming from
$S_\mathrm{tot}^{(1)}$ in \eq{eq:PT_expect_value}, with the improvement coefficients
set to their tree-level values. They are commonly referred to as 
tadpole contributions, and they owe their existence to the non-vanishing background field.
Let us first write ${\mathcal{O}}^{(1)}$ more explicitly
\be
{\mathcal{O}}^{(1)}=\tr\{W_{\vec \ell}^{(1)} \} = \sum_{j=1}^\ell \tr\{  q_{\mu_j} (x^{(j)}) ~ W_{\vec\ell}[V] \}\,.
\ee
The three contributions coming from the order $g_0$ in the expansion (\ref{eq:eS_expansion}) 
of the action are explicitly given in
\eq{eq:S_gh_1}, \eq{eq:gact_1} and \eq{eq:qact_1} for the ghost, gluon and quark cases respectively.
They enable us to find the formal expression\footnote{Due to the notation employed in this section,
the component $T^a_4$ of the tadpole corresponds to $T_0^a$ of \sect{sect:pt_tadpoles}.}
\be
\langle \tr\{W_{\vec\ell}^{(1)}\}  S^{(1)}_\mathrm{tot} \rangle_0 =
- \sum_{a=1}^8 \sum_{\mu=1}^4 \sum_{u_0} \alpha_{\vec\ell,\mu}^a(u_0) ~ T_\mu^a(u_0)\,,
\label{eq:11}
\ee
with
\begin{align}
\alpha_{\vec\ell,\mu}^a(u_0) &=
\tr\{ I^{\bar a} W_{\vec\ell}[V] \}  \sum_{j=1}^\ell \mathrm{sign}(\mu_j)
 \left( \delta_{|\mu_j|,4} + (1-\delta_{|\mu_j|,4}) e^{-i\phi_a(x_0^{(j)})/2}  \right) \nonumber\\
&\quad  \times D_{\mu|\mu_j|}^a\left(\mathbf{0};u_0,x_0^{(j)} - a\delta_{\mu_j+4,0}\right)\,,\label{eq:alpha_ell}
\end{align}
and
\be
T_{\mu}^a(u_0) = T_{\mu,\mathrm{gluon}}^a(u_0) + T_{\mu,\mathrm{ghost}}^a(u_0) + N_fT_{\mu,\mathrm{quark}}^a(u_0)\,.\label{eq:tad}
\ee
Since $T_{\mu}^a$ vanishes for $\mu=4$, due to CP-invariance, and for the color indices $a$ different 
from 3 and 8, due to the structure
of the vertices, we can rewrite \eq{eq:11} as
\be
\langle \tr\{W_{\vec\ell}^{(1)}\}  S^{(1)}_\mathrm{tot} \rangle_0 =
- \sum_{k=1}^3 \sum_{u_0} \left\{\alpha_{\vec\ell,k}^3(u_0) ~ T_k^3(u_0)+\alpha_{\vec\ell,k}^8(u_0) ~ T_k^8(u_0)\right\}\,,
\label{eq:11b}
\ee
and exploit the explicit expressions for the tadpoles $T^a_k$ given in \sect{sect:pt_tadpoles}.

\subsection{Improvement}\label{sect:pt_improvement}

In order to be able to reach the continuum limit with a rate proportional
to $(a/L)^2$ our observable needs to be improved. 
Since there are no operators of dimension 6 with the same symmetries of ${\mathcal{O}}_\mathrm{spin}$,
non-vanishing at one-loop order, and with no valence quarks,
the improvement amounts
to compute the additional contributions stemming from the volume and boundary counterterms
in the action. From the discussion expounded in \sect{sect:Oa_impr} we infer that the
counterterms proportional to $\cs$ and $\tilde{c}_\mathrm{s}$ vanish, because the background
field is purely electric and our observable does not involve relativistic fermion fields. 
The volume term for the quark action has been taken into account
since the very beginning, through the term proportional to $\csw^{(0)}$. The one-loop
expression of $\csw$ is not needed. It amounts to a correction of order $g_0^4$ to our observable.
Analogously for $\tilde{c}_\mathrm{t}$. Inspection of \eq{eq:delta_SFb} reveals that the counterterm
proportional to the one-loop expression of $\tilde{c}_\mathrm{t}$ leads to a correction to our
observable of order $g_0^4$. The only contribution which one needs to take into account
comes from the boundary counterterm of the gauge action (\ref{eq:improved_g_action}) proportional 
to ${c}_\mathrm{t}^{(1)}$. It is convenient to express the corresponding counterterm in the form 
\be\label{eq:impr_diagram}
\langle \tr\{W^{(1)}_{\vec{\ell}}\}\delta S^{(1)}_\mathrm{tot,b}\rangle_0\,,
\ee
where $\delta S^{(1)}_\mathrm{tot,b}$ reads \cite{thesis:skurth}
\be\label{eq:S1_totb}
\delta S^{(1)}_\mathrm{tot,b}=\frac{2}{\sqrt{3}}c_\mathrm{t}^{(1)}
\sum_{k=1}^{3}[q_k^8(\mathbf{0},a)-q_k^8(\mathbf{0},T-a)][\sin(2\gamma)+\sin(\gamma)]\,,
\ee
and the parameter $\gamma$, defined in \eq{eq:gamma_function}, is non-zero only for a non-vanishing
background field.
The coefficient ${c}_\mathrm{t}^{(1)}$ depends on the flavor number \cite{alpha:SU3,pert:1loop}:
\be\label{eq:ct_1loop}
{c}_\mathrm{t}^{(1)}=-0.08900(5)+0.0191410(1)N_\mathrm{f}\,.
\ee
The explicit expression of (\ref{eq:impr_diagram}) can be computed 
similarly to the tadpoles, and reads
\be\label{eq:full_expr}
\langle \tr\{W^{(1)}_{\vec{\ell}}\}\delta S^{(1)}_\mathrm{tot,b}\rangle_0=\frac{2}{\sqrt{3}}c_\mathrm{t}^{(1)}
[\sin(2\gamma)+\sin(\gamma)]\tr\{I^8 W_{\vec{\ell}}[V]\} \sum_{k=1}^{3}M_{\vec{\ell},k}\,,
\ee
with
\begin{align}
&  \hspace{-1cm}M_{\vec{\ell},k}=\sum_{j=1}^{\ell}\mathrm{sign}(\mu_j)\biggr(\delta_{|\mu_j|,4}+(1-\delta_{|\mu_j|,4})
\mathrm{e}^{-i\phi_8(x_0^{(j)})/2}\biggr)\label{eq:Mlk}\\
             &  \hspace{-0.5cm}{\scriptstyle \times}~
\biggr(D^8_{k|\mu_j|}(\mathbf{0},a,x_0^{(j)}-a\delta_{\mu_j+4,0})-D^8_{k|\mu_j|}
(\mathbf{0},T-a,x_0^{(j)}-a\delta_{\mu_j+4,0}) \biggr)\nonumber\,,
\end{align}
where we exploited the fact that $I^8$ and the background
field are diagonal, and that $I^8=I^{\bar{8}}$.

\subsection{Summary}\label{sect:summary_PT}

By collecting the results of this section, we obtain that the expectation value
of the $\Oa$-improved Wilson loop at one-loop order is given by
\begin{align}
&  \hspace{-3cm}\langle \tr\{W_{\vec\ell}\}\rangle_\mathrm{G}=W_{\vec\ell}[V]
+g_0^2\Bigl(\langle \tr\{ W_{\vec\ell}^{(\mathrm{2a})} \} \rangle_0
+\langle \tr\{ W_{\vec\ell}^{(\mathrm{2b})} \} \rangle_0\nonumber\\[-1ex]
&  \label{eq:summary_1loop}\\[-2ex]
&  \hspace{0.3cm}-\langle \tr\{W_{\vec\ell}^{(1)}\}  S^{(1)}_\mathrm{tot} \rangle_0
-\langle \tr\{W^{(1)}_{\vec{\ell}}\}\delta S^{(1)}_\mathrm{tot,b}\rangle_0\Bigr)\,.\nonumber
\end{align}
The one-loop computation of the Polyakov loop with and without insertion of the 
clover leaf operator defined in \sect{sect:SF_Zspin} has been performed 
with the Matlab code \verb|WLINE| described in chapter \ref{chap:PTcode}.
The tadpole loops do not depend on the observable,
and they have been computed and stored on a file. The tadpole contributions 
$\langle \tr\{W_{\vec\ell}^{(1)}\}  S^{(1)}_\mathrm{tot} \rangle_0$ have then been computed 
with very little effort. The improvement counterterms are needed only for
the Polyakov loop with operator insertion, and their computation is cheap too. 
The most time consuming computation is the one of the gluon propagator.

In order to give an idea of the computational cost,  
for $N_\mathrm{f}=2$ and lattice discretization with $L/a=48$ the computation of all diagrams 
and improvement counterterms for the Polyakov loop with insertion
of the clover leaf operator has been carried out in 2 weeks on a PC,
equipped with a single processor Intel Pentium 4 with 2.6 GHz. 
The scaling can be approximated with a polynomial in $L/a$,
and is asymptotically dominated by the highest power, i.e.~$(L/a)^5$.

\section{Two-loop anomalous dimension and cutoff effects}\label{sect:2loop_AD}

In perturbation theory the renormalization factor of the chromo-magnetic operator
in the \SF scheme can be expanded as
\be\label{eq:Z_SF_PTexp}
Z_\mathrm{mag}^\mathrm{SF}(g_0,L/a)=1+Z_\mathrm{mag}^{(1)}(L/a)g_0^2+\mathrm{O}(g_0^2)\,.
\ee
At tree-level it is one, as a consequence of the definition (\ref{eq:ren_cond_rot}),
while the 1-loop coefficient $Z_\mathrm{mag}^{(1)}$ contains a logarithmic divergence.
It is hence natural to decompose it in a constant term, a logarithmic divergent term,
and other terms vanishing in the continuum limit
\be\label{eq:Z_1_exp}
Z_\mathrm{mag}^{(1)}(L/a)=\chi^{(1)}_\mathrm{SF,lat}-\gamma_0\ln(a/L)+\mathrm{O}(a/L)\,.
\ee
These last terms may be written as a linear combination of the form \cite{impr:Sym1}:
\begin{align*}
\sum_{\ell=0}^1\sum_{n=1}^{\infty}c_{\ell n}\cdot(a/L)^n\ln^\ell(a/L)\,.
\end{align*}
After having implemented the $\Oa$-improvement as described in \sect{sect:pt_improvement},
the second sum starts from $n=2$\,.

In view of a non-perturbative computation of the renormalization factor, we define
the step scaling function 
\be\label{eq:SSF_mag}
\Sigma_\mathrm{mag}^\mathrm{SF}(u,a/L)=
\left.\frac{Z_\mathrm{mag}^\mathrm{SF}(2L)}{Z_\mathrm{mag}^\mathrm{SF}(L)}\right|_{\gbsqSF\,(L)=u}\,.
\ee
at vanishing renormalized quark masses. 
The continuum limit 
\be\label{eq:SSF_CL}
\lim_{a/L\to 0}\Sigma_\mathrm{mag}^\mathrm{SF}(u,a/L)=\sigma_\mathrm{mag}^\mathrm{SF}(u)\,,
\ee
conveys the variation experienced by the 
renormalized operator ${\mathcal{O}}^\mathrm{SF}_\mathrm{spin}(\mu)$ 
when the scale $\mu$ is changed by a factor of two,
\be
{\mathcal{O}}^\mathrm{SF}_\mathrm{spin}(\mu)= \sigma_\mathrm{mag}^\mathrm{SF}(\gbsqSF(L)){\mathcal{O}}^\mathrm{SF}_\mathrm{spin}(2\mu)\,,\quad\mu=1/L\,.
\ee

\begin{figure}[t]
\centering
\includegraphics[scale=0.55]{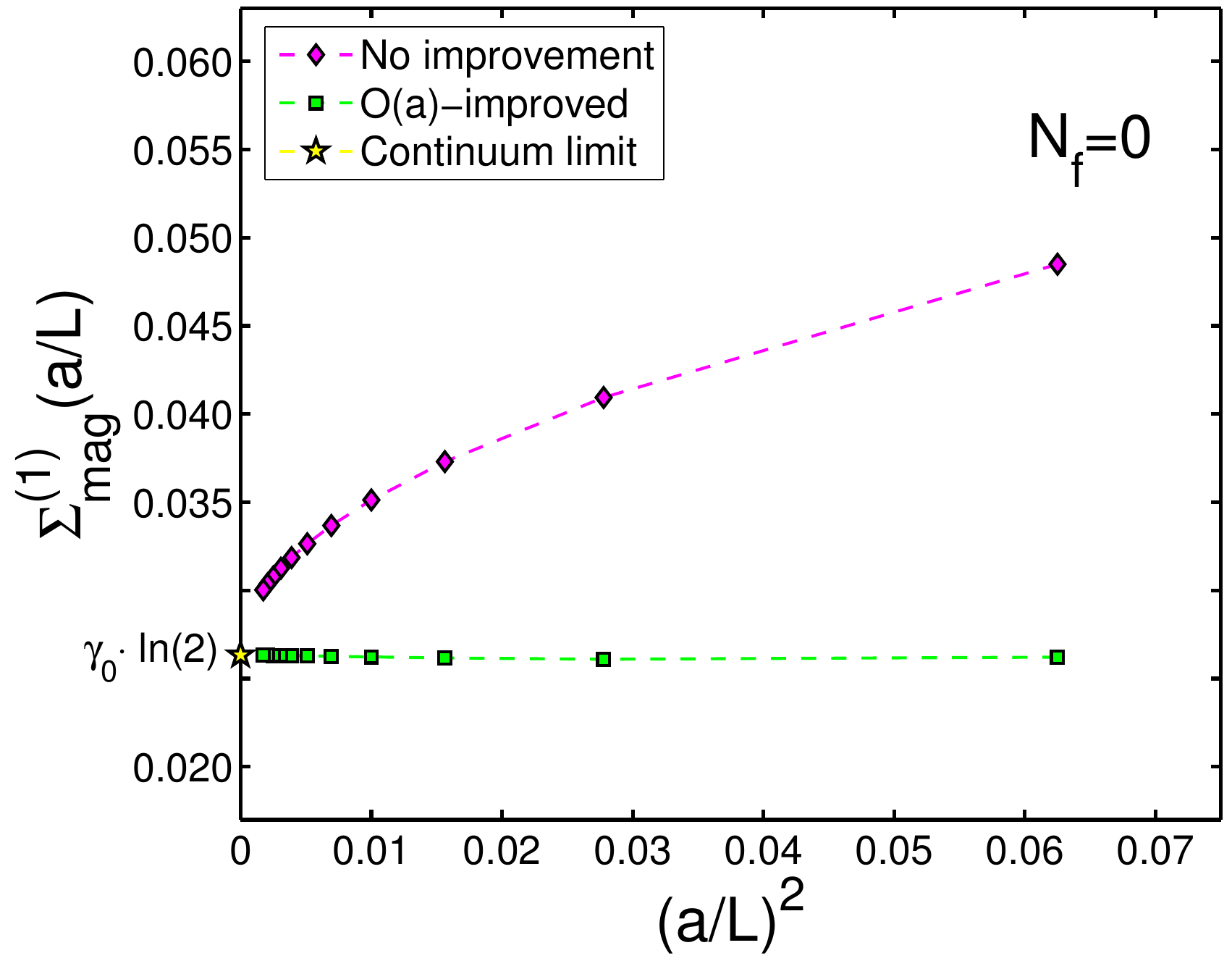}
\mycaption{One-loop contribution $\Sigma_\mathrm{mag}^{(1)}(a/L)$ to the quenched step scaling function
of the chromo-magnetic operator.}\label{fig:Sigma_mag_Nf0}
\end{figure}

{\flushleft The step scaling function (\ref{eq:SSF_mag}) may be expanded in perturbation theory}
\be\label{eq:SSF_mag_exp}
\Sigma_\mathrm{mag}^\mathrm{SF}(u,a/L)=1+\Sigma_\mathrm{mag}^{(1)}(a/L)u+\mathrm{O}(u^2)\,,
\ee
where the one-loop coefficient is given by
\be\label{eq:def_k}
\Sigma_\mathrm{mag}^{(1)}(a/L)=Z_\mathrm{mag}^{(1)}(2L/a)-Z_\mathrm{mag}^{(1)}(L/a)\,,
\ee
whose continuum limit is proportional to the one-loop anomalous dimension
\be\label{eq:CL_k}
\lim_{a/L\to 0} \Sigma_\mathrm{mag}^{(1)}(a/L)\stackrel{(\ref{eq:Z_1_exp})}{=}\gamma_0\ln(2)\,.
\ee
The deviations of $\Sigma_\mathrm{mag}^\mathrm{SF}$ from the continuum limit 
$\sigma_\mathrm{mag}^\mathrm{SF}$ are referred to as cutoff effects, and 
they can be expressed in the form 
\be\label{eq:delta_def}
\delta(u,a/L)=\frac{\Sigma_\mathrm{mag}^\mathrm{SF}(u,a/L)-\sigma_\mathrm{mag}^\mathrm{SF}(u)}{\sigma_\mathrm{mag}^\mathrm{SF}(u)}=
\delta_1(a/L)u+\mathrm{O}(u^2)\,,
\ee
with the one-loop order given by 
\be\label{eq:delta_1_def}
\delta_1(a/L)=\Sigma_\mathrm{mag}^{(1)}(a/L)-\gamma_0\ln(2)\,.
\ee
Here $\delta_1$ can be expanded according to its $N_\mathrm{f}$-dependence,
\be\label{eq:delta_1_Nf_exp}
\delta_1(a/L)=\delta_{1,0}(a/L)+N_\mathrm{f}\,\delta_{1,1}(a/L)\,.
\ee
The computation of $\delta_1$ 
provides important informations upon the cutoff effects
affecting the step scaling function. The results are expected to be of guidance
also for the non-perturbative simulations, at least in the weak coupling regime.
The latter are discussed in the next section.

In \Fig{fig:Sigma_mag_Nf0} the one-loop contribution 
$\Sigma_\mathrm{mag}^{(1)}$ is shown for the quenched case. The lattice
discretization ranges from $L/a=4$ to $L/a=24$. It means that at the finest lattice
we computed $Z_\mathrm{mag}^{(1)}(2L/a=48)$. 
The green squares and the magenta diamonds show the one-loop contribution to 
the step scaling function with and without $\Oa$-improvement counterterms respectively.
The agreement of the continuum limit with the prediction (\ref{eq:CL_k}) is evident,
as well as the effect of the $\Oa$-improvement. 

\begin{figure}[t]
\centering
\includegraphics[scale=0.5]{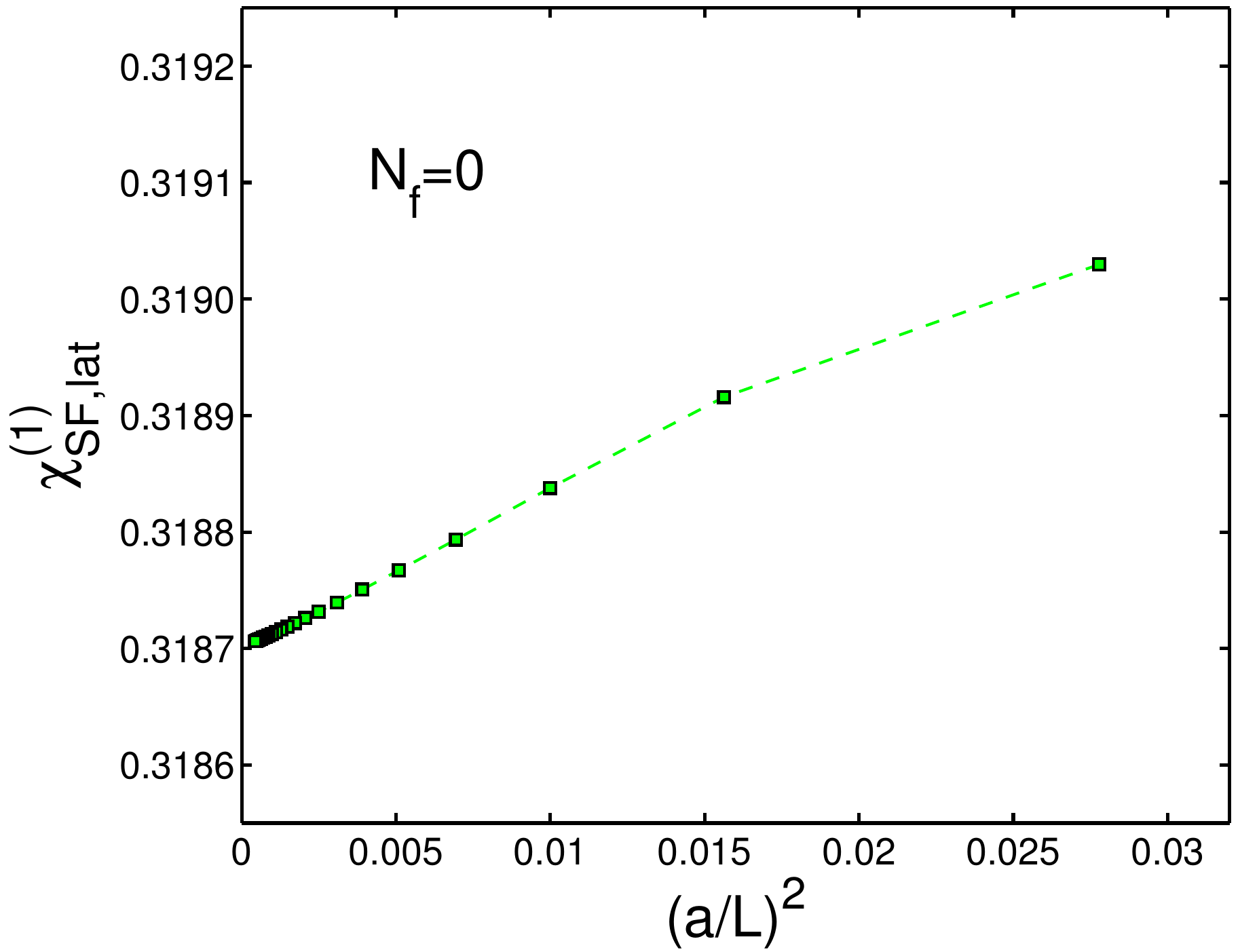}
\mycaption{One-loop order of $Z_\mathrm{mag}$ after removing the logarithmic
divergent part. The plotted points refer to $L/a\geq 6$.}\label{fig:Gamma1_SF_Nf0}
\end{figure}

In \Fig{fig:Gamma1_SF_Nf0} the one-loop order and $\Oa$-improved term $Z_\mathrm{mag}^{(1)}$, after
subtracting the logarithmic divergent part, is shown for the quenched case. 
The lattice discretization ranges from $L/a=4$ to $L/a=48$.
The extrapolated result represents our estimate of $\chi_\mathrm{SF,lat}^{(1)}$
for $N_\mathrm{f}=0$. \newpage

\begin{figure}[t]
\centering
\includegraphics[scale=0.55]{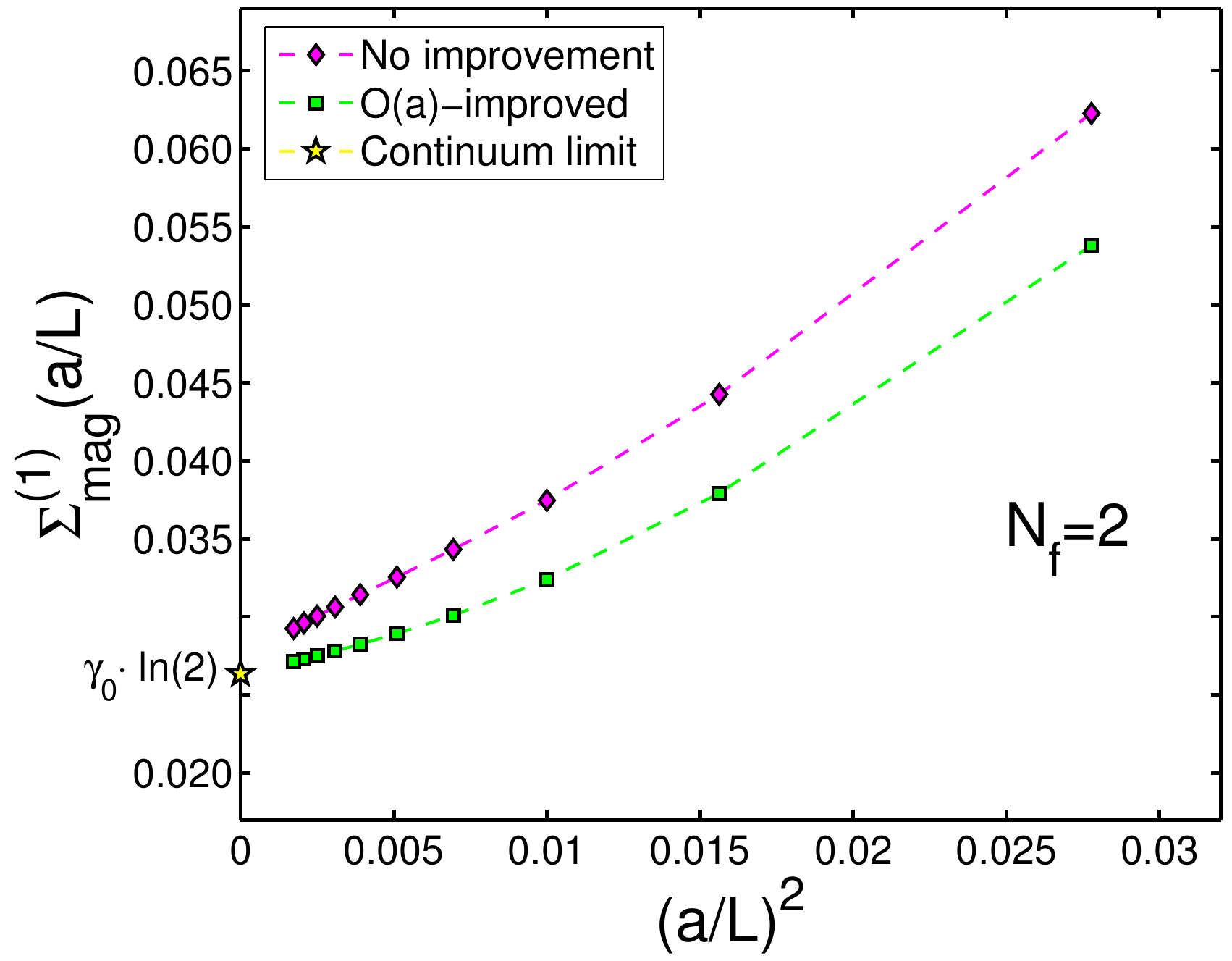}
\mycaption{One-loop contribution $\Sigma_\mathrm{mag}^{(1)}(a/L)$ to the step scaling function
of the chromo-magnetic operator in the $N_\mathrm{f}=2$ case.}\label{fig:Sigma_mag_Nf2}
\end{figure}

The computation with $N_\mathrm{f}\neq 0$ requires, additionally to the quenched case, 
the inclusion of the quark tadpole contributions. 
At this point two remarks are due. The first of them stems from the observation that
the action now includes $N_\mathrm{f}$ mass degenerate light quarks, and our
renormalization condition (\ref{eq:ren_cond_rot}) is intended at 
vanishing quark mass. The latter appears at tree-level in our computation
of the quark tadpoles, which already are one-loop contributions. Therefore
we can fix the mass to zero at tree-level for all lattice discretizations without compromising
the improvement described in \sect{sect:pt_improvement}.
Other choices may be of interest when the comparison with non-perturbative computations
is desired.  The second remark concerns the phase factor $\theta$ appearing in 
the quark boundary conditions (\ref{eq:quark_boundaries4}). For the computation
of the quark tadpoles we take $\theta=-\pi/3$. This choice is motived by the requirement
of having a real renormalization factor $Z_\mathrm{mag}^\mathrm{SF}$. 
According to the analysis of \cite{symm:theta_angle}, the discrete symmetries
of the QCD Sch\"rodinger functional, with the boundary conditions specified in \app{app:Not_Conv},
predict that this can be achieved in \eq{eq:ren_cond_rot} only through the choice $\theta=-\pi/3$.

\begin{figure}[t]
\centering
\vspace{-0.6cm}
\includegraphics[scale=0.5]{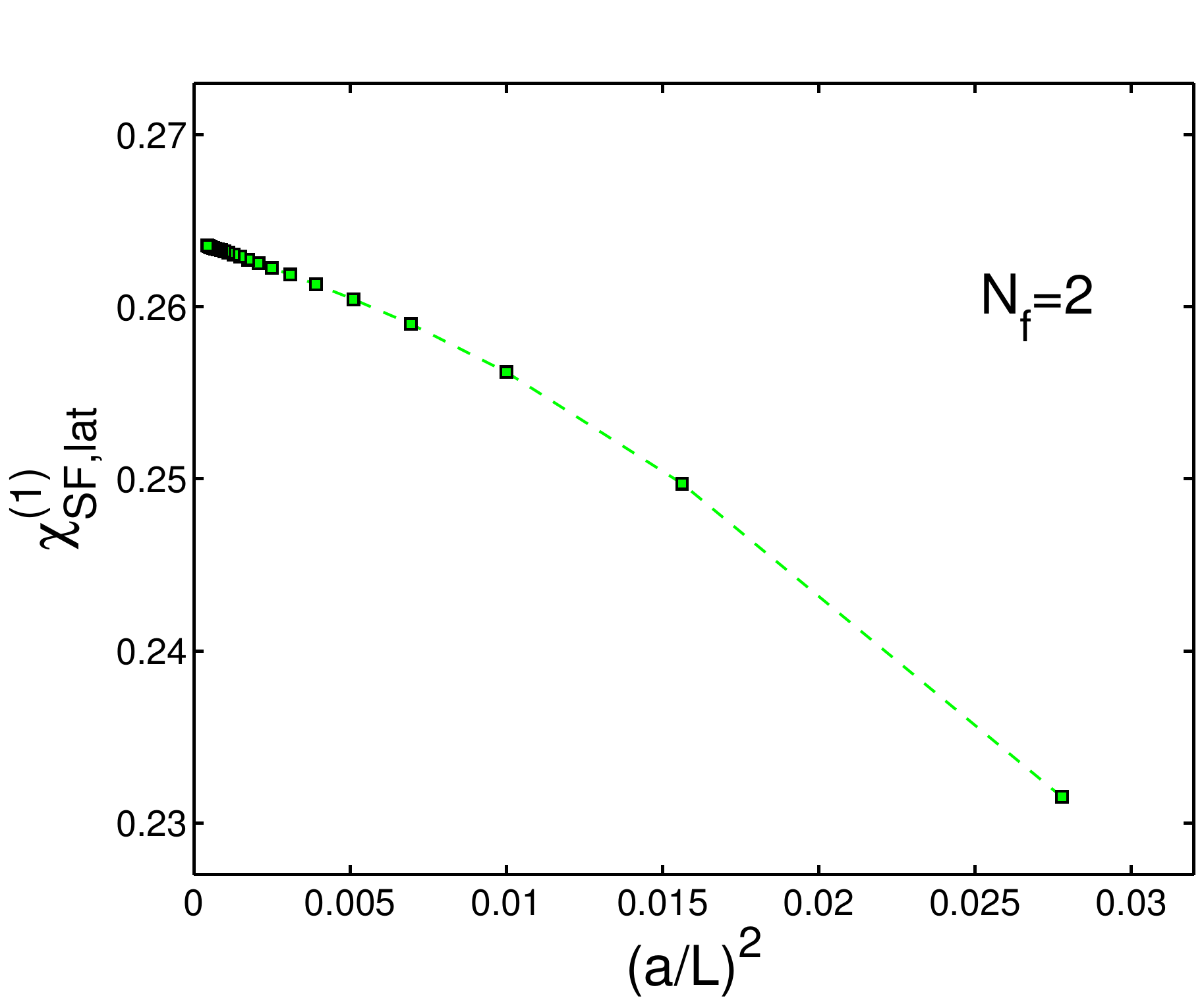}
\mycaption{One-loop order of $Z_\mathrm{mag}$, with $N_\mathrm{f}=2$, after removing the logarithmic
divergent part. The plotted points refer to $L/a\geq 6$.}\label{fig:Gamma1_SF_Nf2}
\vspace{-0.5cm}
\end{figure}

The one-loop contribution to the step scaling 
function in the $N_\mathrm{f}=2$ case is shown in \Fig{fig:Sigma_mag_Nf2}.
Colors and symbols have the same meaning as in the quenched case. The effect of the $\Oa$-improvement
is evident, as well as the fact that, in comparison to the quenched case, the cutoff effects 
are much bigger. Quantitatively, this can be understood by looking at \Tab{tab:cutoff_effects_PT}.\\
\indent The term $Z_\mathrm{mag}^{(1)}$, where the logarithmic divergent part is subtracted,
is shown for $N_\mathrm{f}=2$ in \Fig{fig:Gamma1_SF_Nf2}. There, only the $\Oa$-improved points are plotted.
\newpage

The continuum limits, including the estimate of the uncertainties, are 
performed according to the method described in \cite{pert:2loop_fin}, with
Matlab routines provided by Ulli Wolff. 
The roundoff errors are modelled by assuming that, in double precision,
the rounding amounts to a relative uncertainty of order $10^{-14}$ for the results at $L/a=4$.
This is compatible with the check on the average plaquette described in \sect{sect:av_plaqi}.
As in \cite{pert:2loop_fin}, we assume that the errors grow proportionally to $(L/a)^3$, and
we treat them as a normally distributed superimposed noise, independent
for each lattice discretization. The upshot is that the roundoff errors are negligible
compared to the systematic uncertainties. Another source of error has roots in the improvement
coefficient $\ct$, whose one-loop expression is given in \eq{eq:ct_1loop}. However,
our observables show a very low sensitivity upon variations of $\ct$ within the quoted errors,
and the latter can thus be neglected.\\
\indent The continuum limit for the step scaling functions, both for $N_\mathrm{f}=0$ and for $N_\mathrm{f}=2$,
is in agreement with \eq{eq:CL_k} within a relative systematic error of $\mathrm{O}(10^{-5})$. 
The continuum limit for the one-loop connection between the SF and the lat schemes gives
\be\label{eq:chi_SF_lat_1}
\chi_\mathrm{SF,lat}^{(1)}=0.3187016(1)-0.027448(1)N_\mathrm{f}\,.
\ee
This results can be combined with $\chi_{\MSbar,\mathrm{lat}}^{(1)}$,
appearing in \eq{eq:chi_MSbar_lat_FH}, to get
\begin{align}
\chi_{\mathrm{SF},\MSbar}^{(1)}&= \chi^{(1,0)}_{\mathrm{SF},\MSbar}+\chi^{(1,1)}_{\mathrm{SF},\MSbar}N_\mathrm{f}\,,\\
&  \nonumber\\[-2ex]
\chi^{(1,0)}_{\mathrm{SF},\MSbar}&= -0.0637(3)\,,\quad
\chi^{(1,1)}_{\mathrm{SF},\MSbar}=-0.027448(1)\,,
\end{align}
and finally, through \eq{eq:gamma_1SF},
\be\label{eq:gamma_1SF_result}
\gamma_1^\mathrm{SF}=-0.00236(4)-0.003513(2)N_\mathrm{f}+0.000232 N_\mathrm{f}^2\,.
\ee
The error is dominated by the uncertainty on $\chi_{\MSbar,\mathrm{lat}}^{(1)}$, which
mainly affects the pure gauge contribution. Of course also the coefficient in front of the $N_\mathrm{f}^2$ 
contribution has an error, but it is several orders of magnitude smaller than the others.

\begin{table}[t]
\centering
\begin{tabular}{ccc}
\hline
\hline
& & \\[-1ex]
$L/a$ &$\delta_{1,0}(a/L)$ & $\delta_{1,1}(a/L)$ \\
& & \\[-1ex]
\hline
        & &  \\[-2ex]
4  &  -0.000116&  0.036731\\
        & &  \\[-2ex]
6  &  -0.000236&  0.013742\\
        & &  \\[-2ex]
8  &  -0.000165&  0.005791\\
        & &  \\[-2ex]
10 &  -0.000106&  0.003026\\
        & &  \\[-2ex]
12 &  -0.000072&  0.001876\\
        & &  \\[-2ex]
14 &  -0.000051&  0.001296\\
        & &  \\[-2ex]
16 &  -0.000038&  0.000956\\
        & &  \\[-2ex]
18 &  -0.000029&  0.000738\\
        & &  \\[-2ex]
20 &  -0.000023&  0.000588\\
        & &  \\[-2ex]
22 &  -0.000019&  0.000481\\
        & &  \\[-2ex]
24 &  -0.000016&  0.000400\\
        & &  \\[-2ex]
\hline
\hline
\end{tabular}
\mycaption{Lattice spacing effects of $\Sigma_\mathrm{mag}$ at one-loop order.}
\label{tab:cutoff_effects_PT}
\end{table}


\section{Non-perturbative step scaling functions}\label{sect:NP_ssf}

In order to perform a non-perturbative study of the cutoff 
effects affecting $\Sigma_\mathrm{mag}^\mathrm{SF}$, and com\-pa\-re them with the predictions
of perturbation theory, we computed in the quen\-ched ap\-pro\-xi\-ma\-tion 
two step scaling functions at weak coupling. The simulations parameters
are taken from \cite{mbar:pap1}, and refer to the SF renormalized
couplings $u=\gbsqSF=(0.9944,1.3293)$.

In order to improve the statistical precision, 
we fully exploit translational invariance
and the equivalence of the three spatial coordinates.
In other words, the Polyakov loop is computed in the directions $x,\,y$ and $z$,
and we average over them.
The gauge links building up
the Polyakov loop, but not the inserted clover leaf operator, are evaluated by a 10-hit 
multi-hit procedure, as described in \cite{PPR}. It means that on the time slice $x_0=L/2$,
each link $U_k(x)$ is replaced by
\be\label{eq:ubar_def}
\overline{U}_k(x)=\frac{\int\!\mathrm{d}U_k(x)\,U_k(x)\mathrm{e}^{-S_\mathrm{G}[U_k(x)]}} 
{\int\!\mathrm{d}U_k(x)\,\mathrm{e}^{-S_\mathrm{G}[U_k(x)]}}\,,\quad k=1,2,3\,,
\ee
where the integration is performed over only one SU(3) matrix and depends
on the neighboring links. Each hit consists of one heathbath
update according to the algorithm described in \sect{sect:MC_path_integral}.

\begin{figure}[t]
\centering
\begin{minipage}[t]{0.5\textwidth}
\centering \includegraphics[width=6.5cm,height=5.3cm]{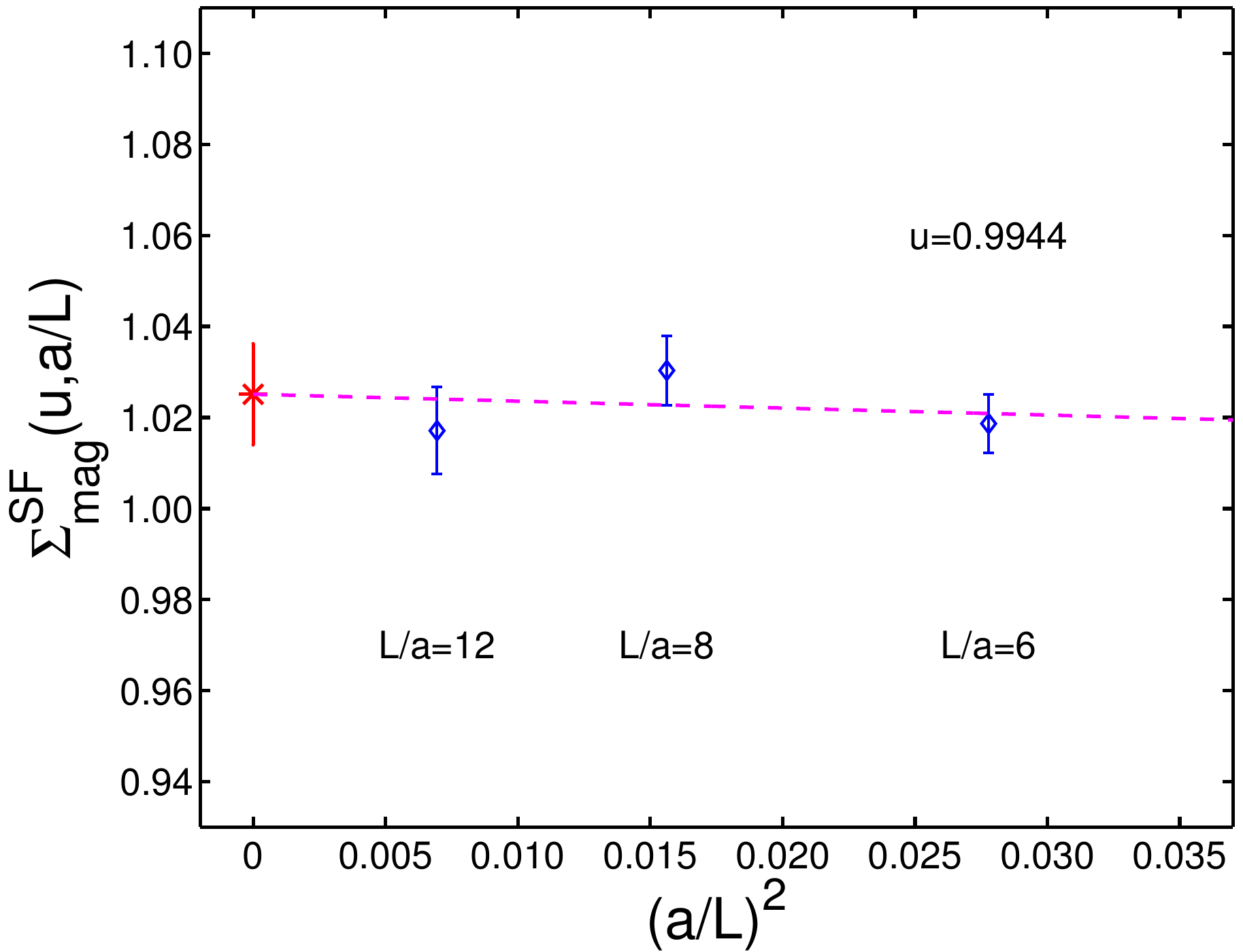}
\end{minipage}%
\begin{minipage}[t]{0.5\textwidth}
\includegraphics[width=6.5cm,height=5.3cm]{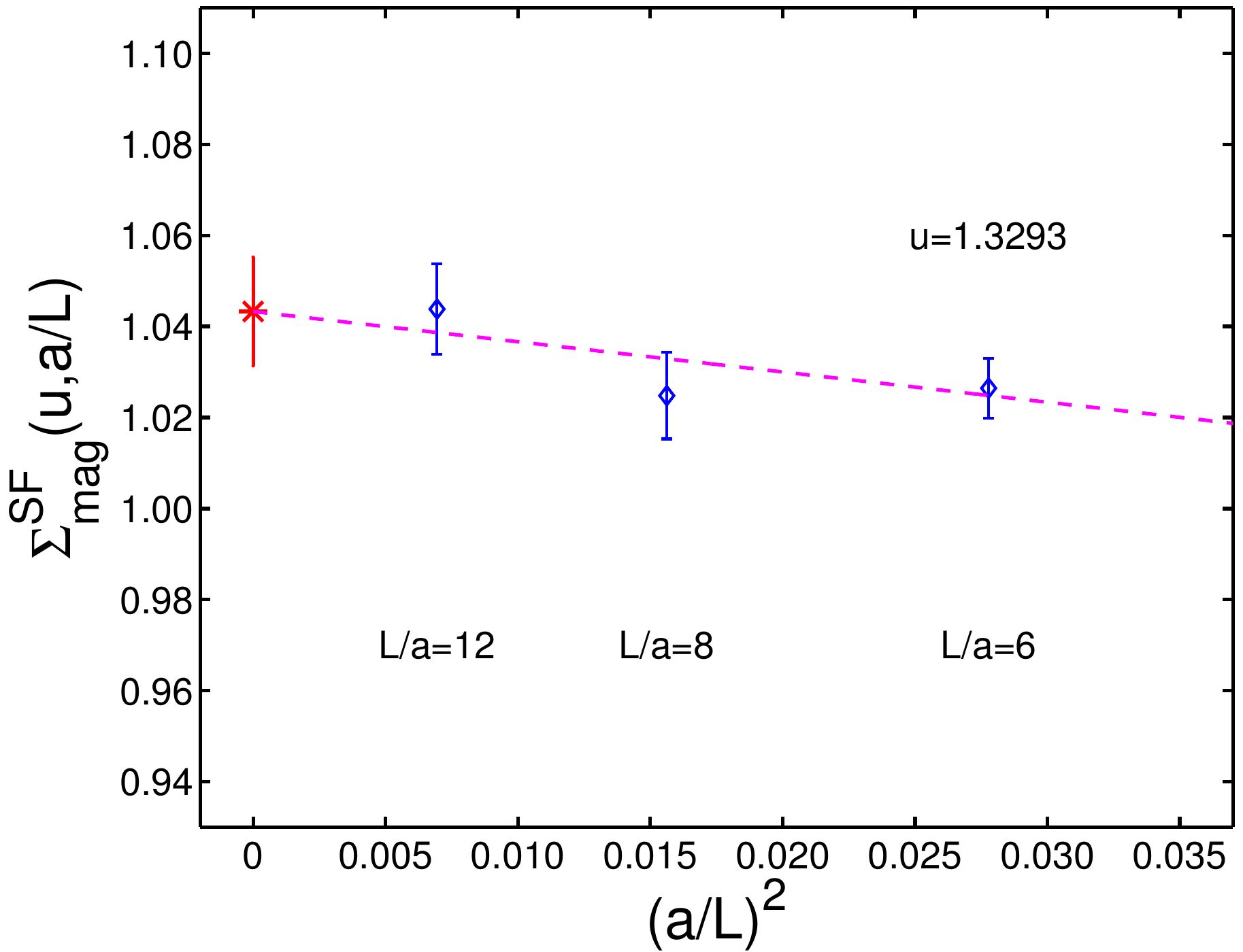}
\end{minipage}
\mycaption{Continuum limit extrapolations for the step scaling function $\Sigma_\mathrm{mag}^\mathrm{SF}$.}\label{fig:NP_SSF}
\end{figure}

The simulation parameters and the results are reported in \Tab{tab:NP_SSF}; the latter are
plotted in \Fig{fig:NP_SSF}. The continuum limit extrapolations are performed
with the ansatz 
\be\label{eq:CL_ansatz}
\Sigma_\mathrm{mag}^\mathrm{SF}(u,a/L)=\sigma_\mathrm{mag}^\mathrm{SF}(u)+\rho(u)({a/L})^2\,,\quad u=\gbsqSF(L)\,,
\ee
and the results are reported in the following table.\vspace{0.3cm}
\begin{center}
\begin{tabular}[h]{ccc}
\hline
\hline
        & &  \\[-1ex]
    $u$ & $\sigma_\mathrm{mag}^\mathrm{SF}(u)$ & $\rho(u)$ \\
        & &  \\[-1ex]
\hline
        & &   \\[-1ex]
    0.9944    &1.025(11) &  -0.15(53) \\
        & &   \\[-1ex]
\hline
        & &   \\[-1ex]
    1.3293    &1.043(12)  &  -0.66(55) \\
        & &   \\[-1ex]
\hline
\hline
\end{tabular}
\end{center}\vspace{0.3cm}
As predicted by perturbation theory, in the quenched approximation the cutoff effects
in the weak coupling regime are indeed invisible, as long as
the statistical precision is around one percent.

The non-perturbative computation
of these step scaling functions is part of a work \cite{Zspin:me, Grozin:2007qv}, whose main result is
the computation of $Z_\mathrm{mag}^\mathrm{RGI}$ and of the $\mathrm{B_{(s)}^\ast}-\mathrm{B_{(s)}^{\phantom{\ast}}}$ mass splitting
in the quenched approximation.

\begin{table}[b]
\centering
\begin{tabular}[h]{crllc}
\hline
\hline
        & & & &\\[-1ex]
 $\beta$ & $L/a$ & $Z_\mathrm{mag}^\mathrm{SF}(g_0,L/a)$ & $Z_\mathrm{mag}^\mathrm{SF}(g_0,2L/a)$ & 
$\Sigma_\mathrm{mag}^\mathrm{SF}(u,a/L)$\\
        & & & &\\[-1ex]
\hline
& & & &\\[-2ex]
\multicolumn{5}{c}{$u=0.9944$} \\
& & & &\\[-2ex]
\hline
        & & & &\\[-1ex]
   10.0500 &6 &1.3651(44) &1.3905(76) &1.0186(64)\\
   10.3000 &8 &1.3514(52) &1.3924(88) &1.0303(76)\\
   10.6086 &12&1.3608(53) &1.384(12) &1.0171(96)\\
        & & & &\\[-1ex]
\hline
& & & &\\[-2ex]
\multicolumn{5}{c}{$u=1.3293$} \\
& & & &\\[-2ex]
\hline
        & & & &\\[-1ex]
    8.6129 &6 &1.4727(57) &1.5116(77)& 1.0264(66)\\
    8.8500 &8 &1.4664(71) &1.503(12) &1.0248(95)\\
    9.1859 &12&1.4528(65) &1.517(13) &1.0438(99)\\
        & & & &\\[-1ex]
\hline
\hline
\end{tabular}
\mycaption{Results for the (quenched) non-perturbative step 
scaling function $\Sigma_\mathrm{mag}^\mathrm{SF}$. The simulations have been performed with 
$\ct$ set to its two-loop value (\ref{eq:ct_2loop}).}\label{tab:NP_SSF}
\end{table}

\clearpage

\section{Summary}\label{sect:Zspin_summary}

The spin splitting in heavy-light quark bound states 
is a key quantity for testing the validity of the 
Heavy Quark Effective Theory, both in the bottom channel 
and the charm one. Lattice computations have started in the quenched approximation, 
providing, so far, results
in disagreement with the experimental predictions. These computations used 
perturbative estimates for the renormalization factor of the chromo-magnetic operator,
introducing a hardly controllable source of uncertainty. 

In this chapter 
a practicable way to non-perturbatively renormalize that operator is introduced. 
The \SF scheme allows
to compute the renormalization factor in a low energy regime, where phenomenology 
directly applies, and to evolve it to high energy scales, where perturbation theory 
can be safely used to finally compute the renormalization group invariant 
renormalization factor. There, a one-loop connection between
the SF and the widely used $\MSbar$ schemes is performed
by exploiting exact relations stemming from the renormalization group equations.
For the latter scheme, the two-loop
anomalous dimension is known, and the here computed connection enables to get 
it for the SF scheme too. The computation
of the two-loop anomalous dimension represents the main result of the chapter, and constitutes
an important ingredient for a precise evaluation of the scale running
of the SF renormalization factor.

In the SF scheme a renormalization condition is imposed by 
introducing correlation functions, which do not involve valence quarks. 
In the quenched approximation, one then 
deals with pure gauge quantities, which considerably simplifies the 
perturbative computations. As light fermions are introduced in the action,
their presence emerges only at the $n$-loop order, with $n\geq 1$.

In order to have a renormalization factor with a non-vanishing tree-level,
a background field has been introduced. Unfortunately, this has the disadvantage
of making the one-loop perturbative computation much more complicated than in the 
vanishing background field case. These difficulties have been faced by developing
a program which automatically generates and computes the required diagrams, as well as 
the necessary $\Oa$-improvement counterterms, for any closed loop and at the perturbative 
one-loop level.
Among the features of the program, it is worth to mention that,
due to the space-time locality
of the involved observables, it is advantageous to compute the gluon loops
in coordinate space. The ghost, gluon and quark propagators are computed through 
recursive methods. The details on those, and more, are explained
in the next chapter.

The computation of the two-loop anomalous dimension does not represent 
the only one result of perturbation theory. 
In view of a non-perturbative computation 
of the renormalization factor $Z_\mathrm{mag}^\mathrm{SF}$ for a wide range
of renormalization scales, \sect{sect:2loop_AD} contains
an interesting one-loop analysis of the cutoff effects affecting 
the step scaling functions. The latter measure the variation which the
factor $Z_\mathrm{mag}^\mathrm{SF}$ undergoes, when the renormalization scale is changed
by a factor of two. The one-loop predictions are expected to provide a guidance for
the non-perturbative computations, at least in the weak coupling regime. 
This has been confirmed in \sect{sect:NP_ssf}, whose results show that, at 
least in the quenched case, the cutoff effects are invisible as long as the relative statistical
precision is around one percent. For the unquenched case,
perturbation theory predicts bigger but still moderate cutoff effects. They may have a strong dependence
on the chosen kinematical parameters, and, unfortunately, a non-perturbative check is still missing.

A final remark has roots in the observation that the connection between the \SF and 
the $\MSbar$ schemes is achieved by exploiting as 
intermediate step the lattice minimal subtraction scheme. The one-loop connection 
between the latter and the $\MSbar$ scheme is known from the literature 
with less precision than achieved here for the SF scheme; it dominates the uncertainty on the two-loop
SF anomalous dimension. As a valuable check,
one may also think of performing a direct two-loop computation in the SF scheme.

\chapter{The Matlab code \texttt{WLINE}}\label{chap:PTcode}

In this chapter we describe the Matlab code 
used for our perturbative computations
in the \SF scheme. We decided to use Matlab in order 
to combine comfortable programming, simplicity and readability
of the code, robustness of the libraries and an acceptable speed for the 
observables and lattice discretizations of chapter \ref{chap:Zspin}.

The first section is devoted to an overview of the
structure of the code \texttt{WLINE} in its released version $1.0$. 
The latter computes
the tree-level and the 1-loop order of the perturbative
expansion of any closed Wilson line
within the \SF scheme on the lattice.
The section provides the necessary informations to enable
the user to compute the desired Wilson line, by specifying
very few input parameters. The generation and the computation of all diagrams,
including the $\Oa$-improvement, 
are then automatically executed. 
With ``automatic generation'' of the diagrams and the improvement counterterms, 
we intend that all of them are already implemented inside the code.  
Furthermore, the program has been also conceived
as an ensemble of several modules. Each propagator, tadpole, improvement counterterm
or diagram is separately accessible.

The second section deals with the free ghost, gluon and quark propagators,
and the attention is focused on the techniques used for their numerical computation.
The presence of a non-vanishing background field prohibits from obtaining
a simple explicit analytical expression for them. A brute force approach would require
the inversion of big sparse matrices, and would be computationally more expensive for fine lattices.
However, clever recursive methods have been developed 
by the authors of \cite{pert:2loop_SU2,pert:2loop_fin}, and they are implemented 
in our code.
To each kind of propagator we dedicate a subsection, which is in turn
divided into three parts. The first part contains the basic definitions and 
the method used for the numerical computation. In the second part we give
the necessary informations which enable to use the implemented subroutines
to compute the propagators. The third part is dedicated to the analytical proof
of the recursive method used for the computation. The user who just intends
to use the program as a black box may skip this part in a first reading.
The analytical proofs are partially based on \cite{propagators:narayanan_notes}.
The computation of the propagators through the recursive method has been successfully
checked, up to machine precision, against the aforementioned method, 
which involves the inversion of big sparse matrices. Further details are given in the 
corresponding subsections.

The third section is mainly dedicated to the ghost, gluon and quark tadpoles.
The latter could be computed through the explicit expressions of the vertices,
provided e.g.~by \cite{thesis:skurth}. However, the formulae given 
in \cite{impr:csw_int_notes}, which have been analytically proven, are 
much simpler to implement and have been found
to significantly reduce the computational effort.
To each kind of tadpole a subsection is dedicated, which is in turn
divided into two parts. While the first part contains the basic definitions, the second
one provides the necessary informations to use the implemented subroutines.
The last subsection is devoted to the $\Oa$-improvement.

The fourth section provides a short report on two tests of the code. 
The first test stems from the simple observation, provided by Creutz in \cite{Creutz:1984mg}, that,
by approximating the gluon action to be quadratic in the gluon fields,
one can derive an exact and analytical expression for the expectation value of the
average plaquette just by counting the number of gluonic degrees of freedom.
This expression has been successfully checked against our code.
The second (successful) test consists in comparing the one-loop results for 
the average plaquette and the Polyakov loops, as defined in chapter \ref{chap:Zspin}, 
produced by \verb|WLINE|,
with the results of the quenched non-perturbative computations, performed at very weak coupling.  
Finally, we remark that the aforementioned observables have been computed
for several values of the gauge fixing parameter $\lambda_0$, appearing in \eq{eq:S_gf},
always giving results consistent within machine precision. 
The consistency does not deteriorate when one increases
the fineness of the lattice. 

The fifth section briefly summarizes the whole chapter, and provides a few observations
and an outlook.  

We remark that throughout the chapter we work in units of the lattice spacing,
and for any unspecified notation we refer to \app{app:Not_Conv}.

\section{The automated code}\label{sect:automatic}

The code is available on request or for download from the webpage
\begin{verbatim}
http://www-zeuthen.desy.de/alpha/
\end{verbatim}
in the directory \verb|Internal/PT software|.
After unpacking the file 
\begin{verbatim}
WLINE.tar.gz
\end{verbatim}
the following directories
are available:

\begin{supertabular}{p{.28\linewidth}p{.65\linewidth}}
\\
        \verb|WLINE_LAUNCH|&\begin{minipage}{1\linewidth} Main directory. It contains the files where
the input parameters have to be specified, and a short script, which launches the program execution.\end{minipage}\\
\\
        \verb|WLINE_SIMPLE|&\begin{minipage}{1\linewidth} The pure gauge 1-loop diagrams, not including the tadpoles,
are generated and computed. Tree-level computation.\end{minipage}\\
\\
        \verb|TADPOLES_GAUGE|&\begin{minipage}{1\linewidth}  Gluon and ghost tadpoles.\end{minipage}\\
\\
        \verb|TG_CONTRIBUTIONS|&\begin{minipage}{1\linewidth} The contribution of the gluon and ghost
tadpoles to the desired observable. \end{minipage}\\
\\

        \verb|TADPOLE_QUARK|&\begin{minipage}{1\linewidth} Quark tadpoles.\end{minipage}\\
\\
        \verb|TQ_CONTRIBUTIONS|&\begin{minipage}{1\linewidth} The contribution of the quark
tadpoles to the desired observable.\end{minipage}\\
\\
\verb|IMPROVEMENT|&\begin{minipage}{1\linewidth} $\Oa$-improvement counterterms.\end{minipage}\\
\\
\end{supertabular}

{\flushleft Inside} the main directory \texttt{WLINE\_LAUNCH} there are six files:
\bi
\item \verb|topo_parameters.dat|\vspace{-0.2cm}
\item \verb|starting_point.dat| \vspace{-0.2cm}
\item \verb|wilson_path.dat|    \vspace{-0.2cm}
\item \verb|wline_automatic.m|\vspace{-0.2cm}
\item \verb|check_path.m|\vspace{-0.2cm}
\item \verb|generateP.m|
\ei
The first of them contains a column vector of length 12, where the lattice topological
parameters and other options are specified. An example is the following
\begin{verbatim}
less topo_parameters.dat

4      %T
4      %L
0.0    %eta
0.0    %nu
1      %if !=0 ==> Activation of the backgroung field
1.0    %lambda_0, gauge fixing parameter
1.0    %1.0=Unix/Linux, 2.0=Windows
1      %Observable: 0, 1 or 2  
-1.04719755119660 %theta_angle=-pi./3 
1.0    %c_sw_0
0.0    %mass_0
2      %Nf
\end{verbatim}
The first two parameters are easy to interpret; they specify the temporal and spatial 
lattice extensions in units of the lattice spacing. The third and fourth parameters
define the non-vanishing background field according to \eqs{eq:angles}. The latter is activated if and only if
the fifth parameter is different from zero. The exact value does not matter at all,
as long as it does not vanish. If it vanishes, the computation is executed
with vanishing background field and boundary fields $C=C'=0$. The sixth parameter specifies the gauge fixing parameter.
The value given in the example, i.e.~$1.0$, corresponds to the Feynman gauge. It can be used to check that
the final result, consisting in the sum of all diagrams, does not depend on the choice of the gauge.
This is possible only in the non-vanishing background field case. For vanishing background field
only the Feynman gauge is available. If another choice is specified, the execution is interrupted.
The value of the seventh parameter has to be chosen according to the kind of operating system.
The difference between Unix/Linux and Windows is for the code important only for the way
the paths are internally specified.
The eighth parameter can be 0, 1 or 2 only. In the first two cases the function \verb|generateP.m| is used
for the computation of Polyakov loops. If it is set to 2, it allows for a general Wilson loop. 
The last four parameters are dedicated to the quark sector. They are of relevance if and only if 
the last of them, specifying the number of mass degenerate light fermions involved, differs from zero. It can
also be a negative integer. The parameter number nine specifies the phase factor $\theta$,
appearing in \eqs{eq:quark_boundaries4}. The following parameter specifies the improvement
coefficient $\csw$ for the fermionic action as in \eq{eq:impr_vol_term}. 
The penultimate parameter corresponds to the mass of all degenerate light fermions.

In all cases, all 12 parameters have to be provided before starting the program. If this
condition is not respected, the execution is interrupted.

Let us now have a look to the file \verb|starting_point.dat|. An example is
\begin{verbatim}
less starting_point.dat

0 0 0 2
\end{verbatim}
It just contains a row vector made of four integers. They specify the
Euclidean coordinates, in lattice units, of the Wilson line starting point with order $(x,y,z,t)$. 
Similarly for the file \verb|wilson_path.dat|, whose content can read
\begin{verbatim}
less wilson_path.dat

3 3 3 3
\end{verbatim}
It is an integer row vector as well, specifying the path covered
by the Wilson line from the starting point. The given example
indicates that the Wilson line is the product of four gauge links
pointing in the $z$-direction. If one wants to insert an electric
plaquette, lying on the $xt$-plane, in the middle of the line, it is sufficient to write the path as
\begin{verbatim}  
3 3 1 4 -1 -4 3 3
\end{verbatim}
Translational invariance can be checked by replacing the aforementioned path by
\begin{verbatim}  
3 1 4 -1 -4 3 3 3
\end{verbatim}
where, given the default starting point (0,0,0,2), the plaquette has been inserted
at the point (0,0,1,2).
The file \verb|generateP.m| contains the function
\begin{verbatim}
function [res] = generateP(option_ins)
\end{verbatim}
which is used if and only if the eighth input parameter in \verb|topo_parameters.dat| is 0 or 1.
According to the definitions of chapter \ref{chap:Zspin},
in the first case the Polyakov loop in the $z$-direction is chosen as observable, while in the
second case the observable is the loop with the insertion of the 
clover leaf operator. In both cases there is no need to specify the content of \verb|wilson_path.dat|.
The latter is produced by the aforementioned function. 

Let us suppose that the program user has chosen his Wilson line,
and coherently specified the input parameters in the three files, which have
just been described. For the computation he can consider the
rest of the program as a black box; he is only asked to start Matlab
and then from the command shell write:
\begin{verbatim}      
>wline_automatic
\end{verbatim}
The user will not interact with the program anymore, no further
specifications are needed. The computation of all 1-loop diagrams, including
the tadpoles, the tree-level and the $\Oa$-improvement are then automatically executed,
and the final result for the 1-loop expression, normalized by the tree-level, is printed out.
The tree-level value is finally accessible by typing 
\begin{verbatim}      
>tree_level
\end{verbatim}
As anticipated at the beginning of the chapter, the expectation value of each 
diagram is accessible:
\begin{verbatim}
>W2a
>W2b
\end{verbatim}
\begin{verbatim}
>Tad_gauge
>Tad_quark
>impr
\end{verbatim}
By referring to \sect{sect:Wilson_loops} for the notation, the commands 
\verb|W2a| and \verb|W2b| produce as output
\begin{displaymath}
\langle \tr\{W^{(2a)}_{\vec\ell}\}\rangle_0\quad 
\mbox{and}\quad \langle \tr\{W^{(2b)}_{\vec\ell}\rangle_0
\end{displaymath}
respectively, \verb|Tad_gauge| the sum of the tadpole contributions
\begin{displaymath} 
\langle \tr\{W^{(1)}_{\vec\ell}S_\mathrm{tot}^{(1)}\}\rangle_{0,\mathrm{g}}
\end{displaymath}
restricted to the ghost and gluon cases,
\verb|Tad_quark| the quark tadpole contribution
\begin{displaymath}
\langle \tr\{W^{(1)}_{\vec\ell}S_\mathrm{tot}^{(1)}\}\rangle_{0,\mathrm{q}}\,,
\end{displaymath}
and \verb|impr| the sum of the $\Oa$-improvement 
counterterms 
\begin{displaymath}
\langle \tr\{W^{(1)}_{\vec\ell}\delta S_\mathrm{tot,b}^{(1)}\}\rangle_{0}\,.
\end{displaymath}
Lower level parts of the computation, e.g.~the propagators, can be accessed
by following the instructions of the next sections.

The file \verb|check_path.m| contains a function which performs some 
checks at the very beginning of the computation. In particular, it checks
whether the Wilson line is really a closed path, and whether the starting point
and the topological parameters have been completely and consistently specified.

\section{The propagators}\label{sect:pt_propagators}

\subsection{The ghost propagator}\label{sect:ghost_propagator}

\subsubsection{Definition}

Given the expansion of the ghost action
\be\label{eq:ghost_action}
S_\mathrm{gh}=\sum_{n=0}^\infty \frac{1}{n!}g_0^nS_\mathrm{gh}^{(n)}\,,
\ee
we consider the $n=0$ term, which is quadratic in the ghost fields, and reads
\be\label{eq:ghost_action_0}
S_\mathrm{gh}^{(0)}=\frac{1}{L^3}\sum_{\vecp}\sum_{t,t'}
\bar{c\kern+0.5pt}^{\bar{a}}(-\vecp,t)F^a(\vecp;t,t'){c\kern+0.5pt}^a(\vecp,t')\,.
\ee
Since in our computations the ghost propagator is present only in connection
with the gluon-ghost-ghost vertex, and the latter is non-vanishing only if it involves
a neutral gluon and the propagator for charged ghost fields, we need $F^a$ only for $a$ different from 3 and 8, 
that means
\be\label{eq:Fa}
F^a(\vecp;t,t')=\delta_{t,t'}[2+\vecs^a(\vecp,t)^2]-\delta_{t+1,t'}-\delta_{t-1,t'}\,.
\ee
The free ghost propagator is given by
\be\label{eq:free_ghost_prop}
\langle {c\kern+0.5pt}^a(\vecp,t) \bar{c\kern+0.5pt}^{b}(\vecpprime,t') \rangle_0=
\delta_{b,\bar{a}}L^3\delta_{\vecp+\vecpprime,\vectn}\,D^a(\vecp;t,t')\,,
\ee
where $D^a$ is the inverse of $F^a$. However, the computation of the propagator
through the inversion of the matrix $F^a$ is computationally not very efficient.
A better way, which exploits recursive relations, is explained in the following.

We fix the color index $a$ and the momentum $\vecp$, and write down the propagator equation
in momentum space with a simplified notation as
\be\label{eq:prop_eq_gh}
-D(t+1,t')+B(t)D(t,t')-D(t-1,t')=\delta_{t,t'}\,,
\ee 
where 
\be\label{eq:prop_eq_gh_rand}
D(0,t')=D(T,t')=0\,,\quad 1\leq t'\leq T-1\,.
\ee
We introduce the forward solution $\psif$, satisfying
\begin{align}
\psif(0)&= 0\,,\quad \psif(1)=1\,,\label{eq:3_gh}\\
\psif(t+1)&= B(t)\psif(t)-\psif(t-1)\,,\quad 1\leq t\leq T-1\,,\label{eq:4_gh}
\end{align}
and the backward solution $\psib$, for which
\begin{align}
\psib(T)&= 0\,,\quad \psib(T-1)=1\,,\label{eq:5_gh}\\
\psib(t-1)&= B(t)\psib(t)-\psib(t+1)\,,\quad 1\leq t\leq T-1\,.\label{eq:6_gh}
\end{align}
Both solutions can be recursively computed and stored. The forward solution
allows to compute $W=\psif(T)$, which is finally used to compute
the solution of
\eq{eq:prop_eq_gh}, with boundary conditions given by \eqs{eq:prop_eq_gh_rand}, as
\begin{align}
\hspace{-0.7cm}D(t,t')
& = 
\left\{
\begin{array}{ll}
\frac{1}{W}\psif(t)\psib(t')\,,\quad&\mbox{for}\,\,t\leq t'\,,\\
\label{eq:solution_gh}\\
\frac{1}{W}\psif(t')\psib(t)\,,\quad&\mbox{for}\,\,t\geq t'\,.
\end{array}
\right. 
\end{align}

\subsubsection{Implementation}

The implementation of the recursive method for the computation of the 
ghost propagator can be found in the program \verb|WLINE|\newpage
as a function in
\begin{verbatim}
TADPOLES_GAUGE/ghost_p.m
\end{verbatim}
Once one is in the directory \verb|TADPOLES_GAUGE|,
it is necessary to load part of the global variables contained in
file \verb|topo_parameters.dat|,
\begin{verbatim}
load_topo_parameters_gtadpole
\end{verbatim}
and initialize the remaining global variables through
the command
\begin{verbatim}
initialization
\end{verbatim}
The function computing the ghost propagator is 
\begin{verbatim}
function [res] = ghost_p(a_su3,p)
\end{verbatim}
The input variable \verb|a_su3| is an integer specifying
the color index, while \verb|p| is a 3-dimensional vector
specifying the spatial momenta $(p_1,p_2,p_3)$ with $p_k=2\pi n_k/L$ and 
$\{n_k\in{\mathbb N}, n_k=0,1,\ldots,L\!-\!1\}$. 
The result
is the ghost propagator (\ref{eq:solution_gh}) written 
in the form of a square matrix with $T\!+\!1$ rows.

\subsubsection{Proof of the recursive method}

First of all we define the function
\be\label{eq:W_gh}
W(t)=\psif(t+1)\psib(t)-\psib(t+1)\psif(t)\,,\quad 0\leq t\leq T-1\,,
\ee
which we now show to be time-independent.
We multiply from the left both sides of \eq{eq:4_gh} by $\psib(t)$ and 
both sides of \eq{eq:6_gh} by $\psif(t)$, and we subtract them
\begin{align}
& \psib(t)\psif(t+1)-\psif(t)\psib(t-1)=-\psib(t)\psif(t-1)+\psif(t)\psib(t+1)\nonumber\\
& \nonumber\\[-2ex]
\Rightarrow&  \psif(t+1)\psib(t)-\psib(t+1)\psif(t)=\psif(t)\psib(t-1)-\psib(t)\psif(t-1)\nonumber\\
&  \nonumber\\[-2ex]
\Rightarrow&  W(t)=W(t-1)\,,\quad 0\leq t\leq T-1\,.
\end{align}
It is convenient to compute $W(t)$ at $t=0$ or $t=T$, where it assumes the 
simple form $W(t)=\psib(0)=\psif(T)=W$.
The correctness of the solution (\ref{eq:solution_gh}) can now be checked with a little 
algebra by verifying that
\begin{align}
(\ref{eq:3_gh}),(\ref{eq:5_gh})&\Rightarrow  (\ref{eq:prop_eq_gh_rand})\nonumber\\
&  \nonumber\\[-2ex]
(\ref{eq:4_gh})&\Rightarrow (\ref{eq:prop_eq_gh})\quad\mbox{for}\quad t<t'\nonumber\\
&  \nonumber\\[-2ex]
(\ref{eq:6_gh})&\Rightarrow (\ref{eq:prop_eq_gh})\quad\mbox{for}\quad t>t'\nonumber
\end{align}
For $t=t'$ we have
\be
-D(t+1,t)+B(t)D(t,t)-D(t-1,t)=1\,,
\ee
which turns out to be
\begin{align}
&  -\frac{1}{W}\psif(t)\psib(t+1)+\frac{1}{W}B(t)\psif(t)\psib(t)-\frac{1}{W}\psif(t-1)\psib(t)=1 \nonumber\\
\Rightarrow& -\frac{1}{W}\psif(t)\psib(t+1)+\frac{1}{W}\psib(t)[B(t)\psif(t)-\psif(t-1)]=1\nonumber\\
\stackrel{(\ref{eq:4_gh})}{\Rightarrow}& -\frac{1}{W}\psif(t)\psib(t+1)+\frac{1}{W}\psib(t)\psif(t+1)=1
\end{align} 
and the last equation can be recognized 
as an identity through the explicit expression of $W$ given in \eq{eq:W_gh}.


\subsection{The gluon propagator}\label{sect:gluon_propagator}

\subsubsection{Definition}

The quadratic part of the pure gluonic action takes the form
\be\label{eq:S0_gluon}
S^{(0)}_\mathrm{G}=\frac{1}{2L^3}\sum_\vecp \sum_{t,t'=0}^{T-1}\sum_a q^{\bar{a}}_\mu(-\vecp,t)
K^a_{\mu\nu}(\vecp;t,t')q_\nu^a(\vecp,t')
\ee
with
\begin{align}
K^a_{kl}(\vecp;t,t') &=  \delta_{t,t'}[
\delta_{kl}\vecs^a(\vecp,t)^2 - s^a_k(\vecp,t)s^a_l(\vecp,t)(1-\lambda_0)]\nonumber\\
\nonumber\\[-2ex]
&\quad   + \delta_{kl}[ 2C_a\delta_{t,t'} - R_a(\delta_{t+1,t'}
+ \delta_{t-1,t'})]\,,
\label{eq:gluonquad1}\\
\nonumber\\[-2ex]
K^a_{k0}(\vecp;t,t') &=  iR_a[\delta_{t,t'}s^a_k(\vecp,t+1)
-\delta_{t-1,t'}s^a_k(\vecp,t')]\nonumber\\
\nonumber\\[-2ex]
&\quad   -i\lambda_0 s^a_k(\vecp,t)[\delta_{t,t'}-\delta_{t-1,t'}]\,,
\label{eq:gluonquad2}\\
\nonumber\\[-2ex]
K^a_{0k}(\vecp;t,t') &=  -K^a_{k0}(\vecp;t',t)\,,
\label{eq:gluonquad3}\\
\nonumber\\[-2ex]
K^a_{00}(\vecp;t,t') &=  R_a\delta_{t,t'}\vecs^a(\vecp,t)\cdot\vecs^a(\vecp,t+1)\nonumber\\
\nonumber\\[-2ex]
&\quad   +\lambda_0[ 2\delta_{t,t'} -\delta_{t+1,t'} -\delta_{t-1,t'}] \nonumber\\
\nonumber\\[-2ex]
&\quad   -\lambda_0\delta_{t,t'}[\delta_{t,0}(1-\chi_a\delta_{\vecp,\vectn})
+\delta_{t,T-1}]\,,\label{eq:gluonquad4}
\end{align}
where $\chi_a=1$ for $a=3,8$ and $\chi_a=0$ otherwise, and the property
\be
K^{\bar{a}}_{\mu\nu}(\vecp;t,t')=K^a_{\nu\mu}(-\vecp;t',t)
\ee
holds. The free gluon propagator
is given by
\be\label{eq:free_gluon_propagator}
\langle q^a_\mu(\vecp,t)q^b_\nu(\vecpprime,t')\rangle_0=\delta_{a,\bar{b}}L^3\delta_{\vecp+\vecpprime,\vectn}
D^a_{\mu\nu}(\vecp;t,t')\,,
\ee
where $D^a$ is the inverse of $K^a$. A brute force inversion of $K^a$ is not a very
efficient way of computing the propagator. A recursive technique can be employed.
First of all we write down the difference equation that we want to solve
\be\label{eq:DK_1}
\sum_{t''}\sum_\nu K^a_{\mu\nu}(\vecp;t,t'')D^a_{\nu\sigma}(\vecp;t'',t')=\delta_{t,t'}\delta_{\mu,\sigma}\,.
\ee
The case $a=3,8$ and $\vecp=\vectn$ is particularly simple, and the solution is explicitly known
\begin{align}
D^a_{\mu\nu}(\vectn;t,t') =\left\{
\begin{array}{ll}
\tfrac{1}{\lambda_0}\bigl(1+\min(t,t')\bigr) & \mbox{if $\mu=\nu=0$},\\
\tfrac{1}{R_a}\bigl(\min(t,t')-\tfrac{t t'}{T}\bigr)
& \mbox{if $\mu=\nu=k$}, \label{eq:D_a38}\\
0 & \mbox{if $ \mu\neq\nu$}.
\end{array}\right.
\end{align}
in all other cases, \eq{eq:DK_1} must be numerically solved. We fix the index $a$
and the momentum $\vecp$, and decompose the operator $K$ according to
\be
K_{\mu\nu}(t,t')=A_{\mu\nu}(t)\delta_{t+1,t'}+B_{\mu\nu}(t)\delta_{t,t'}+A_{\nu\mu}(t-1)\delta_{t-1,t'}\,.
\ee
With this expression, \eq{eq:DK_1} becomes
\begin{align}
&  A_{\mu\lambda}(t)D_{\lambda\nu}(t+1,t')+B_{\mu\lambda}(t)D_{\lambda\nu}(t,t')+A^\dagger_{\mu\lambda}(t-1)
D_{\lambda\nu}(t-1,t')=\delta_{\mu\nu}\delta_{t,t'}\,,\nonumber\\
&  \mbox{with}\quad0\leq t\leq T-1\,,\label{eq:gprop_f}
\end{align}
and $A$, $B$ and $D$ are, at fixed time, $4\!\times\! 4$ matrices. The boundary conditions of the propagator are simply given
by
\be\label{eq:boundaries_D_gprop}
\drvstar{t}D_{0\nu}(t,t')=D_{k\nu}(t,t')=0\,,\quad\mbox{at}\quad t=0\,\;\mbox{and}\,\;t=T\,.
\ee
We construct two solutions $\psif(t)$ and $\psib(t)$ of the homogeneous 
version of \eq{eq:gprop_f} by a two-step recursion forward and 
backward in time. The forward solution $\psif(t)$ starts from
\be
\left. \begin{array}{rr}
\psif_{0\nu}(-1)=\psif_{0\nu}(0)=\,&\!\delta_{0\nu}\\\
                 \psif_{k\nu}(0)=\,&\!0\label{eq:psif123}\\
                 \psif_{k\nu}(1)=\,&\!\delta_{k\nu}
\end{array}\right\}\,,
\ee
while the backward solution $\psib(t)$ starts from
\be
\left. \begin{array}{rr}
\hspace{-0.6cm}\psib_{0\nu}(T)=\psib_{0\nu}(T-1)=\,&\!\delta_{0\nu}\\
                \psib_{k\nu}(T)=\,&\!0\label{eq:psib123}\\
                \psib_{k\nu}(T-1)=\,&\!\delta_{k\nu}
\end{array}\right\}\,,
\ee
where both $\psif$ and $\psib$ are, at fixed time, $4\!\times\! 4$ matrices.
Once one has computed and stored them, the propagator can be expressed in the compact form
\begin{align}
D_{\mu\nu}(t,t') =\left\{
\begin{array}{ll}
\psif_{\mu\sigma}(t)W^{-1}_{\lambda\sigma}\psib_{\mu\lambda}(t') & \mbox{for $t\leq t'$}\,,\\
 & \label{eq:sol_qprop}\\
\psib_{\mu\sigma}(t)W^{-1}_{\sigma\lambda}\psif_{\mu\lambda}(t') & \mbox{for $t\geq t'$}\,.
\end{array}\right.
\end{align}
The Wronskian W is a $4\!\times\! 4$ matrix, and can be computed, after that the recursion
for $\psib$ has been completed, through
\begin{align}
W_{0\nu}&= A_{0\sigma}(0)\psib_{\sigma\nu}(1)+(B_{00}(0)+A^\dagger_{00}(-1))\psib_{0\nu}(0)\,,\\
W_{k\nu}&= -A_{jk}\psib_{j\nu}(0)\,.
\end{align}
For the case of vanishing background field with boundary values $C=C'=0$ the gluon propagator
assumes a simple form if the Feynman gauge, i.e.~$\lambda_0=1$, is 
chosen \cite{impr:axial_int_notes}. For any momentum 
$\vecp$ we define the ``energy'' $\epsilon$ through
\be
\cosh(\epsilon)=1+\tfrac{1}{2}\vecphat^2\,,\quad \hat{p}_k=2\sin(p_k/2)\,.
\ee
For non-zero momenta $\vecp$ the propagator of the spatial components of the gluon field
reads
\begin{align}
&  D_{kj}(\vecp;t,t')=\delta_{kj}\frac{1}{\sinh(\epsilon)\sinh(\epsilon T)}\nonumber\\
&  \hspace{1cm}\times\left\{
\begin{array}{ll}
&\sinh[\epsilon (T-t)]\sinh(\epsilon t')\,,\quad\mbox{if $t\geq t'$}\,,\\[-1ex]
& \\[-1ex]
&\sinh(\epsilon t)\sinh[\epsilon (T-t')]\,,\quad\mbox{if $t\leq t'$}\,,
\end{array}\right.
\end{align}
while for the time component we have
\begin{align}
&  D_{00}(\vecp;t,t')=\frac{1}{\sinh(\epsilon)\sinh(\epsilon T)}\nonumber\\
&  \hspace{1cm}\times\left\{
\begin{array}{ll}
&\cosh[\epsilon (T-t)-\tfrac{1}{2}\epsilon ]\cosh(\epsilon t'+\tfrac{1}{2}\epsilon )\,,\quad\mbox{if $t\geq t'$}\,,\\[-1ex]
& \\[-1ex]
&\cosh(\epsilon t+\tfrac{1}{2}\epsilon )\cosh[\epsilon (T-t')-\tfrac{1}{2}\epsilon ]\,,\quad\mbox{if $t\leq t'$}\,.\\[-1ex]
\end{array}\right.
\end{align}
For $\vecp=\vectn$ the propagator coincides with the one given in \eq{eq:D_a38}
for the non-vanishing background field case; one just has to set $\lambda_0=1$ and $R_a=1$.
The mixed components $D_{k0}$ and $D_{0k}$ vanish for all time coordinates 
and for all momenta $\vecp$.

\subsubsection{Implementation}

The implementation of the recursive method for the computation of the 
gluon propagator in momentum space can be found in the program \verb|WLINE|
as a function in
\begin{verbatim}
WLINE_SIMPLE/gluon_p.m
\end{verbatim}
Once one is in the directory \verb|WLINE_SIMPLE|,
it is necessary to load part of the global variables contained in
file \verb|topo_parameters.dat|,
\begin{verbatim}
load_topo_parameters_wsimple
\end{verbatim}
and initialize the remaining variables through
the command
\begin{verbatim}
initialization
\end{verbatim}
The function computing the gluon propagator is 
\begin{verbatim}
function [res] = gluon_p(a_su3,p,lambda_0)
\end{verbatim}
The input variable \verb|a_su3| is an integer specifying
the color index, while \verb|p| is a 3-dimensional vector
specifying the spatial momenta $(p_1,p_2,p_3)$ with $p_k=2\pi n_k/L$ and 
$\{n_k\in{\mathbb N}, n_k=0,1,\ldots,L\!-\!1\}$. 
The remaining input variable \verb|lambda_0| corresponds to the gauge
fixing parameter $\lambda_0$ appearing in \eq{eq:S_gf}.
For the non-vanishing background field case, the result
is the gluon propagator defined in \eq{eq:DK_1}, and it is written 
in the form of an array of dimension $(4,4,T\!+\!1,T\!+\!1)$. Through
the first two indices one can access the spin structure of the propagator,
while the last two indices specify the temporal coordinates.
Notice that, due to the way Matlab uses to indicate the elements of an array
as well as for convenience of programming, 
the diagonal component $D_{00}$ corresponds to \verb|res[4,4,:,:]| in the program.
The same holds for the mixed components $D_{0k}\Leftrightarrow$ \verb|res[4,k,:,:]|,
with $k=1,2,3$.   

If the background field is deactivated, the gluon propagator is computed 
only for the boundary values $C=C'=0$ and in Feynman gauge using a specific function.
The latter is internally called by \verb|gluon_p| if it is the case.

The gluon propagator in coordinate space, according to the
definition given in \eq{eq:gluon_x}, is computed through the function
\begin{verbatim}
function [res] = gluon_x(a_su3,i_matrixx,lambda_0)
\end{verbatim}
located in the file \verb|WLINE_SIMPLE/gluon_x.m|. The input variables
\verb|a_su3| and \verb|lambda_0| have the same meaning as for the propagator
in momentum space. The variable \verb|i_matrixx| is a matrix with 7 columns,
and a number of rows limited in principle only by the amount of memory 
which can be allocated for the propagator and the available computer time. In order to get
a deeper insight on it,
let us consider the propagator $\Delta^a_{\mu\nu}(\vecx;x_0,y_0)$ in \eq{eq:gluon_x}.
After having initialized the necessary global variables as in the case of \verb|gluon_p|, 
let us suppose that we want to compute $\Delta^5_{33}((0,0,0);2,2)$ in Feynman gauge.
This is achieved through the call
\begin{verbatim}
res = gluon_x(5,[3 3 0 0 0 2 2],1.0)
\end{verbatim}
where \verb|i_matrixx| has only one row. In turn, the function \verb|gluon_x| internally
calls \verb|gluon_p| and performs the required Fourier transform. 

If we now want to 
compute $\Delta^5_{32}((0,0,1);2,2)$, inspection of \eq{eq:gluon_x} reveals 
that it is not needed to call \verb|gluon_p| again. One could use the results of the previous
computation. This suggests to call \verb|gluon_x| at the very beginning in this way
\begin{verbatim}
res = gluon_x(5,[3 3 0 0 0 2 2;3 2 0 0 1 2 2],1.0)
\end{verbatim}
It is now straightforward to extend this feature to 
a path as in \sect{sect:Wilson_loops}. The input 
matrix \verb|i_matrixx| then contains, for fixed color index $a$, 
as many rows as the number of coordinate space propagators one 
wants to compute. 
It is important to remark that, in compliance with the notation of \sect{sect:Wilson_loops},
the temporal direction is denoted by $|\mu|=4$.
The output is given in the form a vector, with as many elements
as the rows of \verb|i_matrixx| and the same coordinate ordering.
Since the computation of the gluon propagators
in momentum space is very expensive, it is clear that
this feature allows to save a large amount of computing time in comparison
to a na\"ive implementation. 

There is another feature which is worth to mention. The authors
of \cite{pert:2loop_SU2} noticed that, since none of the three space directions
is distinguished, there is symmetry under permuting them, which corresponds to certain
discrete rotations and reflections in spin space. Hence one gains a factor of 6 (asymptotically on large lattices)
by computing the gluon propagator in momentum space only 
for the reduced set of momenta $\mathbf{p_1}\leq\mathbf{p_2}\leq\mathbf{p_3}$. Actually, 
a short computation shows that the number
of needed momenta amounts to
\be
\frac{L(L+1)(L+2)}{6}\,,\nonumber
\ee 
to be compared to $L^3$ of a na\"ive Fourier transform. The time needed to perform
the Fourier transforms is, for the observables in chapter \ref{chap:Zspin}, negligible 
in comparison to the computation of the gluon 
propagators in momentum space, even for the reduced set of momenta. Hence we found no need
of employing a FFT.
The function \verb|gluon_x| internally calls another function, called \verb|ppermute| 
and located in the same directory in the file \verb|ppermute.m|, which 
calls \verb|gluon_p| only for the reduced set of momenta, and, through matrix 
manipulations optimized for Matlab and of negligible computational cost, produces 
as output the gluon propagator for all momenta.

\begin{figure}[t]
\centering
\includegraphics[scale=0.5]{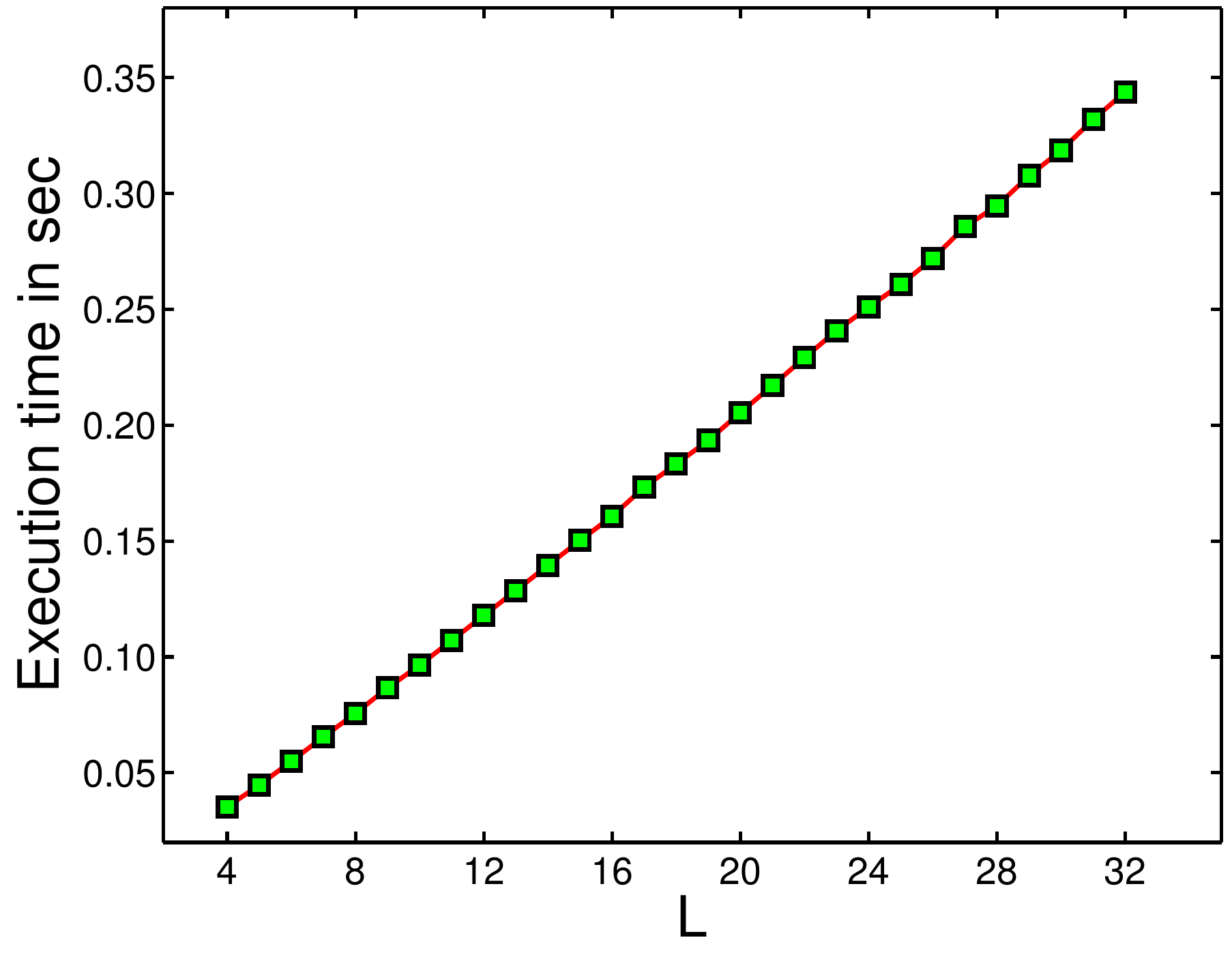}
\mycaption{Computational cost for the gluon propagator for a fixed momentum. Test performed
on a PC with a processor Intel Pentium 4 with 2.6 GHz.}\label{fig:gluon_time}
\end{figure}

Finally, we show in \Fig{fig:gluon_time} the cost
for the computation of the gluon propagator for a fixed momentum
and non-trivial color, i.e.~$a\neq 3,8$, in the range $4\leq L\leq 32$
and for $T=L$. The expected linear scaling is evident. On a computer with
processor Intel Pentium 4 with 2.6 GHz, we found
\be\label{eq:gluon_prop_cost}
\mbox{cost in seconds} = k_1+k_2L\,,
\ee
with $k_2\simeq 0.011$.

\subsubsection{Proof of the recursive method}
 
So far we listed the basic properties of the recursive method and gave
the necessary tools to implement it on a computer. 
However, we did not demonstrate that (\ref{eq:sol_qprop})
is the correct and unique solution \eq{eq:DK_1}. This is the subject of the rest
of this subsection.

We start from the forward solution $\psif$ and observe that
\be\label{eq:gprop_4}
A_{0\lambda}\psif_{\lambda\nu}(1)+[B_{00}(0)+A^\dagger_{00}(-1)]\psif_{0\nu}(0)=0\,,
\ee
which will be our starting point in the implementation of the recursive method.
In fact, together with \eqs{eq:psif123}, we notice that
\begin{align}
\nu=0\,,&\quad \psif_{00}(1)=-A^{-1}_{00}(0)[B_{00}(0)+A^\dagger_{00}(-1)]\,,\\
\nu=j\,,&\quad \psif_{0j}(1)=-A^{-1}_{00}(0)A_{0j}(0)\,,
\end{align}
and we have defined all components of $\psif$ for $t=0,1$.
The matrix $A$ has been checked to be invertible for all times
by inspecting its explicit expression. The invertibility
holds as long as the gauge fixing parameter $\lambda_0> 0$ and 
the constant electric background field does not vanish. 
Further, we use the following equation 
\be
A(t)\psif(t+1)+B(t)\psif(t)+A^\dagger(t-1)\psif(t-1)=0\,,\label{eq:gprop_5}
\ee
written in a looser notation, to recursively compute $\psif(t)$
in the interval $2\leq t\leq T$. Analogously, we let the backward solution $\psib$
satisfy \eqs{eq:psib123} and
\be
A(t)\psib(t+1)+B(t)\psib(t)+A^\dagger(t-1)\psib(t-1)=0\,.\label{eq:gprop_7}
\ee
The latter is used to compute $\psib(t)$ for $0\leq t\leq T\!-\!2a$. We look for a solution of
\eq{eq:DK_1} having the form
\begin{align}
D_{\mu\nu}(t,t') =\left\{
\begin{array}{ll}
\psif(t)U(t')& \mbox{for $t\leq t'$}\,,\\
 & \label{eq:sol_gprop_UV}\\
\psib(t)V(t') & \mbox{for $t\geq t'$}\,,
\end{array}\right.
\end{align}
which is compatible with the boundary conditions in \eqs{eq:boundaries_D_gprop}
and satisfies \eq{eq:DK_1} for $t\neq t'$ from the very construction. 
At $t=t'$
we first require the consistency of the solution (\ref{eq:sol_gprop_UV}), i.e.~
\be
\psif(t)U(t)=\psib(t)V(t)\,,\label{eq:gprop_9}
\ee
and then require to satisfy \eq{eq:DK_1}; this can be rewritten as
\be
A(t)\psib(t+1)V(t)+B(t)\psif(t)U(t)+A^\dagger(t-1)\psif(t)U(t)=1\,,
\ee
where we insert \eq{eq:gprop_5} to be able to get
\be
A(t)\psib(t+1)V(t)-A(t)\psif(t+1)U(t)=1\,.\label{eq:gprop_10}
\ee
The latter equation is used together with \eq{eq:gprop_9} to solve for $U$ and $V$:
\begin{align}
(\ref{eq:gprop_9})&\Rightarrow \! V(t)=[\psib(t)]^{-1}\psif(t)U(t)\label{eq:gprop_11}\\
&  \nonumber\\[-2ex]
(\ref{eq:gprop_10}),(\ref{eq:gprop_11})&\Rightarrow \! 
A(t)\{\psib(t+1)[\psib(t)]^{-1}\psif(t)-\psif(t+1)\}U(t)=1\nonumber\\
&  \nonumber\\[-2ex]
&\Rightarrow \! A(t)\{\psib(t+1)[\psib(t)]^{-1}-\psif(t+1)[\psif(t)]^{-1}\}\psif(t)U(t)=1\nonumber
\end{align}
and get
\begin{align}
\left[U(t)\right]^{-1}&= A(t)\{\psib(t+1)[\psib(t)]^{-1}-\psif(t+1)[\psif(t)]^{-1}\}\psif(t)\,,\label{eq:gprop_121}\\
&\,\,\Downarrow  {\scriptstyle(\ref{eq:gprop_11})}\nonumber\\
\left[V(t)\right]^{-1}&= A(t)\{\psib(t+1)[\psib(t)]^{-1}-\psif(t+1)[\psif(t)]^{-1}\}\psib(t)\,.\label{eq:gprop_122}
\end{align}
With these two expressions we have determined the solution in the interval $1\leq t'\leq T-1$. For $t'=0$
we proceed as follows. With
\be\label{eq:gprop_13}
D(t,0)=\psib(t)V(0)\,,\quad 0\leq t \leq T\,,
\ee
and the boundary condition $D_{k\nu}(0,0)=0$ we get 
\be\label{eq:gprop_14}
\psib_{k\nu}(0)V_{\lambda\nu}(0)=0\,, 
\ee
and consider \eq{eq:gprop_f} for $t=0$, which is for $\mu=0$
\be
A_{0\lambda}(0)\psib_{\lambda\sigma}(1)V_{\sigma\nu}(0)+[B_{00}(0)+A^\dagger_{00}(-1)]
\psib_{0\sigma}(0)V_{\sigma\nu}(0)=\delta_{0\nu}\,.\label{eq:gprop_15}
\ee
Finally we can gather \eq{eq:gprop_14} and \eq{eq:gprop_15} onto a the compact 
\begin{align}
&  \hspace{-1.03cm}NV=R\,,\,R_{k\nu}=0\,,\, R_{0\nu}=\delta_{0\nu}\,.\\
&  \hspace{-1cm}N_{k\nu}=\psib_{k\nu}(0)\,,\, N_{0\nu}=A_{0\lambda}(0)\psib_{\lambda\nu}(1)
+[B_{00}(0)+A^\dagger_{00}(-1)]\psib_{0\nu}(0)\,.
\end{align}
The final step is the construction of the Wronskian matrix. We multiply \eq{eq:gprop_5}
by $[\psib(t)]^\dagger$ from the left, and the hermitian conjugate of
both sides of \eq{eq:gprop_7} by $\psif(t)$ from the right. The subtraction of the l.h.s.~of 
the two resulting equations gives
\begin{align}
&  \hspace{-1cm}[\psib(t)]^\dagger A(t)\psif(t+1)-[\psib(t+1)]^\dagger A^\dagger(t)\psif(t)\nonumber\\
&  =[\psib(t-1)]^\dagger A(t-1)\psif(t)- [\psib(t)]^\dagger A^\dagger(t-1) \psif(t-1)\,,\label{eq:gprop_W21}     
\end{align}
valid in the interval $1\leq t\leq T-1$. We define the $4\!\times\!4$ matrix
\be\label{eq:gprop_16}
W_{21}(t)=-[\psib(t)]^\dagger A(t)\psif(t+1)+[\psib(t+1)]^\dagger A^\dagger(t)\psif(t)\,,
\ee
in the interval $0\leq t\leq T-1$, and observe that, thanks to \eq{eq:gprop_W21}, $W_{21}$ is time-independent
\be
W_{21}(t)=W_{21}(t-1)\,,\quad 1\leq t\leq T-1\,.
\ee
We repeat the procedure and multiply both sides of \eq{eq:gprop_5} by $[\psif(t)]^\dagger$
and then subtract their hermitian conjugates. The result
\begin{align}
&  \hspace{-1cm}[\psif(t)]^\dagger A(t)\psif(t+1)-[\psif(t+1)]^\dagger A^\dagger(t)\psif(t)\nonumber\\
&  =[\psif(t-1)]^\dagger A(t-1)\psif(t)-[\psif(t)]^\dagger A^\dagger(t-1)\psif(t-1)\,,
\end{align}
valid in the interval $1\leq t\leq T-1$, implies that the $4\!\times\!4$ matrix
\be\label{eq:gprop_18}
W_{11}(t)=[\psif(t)]^\dagger A(t)\psif(t+1)-[\psif(t+1)]^\dagger A^\dagger(t)\psif(t)\,,
\ee
is time-independent. The same procedure but by starting with \eq{eq:gprop_7} instead of \eq{eq:gprop_5}
let us conclude that the matrix
\be\label{eq:gprop_20}
W_{22}(t)=[\psib(t)]^\dagger A(t)\psib(t+1)-[\psib(t+1)]^\dagger A^\dagger(t)\psib(t)\,,
\ee
is time independent too. The matrix $W_{ij}$, with $W_{12}=W_{21}^\dagger$, is the Wronskian matrix
and it is time-independent. Inspection of the explicit expression of $W_{11}$ at $t=0$
and of $W_{22}$ at $t=T-1$ reveals that
\be\label{eq:gprop_22}
W_{11}=W_{22}=0\,.
\ee
To evaluate $W_{21}$ we can set $t=0$ in \eq{eq:gprop_16} and get
\be\label{eq:gprop_24}
W_{21}=-[\psib(0)]^\dagger A(0)\psif(1)+[\psib(1)]^\dagger A^\dagger(0)\psif(0)\,.
\ee
Now we go back to \eq{eq:gprop_121}
\begin{align}
[\psib(t)]^\dagger\left[U(t)\right]^{-1}&= [\psib(t)]^\dagger A(t)\psib(t+1)[\psib(t)]^{-1}\psif(t)-
[\psib(t)]^\dagger A(t)\psif(t+1)\nonumber\\
&  \nonumber\\[-2ex]
&\hspace{-0.7cm}\stackrel{(\ref{eq:gprop_20}),(\ref{eq:gprop_22})}{=}  [\psib(t+1)]^\dagger A^\dagger(t)\psif(t)-
[\psib(t)]^\dagger A(t) \psif(t+1)\nonumber\\
&  \nonumber\\[-2ex]
&\hspace{-0.25cm}\stackrel{(\ref{eq:gprop_16})}{=} W_{21}
\end{align}
We consider now \eq{eq:gprop_122}, and proceed in the same way
\begin{align}
[\psif(t)]^\dagger\left[V(t)\right]^{-1}&=
[\psif(t)]^\dagger A(t)\psib(t+1)\nonumber\\
&  \nonumber\\[-2ex]
&\quad  -[\psif(t)]^\dagger A(t)\psif(t+1)[\psif(t)]^{-1}\psib(t)\nonumber\\
&  \nonumber\\[-2ex]
&\hspace{-0.7cm}\stackrel{(\ref{eq:gprop_18}),(\ref{eq:gprop_22})}{=} [\psif(t)]^\dagger A(t)\psib(t+1)
-[\psif(t+1)]^\dagger A^\dagger(t)\psib(t)\nonumber\\
&  \nonumber\\[-2ex]
&\hspace{-0.25cm}\stackrel{(\ref{eq:gprop_16})}{=} W_{21}^\dagger=W_{12}
\end{align}
This allows us to write the solution (\ref{eq:sol_gprop_UV}) as
\begin{align}
D_{\mu\nu}(t,t') =\left\{
\begin{array}{ll}
\psif(t)W_{21}^{-1}[\psib(t')]^\dagger&\quad \mbox{for $t\leq t'$}\,,\\
 & \label{eq:sol_gprop_W}\\
\psib(t)[W_{21}^\dagger]^{-1}[\psif(t')]^\dagger &\quad \mbox{for $t\geq t'$}\,,
\end{array}\right.
\end{align}
holding for $t=t'$ too. Now we work out $W_{12}$ in detail
\be
W_{12}=[\psif(0)]^\dagger A(0)\psib(1)-[\psif(1)]^\dagger A^\dagger(0)\psib(0)\,,
\ee
and have a look at 
\begin{align}
[W_{12}]_{0\nu}&= \psif_{\sigma 0}(0)A_{\sigma\lambda}(0)\psib_{\lambda\nu}(1)
-\psif_{\sigma 0}(1)A_{\lambda\sigma}(0)\psib_{\lambda\nu}(0)\\
&  \nonumber\\[-2ex]
&\hspace{-0.25cm}\stackrel{(\ref{eq:psif123})}{=} A_{0\lambda}(0)
\psib_{\lambda\nu}(1)-\psif_{00}(1)A_{\lambda 0}(0)\psib_{\lambda\nu}(0)\,.
\end{align}
From \eq{eq:gprop_4} we have that
\begin{align}
&  \hspace{-0.7cm}A_{00}(0)\psif_{00}(1)+A_{0k}(0)\psif_{k0}(1)+[B_{00}(0)+A^\dagger_{00}(-1)]\psif_{00}(0)=0\,,\\
&  \nonumber\\[-2ex]
&  \hspace{-0.7cm}(\ref{eq:psif123})
\Rightarrow A_{00}(0)\psif_{00}(1)=-[B_{00}(0)+A^\dagger_{00}(-1)]\,,
\end{align}
and consequently
\be
[W_{12}]_{0\nu}=A_{0\lambda}(0)\psib_{\lambda\nu}(1)+\frac{B_{00}(0)+A^\dagger_{00}(-1)}{A_{00}(0)}
A_{\lambda 0}(0)\psib_{\lambda\nu}(0)\,.
\ee
Since $A_{k0}(0)=0$ we have
\be
[W_{12}]_{0\nu}=A_{0\lambda}(0)\psib_{\lambda\nu}(1)+[B_{00}(0)+A^\dagger_{00}(-1)]\psib_{0\nu}(0)\,.
\ee
The last step consists in considering
\be
[W_{12}]_{k\nu}=\psif_{\sigma k}(0)A_{\sigma\lambda}(0)\psib_{\lambda\nu}(1)-
\psif_{\sigma k}(1)A_{\lambda \sigma}(0)\psib_{\lambda \nu}(0)\,.
\ee
The boundary conditions on $\psif$ let us write that
\be
\psif_{\sigma k}(1)A_{\lambda \sigma}(0)\psib_{\lambda \nu}(0)=
A_{\lambda k}(0)\psib_{\lambda\nu}(0)+\psif_{0k}(1)A_{\lambda 0}(0)\psib_{\lambda\nu}(0)\,,
\ee
and from \eq{eq:gprop_4} we have that
\be
A_{00}(0)\psif_{0k}(1)+A_{0j}(0)\psi_{jk}(a)+[B_{00}(0)+A^\dagger_{00}(-1)]\psif_{0k}(0)=0\,,
\ee
which can be simplified by using again the boundary conditions on $\psif$:
\be
A_{00}(0)\psif_{0k}(1)+A_{0k}(0)=0\Rightarrow \psif_{0k}(1)=-\frac{A_{0k}(0)}{A_{00}(0)}\,.
\ee
This lets us write
\begin{align}
[W_{12}]_{k\nu}&= -A_{\lambda k}(0)\psib_{\lambda\nu}(0)+\frac{A_{0k}(0)}{A_{00}(0)}
A_{\lambda 0}(0)\psib_{\lambda\nu}(0)\nonumber\\
&  \nonumber\\
&\hspace{-0.45cm}\stackrel{A_{k0}(0)=0}{=} -A_{\lambda k}(0)\psib_{\lambda\nu}(0)+A_{0k}(0)\psib_{0\nu}(0)\\
&  \nonumber\\
&= -A_{jk}(0)\psib_{j\nu}(0)\,.
\end{align}
If we now rename the matrix $W_{12}$ simply $W$, we demonstrated that (\ref{eq:sol_qprop})
is solution of \eq{eq:DK_1}.

\subsection{The quark propagator}\label{sect:quark_propagator}

\subsubsection{Definition}

We introduce the Fourier transformed quark fields according to
\begin{align}
\psibar(\vecx,t)&= \frac{1}{L^3}\sum_\vecp \mathrm{e}^{i\vecp\vecx}\psibar(\vecp,t)\,,\\
\psi(\vecx,t)&= \frac{1}{L^3}\sum_\vecp \mathrm{e}^{i\vecp\vecx}\psi(\vecp,t)\,,
\end{align}
whose representation in momentum space is used to express the free part of the quark action as
\be
S_\mathrm{F}^{(0)}=\frac{1}{L^3}\sum_\vecp \sum_{t,t'}\psibar(-\vecp,t)\tilde{D}(\vecp;t,t')\psi(\vecp,t')\,,
\ee
where $\tilde{D}$ is the improved Dirac-Wilson operator, including the mass term,
at lowest order in $g_0$. It is diagonal in momentum and color space 
for the background field under consideration. It can be written as
\be
\tilde{D}(\vecp;t,t')=-P_-\delta_{t+1,t'}+B(\vecpplus,t)\delta_{t,t'}-P_+\delta_{t-1,t'}\,,
\ee 
where the matrices $P_\pm$ are defined in \app{app:Not_Conv}, $p^+_k=p_k+\theta/L$, and the matrix
$B$ reads
\begin{align}
B(\vecp,t)&= 4+m_0-\sum_k \left[\frac{1}{2}(1+\gamma_k)\mathrm{e}^{-ip_k}V^\dagger(t)
+\frac{1}{2}(1-\gamma_k)\mathrm{e}^{ip_k}V(t)\right]\nonumber\\
              &\quad  +iH\gamma_0 \sum_k \gamma_k\,.\label{eq:B_quark_def}
\end{align}
The matrix $H$ is proportional to the tree-level value of $\csw$, and is diagonal in color space 
with the elements
\be
H_{\alpha\alpha}=-\frac{1}{2}\csw^{(0)}\sin \mathcal{E}_\alpha\,.
\ee
The propagator $S$ is the inverse of the Dirac-Wilson operator,
\be\label{eq:diff_eq_qprop}
\sum_{t''}\tilde{D}(\vecp;t,t'')S(\vecp;t'',t')=\delta_{t,t'}\,,\quad 0<t,t'<T\,,
\ee
and is pseudo-hermitian,
\be\label{eq:S_g5h}
\gamma_5 S(\vecp;t,t')\gamma_5=S(\vecp;t',t)^\dagger\,,
\ee
as well as the operator $\tilde{D}$. The boundary conditions are given
by
\begin{align}
P_+S(\vecp;t,t')|_{t=0}&= P_-S(\vecp;t,t')|_{t=T}=0\,,\\
S(\vecp;t,t')P_-|_{t'=0}&= S(\vecp;t,t')P_+|_{t'=T}=0\,.
\end{align}
The computation of $S$ by a brute force inversion of $\tilde{D}$ can become
computationally very expensive for fine lattices. We prefer to proceed
in analogy to the ghost and gluon cases, by fixing the momentum $\vecp$,
and by constructing the solutions of the homogeneous
equation by forward and backward recursion. 
We call these solutions $\psif$ and $\psib$ respectively; they
are $4\!\times\!2$ matrices, and satisfy
\begin{align}
\sum_{t'}\tilde{D}(t,t')\psif(t')&= 0\,,\quad P_+\psif(0)=\psif_+(0)=0\,,\label{eq:qprop_f}\\
\sum_{t'}\tilde{D}(t,t')\psib(t')&= 0\,,\quad P_-\psib(T)=\psib_-(T)=0\,.\label{eq:qprop_b}
\end{align}
The recursion is simpler than in the gluon case, because here one has to solve
a first order difference equation. To see this, we define
\be
F^{f/b}(t)=P_-\psi^{f/b}(t)+P_+\psi^{f/b}(t-1)\,,
\ee
which enables us to write the homogeneous equation as
\be\label{eq:B27}
[B(t)P_+-P_-]F^{f/b}(t+1)+[B(t)P_--P_+]F^{f/b}(t)=0\,.
\ee
Once the forward and backward solutions have been computed, the quark propagator
can be constructed through
\be\label{eq:quark_p_sol}
S(t,t') = \left\{
\begin{array}{ll}
\psi^f(t)\vcal[\psi^b(t')]^{\dagger}\gamma_5
& \mbox{for $t<t'$},\\
\psi^b(t)\vcal^{\dagger}[\psi^f(t')]^{\dagger}\gamma_5
& \mbox{for $t>t'$},\\
\psi^f_-(t)\vcal[\psi^b(t)]^{\dagger}\gamma_5 & \\
\;\; + \psi^b_+(t)\vcal^{\dagger}[\psi^f(t)]^{\dagger}\gamma_5
& \mbox{for $t=t'$}.
\end{array}\right.
\ee
The $2\!\times\!2$ matrix $\vcal$ can be computed from
\be\label{eq:vcal_ta}
[\vcal^\dagger]^{-1}=[\psif_-(1)]^\dagger\gamma_5\psib_+(0)\,.
\ee

\subsubsection{Implementation}

The implementation of the recursive method for the computation of the
quark propagator can be found in the program \verb|WLINE|
as a function in
\begin{verbatim}
TADPOLE_QUARK/quark_p.m
\end{verbatim}
Once one is in the directory \verb|TADPOLES_GAUGE|,
it is necessary to load part of the global variables contained in
file \verb|topo_parameters.dat|,
\begin{verbatim}
load_topo_parameters_qtadpole
\end{verbatim}
and initialize the remaining global variables through
the command
\begin{verbatim}
initialization
\end{verbatim}
The function computing the quark propagator is
\begin{verbatim}
function [res] = quark_p(p,theta_angle,csw_0,mass_0)
\end{verbatim}
The variable \verb|p| is a 3-dimensional vector
specifying the spatial momenta $(p_1,p_2,p_3)$ with $p_k=2\pi n_k/L$ and 
$\{n_k\in{\mathbb N}, n_k=0,1,\ldots,L\!-\!1\}$,
\verb|theta_angle| 
specifies the phase appearing in \eqs{eq:quark_boundaries4},
\verb|csw_0| is the tree-level value of $\csw$ entering in the
improvement counterterm to the fermion action defined in \eq{eq:impr_vol_term},
and \verb|mass_0| is the quark mass according to \eq{eq:B_quark_def}.
The output is the propagator (\ref{eq:quark_p_sol}) in the form of an
array of dimension $(3,T\!+\!1,T\!+\!1,4,4)$. Through the first index one accesses
the color structure. The propagator is diagonal in color space,
and that index selects the diagonal elements. The second and third indices
refer to the temporal coordinates $t$ and $t'$, while the last two indices 
specify the Dirac structure. 

\subsubsection{Proof of the recursive method}

We now show that (\ref{eq:quark_p_sol}) is the solution of 
the difference equation (\ref{eq:diff_eq_qprop}). We proceed in analogy
to the gluon case, and, first of all, look for a solution of the form 
\begin{align}
S(t,t') =\left\{
\begin{array}{ll}
\psif(t)N^f(t')\gamma_5& \mbox{for $t<t'$}\,,\\
 & \label{eq:sol_qprop_NN}\\
\psib(t)N^b(t')\gamma_5 & \mbox{for $t>t'$}\,,
\end{array}\right.
\end{align}
To determine $N^f$ and $N^b$ we exploit the pseudo-hermicity of the 
propagator given in \eq{eq:S_g5h}, to get
\be
N^f(t')=\vcal[\psib(t')]^\dagger\,,\quad N^b(t')=\vcal^\dagger[\psif(t')]^\dagger\,.
\ee
We now inspect \eq{eq:diff_eq_qprop} for $t'=t+1$,
\be\label{eq:diff_eq_qprop1}
P_-S(t+1,t+1)=-P_+S(t-1,t+1)+B(t)S(t,t+1)\,.
\ee
The identity
\be
B(t)S(t,t+1)=B(t)(P_++P_-)\psif(t)\vcal[\psib(t+1)]^\dagger\gamma_5\,,
\ee
together with the forward version of \eq{eq:B27} in the form 
\be\label{eq:B27f}
B(t)(P_++P_-)\psif(t)=P_-\psif(t+1)+P_+\psif(t-1)\,,
\ee
let us rewrite \eq{eq:diff_eq_qprop1} as
\be\label{eq:P-tt}
P_-S(t+1,t+1)=P_-\psif(t+1)\vcal[\psib(t+1)]^\dagger\gamma_5\,, 
\ee
Analogously, we inspect \eq{eq:diff_eq_qprop} for $t'=t-1$,
\be\label{eq:diff_eq_qprop2}
P_+S(t-1,t-1)=-P_-S(t+1,t-1)+B(t)S(t,t-1)\,.
\ee
The identity
\be
B(t)S(t,t-1)=B(t)(P_++P_-)\psib(t)\vcal^\dagger [\psif(t-1)]^\dagger \gamma_5\,,
\ee
and the backward version of \eq{eq:B27} in the form
\be\label{eq:B27b}
B(t)(P_+-P_-)\psib(t)=P_-\psib(t+1)+P_+\psib(t-1)\,,
\ee
let us rewrite \eq{eq:diff_eq_qprop2} as
\be\label{eq:P+tt}
P_+S(t-1,t-1)=P_+\psib(t-1)\vcal^\dagger [\psif(t-1)]^\dagger \gamma_5\,.
\ee
It is now clear that eqs.~(\ref{eq:P-tt}, \ref{eq:P+tt}) prove the correctness
of the solution (\ref{eq:quark_p_sol}) for $t=t'$. It remains now to determine the 
explicit expression of $\vcal$. First of all we notice that the pseudo-hermicity
of the propagator implies the following useful relations
\begin{align}
\psif_-(t)\vcal[\psib_-(t)]^\dagger\gamma_5&= \psib_-(t)\vcal^\dagger[\psif_-(t)]^\dagger\gamma_5\,,\label{eq:B6}\\
&  \nonumber\\[-2ex]
\psif_+(t)\vcal[\psib_+(t)]^\dagger\gamma_5&= \psib_+(t)\vcal^\dagger[\psif_+(t)]^\dagger\gamma_5\,,\label{eq:B7}
\end{align}
and we then proceed by checking that the equation
\begin{align}
1&= -\psif_+(t-1)\vcal [\psib_-(t)]^\dagger\gamma_5+\psib_+(t-1)\vcal^\dagger [\psif_-(t)]^\dagger\gamma_5\nonumber\\[-1ex]
 &  \label{eq:B8}\\[-1ex]
 &\quad  +\psif_-(t+1)\vcal [\psib_+(t)]^\dagger\gamma_5-\psib_-(t+1)\vcal^\dagger [\psif_+(t)]^\dagger\gamma_5\nonumber
\end{align}
holds for $a\leq t< T$. We write down \eq{eq:diff_eq_qprop} for $t=t'$,
\begin{align}
1&= -\psib_-(t+1)\vcal^\dagger[\psif(t)]^\dagger\gamma_5-\psif_+(t-1)\vcal[\psib(t)]^\dagger \gamma_5\nonumber\\[-1ex]
 &  \label{eq:diff_eq_qprop0}\\[-1ex]
 &\quad  +B(t)\left\{\psif_-(t)\vcal[\psib(t)]^\dagger\gamma_5+\psib_+(t)\vcal^\dagger [\psif(t)]^\dagger\gamma_5 \right\}\,,\nonumber
\end{align}
and multiply it by $P_-$ from the right
\begin{align}
P_-&= -\psib_-(t+1)\vcal^\dagger[\psif(t)_+]^\dagger\gamma_5-\psif_+(t-1)\vcal[\psib(t)_+]^\dagger \gamma_5\nonumber\\
 &  \nonumber\\[-2ex]
 &\quad  +B(t)\left\{\psif_-(t)\vcal[\psib(t)_+]^\dagger\gamma_5+\psib_+(t)\vcal^\dagger [\psif(t)_+]^\dagger\gamma_5\right\}\,.\nonumber
\end{align}
Then we observe that 
\begin{align}
B(t)& \left\{\psif_-(t)\vcal [\psib_+(t)]^\dagger\gamma_5
+\psib_+(t)\vcal^\dagger [\psif_+(t)]^\dagger\gamma_5\right\}\nonumber\\
& \nonumber\\[-1ex]
&\hspace{-0.25cm}\stackrel{(\ref{eq:B7})}{=} B(t)\left\{\psif_-(t)\vcal [\psib_+(t)]^\dagger\gamma_5+
\psif_+(t)\vcal[\psib_+(t)]^\dagger\gamma_5\right\}\nonumber\\
&  \nonumber\\[-1ex]
&= B(t)(P_-+P_+)\psif(t)\vcal[\psib_+(t)]^\dagger\gamma_5\nonumber\\
&  \nonumber\\[-1ex]
&\hspace{-0.25cm}\stackrel{(\ref{eq:B27f})}{=} \left\{\psif_-(t+1)+\psif_+(t-1)\right\}\vcal[\psib_+(t)]^\dagger\gamma_5\,,\nonumber
\end{align}
which yields
\be\label{eq:B81}
P_-=-\psib_-(t+1)\vcal^\dagger[\psif_+(t)]^\dagger\gamma_5+\psif_-(t+1)\vcal[\psib_+(t)]^\dagger\gamma_5\,.
\ee
By multiplying \eq{eq:diff_eq_qprop0} by $P_+$ from the right and proceeding analogously,
we get
\be\label{eq:B82}
P_+=-\psif_+(t-1)\vcal [\psib_-(t)]^\dagger\gamma_5+\psib_+(t-1)\vcal^\dagger [\psif_-(t)]^\dagger\gamma_5\,.
\ee
It is now very easy to see that eqs.~(\ref{eq:B81}, \ref{eq:B82}) imply \eq{eq:diff_eq_qprop0}\,.
The next step consists in showing that 
\be\label{eq:B9}
[\psif(t-1)]^\dagger\gamma_5\psif_-(t)-[\psif(t)]^\dagger\gamma_5\psif_+(t-1)=0\,,
\ee
holds for $1\leq t\leq T$. First of all, we demonstrate that the l.h.s.~is time-independent.
We notice that
\be\label{eq:Bt_g5}
B(t)=\gamma_5 [B(t)]^\dagger \gamma_5\,,
\ee
and write the forward homogeneous equation in two equivalent forms
\begin{align}
&B(t)\psif(t)=\psif_-(t+1)+\psif_+(t-1)\,,\label{eq:Bpsif}\\[-1ex]
&\nonumber\\[-1ex]
&[\psif(t)]^\dagger [B(t)]^\dagger=[\psif_-(t+1)]^\dagger+[\psif_+(t-1)]^\dagger\,.\label{eq:Bpsifdagger}
\end{align}
We multiply the first equation by $[\psif(t)]^\dagger\gamma_5$ from the right
\begin{align}
[\psif(t)]^\dagger\gamma_5B(t)\psif(t)&= \underbrace{[\psif(t)]^\dagger\gamma_5\psif_-(t+1)}_{\oplus}+
\underbrace{[\psif(t)]^\dagger\gamma_5\psif_+(t-1)}_{\ominus}\nonumber\\
&\nonumber\\[-1ex]
&\hspace{-0.35cm}\stackrel{(\ref{eq:Bt_g5})}{=}  [\psif(t)]^\dagger [B(t)]^\dagger \gamma_5\psif(t)\nonumber\\
&\nonumber\\[-1ex]           
&\hspace{-0.35cm}\stackrel{(\ref{eq:Bpsifdagger})}{=} \underbrace{[\psif_-(t+1)]^\dagger\gamma_5\psif(t)}_{\odot}+
\underbrace{[\psif_+(t-1)]^\dagger\gamma_5\psif(t)}_{\otimes}\,,\nonumber
\end{align}
and by noticing that $[\psif_\pm]^\dagger\gamma_5\psif=[\psif]^\dagger\gamma_5\psif_\mp$, we have that
the equation $\otimes - \ominus=\oplus-\odot$ reads
\begin{align}
[\psif(t-1)]^\dagger\gamma_5\psif_-(t)&- [\psif(t)]^\dagger\gamma_5\psif_+(t-1)\\
&  \nonumber\\[-2ex]
&\quad  =[\psif(t)]^\dagger\gamma_5\psif_-(t+1)-[\psif(t+1)]^\dagger\gamma_5\psif_+(t)\,,\nonumber
\end{align}
which proves that the l.h.s.~of \eq{eq:B9} is time-independent.
Furthermore we observe that, due to the boundary condition $\psif_+(0)=0$,
it vanishes at $t=1$. The time-independence implies that \eq{eq:B9} is valid in the 
whole interval $1\leq t\leq T$.

We multiply \eq{eq:B8} by $P_+$ from the left and by $\psif(t-1)$ from the right
\begin{align}
P_+\psif(t-1)\!&= \!-\psif_+(t-1)\vcal[\psib_-(t)]^\dagger\gamma_5 \psif(t-1)\nonumber\\
             &\quad  +\psib_+(t-1)\vcal^\dagger[\psif_-(t)]^\dagger\gamma_5\psif(t-1)\nonumber\\
             &\hspace{-0.35cm}\stackrel{(\ref{eq:B9})}{=} \!-\psif_+(t-1)\vcal[\psib_-(t)]^\dagger\gamma_5 \psif(t-1)\nonumber\\
             &\qquad \hspace{-0.15cm}  +\psib_+(t-1)\vcal^\dagger[\psif_+(t-1)]^\dagger\gamma_5\psif(t)\nonumber\\
             &\hspace{-0.35cm}\stackrel{(\ref{eq:B7})}{=} \!-\psif_+(t-1)\vcal[\psib_-(t)]^\dagger\gamma_5 \psif(t-1)\nonumber\\
             &\qquad \hspace{-0.15cm}  +\psif_+(t-1)\vcal[\psib_+(t-1)]^\dagger\gamma_5\psif(t)\nonumber\\
             &= \!\psif_+(t-1)\vcal\left\{[\psib_+(t-1)]^\dagger\gamma_5\psif(t)
                \!-[\psib_-(t)]^\dagger\gamma_5 \psif(t-1)\right\}\,,\nonumber
\end{align}
{\flushleft and conclude that}
\be
\vcal^{-1}=[\psib_+(t-1)]^\dagger\gamma_5\psif(t)-[\psib_-(t)]^\dagger\gamma_5 \psif(t-1)\,,
\ee
or equivalently
\be\label{eq:vcal_dagger_inv}
[\vcal^\dagger]^{-1}=[\psif(t)]^\dagger\gamma_5\psib_+(t-1)
-[\psif(t-1)]^\dagger\gamma_5\psib_-(t)\,.
\ee
Finally we show that the r.h.s.~of the above equation is time-independent.
To achieve that, we multiply \eq{eq:B27b} by $[\psif(t)]^\dagger \gamma_5$ from the left
\be
[\psif(t)]^\dagger \gamma_5 B(t)\psib(t)\stackrel{(\ref{eq:B27b})}{=}[\psif(t)]^\dagger \gamma_5 \psib_-(t+1)
+[\psif(t)]^\dagger \gamma_5 \psib_+(t-1)\,,
\ee
and \eq{eq:Bpsifdagger} by $\gamma_5\psib(t)$ from the right
\begin{align}
[\psif(t)]^\dagger [B(t)]^\dagger \gamma_5 \psib(t)&= [\psif(t)]^\dagger \gamma_5 B(t)\psib(t)\\
&  \nonumber\\[-2ex]
&\hspace{-0.35cm}\stackrel{(\ref{eq:B27f})}{=} [\psif_-(t+1)]^\dagger\gamma_5\psib(t)+[\psif_+(t-1)]^\dagger\gamma_5\psib(t)\,,\nonumber
\end{align}
and subtract them. It follows that
\begin{align}
&  [\psif(t)]^\dagger \gamma_5 \psib_-(t+1)+[\psif(t)]^\dagger \gamma_5 \psib_+(t-1)\nonumber\\
&  \nonumber\\[-2ex]
&  \,\,=[\psif_-(t+1)]^\dagger\gamma_5\psib(t)+[\psif_+(t-1)]^\dagger\gamma_5\psib(t)\nonumber\\
&  \nonumber\\[-2ex]
&  \,\,=[\psif(t+1)]^\dagger \gamma_5 \psib_+(t)+[\psif(t-1)]^\dagger \gamma_5 \psib_-(t)\,,\nonumber
\end{align}
which can be rewritten as
\begin{align}
[\psif(t)]^\dagger \gamma_5 \psib_+(t-1)&- [\psif(t-1)]^\dagger \gamma_5 \psib_-(t)\\
&  \nonumber\\[-2ex]
&\quad  =[\psif(t+1)]^\dagger \gamma_5 \psib_+(t)-[\psif(t)]^\dagger \gamma_5 \psib_-(t+1)\,,\nonumber
\end{align}
and shows the time independence of the r.h.s.~of \eq{eq:vcal_dagger_inv}. The latter,
due to the boundary conditions, becomes
particularly simple for $t=1$, that is the expression in \eq{eq:vcal_ta}.

%
%

\section{Tadpoles and improvement}\label{sect:pt_tadpoles}

In this subsection we provide the definitions relative 
to the tadpoles as well as their implementation in \verb|WLINE|.
It is important to stress that if one wants to compute several observables,
it is not necessary to compute the tadpoles for each of them. 
The tadpole loops $T^a_{k,\mbox{\scriptsize ghost}}(u_0)$, $T^a_{k,\mbox{\scriptsize gluon}}(u_0)$
and $T^a_{k,\mbox{\scriptsize quark}}(u_0)$ do not depend on the observable, but only on the 
kinematics. They can thus be computed and stored. Then, for each observable,
the tadpole contributions can be calculated according to 
the formulae given in \sect{sect:tadpole_contributions}. This latter computation
is much cheaper than the tadpole loops.

The last subsection is devoted to the $\Oa$-improvement counterterms. 
Under the point of view of the theory, there is nothing to add 
to \sect{sect:pt_improvement}, and we describe only the Matlab implementation.

\subsection{Ghost tadpole}\label{sect:ghost_tadpole}

\subsubsection{Definition}

The $\mathrm{O}(g_0)$-term of the
perturbative expansion of the ghost action is given by
\begin{align}
S_{\mathrm{gh}}^{(1)} &=  \frac{1}{L^6}\sum_{\vecp,\vecpprime,\vecq}
\delta_{\mathrm{P}}(\vecp+\vecpprime+\vecq)\sum_{s_0,t_0,u_0}\sum_{\mu}\label{eq:S_gh_1}\\
&\quad  \sum_{a,b,c} {\bar{c}}^a(-\vecpprime,s_0)
F^{abc}_{\mu}(\vecpprime,\vecp,\vecq;s_0,t_0,u_0){c}^b(-\vecp,t_0)
{q}^c_{\mu}(-\vecq,u_0)\,.\nonumber
\end{align}
where $\delta_{\mathrm{P}}$ indicates the periodic delta function, i.e.~the delta function 
modulo $2\pi$, and $F$ is the gluon-ghost-ghost vertex, whose explicit expression
is readable e.g.~in \cite{thesis:skurth}. It enters in the ghost tadpole,
formally given by 
\be
T^a_{\mu,\mbox{\scriptsize ghost}}(u_0)
= \frac{1}{L^3}\sum_{\vecq}\sum_{s_0,t_0}\sum_c
F^{\bar{c}ca}_{\mu}(\vecq,-\vecq,\vectn;s_0,t_0,u_0)D^c(\vecq;t_0,s_0)\,,
\ee 
which is non-vanishing only for $\mu\neq 0$ and $a=3,8$. The non-vanishing components
can be computed through \cite{impr:csw_int_notes}:
\be\label{eq:ghost_tadpole}
T^a_{k,\mbox{\scriptsize ghost}}(u_0)
= \frac{1}{L^3}\sum_{\vecq}\sum_c
c_{\bar{c}ca}\sin[q_k+\phi_c(u_0)]D^c(\vecq;u_0,u_0)\,.
\ee

\subsubsection{Implementation}

The implementation of the ghost tadpole according to \eq{eq:ghost_tadpole}
can be found in the program \verb|WLINE| as a function in
\begin{verbatim}
TADPOLES_GAUGE/ghost_tadpole.m
\end{verbatim}
Once one is in the directory \verb|TADPOLES_GAUGE|,
it is necessary to load the global variables,
\begin{verbatim}
T L eta nu
\end{verbatim}
contained in
file \verb|topo_parameters.dat|, through
the command
\begin{verbatim}
load_topo_parameters_gtadpole
\end{verbatim}
and initialize the remaining global variables through
\begin{verbatim}
initialization
\end{verbatim}
The function computing the ghost tadpole is 
\begin{verbatim}
function [res] = ghost_tadpole()
\end{verbatim}
The output has dimension
\begin{verbatim}
size(res) = 2 3 T+1
\end{verbatim}
The first index refers to the color; if it is 1, it corresponds 
to the color $a=3$, if 2 to $a=8$. The second index refers
to $k$, and the last one to the temporal coordinate $u_0$.

\subsection{Gluon tadpole}\label{sect:gluon_tadpole}

\subsubsection{Definition}
At order $g_0$ the gluon action can be expressed in the compact form
\begin{align}
S_\mathrm{G}^{(1)} &=  \frac{1}{6L^6 }\sum_{\vecq_1,\vecq_2,\vecq_3}
\delta_{\mathrm{P}}(\vecq_1 + \vecq_2 + \vecq_3)\label{eq:gact_1}\\
& \quad  \times\sum_{\mu_1,\mu_2,\mu_3}\sum_{t_1,t_2,t_3}\sum_{a_1,a_2,a_3}
V^{a_1 a_2 a_3}_{\mu_1\mu_2\mu_3}(\vecq_1,\vecq_2,\vecq_3;t_1,t_2,t_3)
\prod_j {q}^{a_j}_{\mu_j}(-\vecq_j,t_j)\,,\nonumber
\end{align}
where $V$ is the triple gluon vertex. The latter has a quite involved 
expression, and is readable e.g.~in \cite{thesis:skurth}. It is needed to compute
the gluon tadpole, which reads
\be
T_{\mu,\mathrm{gluon}}^a(u_0)=-\frac{1}{2L^3}\sum_{\vecq}
\sum_{\nu,\rho}\sum_{s_0,t_0}\sum_{c}
V^{\bar{c}ca}_{\nu\rho\mu}(\vecq,-\vecq,\vectn;s_0,t_0,u_0)
D^c_{\rho\nu}(\vecq;t_0,s_0)\,.
\ee
As mentioned at the beginning of this chapter, the tadpole is non-vanishing
only for $\mu\neq 0$ and $a=3,8$. It assumes 
the expression \cite{impr:csw_int_notes}:
\begin{align}
&  T_{k,\mathrm{gluon}}^a(u_0)=-\frac{1}{2L^3}\sum_{\vecq}\sum_c\Biggl\{\nonumber\\
&  \nonumber\\[-2ex]
&  \hspace{1cm}-c_{ac\bar{c}}\,c_k^c(\vecq,u_0)\sum_j\Bigl(D_{jj}^c(\vecq;u_0,u_0)s_k^c(\vecq,u_0)
-D_{jk}^c(\vecq;u_0,u_0)s_j^c(\vecq,u_0)\Bigr)\nonumber\\
&  \nonumber\\[-2ex]
&  \hspace{1cm}-\tfrac{1}{2}f_{ac}^+\Bigl(D_{kk}^c(\vecq;u_0-1,u_0-1)-D_{kk}^c(\vecq;u_0+1,u_0+1)\Bigr)\nonumber\\
&  \nonumber\\[-2ex]
&  \hspace{1cm}-e_{ac}^+\Bigl(D_{kk}^c(\vecq;u_0,u_0+1)-D_{kk}^c(\vecq;u_0,u_0-1)\Bigr)\nonumber\\
&  \nonumber\\[-2ex]
&  \hspace{1cm}-ie_{ac}^+\biggl[-D_{0k}^c(\vecq;u_0,u_0)s_k^c(\vecq,u_0+1)-D_{0k}^c(\vecq;u_0-1,u_0)s_k^c(\vecq,u_0-1)\nonumber\\
&  \nonumber\\[-2ex]
&  \hspace{1cm}+\Bigl(D_{0k}^c(\vecq;u_0-1,u_0-1)+D_{0k}^c(\vecq;u_0,u_0+1)\Bigr)s_k^c(\vecq,u_0) \biggr]\nonumber\\
&  \nonumber\\[-2ex]
&  \hspace{1cm}-ie_{ac}^-\Bigl(D_{0k}^c(\vecq;u_0-1,u_0-1)-D_{0k}^c(\vecq;u_0,u_0+1)\Bigr)c_k^c(\vecq,u_0)\nonumber\\
&  \nonumber\\[-2ex]
&  \hspace{1cm}-\tfrac{1}{2}f_{ac}^+\Bigl(D_{00}^c(\vecq;u_0-1,u_0-1)s_k^c(\vecq,u_0-1)^2\Bigr.\nonumber\\
&  \nonumber\\[-2ex]
&  \hspace{1cm}\Bigl.-D_{00}^c(\vecq;u_0,u_0)s_k^c(\vecq,u_0+1)^2\Bigr)\nonumber\\
&  \nonumber\\[-2ex]
&  \hspace{1cm}-\tfrac{1}{2}f_{ac}^-\Bigl(D_{00}^c(\vecq;u_0-1,u_0-1)s_k^c(\vecq,u_0-1)c_k^c(\vecq,u_0-1)\nonumber\\
&  \nonumber\\[-2ex]
&  \hspace{1cm}+D_{00}^c(u_0,u_0)s_k^c(\vecq,u_0+1)c_k^c(\vecq,u_0+1)\Bigr)\Biggr\}\,,\label{eq:gluon_tadpole}
\end{align}
where it is understood that at the temporal boundaries
\be
D(\vecp;t_0,v_0)=0\,,\qquad\mbox{if $t_0$ or $v_0$ $=-1,\, T+1$}\,.
\ee

\subsubsection{Implementation}

The implementation of the gluon tadpole according to \eq{eq:gluon_tadpole}
can be found in the program \verb|WLINE| as a function in
\begin{verbatim}
TADPOLES_GAUGE/gluon_tadpole.m
\end{verbatim}
Once one is in the directory \verb|TADPOLES_GAUGE|,
it is necessary to load the global variables,
\begin{verbatim}
T L eta nu lambda_0
\end{verbatim}
contained in
file \verb|topo_parameters.dat|, through
the command
\begin{verbatim}
load_topo_parameters_gtadpole
\end{verbatim}
and initialize the remaining global variables through
\begin{verbatim}
initialization
\end{verbatim}
The function computing the ghost tadpole is 
\begin{verbatim}
function [res] = gluon_tadpole()
\end{verbatim}
The output has dimension
\begin{verbatim}
size(res) = 2 3 T+1
\end{verbatim}
The first index refers to the color; if it is 1, it corresponds 
to the color $a=3$, if 2 to $a=8$. The second index refers
to $k$, and the last one to the temporal coordinate $u_0$.

\subsection{Quark tadpole}\label{sect:quark_tadpole}

\subsubsection{Definition}

We proceed in analogy with the ghost and gluon cases, and study
the term of the perturbative expansion of the quark action appearing at order $g_0$.
The Wilson part takes the form
\begin{align}
S_{\mathrm{F,Wilson}}^{(1)} &= 
\frac{1}{L^6}\sum_{\vecp,\vecpprime,\vecq}
\delta_{\mathrm{P}}(\vecp + \vecpprime + \vecq)
\sum_{\mu}\sum_{x_0,y_0,z_0}\label{eq:qact_1}\\
&\quad   \psibar(-\vecpprime,x_0)
V^{a}_{\mu}(\vecs;x_0,y_0,z_0)\psi(-\vecp,y_0)
{q}^{a}_{\mu}(-\vecq,z_0)\,,\nonumber
\end{align}
with $\vecs=(\vecpprime-\vecp)/2$. It comes together with the Sheikholeslami-Wohlert
part 
\begin{align}
\delta S_{\mathrm{v}}^{(n)} &=  \frac{\csw^{(0)}}{L^6}
\sum_{\vecp,\vecpprime,\vecq}\delta_{\mathrm{P}}
(\vecp + \vecpprime + \vecq)
\sum_{x_0,z_{0}}\sum_{\mu}\sum_{a}\\
&\quad   \psibar(-\vecpprime,x_0)S^{a}_{\mu}
(\vecq;x_0,z_{0})\psi(-\vecp,x_0){q}^{a}_{\mu}(-\vecq,z_{0})\,.\nonumber
\end{align}
It follows that the complete gluon-quark-quark vertex is given by
\be
V^a_\mu(\vecpprime,\vecp,\vecq;s_0,t_0,u_0)=V^a_\mu(\vecs;s_0,t_0,u_0)+\csw^{(0)}S^{a}_{\mu}
(\vecq;s_0,u_{0})\delta_{s_0,t_0}\,,
\ee
entering in the quark tadpole as
\be
T^a_{\mu,\mathrm{quark}}(u_0) = \frac{1}{L^3}\sum_{\vecq}\sum_{s_0,t_0}
\Tr\Bigl\{V^a_{\mu}(\vecq,-\vecq,\vectn;s_0,t_0,u_0)S(\vecq;t_0,s_0)\Bigr\}\,.
\ee
The latter is non-vanishing only for $\mu\neq 0$ and $a=3,8$;
the explicit expression reads
\begin{align}
&  T^{a}_{k,\mathrm{quark}}(u_0)=\label{eq:Tq_final}\\
&  \hspace{1cm}-\frac{1}{L^3}\sum_\vecq\Tr\, I^a
\Biggl\{\Bigl(i\sin[\alpha_k(\vecq,u_0)]
-\gamma_k\cos[\alpha_k(\mathbf{q},u_0)]\Bigl)
S(\mathbf{q};u_0,u_0)\nonumber\\
&  \hspace{1cm}+\frac{i}{4}c_\mathrm{SW}^{(0)}\cos(\mathcal{E})
\sigma_{0k}\Big(S(\mathbf{q};u_0+1,u_0+1)-S(\mathbf{q};u_0-1,u_0-1)\Bigl)\Biggr\}\,.\nonumber
\end{align}

\subsubsection{Implementation}

The implementation of the quark tadpole according to \eq{eq:Tq_final}
can be found in the program \verb|WLINE| as a function in
\begin{verbatim}
TADPOLE_QUARK/quark_tadpole.m
\end{verbatim}
Once one is in the directory \verb|TADPOLE_QUARK|,
it is necessary to load the global variables,
\begin{verbatim}
T L eta nu theta_angle csw_0 mass_0
\end{verbatim}
contained in
file \verb|topo_parameters.dat|, through
the command
\begin{verbatim}
load_topo_parameters_qtadpole
\end{verbatim}
and initialize the remaining global variables through
\begin{verbatim}
initialization
\end{verbatim}
The function computing the quark tadpole is 
\begin{verbatim}
function [res] = quark_tadpole()
\end{verbatim}
The output has dimension
\begin{verbatim}
size(res) = 2 3 T+1
\end{verbatim}
The first index refers to the color; if it is 1, it corresponds 
to the color $a=3$, if 2 to $a=8$. The second index refers
to $k$, and the last one to the temporal coordinate $u_0$.

\subsection{Improvement}\label{sect:code_improvement}

\subsubsection{Implementation}

The implementation of the $\Oa$-improvement counterterms according to \eq{eq:full_expr}
can be found in the program \verb|WLINE| as a function in
\begin{verbatim}
IMPROVEMENT/wline_impr.m
\end{verbatim}
Once one is in the directory \verb|IMPROVEMENT|,
it is necessary to load the global variables,
\begin{verbatim}
T L eta nu back_switch lambda_0 Nf
\end{verbatim}
contained in
file \verb|topo_parameters.dat|, through
the command
\begin{verbatim}
load_topo_parameters_impr
\end{verbatim}
and initialize the remaining global variables through
\begin{verbatim}
initialization
\end{verbatim}
The function computing the improvement counterterm is 
\begin{verbatim}
function [res] = wline_impr(starting_point,wilson_path)
\end{verbatim}
which needs the starting point and the path describing the Wilson loop
as is explained in \sect{sect:automatic}.
The output is a complex number, and precisely the expectation value in \eq{eq:full_expr}. 

\section{Tests of the code}\label{sect:tests_code}

\subsection{The average plaquette}\label{sect:av_plaqi}

We consider the unimproved gauge action
\be\label{eq:g_act_noimpr}
S_\mathrm{G}[U]=\frac{\beta}{3}\sum_p \Re\tr\{1-U(p)\}\,,
\ee  
where the sum is over all unoriented plaquettes,
and rewrite the parametrization (\ref{eq:U_parametrization}) of the gauge links in the form
\be
U_\mu(x)=\left\{1+\sigma_\mu(x)+\mathrm{O}(\sigma^2)\right\}V_\mu(x)\,,\quad \sigma=\mathrm{O}(\beta^{-1/2})\,,\label{eq:U_sigma_param}
\ee
where the fields $\sigma_\mu$ are nothing but a rescaling of the gluon fields $q_\mu$.
In the following we assume that the gauge has been fixed. Although the details of the gauge fixing
are not relevant for the present discussion, one may assume that we have followed 
the procedure described in \sect{sect:gauge_fix}.
In the weak coupling regime we expand the action in powers of $1/\beta$,
and write
the partition function in the simplified form
\be\label{eq:part_f_plaqi}
\mathcal{Z}=K\int \mathrm{D}[\sigma]\mathrm{exp}\left\{-\tfrac{1}{2}\beta \sigma D^{-1}\sigma+\mbox{higher orders}\right\}\,.
\ee
Here $K$ is an overall constant factor and $D$ is the gluon propagator defined in \eq{eq:DK_1}. 
Since the integral in \eq{eq:part_f_plaqi} is Gaussian, its value is a determinant
\be
\mathcal{Z}=K'|D/\beta|^{1/2}(1+\mathrm{O}(\beta^{-1}))\,.
\ee 
The matrix $D$ has the dimensionality of the parameter space after gauge fixing.
In other words, it is a square matrix with $N_\mathrm{g}$ rows, where $N_\mathrm{g}$
is the number of physical gluonic degrees of freedom. Removing a factor of $\beta$
from each row of the matrix, we find
\be
\mathcal{Z}=K'|D|^{1/2}\beta^{-N_\mathrm{g}/2}(1+\mathrm{O}(\beta^{-1}))\,.
\ee
We now exploit this result to obtain the following 1-loop prediction for 
the average plaquette 
\begin{align}
\frac{1}{3N_p}\langle \sum_p \Re\tr\{1-U(p)\}\rangle&= -\frac{1}{N_p}\frac{\partial}{\partial \beta}\ln \mathcal{Z}\nonumber\\
&= -\frac{1}{N_p}\frac{-\tfrac{N_\mathrm{g}}{2}\beta^{-1}(1+\mathrm{O}(\beta^{-1}))
+\mathrm{O}(\beta^{-2})}{1+\mathrm{O}(\beta^{-1})}\nonumber\\
&= \frac{N_\mathrm{g}}{2N_p\beta}+\mathrm{O}(\beta^{-2})\,,\label{eq:av_plaqi}
\end{align}
where we have summed only over the $N_p=3L^3(2T-1)$ dynamical plaquettes,
and we have kept the factor $1/3$ in order to normalize the 
trace of the unity matrix. The number
of degrees of freedom is given by
\be
N_\mathrm{g}=8[L^3(4T-3)-L^3(T-1)]-\nu\,.
\ee 
The factor 8 in front stems from the fact that
the Lie algebra $su(3)$ of SU(3) is a real vector space 
of dimension 8. The first factor in square brackets counts
the lattice variables $q_\mu(x)$, and the second factor subtracts
the gauge fixed ones. The volume term $\nu$ is given by
\begin{align}
\nu=\left\{
\begin{array}{ll}
&8\,,\quad\mbox{for irreducible background field}\,,\\[-1ex]
&\label{eq:nu_bk_dep}\\[-1ex]
&2\,,\quad\mbox{for non-vanishing Abelian background field}\,.
\end{array}\right.
\end{align}
For irreducible background field we intend e.g.~a setup as in chapter \ref{chap:SSM},
with boundary values $C=C'=0$. The value $\nu=8$ stems from the zero momentum $q_0$
gluons at the temporal boundary $x_0=0$. We have eight of them, because dim$(su(3))=8$. 
They obey Dirichlet boundary conditions, and
are associated with spatially constant diagonal modes. The latter are not fixed
by the gauge fixing procedure, as pointed out in \sect{sect:gauge_fix}, 
and do not belong to the degrees of freedom. They just survive as a global
symmetry of the theory. 
For the non-vanishing background field case the situation is analogous, with the only
difference that at the lower temporal boundary we have a mixture of Dirichlet and Neumann
boundary conditions. Inspection of \eqs{eq:boundaries_q0} reveals that only the two gluons
obeying Dirichlet boundary conditions are associated with spatially
constant diagonal modes; this implies $\nu=2$.

In order to test the code, we have reproduced, for various choices
of $L$ and $T$ up to 32, the result (\ref{eq:av_plaqi}),
for both vanishing and non-vanishing background field, with a numerical
precision up to 12 digits.

\subsection{Comparison with Monte Carlo results}\label{sect:NP_vs_PT}

\begin{figure}[ht]
     \begin{minipage}{0.5\textwidth}
      \centering
       \includegraphics[width=0.9\textwidth]{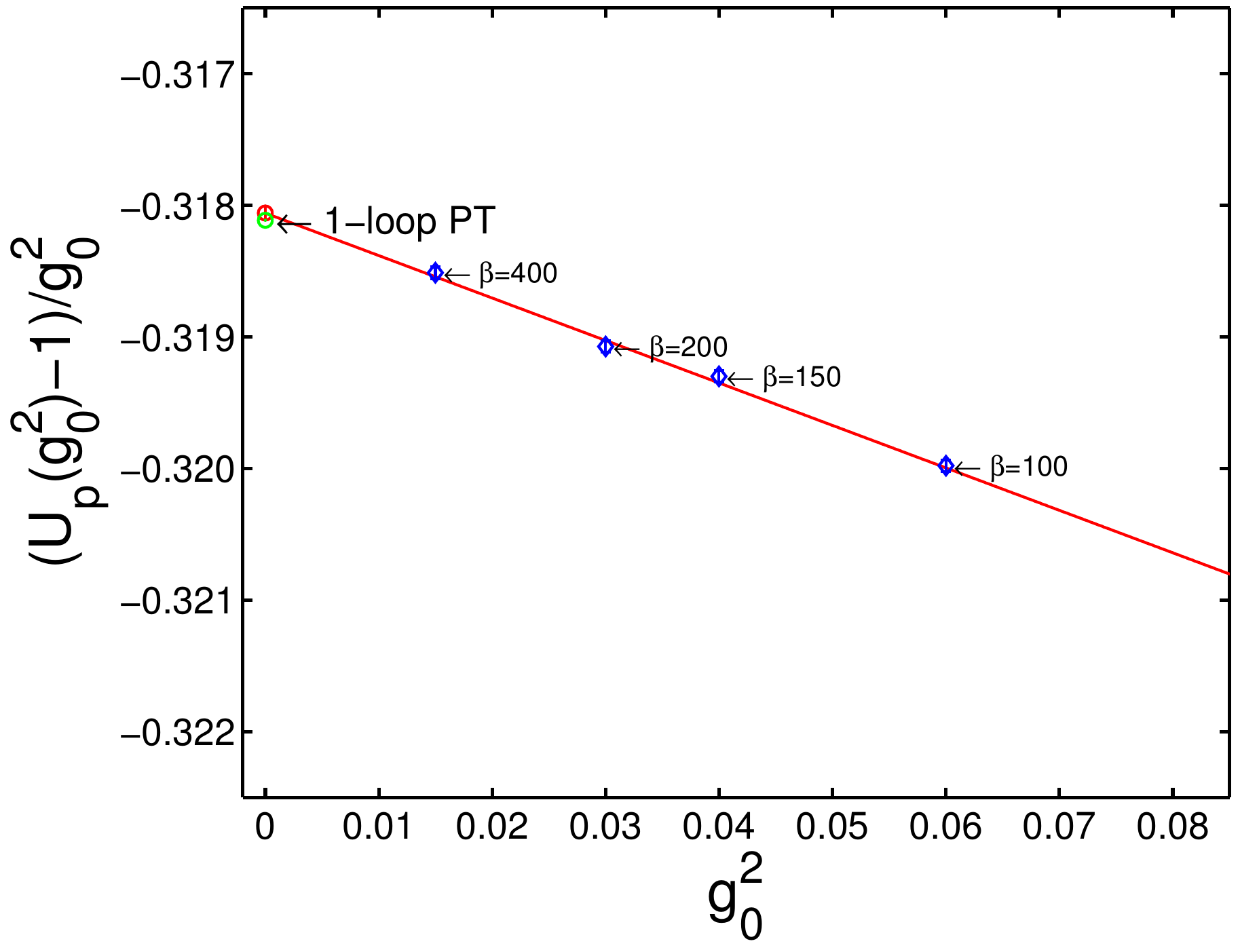}
     \end{minipage}\hfill
     \begin{minipage}{0.5\textwidth}
      \centering
       \includegraphics[width=0.9\textwidth]{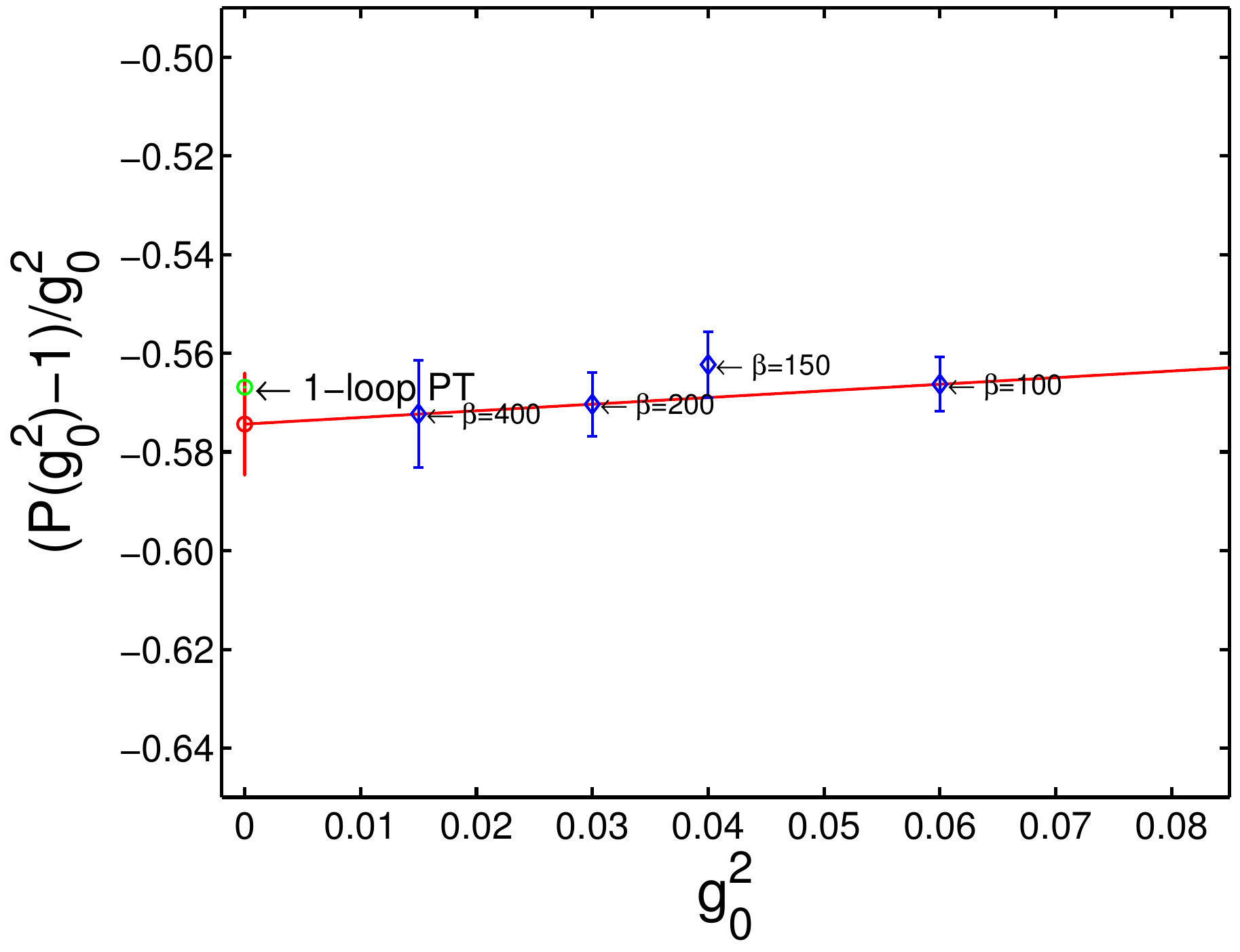}\\
       \vspace{0.5cm}
       \includegraphics[width=0.9\textwidth]{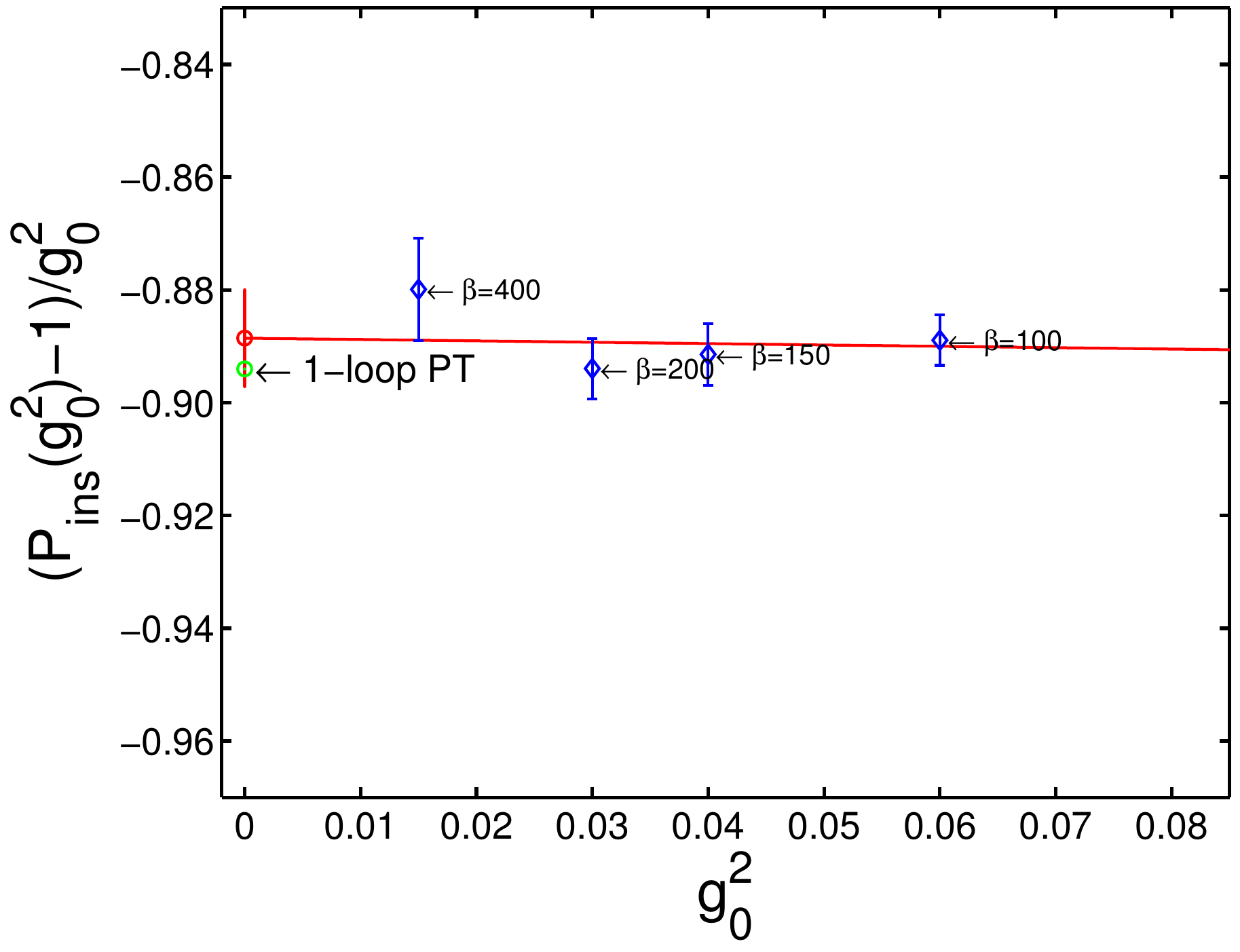}
     \end{minipage}
\mycaption{Comparison of the perturbative results with the 
non-perturbative ($N_\mathrm{f}=0$) ones.}\label{fig:NP_vs_PT}
   \end{figure}

The perturbative results for the plaquette and the Polyakov loops with 
and without insertion of the electric operator have been checked
against the corresponding non-perturbative (quenched) computations.
The latter have been performed at small bare couplings, $0.015\leq g_0^2\leq 0.06$,
setting the improvement coefficient $\ct$ to its tree-level value. 
In all cases $L=T=4$. The 1-loop contribution for the 
observable ${\mathcal{O}}$ is extracted from the non-perturbative result $\langle{\mathcal{O}}\rangle$
according to
\be
\langle{\mathcal{O}}^{(1)}\rangle=\lim_{g_0^2\to 0} \left(\langle{\mathcal{O}}(g_0^2)\rangle-1\right)/g_0^2\,,
\ee
where the observable is normalized by its tree-level value. The plot on the left of \Fig{fig:NP_vs_PT}
shows the case where the observable is the average plaquette. 
The upper and lower plots on the right 
show the simple Polyakov loop and the Polyakov loop with insertion of the clover leaf
operator respectively. The extrapolation
is achieved through a simple linear fit.
The agreement is found within the statistical uncertainty.


\section{Summary}\label{sect:matlab_summary}

The program \verb|WLINE| has been conceived 
as a double-purpose project. The first purpose is to provide
a program able to compute the expectation value, at one-loop order 
of perturbation theory, of any QCD Wilson line in the \SF scheme 
with vanishing or non-vanishing background field,
by generating and computing all needed diagrams
and the improvement counterterms. The user is asked
to specify only the path describing the Wilson line and
a few kinematical parameters. The rest of the program
can be seen as a black box, and the execution does not require
any further effort from the user's side.
The second purpose is to provide a 
structure with several modules. Each module has its own task, and
can easily be accessed and used outside the program.
Each function computing the propagators, the tadpoles or the improvement 
counterterms shares with the rest of program only the necessary 
kinematical variables. 

The presence of a non-vanishing background field precludes from finding
a simple analytical expression for the ghost, gluon and quark propagators.
By starting from the very definition of the latter, one is found to
invert big sparse matrices. As a consequence, the computational cost increases
very rapidly with the fineness of the lattice. A much more viable way is provided
in \sect{sect:pt_propagators}, where recursive methods are introduced. They allow
to considerably reduce the computational cost, which is found to linearly scale with
the lattice extension. For each propagator, a detailed proof of the recursive method
is provided.

The non-vanishing background field is also the cause 
of the presence of the tadpole diagrams. Although they could be computed
from the complete expression of the vertices, the formulae provided by Peter Weisz
allow to significantly reduce the programming as well as the computational efforts.
The tadpole loops are observable independent. They can be separately computed and stored
once the kinematics is defined. Of course, their contribution to the one-loop expansion of 
the desired observable depends on the observable itself, but this last calculation
is computationally not very time consuming if the loops are already stored.

The improvement is restricted to the computation of the terms
stemming from the one-loop expansion of the boundary counterterms
of the gauge action, and in particular to the time-like plaquettes.
They owe their existence to the
presence of a non-vanishing
background field, and
are not computationally demanding, even in the $N_\mathrm{f}\neq 0$ case.

Among all tests performed to check the correctness of the code,
it is worth mentioning the computation of the average plaquette. 
It is a simple test comparing the result of the program with the predictions
of the theory, in which only the counting of the gluonic degrees of freedom enters.
This kind of test can be straightforwardly extended to other actions and choices of boundary
conditions.

The program \verb|WLINE| is available in its released version 1.0. Plans
for a version 1.1 are already established. Apart from a few small improvements,
mainly addressed to speed up the computation, the main purpose is to remove the restriction
of an observable consisting of a closed loop of gauge links, and admit a more general class
of observables, including, among others, two separated Wilson loops. 
In a next future, a version 2.0, allowing for smeared gauge links, is expected to
enhance the support given by perturbation theory to non-perturbative HQET computations.

\chapter{Conclusions}\label{chap:conclusions}

The subject of the first part of this work is the combination of the 
Tor Vergata Step Scaling Method, involving the relativistic QCD Lagrangian, 
and HQET. The final result is a very precise and controlled lattice 
determination, in the continuum limit, of 
the b-quark mass and $\mathrm{B_s}$-meson decay constant. The 
former is affected
by an uncertainty of 1.5\%, while the decay constant is extracted with a precision 
slightly worse than 3\%. The meticulous error analysis attending the
simulations ensures that these results do not
suffer from any systematic errors, apart from the use of the quenched approximation.
Neither extrapolations in the heavy quark mass, nor other approximations on the kinematics
of the quark-antiquark pair are introduced.
  
The results of this study are not restricted only to these two numbers. The interplay 
of relativistic QCD and HQET shows that the static limit alone provides high precision results,
which lie very close to the interpolated final numbers, with only one exception. The latter
is the decay constant in the small volume, where the finite quark mass corrections amount to
roughly 5\%. On the other side, the Step Scaling Method is shown to be a successful and 
powerful tool to investigate two-scale problems. However, without the support of the effective 
theory the extrapolation to the b-region is difficult to control, especially when the simulated quark
masses are much lighter than the extrapolated one.

The second part of this thesis deals with the renormalization of the
chromo-ma\-gne\-tic operator, appearing at order $1/m$ in the HQET Lagrangian. It is 
the leading order operator, in the effective theory, responsible for the spin splitting in heavy-light
quark bound states. A precise computation of its renormalization group invariant expression, 
along the guidelines of the ALPHA collaboration, requires in turn an accurate knowledge
of the scale running of the renormalization factor in the \SF scheme. The main result
for this purpose is the two-loop anomalous dimension of the operator, as well as a both
perturbative and non-perturbative (in the quenched approximation) study
of the cutoff effects affecting the renormalization factor and the corresponding step scaling functions.
The latter measure the variation which the former undergoes, when the renormalization 
scale is changed by a factor of two. 

The anomalous dimension is computed by starting from the known two-loop expression in the $\MSbar$ scheme
as well as the one-loop connection between the latter and an intermediate scheme, the lattice 
minimal subtraction one. By computing the one-loop connection between the latter and the Schr\"odinger functional,
one ends up with the two-loop SF anomalous dimension. 

Perturbation theory, at one-loop order, and the non-perturbative
simulations, without sea quarks, agree with the fact that the cutoff effects
associated with the step scaling functions are invisible as long as the relative statistical precision
is around one percent. In presence of sea quarks, perturbation theory predicts
bigger but still moderate deviations from the continuum limit. Unfortunately, a non-perturbative confirmation
is still missing.

The perturbative computations in the \SF scheme are complicated by the presence
of a non-vanishing background field. With respect to the vanishing case, it introduces new diagrams
and forbids to have a simple analytical expression for the propagators. Furthermore,
it requires the addition of the $\Oa$-improvement counterterms in order to reach the continuum limit
with a rate proportional to $(a/L)^2$. 
The problem is faced by carrying out the perturbative expansion in position space. This yields
a considerable speed-up in comparison with a momentum space computation. 
This is part of the background of the Matlab code \verb|WLINE|,
which joins the requirements of a robust, efficient and reasonably fast program, with a very 
user-friendly interface.

In an ideal to-do list, we would write the following points: 
\bi
\item{\textit{Reduction of the uncertainty associated with the renormalization factor of the quark mass.}}\\[-2ex]
\item{\textit{Improvement of the precision of the large volume computations both in HQET and in relativistic QCD.}}\\[-2ex]
\item{\textit{Computation of $Z_\mathrm{mag}^\mathrm{RGI}$ via matching between HQET and QCD. At first in the quenched
approximation.}}\\[-2ex]
\item{\textit{Dynamical fermions.}}\\[-2ex]
\item{\textit{Development of the code} \verb|WLINE|. }
\ei

The first point has roots in the observation 
that the uncertainty on the final result for the b-quark mass
is dominated by the one on the total renormalization factor $Z_\mathrm{M}$
of the RGI quark mass. 

The second point originates from the importance that large volume computations already
in the static approximation have in a precise determination of the bottom-light 
decay constant. The use of actions with smeared gauge links as well as refined
fitting techniques \cite{Estat:me} are helpful, but, so far, not sufficient to overcome
the problem of the degradation of the signal-to-noise ratio in the heavy-light and heavy-heavy correlators.
Especially in view of the inclusion of sea quarks, the most promising way is probably represented
by the all-to-all propagators \cite{Foley:2005ac}.

About the third point it is worth mentioning an alternative approach for the renormalization
of the chromo-magnetic operator. It can be achieved by following the procedure \cite{hqet:pap1,mb:nf0},
consisting in a non-perturbative matching between HQET and QCD, where, unlike in chapter \ref{chap:Zspin},
valence light quarks appear.

The fourth point certainly represents the next future. For the combination of HQET and relativistic
QCD the challenge is to perform simulations in a large volume ($\gtrsim\!1.5$ fm) with a lattice
resolution, where quarks like the charm and heavier can be handled with confidence.
For the renormalization of the chromo-magnetic operator the two-loop anomalous dimension
is now known also for non-vanishing $N_\mathrm{f}$. However, the renormalization procedure depicted 
in chapter \ref{chap:Zspin}
has shown in \cite{Zspin:me}, for $N_\mathrm{f}\!=\! 0$, to require a large amount 
of gauge configurations to achieve a good statistical precision.
This is of course an unpleasant situation in presence of dynamical fermions, for which the
production of statistically uncorrelated gauge samples is computationally very expensive.
The method mentioned in the third point may provide a more viable approach. 

The code \verb|WLINE| has been successfully applied for the computations 
reported in chapter \ref{chap:Zspin}. Its double-purpose structure will be helpful in the future
for testing other perturbative and non-perturbative codes,
as well as to perturbatively compute other observables and study their cutoff effects.



\appendix


\bibliography{bibliography}

\begin{thebibliography}{102}
\providecommand{\natexlab}[1]{#1}
\providecommand{\url}[1]{\texttt{#1}}
\expandafter\ifx\csname urlstyle\endcsname\relax
  \providecommand{\doi}[1]{doi: #1}\else
  \providecommand{\doi}{doi: \begingroup \urlstyle{rm}\Url}\fi

\bibitem[Della~Morte et~al.(2007{\natexlab{a}})]{Estat:me}
Michele Della~Morte et~al.
\newblock Heavy-strange meson decay constants in the continuum limit of
  quenched {Q}{C}{D}.
\newblock 2007{\natexlab{a}}.
\newblock arXiv:0710.2201 [hep-lat].

\bibitem[Heitger et~al.(2003)Heitger, Kurth, and Sommer]{zastat:pap3}
Jochen Heitger, Martin Kurth, and Rainer Sommer.
\newblock Non-perturbative renormalization of the static axial current in
  quenched {Q}{C}{D}.
\newblock \emph{Nucl. Phys.}, B669:\penalty0 173--206, 2003.

\bibitem[Politzer(1973)]{Politzer:1973fx}
H.~David Politzer.
\newblock Reliable perturbative results for strong interactions?
\newblock \emph{Phys. Rev. Lett.}, 30:\penalty0 1346--1349, 1973.

\bibitem[Gross and Wilczek(1973)]{Gross:1973id}
D.~J. Gross and Frank Wilczek.
\newblock Ultraviolet behavior of non-abelian gauge theories.
\newblock \emph{Phys. Rev. Lett.}, 30:\penalty0 1343--1346, 1973.

\bibitem[Wilson(1974)]{Wilson}
Kenneth~G. Wilson.
\newblock Confinement of quarks.
\newblock \emph{Phys. Rev.}, D10:\penalty0 2445--2459, 1974.

\bibitem[Eichten(1988)]{stat:eichten}
E.~Eichten.
\newblock Heavy quarks on the lattice.
\newblock \emph{Nucl. Phys. Proc. Suppl.}, 4:\penalty0 170, 1988.

\bibitem[Guagnelli et~al.(2002{\natexlab{a}})Guagnelli, Palombi, Petronzio, and
  Tantalo]{Guagnelli:2002jd}
Marco Guagnelli, Filippo Palombi, Roberto Petronzio, and Nazario Tantalo.
\newblock f({B}) and two scales problems in lattice {Q}{C}{D}.
\newblock \emph{Phys. Lett.}, B546:\penalty0 237--246, 2002{\natexlab{a}}.

\bibitem[Eichten and Hill(1990{\natexlab{a}})]{stat:eichhill2}
Estia Eichten and Brian Hill.
\newblock Static effective field theory: 1/m corrections.
\newblock \emph{Phys. Lett.}, B243:\penalty0 427--431, 1990{\natexlab{a}}.

\bibitem[Falk et~al.(1991)Falk, Grinstein, and Luke]{Falk:1990pz}
Adam~F. Falk, Benjamin Grinstein, and Michael~E. Luke.
\newblock Leading mass corrections to the heavy quark effective theory.
\newblock \emph{Nucl. Phys.}, B357:\penalty0 185--207, 1991.

\bibitem[{L\"uscher} et~al.(1991){L\"uscher}, Weisz, and Wolff]{alpha:sigma}
Martin {L\"uscher}, Peter Weisz, and Ulli Wolff.
\newblock A numerical method to compute the running coupling in asymptotically
  free theories.
\newblock \emph{Nucl. Phys.}, B359:\penalty0 221--243, 1991.

\bibitem[Bode et~al.(2001)]{alpha:letter}
Achim Bode et~al.
\newblock First results on the running coupling in {{Q}{C}{D}} with two
  massless flavors.
\newblock \emph{Phys. Lett.}, B515:\penalty0 49--56, 2001.

\bibitem[Capitani et~al.(1999)Capitani, {L\"uscher}, Sommer, and
  Wittig]{mbar:pap1}
Stefano Capitani, Martin {L\"uscher}, Rainer Sommer, and Hartmut Wittig.
\newblock Nonperturbative quark mass renormalization in quenched lattice
  {{Q}{C}{D}}.
\newblock \emph{Nucl. Phys.}, B544:\penalty0 669, 1999.

\bibitem[Isgur and Wise(1989)]{Isgur:1989vq}
Nathan Isgur and Mark~B. Wise.
\newblock Weak decays of heavy mesons in the static quark approximation.
\newblock \emph{Phys. Lett.}, B232:\penalty0 113, 1989.

\bibitem[Korner and Thompson(1991)]{Korner:1991kf}
J.~G. Korner and George Thompson.
\newblock The heavy mass limit in field theory and the heavy quark effective
  theory.
\newblock \emph{Phys. Lett.}, B264:\penalty0 185--192, 1991.

\bibitem[Sommer(2006)]{Sommer:Nara}
Rainer Sommer.
\newblock Non-perturbative {Q}{C}{D}: Renormalization, {O}(a)-improvement and
  matching to heavy quark effective theory, 2006.
\newblock hep-lat/0611020.

\bibitem[Georgi(1990)]{Georgi:1990um}
Howard Georgi.
\newblock An effective field theory for heavy quarks at low-energies.
\newblock \emph{Phys. Lett.}, B240:\penalty0 447--450, 1990.

\bibitem[Isgur and Wise(1990)]{Isgur:1989ed}
Nathan Isgur and Mark~B. Wise.
\newblock Weak transition form-factors between heavy mesons.
\newblock \emph{Phys. Lett.}, B237:\penalty0 527, 1990.

\bibitem[Kurth and Sommer(2001{\natexlab{a}})]{Kurth:2000ki}
Martin Kurth and Rainer Sommer.
\newblock Renormalization and {O}(a)-improvement of the static axial current.
\newblock \emph{Nucl. Phys.}, B597:\penalty0 488--518, 2001{\natexlab{a}}.

\bibitem[Heitger and Sommer(2004)]{hqet:pap1}
Jochen Heitger and Rainer Sommer.
\newblock Non-perturbative heavy quark effective theory.
\newblock \emph{JHEP}, 02:\penalty0 022, 2004.

\bibitem[Collins(1984)]{Collins:1984xc}
John~C. Collins.
\newblock \emph{Renormalization. An introduction to renormalization, the
  renormalization group, and the operator product expansion}.
\newblock Cambridge, Uk: Univ. Pr., 1984.

\bibitem[Maiani et~al.(1992)Maiani, Martinelli, and Sachrajda]{stat:MMS}
L.~Maiani, G.~Martinelli, and C.~T. Sachrajda.
\newblock Nonperturbative subtractions in the heavy quark effective field
  theory.
\newblock \emph{Nucl. Phys.}, B368:\penalty0 281--292, 1992.

\bibitem['t~Hooft(1973)]{ms:thooft}
G.~'t~Hooft.
\newblock Dimensional regularization and the renormalization group.
\newblock \emph{Nucl. Phys.}, B61:\penalty0 455--468, 1973.

\bibitem[Bardeen et~al.(1978)Bardeen, Buras, Duke, and Muta]{msbar:gen}
William~A. Bardeen, A.~J. Buras, D.~W. Duke, and T.~Muta.
\newblock Deep inelastic scattering beyond the leading order in asymptotically
  free gauge theories.
\newblock \emph{Phys. Rev.}, D18:\penalty0 3998, 1978.

\bibitem[{L\"uscher} et~al.(1992){L\"uscher}, Narayanan, Weisz, and
  Wolff]{SF:LNWW}
Martin {L\"uscher}, Rajamani Narayanan, Peter Weisz, and Ulli Wolff.
\newblock The {S}chr{\"o}dinger functional: A renormalizable probe for
  nonabelian gauge theories.
\newblock \emph{Nucl. Phys.}, B384:\penalty0 168--228, 1992.

\bibitem[Sint(1994)]{SF:stefan1}
Stefan Sint.
\newblock On the {Schr\"odinger} functional in {{Q}{C}{D}}.
\newblock \emph{Nucl. Phys.}, B421:\penalty0 135--158, 1994.

\bibitem[Eichten and Hill(1990{\natexlab{b}})]{stat:eichhill1}
Estia Eichten and Brian Hill.
\newblock An effective field theory for the calculation of matrix elements
  involving heavy quarks.
\newblock \emph{Phys. Lett.}, B234:\penalty0 511, 1990{\natexlab{b}}.

\bibitem[Eichten and Hill(1990{\natexlab{c}})]{stat:eichhill_za}
Estia Eichten and Brian Hill.
\newblock Renormalization of heavy-light bilinears and f({B}) for {W}ilson
  fermions.
\newblock \emph{Phys. Lett.}, B240:\penalty0 193, 1990{\natexlab{c}}.

\bibitem[Della~Morte et~al.(2004)]{stat:letter}
Michele Della~Morte et~al.
\newblock Lattice {HQET} with exponentially improved statistical precision.
\newblock \emph{Phys. Lett.}, B581:\penalty0 93--98, 2004.

\bibitem[Della~Morte et~al.(2005{\natexlab{a}})Della~Morte, Shindler, and
  Sommer]{stat:actpaper}
Michele Della~Morte, Andrea Shindler, and Rainer Sommer.
\newblock On lattice actions for static quarks.
\newblock \emph{JHEP}, 08:\penalty0 051, 2005{\natexlab{a}}.

\bibitem[Hasenfratz and Knechtli(2001)]{HYP}
Anna Hasenfratz and Francesco Knechtli.
\newblock Flavor symmetry and the static potential with hypercubic blocking.
\newblock \emph{Phys. Rev.}, D64:\penalty0 034504, 2001.

\bibitem[de~Divitiis et~al.(1995)de~Divitiis, Frezzotti, Guagnelli, and
  Petronzio]{deDivitiis:1995yp}
G.~M. de~Divitiis, R.~Frezzotti, M.~Guagnelli, and R.~Petronzio.
\newblock Nonperturbative determination of the running coupling constant in
  quenched {SU}(2).
\newblock \emph{Nucl. Phys.}, B433:\penalty0 390--402, 1995.

\bibitem[Creutz(1980)]{pot:creutz}
M.~Creutz.
\newblock Monte carlo study of quantized {SU}(2) gauge theory.
\newblock \emph{Phys. Rev.}, D21:\penalty0 2308, 1980.

\bibitem[Fabricius and Haan(1984)]{Fabricius:1984wp}
K.~Fabricius and O.~Haan.
\newblock Heat bath method for the twisted {E}guchi-{K}awai model.
\newblock \emph{Phys. Lett.}, B143:\penalty0 459, 1984.

\bibitem[Kennedy and Pendleton(1985)]{Kennedy:1985nu}
A.~D. Kennedy and B.~J. Pendleton.
\newblock Improved heat bath method for {M}onte {C}arlo calculations in lattice
  gauge theories.
\newblock \emph{Phys. Lett.}, B156:\penalty0 393--399, 1985.

\bibitem[Creutz(1987)]{HOR1}
Michael Creutz.
\newblock Overrelaxation and {M}onte {C}arlo simulation.
\newblock \emph{Phys. Rev.}, D36:\penalty0 515, 1987.

\bibitem[Cabibbo and Marinari(1982)]{Cabibbo:1982zn}
N.~Cabibbo and E.~Marinari.
\newblock A new method for updating {SU}({N}) matrices in computer simulations
  of gauge theories.
\newblock \emph{Phys. Lett.}, B119:\penalty0 387--390, 1982.

\bibitem[{L\"uscher} et~al.(1994){L\"uscher}, Sommer, Weisz, and
  Wolff]{alpha:SU3}
Martin {L\"uscher}, Rainer Sommer, Peter Weisz, and Ulli Wolff.
\newblock A precise determination of the running coupling in the {SU(3)}
  {Y}ang-{M}ills theory.
\newblock \emph{Nucl. Phys.}, B413:\penalty0 481--502, 1994.

\bibitem[Kennedy(2006)]{Kennedy:2006ax}
A.~D. Kennedy.
\newblock Algorithms for dynamical fermions, 2006.
\newblock hep-lat/0607038.

\bibitem[Giusti(2007)]{Giusti:2007hk}
Leonardo Giusti.
\newblock Light dynamical fermions on the lattice: Toward the chiral regime of
  {Q}{C}{D}.
\newblock \emph{PoS.}, LAT2006, 2007.

\bibitem[Symanzik(1982)]{Symanzik:1982}
K.~Symanzik, 1982.
\newblock Some topics in quantum field theory, in Mathematical problems in
  theoretical physics, eds. R. Schrader et al., Lecture Notes in Physics. Vol.
  153 (Springer, New York, 1982).

\bibitem[Symanzik(1983{\natexlab{a}})]{impr:Sym1}
K.~Symanzik.
\newblock Continuum limit and improved action in lattice theories. 1.
  {P}rinciples and $\phi^4$ theory.
\newblock \emph{Nucl. Phys.}, B226:\penalty0 187, 1983{\natexlab{a}}.

\bibitem[Symanzik(1983{\natexlab{b}})]{impr:Sym2}
K.~Symanzik.
\newblock Continuum limit and improved action in lattice theories. 2. {O($N$)}
  nonlinear sigma model in perturbation theory.
\newblock \emph{Nucl. Phys.}, B226:\penalty0 205, 1983{\natexlab{b}}.

\bibitem[{L\"uscher} et~al.(1996){L\"uscher}, Sint, Sommer, and
  Weisz]{impr:pap1}
Martin {L\"uscher}, Stefan Sint, Rainer Sommer, and Peter Weisz.
\newblock Chiral symmetry and {O($a$)} improvement in lattice {{Q}{C}{D}}.
\newblock \emph{Nucl. Phys.}, B478:\penalty0 365--400, 1996.

\bibitem[Sheikholeslami and Wohlert(1985)]{impr:SW}
B.~Sheikholeslami and R.~Wohlert.
\newblock Improved continuum limit lattice action for {{Q}{C}{D}} with {W}ilson
  fermions.
\newblock \emph{Nucl. Phys.}, B259:\penalty0 572, 1985.

\bibitem[{L\"uscher} et~al.(1997{\natexlab{a}}){L\"uscher}, Sint, Sommer,
  Weisz, and Wolff]{impr:pap3}
Martin {L\"uscher}, Stefan Sint, Rainer Sommer, Peter Weisz, and Ulli Wolff.
\newblock Non-perturbative {O($a$)} improvement of lattice {{Q}{C}{D}}.
\newblock \emph{Nucl. Phys.}, B491:\penalty0 323--343, 1997{\natexlab{a}}.

\bibitem[Sint and Weisz(1997)]{impr:pap5}
Stefan Sint and Peter Weisz.
\newblock Further results on {O}($a$) improved lattice {{Q}{C}{D}} to one loop
  order of perturbation theory.
\newblock \emph{Nucl. Phys.}, B502:\penalty0 251, 1997.

\bibitem[Bode et~al.(2000)Bode, Weisz, and Wolff]{pert:2loop_fin}
Achim Bode, Peter Weisz, and Ulli Wolff.
\newblock Two loop computation of the {Schr\"odinger} functional in lattice
  {{Q}{C}{D}}.
\newblock \emph{Nucl. Phys.}, B576:\penalty0 517--539, 2000.

\bibitem[Kurth and Sommer(2001{\natexlab{b}})]{zastat:pap1}
Martin Kurth and Rainer Sommer.
\newblock Renormalization and {O}($a$)-improvement of the static axial current.
\newblock \emph{Nucl. Phys.}, B597:\penalty0 488--518, 2001{\natexlab{b}}.

\bibitem[Guagnelli et~al.(1999)Guagnelli, Heitger, Sommer, and
  Wittig]{mbar:pap2}
Marco Guagnelli, Jochen Heitger, Rainer Sommer, and Hartmut Wittig.
\newblock Hadron masses and matrix elements from the {{Q}{C}{D}}
  {Schr\"odinger} functional.
\newblock \emph{Nucl. Phys.}, B560:\penalty0 465, 1999.

\bibitem[de~Divitiis et~al.(2003{\natexlab{a}})de~Divitiis, Guagnelli,
  Petronzio, Tantalo, and Palombi]{romeII:mb}
Giulia~Maria de~Divitiis, Marco Guagnelli, Roberto Petronzio, Nazario Tantalo,
  and Filippo Palombi.
\newblock Heavy quark masses in the continuum limit of lattice {Q}{C}{D}.
\newblock \emph{Nucl. Phys.}, B675:\penalty0 309--332, 2003{\natexlab{a}}.

\bibitem[de~Divitiis et~al.(2003{\natexlab{b}})de~Divitiis, Guagnelli, Palombi,
  Petronzio, and Tantalo]{romeII:fb}
G.~M. de~Divitiis, M.~Guagnelli, F.~Palombi, R.~Petronzio, and N.~Tantalo.
\newblock Heavy-light decay constants in the continuum limit of lattice
  {Q}{C}{D}.
\newblock \emph{Nucl. Phys.}, B672:\penalty0 372--386, 2003{\natexlab{b}}.

\bibitem[Guazzini et~al.(2007{\natexlab{a}})Guazzini, Sommer, and
  Tantalo]{Guazzini:2007ja}
Damiano Guazzini, Rainer Sommer, and Nazario Tantalo.
\newblock Precision for {B}-meson matrix elements.
\newblock 2007{\natexlab{a}}.
\newblock arXiv:0710.2229 [hep-lat].

\bibitem[Sommer(1994)]{pot:r0}
R.~Sommer.
\newblock A new way to set the energy scale in lattice gauge theories and its
  applications to the static force and $\alpha_s$ in {SU(2) Yang-Mills} theory.
\newblock \emph{Nucl. Phys.}, B411:\penalty0 839, 1994.

\bibitem[Guagnelli et~al.(1998)Guagnelli, Sommer, and Wittig]{pot:r0_SU3}
Marco Guagnelli, Rainer Sommer, and Hartmut Wittig.
\newblock Precision computation of a low-energy reference scale in quenched
  lattice {{Q}{C}{D}}.
\newblock \emph{Nucl. Phys.}, B535:\penalty0 389, 1998.

\bibitem[Necco and Sommer(2002)]{pot:intermed}
Silvia Necco and Rainer Sommer.
\newblock The {N}(f) = 0 heavy quark potential from short to intermediate
  distances.
\newblock \emph{Nucl. Phys.}, B622:\penalty0 328--346, 2002.

\bibitem[Guagnelli et~al.(2002{\natexlab{b}})Guagnelli, Petronzio, and
  Tantalo]{Guagnelli:2002ia}
Marco Guagnelli, Roberto Petronzio, and Nazario Tantalo.
\newblock The lattice scale at large beta in quenched {Q}{C}{D}.
\newblock \emph{Phys. Lett.}, B548:\penalty0 58--62, 2002{\natexlab{b}}.

\bibitem[Gasser and Leutwyler(1982)]{Gasser:1982ap}
J.~Gasser and H.~Leutwyler.
\newblock Quark masses.
\newblock \emph{Phys. Rept.}, 87:\penalty0 77--169, 1982.

\bibitem[Gasser and Leutwyler(1984)]{chir:GaLe1}
J.~Gasser and H.~Leutwyler.
\newblock Chiral perturbation theory to one loop.
\newblock \emph{Ann. Phys.}, 158:\penalty0 142, 1984.

\bibitem[Gasser and Leutwyler(1985)]{Gasser:1984gg}
J.~Gasser and H.~Leutwyler.
\newblock Chiral perturbation theory: Expansions in the mass of the strange
  quark.
\newblock \emph{Nucl. Phys.}, B250:\penalty0 465, 1985.

\bibitem[Guazzini et~al.(2006)Guazzini, Sommer, and Tantalo]{lat06:damiano}
Damiano Guazzini, Rainer Sommer, and Nazario Tantalo.
\newblock m(b) and f({B}/s) from a combination of {HQET} and {Q}{C}{D}.
\newblock \emph{PoS}, LAT2006:\penalty0 084, 2006.

\bibitem[{L\"uscher} et~al.(1997{\natexlab{b}}){L\"uscher}, Sint, Sommer, and
  Wittig]{impr:pap4}
Martin {L\"uscher}, Stefan Sint, Rainer Sommer, and Hartmut Wittig.
\newblock Nonperturbative determination of the axial current normalization
  constant in {O}($a$) improved lattice {{Q}{C}{D}}.
\newblock \emph{Nucl. Phys.}, B491:\penalty0 344--364, 1997{\natexlab{b}}.

\bibitem[Guagnelli et~al.(2001)]{impr:babp}
Marco Guagnelli et~al.
\newblock Non-perturbative results for the coefficients $b_m$ and {$b_A-b_P$}
  in {O($a$)} improved lattice {{Q}{C}{D}}.
\newblock \emph{Nucl. Phys.}, B595:\penalty0 44--62, 2001.

\bibitem[Heitger and Wennekers(2004)]{HQET:pap2}
Jochen Heitger and Jan Wennekers.
\newblock Effective heavy-light meson energies in small-volume quenched
  {Q}{C}{D}.
\newblock \emph{JHEP}, 02:\penalty0 064, 2004.

\bibitem[Garden et~al.(2000)Garden, Heitger, Sommer, and Wittig]{mbar:pap3}
Joyce Garden, Jochen Heitger, Rainer Sommer, and Hartmut Wittig.
\newblock Precision computation of the strange quark's mass in quenched
  {{Q}{C}{D}}.
\newblock \emph{Nucl. Phys.}, B571:\penalty0 237--256, 2000.

\bibitem[van Ritbergen et~al.(1997)van Ritbergen, Vermaseren, and
  Larin]{MS:4loop1}
T.~van Ritbergen, J.~A.~M. Vermaseren, and S.~A. Larin.
\newblock The four loop beta function in quantum chromodynamics.
\newblock \emph{Phys. Lett.}, B400:\penalty0 379--384, 1997.

\bibitem[Vermaseren et~al.(1997)Vermaseren, Larin, and van
  Ritbergen]{MS:4loop3}
J.~A.~M. Vermaseren, S.~A. Larin, and T.~van Ritbergen.
\newblock The four loop quark mass anomalous dimension and the invariant quark
  mass.
\newblock \emph{Phys. Lett.}, B405:\penalty0 327--333, 1997.

\bibitem[Della~Morte et~al.(2007{\natexlab{b}})Della~Morte, Garron, Papinutto,
  and Sommer]{mb:nf0}
Michele Della~Morte, Nicolas Garron, Mauro Papinutto, and Rainer Sommer.
\newblock Heavy quark effective theory computation of the mass of the bottom
  quark.
\newblock \emph{JHEP}, 01:\penalty0 007, 2007{\natexlab{b}}.

\bibitem[Martinelli and Sachrajda(1999)]{mbstat:MaSa}
G.~Martinelli and C.~T. Sachrajda.
\newblock Computation of the b-quark mass with perturbative matching at the
  next-to-next-to-leading order.
\newblock \emph{Nucl. Phys.}, B559:\penalty0 429, 1999.

\bibitem[Gimenez et~al.(2000)Gimenez, Giusti, Martinelli, and
  Rapuano]{mb:Gimenez2}
V.~Gimenez, L.~Giusti, G.~Martinelli, and F.~Rapuano.
\newblock {N}{N}{L}{O} unquenched calculation of the b quark mass.
\newblock \emph{JHEP}, 03:\penalty0 018, 2000.

\bibitem[Collins(2000)]{mb:NRQCD}
S.~Collins.
\newblock The mass of the b quark from lattice {NRQCD}.
\newblock \emph{Wien 2000, Quark confinement and the hadron spectrum}, pages
  325--327, 2000.
\newblock hep-lat/0009040.

\bibitem[Onogi(2006)]{Onogi:2006km}
Tetsuya Onogi.
\newblock Heavy flavor physics from lattice {Q}{C}{D}.
\newblock \emph{PoS}, LAT2006:\penalty0 017, 2006.

\bibitem[Chetyrkin and Grozin(2003)]{ChetGrozin}
K.~G. Chetyrkin and A.~G. Grozin.
\newblock Three-loop anomalous dimension of the heavy-light quark current in
  {HQET}.
\newblock \emph{Nucl. Phys.}, B666:\penalty0 289--302, 2003.

\bibitem[Heitger et~al.(2004)Heitger, {J\"uttner}, Sommer, and
  Wennekers]{hqet:pap3}
Jochen Heitger, Andreas {J\"uttner}, Rainer Sommer, and Jan Wennekers.
\newblock Non-perturbative tests of heavy quark effective theory.
\newblock \emph{JHEP}, 11:\penalty0 048, 2004.

\bibitem[Rolf et~al.(2004)]{lat03:juri}
J.~Rolf et~al.
\newblock Towards a precision computation of f({B}/s) in quenched {Q}{C}{D}.
\newblock \emph{Nucl. Phys. Proc. Suppl.}, 129:\penalty0 322--324, 2004.

\bibitem[Ali~Khan et~al.(2007)]{AliKhan:2007tm}
A.~Ali~Khan et~al.
\newblock Decay constants of charm and beauty pseudoscalar heavy- light mesons
  on fine lattices, 2007.
\newblock hep-lat/0701015.

\bibitem[Bowler et~al.(2001)]{Bowler:2000xw}
K.~C. Bowler et~al.
\newblock Decay constants of {B} and {D} mesons from non-perturbatively
  improved lattice {Q}{C}{D}.
\newblock \emph{Nucl. Phys.}, B619:\penalty0 507--537, 2001.

\bibitem[Ali~Khan et~al.(2001)]{AliKhan:2000eg}
A.~Ali~Khan et~al.
\newblock Decay constants of {B} and {D} mesons from improved relativistic
  lattice {Q}{C}{D} with two flavours of sea quarks.
\newblock \emph{Phys. Rev.}, D64:\penalty0 034505, 2001.

\bibitem[Bernard et~al.(2002)]{Bernard:2002pc}
C.~Bernard et~al.
\newblock Lattice calculation of heavy-light decay constants with two flavors
  of dynamical quarks.
\newblock \emph{Phys. Rev.}, D66:\penalty0 094501, 2002.

\bibitem[Neubert(1994)]{Neubert:1993mb}
Matthias Neubert.
\newblock Heavy quark symmetry.
\newblock \emph{Phys. Rept.}, 245:\penalty0 259--396, 1994.

\bibitem[Bochicchio et~al.(1992)Bochicchio, Martinelli, Allton, Sachrajda, and
  Carpenter]{stat:bbstar}
M.~Bochicchio, G.~Martinelli, C.~R. Allton, Christopher~T. Sachrajda, and D.~B.
  Carpenter.
\newblock Heavy quark spectroscopy on the lattice.
\newblock \emph{Nucl. Phys.}, B372:\penalty0 403--420, 1992.

\bibitem[Gimenez et~al.(1997)Gimenez, Martinelli, and
  Sachrajda]{Gimenez:1996av}
V.~Gimenez, G.~Martinelli, and Christopher~T. Sachrajda.
\newblock A high-statistics lattice calculation of lambda(1) and lambda(2) in
  the {B}-meson.
\newblock \emph{Nucl. Phys.}, B486:\penalty0 227--244, 1997.

\bibitem[Aoki et~al.(2004)]{Aoki:2003jf}
S.~Aoki et~al.
\newblock Heavy quark expansion parameters from lattice {NRQCD}.
\newblock \emph{Phys. Rev.}, D69:\penalty0 094512, 2004.

\bibitem[Amoros et~al.(1997)Amoros, Beneke, and Neubert]{Amoros:1997rx}
G.~Amoros, M.~Beneke, and M.~Neubert.
\newblock Two-loop anomalous dimension of the chromo-magnetic moment of a heavy
  quark.
\newblock \emph{Phys. Lett.}, B401:\penalty0 81--90, 1997.

\bibitem[Czarnecki and Grozin(1997)]{Czarnecki:1997dz}
A.~Czarnecki and A.~G. Grozin.
\newblock {HQET} chromomagnetic interaction at two loops.
\newblock \emph{Phys. Lett.}, B405:\penalty0 142--149, 1997.

\bibitem[Sint and Weisz(1999)]{mbar:pert}
Stefan Sint and Peter Weisz.
\newblock The running quark mass in the {SF} scheme and its two loop anomalous
  dimension.
\newblock \emph{Nucl. Phys.}, B545:\penalty0 529, 1999.

\bibitem[{L\"uscher} and Weisz(1996)]{impr:pap2}
M.~{L\"uscher} and P.~Weisz.
\newblock O($a$) improvement of the axial current in lattice {{Q}{C}{D}} to one
  loop order of perturbation theory.
\newblock \emph{Nucl. Phys.}, B479:\penalty0 429--260, 1996.

\bibitem[Sint and Sommer(1996)]{pert:1loop}
Stefan Sint and Rainer Sommer.
\newblock The running coupling from the {{Q}{C}{D}} {Schr\"odinger} functional:
  A one loop analysis.
\newblock \emph{Nucl. Phys.}, B465:\penalty0 71--98, 1996.

\bibitem[Flynn and Hill(1991)]{Flynn:1991kw}
Jonathan~M. Flynn and Brian~Russell Hill.
\newblock {B} - {B}* splitting: A test of heavy quark methods.
\newblock \emph{Phys. Lett.}, B264:\penalty0 173--177, 1991.

\bibitem[{L\"uscher}(1988)]{FSE:martin1}
M.~{L\"uscher}.
\newblock Selected topics in lattice field theory, 1988.
\newblock Lectures given at Summer School 'Fields, Strings and Critical
  Phenomena', Les Houches, France, Jun 28 - Aug 5, 1988.

\bibitem[Narayanan and Wolff(1995)]{pert:2loop_SU2}
Rajamani Narayanan and Ulli Wolff.
\newblock Two loop computation of a running coupling in lattice {Y}ang-{M}ills
  theory.
\newblock \emph{Nucl. Phys.}, B444:\penalty0 425--446, 1995.

\bibitem[Kurth(2002)]{thesis:skurth}
Stefan Kurth.
\newblock \emph{The renormalised quark mass in the {Schr\"odinger} functional
  of lattice {{Q}{C}{D}}: A one-loop calculation with a non-vanishing
  background field}.
\newblock PhD thesis, Humboldt U., Berlin, 2002.
\newblock hep-lat/0211011.

\bibitem[Della~Morte et~al.(2005{\natexlab{b}})Della~Morte, Hoffmann, and
  Knechtli]{symm:theta_angle}
M.~Della~Morte, R.~Hoffmann, and F.~Knechtli.
\newblock Discrete symmetries of lattice {Q}{C}{D}.
\newblock \emph{Internal notes of the ALPHA collaboration}, 2005{\natexlab{b}}.

\bibitem[Parisi et~al.(1983)Parisi, Petronzio, and Rapuano]{PPR}
G.~Parisi, R.~Petronzio, and F.~Rapuano.
\newblock A measurement of the string tension near the continuum limit.
\newblock \emph{Phys. Lett.}, 128B:\penalty0 418, 1983.

\bibitem[Guazzini et~al.(2007{\natexlab{b}})Guazzini, Meyer, and
  Sommer]{Zspin:me}
Damiano Guazzini, Harvey~B. Meyer, and Rainer Sommer.
\newblock Non-perturbative renormalization of the chromo-magnetic operator in
  heavy quark effective theory and the {B}* - {B} mass splitting.
\newblock \emph{JHEP}, 10:\penalty0 081, 2007{\natexlab{b}}.

\bibitem[Grozin et~al.(2007)]{Grozin:2007qv}
A.~G. Grozin et~al.
\newblock The {B}-meson mass splitting from non-perturbative quenched lattice
  {Q}{C}{D}.
\newblock \emph{PoS}, LAT2007:\penalty0 100, 2007.

\bibitem[Narayanan and Wolff(1992)]{propagators:narayanan_notes}
R.~Narayanan and U.~Wolff, 1992.
\newblock Unpublished notes.

\bibitem[Weisz(1996)]{impr:csw_int_notes}
Peter Weisz.
\newblock Computation of the improvement coefficient c(sw) to 1-loop.
\newblock \emph{Internal notes of the ALPHA collaboration}, 1996.

\bibitem[Creutz(1984)]{Creutz:1984mg}
M.~Creutz.
\newblock \emph{Quarks, gluons and lattices}.
\newblock Cambridge, Uk: Univ. Pr. 169 P. ( Cambridge Monographs On
  Mathematical Physics), 1984.

\bibitem[{L\"uscher}(1996)]{impr:axial_int_notes}
Martin {L\"uscher}.
\newblock Improved axial current at one-loop order of perturbation theory.
\newblock \emph{Internal notes of the ALPHA collaboration}, 1996.

\bibitem[Foley et~al.(2005)]{Foley:2005ac}
Justin Foley et~al.
\newblock Practical all-to-all propagators for lattice {Q}{C}{D}.
\newblock \emph{Comput. Phys. Commun.}, 172:\penalty0 145--162, 2005.

\bibitem[Wolff(2004)]{Wolff:2003sm}
Ulli Wolff.
\newblock {M}onte {C}arlo errors with less errors.
\newblock \emph{Comput. Phys. Commun.}, 156:\penalty0 143--153, 2004.

\bibitem[Press et~al.(1992)Press, Flannery, Teukolsky, and Vetterling]{NumRec}
William~H. Press, Brian~P. Flannery, Saul~A. Teukolsky, and William~T.
  Vetterling.
\newblock \emph{Numerical Recipes in C : The Art of Scientific Computing}.
\newblock {Cambridge University Press}, October 1992.
\newblock ISBN 0521431085.
\newblock URL
  \url{http://www.amazon.co.uk/exec/obidos/ASIN/0521431085/citeulike-21}.

\end{thebibliography}
\bibliographystyle{unsrtnat} 
\chapter{Notations and conventions}\label{app:Not_Conv}

\section{The Dirac matrices}\label{asect:Dirac_matrices}

We employ the hermitian euclidean Dirac matrices $\gamma_\mu$ with
\be
\gamma_0=\left(
\begin{array}{cr}
1 & 0 \\
0 & -1 \\
\end{array}
\right)\,,\quad
\gamma_k=\left(
\begin{array}{cc}
0 & -i\sigma_k \\
i\sigma_k & 0 \\
\end{array}
\right)\,,\quad k=1,2,3\,,
\ee
where the Pauli matrices
\be
\sigma_1=\left(
\begin{array}{cc}
0 & \phantom{-}1 \\
1 & \phantom{-}0 \\
\end{array}
\right)\,,\quad
\sigma_2=\left(
\begin{array}{cc}
0 & -i \\
i & \phantom{-}0 \\
\end{array}
\right)\,,\quad
\sigma_3=\left(
\begin{array}{cr}
1 & 0 \\
0 & -1 \\
\end{array}
\right)\,,
\ee
appear. In this representation $\gamma_5=\gamma_0\gamma_1\gamma_2\gamma_3$
is not diagonal
\be
\gamma_5=\left(
\begin{array}{cc}
0 & -1 \\
-1 & \phantom{-}0 \\
\end{array}
\right)\,,
\ee
The anticommutation relations are given by
\be
\{\gamma_\mu,\gamma_\nu\}=2\delta_{\mu\nu}\,,\quad \{\gamma_5,\gamma_\nu\}=0\,.
\ee
Finally the projectors
\be
P_+=\tfrac{1}{2}(1+\gamma_0)=\left(
\begin{array}{cc}
1 & 0 \\
0 & 0 \\
\end{array}
\right)\,,\quad
P_-=\tfrac{1}{2}(1-\gamma_0)=\left(
\begin{array}{cc}
0 & 0 \\
0 & 1 \\
\end{array}
\right)\,,
\ee
are used.

\section{The basis of the Lie algebra}\label{asect:Lie_algebra}

The Lie algebra $su(3)$ of SU(3) can be identified with the space of
complex $3\!\times\!3$ matrices constructed as in \cite{impr:csw_int_notes} from the modified Gell-Mann matrices
\begin{align}
\tilde{\lambda}_1&= \left(
\begin{array}{ccc}
\,0\, & \,1\, & \,0\, \\
\,1\, & \,0\, & \,0\, \\
\,0\, & \,0\, & \,0\, \\
\end{array}
\right)\,,\quad
\tilde{\lambda}_2=\left(
\begin{array}{ccc}
\,0\, & -i & \,0\, \\
\,i\, & 0 & \,0\, \\
\,0\, & 0 & \,0\, \\
\end{array}
\right)\,,\nonumber\\
\tilde{\lambda}_4&= \left(
\begin{array}{ccc}
\,0\, & \,0\, & \,1\, \\
\,0\, & \,0\, & \,0\, \\
\,1\, & \,0\, & \,0\, \\
\end{array}
\right)\,,\quad
\tilde{\lambda}_5=\left(
\begin{array}{ccc}
\,0\, & \,0\, & -i \\
\,0\, & \,0\, & 0 \\
\,i\, & \,0\, & 0 \\
\end{array}
\right)\,,\nonumber\\[-1.5ex]
&  \label{eq:ILie_matrices}\\[-1.5ex]
\tilde{\lambda}_6&= \left(
\begin{array}{ccc}
\,0\, & \,0\, & \,0\, \\
\,0\, & \,0\, & \,1\, \\
\,0\, & \,1\, & \,0\, \\
\end{array}
\right)\,,\quad
\tilde{\lambda}_7=\left(
\begin{array}{ccc}
\,0\, & \,0\, & 0 \\
\,0\, & \,0\, & -i \\
\,0\, & \,i\, & 0 \\
\end{array}
\right)\,,\nonumber\\
\tilde{\lambda}_3&= \left(
\begin{array}{ccc}
\,0\, & \,0\, & 0 \\
\,0\, & \,1\, & 0 \\
\,0\, & \,0\, & -1 \\
\end{array}
\right)\,,\quad
\tilde{\lambda}_8=\frac{1}{\sqrt{3}}\left(
\begin{array}{ccc}
\,2\, & 0 & 0 \\
\,0\, & -1 & 0 \\
\,0\, & 0 & \!-1 \\
\end{array}
\right)\,.\nonumber
\end{align}
For $a=1,\ldots,8$ we define
\be
T_a=\frac{1}{2i}\tilde{\lambda}_a\,,
\ee
and finally the basis
\begin{align}
I^1&= \tfrac{1}{\sqrt{2}}(T_1+iT_2)\,,\quad I^2=\tfrac{1}{\sqrt{2}}(T_1-iT_2)\,,\nonumber\\
 &  \nonumber\\[-1ex]
I^4&= \tfrac{1}{\sqrt{2}}(T_4+iT_5)\,,\quad I^5=\tfrac{1}{\sqrt{2}}(T_4-iT_5)\,,\nonumber\\[-0.5ex]
 &   \\[-0.5ex]
I^6&= \tfrac{1}{\sqrt{2}}(T_6+iT_7)\,,\quad I^7=\tfrac{1}{\sqrt{2}}(T_6-iT_7)\,,\nonumber\\
 &  \nonumber\\[-1ex]
I^3&= T_3\,,\quad\hspace{1.87cm} I^8=T_8\,,\nonumber
\end{align}
where only the last two matrices are diagonal. With this choice
\be
[I^a]^\dagger=I^{\bar{a}}\,,\quad\tr\{I^aI^b\}=-\frac{1}{2}\delta_{b\bar{a}}\,
\ee
where
\be
\bar{1}=2\,,\quad\bar{4}=5\,,\quad\bar{6}=7\,,
\ee
and vice versa. For the diagonal matrices $\bar{3}=3$ and $\bar{8}=8$.

\section{The non-vanishing background field}\label{asect:bk_field}

We work in units of the lattice spacing and remark that 
the non-vanishing background field, appearing in chapters \ref{chap:Zspin} and \ref{chap:PTcode},
is induced by the spatially constant Abelian boundary fields \cite{SF:LNWW}:
\be\label{eq:defs_C_Cprime}
C=\frac{i}{L}\mathrm{diag}\left(\angle_1,\angle_2,\angle_3\right)\,,\quad
C'=\frac{i}{L}\mathrm{diag}\left(\angleprime_1,\angleprime_2,\angleprime_3\right)\,,
\ee
where the angles $\angle_m$ and $\angleprime_m$ are given in terms of the parameters $\eta$
and $\nu$
\begin{align}
\angle_1&= \eta-\tfrac{\pi}{3}\,,\qquad\hspace{1.38cm}\angleprime_1=-\eta-\pi\,,\nonumber\\
&  \nonumber\\[-1ex]
\angle_2&= \eta(-\tfrac{1}{2}+\nu)\,,\qquad\hspace{0.525cm}\angleprime_2=\eta(\tfrac{1}{2}+\nu)+\tfrac{\pi}{3}\,,\label{eq:angles}\\
&  \nonumber\\[-1ex]
\angle_3&= -\eta(\tfrac{1}{2}+\nu)+\tfrac{\pi}{3}\qquad\angleprime_3=\eta(\tfrac{1}{2}-\nu)+\tfrac{2\pi}{3}\,.\nonumber
\end{align}
The ``point A'' is defined as in \cite{alpha:SU3} by the choice $\eta=\nu=0$. The 
renormalization condition in \eq{eq:ren_cond_rot} is given with Dirichlet boundary conditions
in time, i.e.
\be\label{eq:U_x0}
  U_k(x)|_{x_0=0} = \exp(C)\,, \quad
  U_k(x)|_{x_0=T} = \exp(C')\,, \quad k=1,2,3\,,
\ee
and $U_0(x)|_{x_0=0}$ unconstrained.
The background field
represents a constant electric field
\be\label{eq:electric_E}
\mathcal{E}=-i[C'-C]/T\,,
\ee
which can be rewritten in terms of the matrix $I^8$, 
appearing in \eqs{eq:ILie_matrices}, as 
\be
\mathcal{E}=-\gamma\left(
\begin{array}{ccc}
\,2\, & 0 & 0 \\
\,0\, & -1 & 0 \\
\,0\, & 0 & \!-1 \\
\end{array}
\right)\,,
\ee
with
\be\label{eq:gamma_function}
\gamma=\frac{1}{LT}\left(\eta+\frac{\pi}{3}\right)\,.
\ee
The configuration $V$ of least action appearing 
in the parametrization (\ref{eq:U_parametrization}) is
\be\label{eq:def_V_bk}
V_0(x)=1\,,\quad V_k(x)=V(x_0)\,,
\ee 
with
\be
V(x_0)=\exp\{i[\mathcal{E}x_0-iC]\}\,.
\ee
An equivalent parametrization of $V$ used in chapters \ref{chap:Zspin} and \ref{chap:PTcode} is 
\be\label{eq:def_V_bk_exp}
V_\mu(x)=\exp\{ib_\mu(x)\}\,,
\ee
\begin{table}[t]
\vspace{-0.6cm}
\centering
\begin{tabular}{cccc}
\hline
\hline
& & & \\[-1.5ex]
$a$&$C_a$&$R_a$ &$\phi_a(x_0)$ \\
& & & \\[-1.5ex]
\hline
& & & \\[-1.5ex]
$1$&$\tfrac{1}{2}(\cos 2\gamma+\cos\gamma)$&$\cos\tfrac{\gamma}{2}$ & $-3\gamma x_0+\tfrac{1}{L}(\eta[\tfrac{3}{2}-\nu]-\tfrac{\pi}{3})$\\
& & & \\[-2ex]
$3$&$\cos\gamma$&$\cos\gamma$ & $0$\\
& & & \\[-2ex]
$4$&$\tfrac{1}{2}(\cos 2\gamma+\cos\gamma)$&$\cos\tfrac{\gamma}{2}$ & $-3\gamma x_0+\tfrac{1}{L}(\eta[\tfrac{3}{2}+\nu]-\tfrac{2\pi}{3})$\\
& & & \\[-2ex]
$6$&$\cos\gamma$&$\cos\gamma$ & $\tfrac{1}{L}(2\eta\nu+\tfrac{\pi}{3})$\\
& & & \\[-2ex]
$8$&$\tfrac{1}{3}(2\cos 2\gamma+\cos\gamma)$&$\tfrac{1}{3}(2\cos 2\gamma+\cos\gamma)$ & $0$\\
& & & \\[-1.5ex]
\hline
\hline
\end{tabular}
\mycaption{$C_a$, $R_a$ and $\phi_a$ for the gauge group SU(3). The other coefficients are
$C_2=C_1$, $C_5=C_4$, $C_7=C_6$, $R_2=R_1$, $R_5=R_4$, $R_7=R_6$, $\phi_2=-\phi_1$, $\phi_5=-\phi_4$ and $\phi_7=-\phi_6$.}
\label{tab:Ca_Ra_phi_a}
\vspace{-0.3cm}
\end{table}
with
\be\label{deq:def_cal_B}
b_0(x)=0\,,\quad\mbox{and}\quad b_k(x)=b_k(x_0)=\mathcal{E}x_0-iC\,.
\ee
Other functions appearing in chapter \ref{chap:PTcode} are
given in \Tab{tab:Ca_Ra_phi_a}, and
\begin{align}
s_k^a(\vecp,x_0)&= 2\sin\left[\tfrac{1}{2}\Bigl(p_k+\phi_a(x_0)\Bigr)\right]\,,\label{eq:def_small_s}\\
&  \nonumber\\[-1ex]
c_k^a(\vecp,x_0)&= 2\cos\!\left[\tfrac{1}{2}\Bigl(p_k+\phi_a(x_0)\Bigr)\right]\,,\label{eq:def_small_c}\\
&  \nonumber\\[-1ex]
c_{abc}&= -2i\tr\Bigl\{I^a[I^b,I^c]\Bigr\}\,,\label{eq:def_cabc}\\
&  \nonumber\\[-1ex]
e_{abc}&= -2i\tr\Bigl\{\mathrm{e}^{i\mathcal{E}}I^aI^bI^c-\mathrm{e}^{-i\mathcal{E}}I^aI^cI^b\Bigr\}\,,\label{eq:def_eabc}\\
&  \nonumber\\[-1ex]
\alpha_k(\vecp,x_0)&= b_k(x_0)+p^+_k\,.\label{eq:def_alpha_k_b}
\end{align}
The following formulae are used to compute the tadpoles and are valid 
only for $a$ neutral label ($\Leftrightarrow a=3,8$)
\begin{align}
f^+_{ac}&= \tfrac{i}{2}\Bigl(e_{ac\bar{c}}+e_{a\bar{c}c}\Bigr)\,,\label{eq:f+}\\
&  \nonumber\\[-1ex]
f^-_{ac}&= \tfrac{1}{2}\Bigl(e_{ac\bar{c}}-e_{a\bar{c}c}\Bigr)\,,\label{eq:f-}\\
&  \nonumber\\[-1ex]
e^+_{ac}&= \tfrac{i}{2}\Bigl(e_{ac\bar{c}}\,\mathrm{e}^{-\tfrac{i}{2}\breve{\phi}_c}
+e_{a\bar{c}c}\,\mathrm{e}^{\tfrac{i}{2}\breve{\phi}_c} \Bigr)\,,\label{eq:e+}\\
&  \nonumber\\[-1ex]
e^-_{ac}&= \tfrac{1}{2}\Bigl(e_{ac\bar{c}}\,\mathrm{e}^{-\tfrac{i}{2}\breve{\phi}_c}
-e_{a\bar{c}c}\,\mathrm{e}^{\tfrac{i}{2}\breve{\phi}_c} \Bigr)\,,\label{eq:e-}
\end{align}
where $\breve{\phi}_c=\partial_0 \phi_c(x_0)$ which is independent of $x_0$.

\chapter{Statistical uncertainties in Monte Carlo simulations}\label{MC_errors}


\section{Thermalization}\label{sect:therm}

The initial configuration of gauge fields
can be arbitrarily chosen, but the average $\overline{{\mathcal{O}}}$ 
over many configurations cannot sensibly depend on the initial one.
For this reason one starts from a fixed configuration, e.g. where
$U_\mu(x)=1_{3\times 3}$ for all dynamical links (eventually with some random noise),
and the updating algorithm proceeds as discussed above. The number $N_\mathrm{UP}$
of updates must be chosen to be large enough to let the system ``forget'' the initial configuration.
At the end of the computation, one calculates several averages $\overline{{\mathcal{O}}}$
by discarding the first $N_\mathrm{disc}$ configurations, and checks that, as a function
of $N_\mathrm{disc}$, the fluctuations
on these averages are much smaller (order of magnitudes) than the statistical error $\overline{{\mathcal{O}}}$.
The latter depends of course on $N_\mathrm{disc}$, but the dependence is negligible as long as $N_\mathrm{disc}\ll N$.
If the check is successful, one can assume
that the thermal equilibrium has been reached, and the analysis is performed only on the last $N-N_\mathrm{disc}$
measurements.  

\section{Autocorrelation and errors estimate}\label{sect:autocorr} 

In this section we explain how the statistical errors in our Monte Carlo
simulations have been estimated. The whole section is far from being original,
and it is mainly based on \cite{Wolff:2003sm}. 

The first subsection is devoted to a short explanation of the $\Gamma$-method,
which has been used for the error analysis of our HQET data. The second
subsection reports the key formulae of the jackknife error analysis,
which has been used by the Tor Vergata group for the analysis of the relativistic QCD data. 
In \Ref{Wolff:2003sm} it is shown that the first method leads to more reliable
and robust error estimate. On the other side, the jackknife allows to compute the correlation
between derived observable in a simple way.

\subsection{The Gamma method}\label{sect:Gamma_method}

We assume that the thermal equilibrium has been reached, and to have a number 
of primary observables
labelled by a Greek index with exact statistical mean values $A_{\alpha}$.
Our derived observable is a function of these,
\be\label{eq:deriv_funct}
{\mathcal{O}}\equiv f(A_1,A_2,\ldots)\equiv f(A_{\alpha})\,.
\ee
The Monte Carlo estimates of the primary observables
are labelled by $a^{i,r}_{\alpha}$, where $i=1,\ldots,N_r$
counts the measurements for each replicum, and $r=1,\ldots,R$ the number of statistically
independent replica. From them one can compute the autocorrelation function $\Gamma_{\alpha\beta}$,
defined by
\be
\langle (a^{i,r}_{\alpha}-A_{\alpha})(a^{j,s}_{\beta}-A_{\beta})\rangle=
\delta_{r,s}\Gamma_{\alpha\beta}(j-i)\,,\quad\Gamma_{\alpha\beta}(n)=\Gamma_{\beta\alpha}(-n)\,.
\ee
Here and in the following of this section the averages $\langle\cdot\rangle$
refer to an ensemble of identical numerical experiments with independent
random numbers and initial states. 
By means of the per replicum means
\be
\bar{a}_\alpha^r=\frac{1}{N_r}\sum_{i=1}^{N_r}a^{i,r}_{\alpha}\,,
\ee
we define the natural estimator of the primary quantities $A_{\alpha}$ as
\be
\bar{\bar{a}}_\alpha=\frac{1}{N}\sum_{r=1}^{R}N_r \bar{a}_\alpha^r\,,\quad N=\sum_{r=1}^{R}N_r\,.
\ee
The covariance matrix assumes the simple form
\be
\mathrm{Cov}(\bar{\bar{a}}_\alpha,\bar{\bar{a}}_\beta)=\frac{1}{N}C_{\alpha\beta}\times\{1+\mathrm{O}(R\tau/N)\}\,,
\ee
where 
\be
C_{\alpha\beta}=\sum_{t=-\infty}^\infty \Gamma_{\alpha\beta}(t)\,,
\ee
and the finite scale $\tau$ characterizes the asymptotic exponential decay of $\Gamma_{\alpha\beta}$
\be
\Gamma_{\alpha\beta}(t)\stackrel{|t|\to\infty}{\sim}\mathrm{e}^{-|t|/\tau}\,.
\ee
To determine ${\mathcal{O}}$ we consider the estimator
\be\label{eq:Obarbar}
\bar{\bar{\!{\mathcal{O}}}}=f(\bar{\bar{a}}_\alpha)\,,
\ee
and the derivatives $f_\alpha=\partial f/\partial A_\alpha$, taken at the exact values $A_1,A_2,\ldots$.
One can then write the variance of ${\mathcal{O}}$ through the expression
\be
\sigma_{\mathcal{O}}^2=\frac{2\tau_{\mathrm{int},{\mathcal{O}}}}{N}v_{\mathcal{O}}\,.
\ee
Here $v_{\mathcal{O}}$ is the na\"ive, i.e.~disregarding autocorrelations, variance
\be
v_{\mathcal{O}}=\sum_{\alpha\beta}f_\alpha f_\beta \Gamma_{\alpha\beta}(0)\,,
\ee
and the integrated autocorrelation time for ${\mathcal{O}}$ reads
\be
\tau_{\mathrm{int},{\mathcal{O}}}=\frac{1}{2v_{\mathcal{O}}}\sum_{t=-\infty}^\infty\sum_{\alpha\beta}f_\alpha f_\beta 
\Gamma_{\alpha\beta}(t)\,.
\ee
Our best estimate for ${\mathcal{O}}$ is given by \eq{eq:Obarbar}, and, in absence of autocorrelations,
we would have $\Gamma_{\alpha\beta}(t)\propto \delta_{t,0}$, which implies $\tau_{\mathrm{int},{\mathcal{O}}}=0.5$.
The integrated autocorrelation time thus 
represents an estimate of the efficiency of the algorithm in use
for the determination of ${\mathcal{O}}$. 

The next step consists in the computation of the gradient $f_\alpha$
and of $\Gamma_{\alpha\beta}$. For the former we define the estimator $\,\bar{\bar{f}}_\alpha$,
where the derivative is evaluated at arguments $\bar{\bar{a}}_1,\bar{\bar{a}}_2,\ldots$, while for
the autocorrelation function we have
\be
\bar{\bar{\Gamma}}_{\alpha\beta}(t)=\frac{1}{N-Rt}\sum_{r=1}^{R}
\left[\sum_{i=1}^{N_r-t}(a^{i,r}_{\alpha}-\bar{\bar{a}}_\alpha)(a^{i+t,r}_{\beta}-\bar{\bar{a}}_\beta)\right]\,,
\ee
from which one computes the projected autocorrelation function
\be
\bar{\bar{\Gamma}}_{\mathcal{O}}(t)=\sum_{\alpha\beta}\,\bar{\bar{f}}_{\alpha}\,\hspace{0.1cm}\bar{\bar{f}}_{\beta}
\,\bar{\bar{\Gamma}}_{\alpha\beta}(t)\,.
\ee
The latter is finally used to compute
\begin{align}
\bar{\hspace{-0.02cm}\bar{v}}_{\mathcal{O}}&= \bar{\bar{\Gamma}}_{\mathcal{O}}(0)\,,\quad\mbox{and}\\
\bar{\bar{\sigma}}_{\mathcal{O}}^2&= \frac{1}{N}\left[\bar{\bar{\Gamma}}_{\mathcal{O}}(0)+2\sum_{t=1}^W\bar{\bar{\Gamma}}_{\mathcal{O}}(t)\right]\,,
\end{align}
where the last equation defines our best estimate of the variance. There 
we have introduced the summation window $W$, in order to get a good estimator of the autocorrelation time.
This cut-off is necessary because the signal for $\Gamma_{\mathcal{O}}(t)/\Gamma_{\mathcal{O}}(0)$
is overwhelmed by the noise for $|t|\gg \tau$. At the same time, the summation should not be truncated
too early, because it would compromise the accuracy of $\bar{\bar{\sigma}}_{\mathcal{O}}^2$.

A Matlab routine, called \texttt{UWerr.m}, that implements the whole method, including an automatic
windowing procedure, the necessary plots and much more, is documented and offered by Ulli Wolff for
download from the webpage:\\ $ $\\ \texttt{www-com.physik.hu-berlin.de/ALPHAsoft}.

\subsection{Jackknife}\label{sect:jackknife}

We borrow the notation of the previous subsection,
and give the basic formulae for error estimation by jackknife.
We assume data $a_\alpha^i$ where possible replica are sewed to one history
\be
a_\alpha^{j+\sum_{k=1}^{r-1}N_k}=a_\alpha^{j,r}\,,
\ee 
of length $N=BN_B$ which we divide into $N_B$ sections of $B$ consecutive measurements
each. Through the average 
\be
\bar{a}_\alpha=\frac{1}{N}\sum_{i=1}^N a_\alpha^i\,,  
\ee
we take as best estimate for ${\mathcal{O}}$
\be
\bar{\mathcal{O}}=f(\bar{a}_\alpha)\,,
\ee
where it is understood that $\bar{\mathcal{O}}=\bar{\bar{\!{\mathcal{O}}}}$, the latter appearing in \eq{eq:Obarbar}.
We further proceed by forming the blocked measurements
\be
b_\alpha^k=\frac{1}{B}\sum_{i=1}^B a_\alpha^{(k-1)B+i}\,,\quad k=1,\ldots,N_B\,,
\ee
and the jackknife bins
\be
c_\alpha^k=\frac{1}{N-B}\left(\sum_{i=1}^N a_\alpha^i-Bb_\alpha^k\right)\,.
\ee
The resulting jackknife error estimator is given by
\be\label{eq:jack_variance}
\bar{\sigma}^2_{{\mathcal{O}},\mathrm{Jack}}=\frac{N_B-1}{N_B}\sum_{k=1}^{N_B}(f(c_\alpha^k)-\bar{\mathcal{O}})^2\,.
\ee
Finally we consider the case where we want to compute another observable $\mathcal{Q}$, sharing
with ${\mathcal{O}}$ the same gauge configurations, and eventually expressed as function
of primary observables, whose a group may be in common with ${\mathcal{O}}$. 
We thus have ${\mathcal{O}}\equiv f(A_\alpha)$ and $\mathcal{Q}\equiv g(A_\alpha)$, and the correlation between
the two observables can be easily computed by jackknife 
\be\label{eq:cov_jack}
\mathrm{Cov}({\mathcal{O}},\mathcal{Q})_\mathrm{Jack}=\frac{N_B-1}{N_B}\sum_{k=1}^{N_B}(f(c_\alpha^k)-\bar{\mathcal{O}})(g(c_\alpha^k)-\bar{\mathcal{Q}})\,.
\ee
By performing the whole analysis with different choices of the bin length $B$, one can
check the stability of the estimated correlation.

\chapter{Simulation results}\label{app:CL_SSM}

This appendix is devoted to collect the results
of the simulations performed to apply the Step Scaling
Method. More emphasis and details are given for the
static simulations, which have been performed 
after the relativistic QCD results 
were published \cite{romeII:mb,romeII:fb}. They
have been reanalyzed, taking into account the correlation between
observables computed on the same gauge configurations. The statistical
uncertainties on the regularization dependent
part of
the renormalization constants and the lattice spacing 
are included before performing the continuum limit extrapolations;
they do not appear as a separate uncertainty.

\section{Static data}\label{asect:CL_static}

The parameters for the HQET simulations exploited
in the Step Scaling Method are reported in \tab{tab:det_param_static}. 
The method employed to compute them is explained in \sect{sect:SSM_simu_para}.
The first column of the table indicates
on which volume the parameters are determined. For the volumes indicated by
$L_0^\star$ the values of $\kappa_\mathrm{s}$ are obtained through an interpolation of
the data in \cite{romeII:mb}. The
errors on the last column for the RGI strange quark mass
include the uncertainties on $k_\mathrm{c}$ and on the PCAC mass (\ref{eq:mbar3}). An additional error is added in quadrature
to take into account the uncertainty on the regularization dependent part of the renormalization constants of the RGI quark mass
and on $r_0/a$ according to \sect{sect:SSM_simu_para}.

In the following tables and figures the observables at non-vanishing
lattice spacing borrow the continuum notation. The uncertainty stemming
from the perturbative determination
of $\castat$ and $\bastat$ is negligible compared to the statistical errors
on all observables.

\clearpage

\begin{table}
\centering
\footnotesize
\begin{tabular}{cccccc}
\hline
\hline
& & & & &\\
Volume&$L/a$ &$\beta$ & $\kappa_\mathrm{crit}$ & $\kappa_\mathrm{s}$ & $M_\mathrm{s}$ (GeV)\\
& & & & &\\
\hline
& & & & &\\[-2ex]
$L_0$ &    6 &  6.2110& 0.135625(15)&0.134766&0.1381(20)\\
& & & & &\\[-2ex]
\hline
& & & & &\\[-2ex]
$L_0$ &    8 &  6.4200& 0.135616(13)&0.135015&0.1365(20)\\
& & & & &\\[-2ex]
\hline
& & & & &\\[-2ex]
$L_0^\star$ &    12& 6.7370 &0.135235(5)&0.134801&0.1367(20)\\
& & & & &\\[-2ex]
\hline
& & & & &\\[-2ex]
$L_0^\star$ &    16& 6.9630 &0.134832(4)&0.134526&0.1343(21)\\
& & & & &\\[-2ex]
\hline
& & & & &\\[-2ex]
$L_0^\star$ &    24& 7.3000 &0.134235(3)&0.134041&0.1396(18)\\
& & & & &\\[-2ex]
\hline
\hline
& & & & &\\[-2ex]
$L_1$ &    8 &  5.9598&0.134700(18)&0.133274 &0.1373(16) \\
& & & & &\\[-2ex]
\hline
& & & & &\\[-2ex]
$L_1$ &   10 &  6.0914& 0.135494(23)&0.134476&0.1362(15)\\
& & & & &\\[-2ex]
\hline
& & & & &\\[-2ex]
$L_1$ &   12 &  6.2110& 0.135772(8)&0.134908&0.1352(15)\\
& & & & &\\[-2ex]
\hline
& & & & &\\[-2ex]
$L_1$ &   16 &  6.4200& 0.135687(8)&0.135087&0.1322(15)\\
& & & & &\\[-2ex]
\hline
& & & & &\\[-2ex]
$L_1$ &   24 &  6.7370& 0.135221(7)&0.134832&0.1340(17)\\
& & & & &\\[-2ex]
\hline
\hline
\end{tabular}
\mycaption{Collection of all simulation parameters for the HQET part of the Step Scaling Method.}
\label{tab:det_param_static}
\end{table}

\begin{table}
\centering
\footnotesize
\begin{tabular}{cccc}
\hline
\hline
& & & \\
Observable&C.L.&$\castat$ & $\bastat$ \\
& & & \\
\hline
& & & \\[-2ex]
$\sigma_\mathrm{m}^\mathrm{stat}(L_2)$ & 1.561(53)    & 1-loop & - \\
& & & \\[-2ex]
\hline
& & & \\[-2ex]
 &  0.233(36)     & 1-loop  & - \\[-1.5ex]
$\sigma_\mathrm{m}^\mathrm{stat}(L_1)$ & & & \\[-1.5ex]
 &  0.240(36) &tree-level &- \\
\hline
& & & \\[-2ex]
$Y_\mathrm{SF}(L_2,\mu)$ & 5.06(21)& 1-loop & 1-loop \\
& & & \\[-2ex]
\hline
& & & \\[-2ex]
 & -1.754(21) & 1-loop  & 1-loop  \\
 & -1.752(21) &1-loop & tree-level\\[-1.3ex]
$Y_\mathrm{SF}(L_1,1/L_1)$ & &   &  \\[-1.3ex]
 &-1.749(20) &tree-level   &1-loop \\
 &-1.746(20) &tree-level   &tree-level   \\
\hline
& & & \\[-2ex]
 & 0.4337(44) & 1-loop  & - \\[-1.5ex]
$\sigma_\mathrm{f}^\mathrm{stat}(L_1)$ & & & \\[-1.5ex]
 & 0.4339(44) &tree-level &- \\
\hline
& & & \\[-2ex]
 & -1.592(11) & 1-loop  & 1-loop  \\
 & -1.591(10) &1-loop & tree-level\\[-1.3ex]
$Y_\mathrm{SF}(L_0,1/L_0)$ & &   &  \\[-1.3ex]
 & -1.590(10) &tree-level   &1-loop \\
 & -1.589(10) &tree-level   &tree-level   \\
\hline
\hline
\end{tabular}
\mycaption{Collection of all continuum limit extrapolations for the HQET part of the Step Scaling Method.
The result $Y_\mathrm{SF}(L_2,\mu)$ is taken from \cite{Estat:me}, and is intended to be 
multiplied by $I^\mathrm{stat}(\mu=(1.436r_0)^{-1})=0.9191(83)$ computed in \cite{zastat:pap3} to get \eq{eq:YRGI_L2}.}
\label{tab:static_CL}
\end{table}

\clearpage

\begin{table}
\centering
\footnotesize
\begin{tabular}{ccccccc}
\hline
\hline
& & & & & & \\
$\castat$ & $\beta$ & \multicolumn{2}{c}{$L_1/a$}  & $\kappa_\mathrm{s}$ & \multicolumn{2}{c}{$a\Gamma_\mathrm{stat}(L_1)$}  \\
& & & & & & \\
\hline
& & & & & & \\[-2ex]
1-loop& & & &  &\multicolumn{2}{c}{0.31830(84)}  \\[-1.6ex]
 &5.9598 & \multicolumn{2}{c}{8}  &0.133274 &  & \\[-1.6ex]
tree-level& & & & & \multicolumn{2}{c}{0.31823(85)}  \\
& & & & & & \\[-2ex]
\hline
& & & & & & \\[-2ex]
1-loop& & & & &\multicolumn{2}{c}{0.28053(63)}  \\[-1.6ex]
 &6.0914 & \multicolumn{2}{c}{10}  &0.134476 & & \\[-1.6ex]
tree-level& & & &  &\multicolumn{2}{c}{0.28051(63)}  \\
& & & & & & \\[-2ex]
\hline
& & & & & & \\[-2ex]
1-loop& & & & &\multicolumn{2}{c}{0.25328(58)}  \\[-1.6ex]
 &6.2110 & \multicolumn{2}{c}{12}  & 0.134908 & &  \\[-1.6ex]
tree-level& & & & &\multicolumn{2}{c}{0.25326(59)}  \\
& & & & & &\\[-2ex]
\hline
& & & & & &\\[-2ex]
1-loop& & & &  &\multicolumn{2}{c}{0.21140(74)}  \\[-1.6ex]
 &6.4200 & \multicolumn{2}{c}{16}  &0.135087 & &  \\[-1.6ex]
tree-level& & & & & \multicolumn{2}{c}{0.21139(75)}  \\
& & & & & &\\[-2ex]
\hline
& & & & & &\\[-2ex]
1-loop& & & & & \multicolumn{2}{c}{0.1722(16)}  \\[-1.6ex]
 &6.7370 & \multicolumn{2}{c}{24}  &0.134832 & & \\[-1.6ex]
tree-level& & & & & \multicolumn{2}{c}{0.1722(16)}  \\
& & & & & &\\[-2ex]
\hline
\hline
& & & & & \multicolumn{2}{c}{Interpolated} \\
$\castat$&$\beta$ &\multicolumn{2}{c}{$L_2/a$} &  $aE_\mathrm{stat}$ & \multicolumn{2}{c}{results for} \\
& & & & & \multicolumn{2}{c}{$a\Gamma_\mathrm{stat}(L_1)$} \\
\hline
& & & & & &\\[-2ex]
1-loop& & & &0.4053(49) &\multicolumn{2}{c}{0.30003(51)} \\[-1.6ex]
&6.0219 &\multicolumn{2}{c}{16} & & & \\[-1.6ex]
tree-level& & & & & \multicolumn{2}{c}{0.29998(51)} \\
& & & & & & \\[-2ex]
\hline
& & & & & & \\[-2ex]
1-loop& & & & 0.3011(33) &\multicolumn{2}{c}{0.23602(47)}  \\[-1.6ex]
&6.2885 &\multicolumn{2}{c}{24} & & &\\[-1.6ex]
tree-level& & & & & \multicolumn{2}{c}{0.23601(47)} \\
& & & & & &\\[-2ex]
\hline
& & & & & &\\[-2ex]
1-loop& & & & 0.2564(9) &\multicolumn{2}{c}{0.20671(57)}  \\[-1.6ex]
&6.4500 &\multicolumn{2}{c}{32} & & &\\[-1.6ex]
tree-level& & & & & \multicolumn{2}{c}{0.20671(57)} \\
& & & & & &\\[-2ex]
\hline
& & & & & &\\[-2ex]
1-loop& & & & 0.2461(14) &\multicolumn{2}{c}{0.19972(63)} \\[-1.6ex]
&6.4956 &\multicolumn{2}{c}{32} & & &\\[-1.6ex]
tree-level& & & & & \multicolumn{2}{c}{0.19973(63)} \\
& & & & & & \\[-2ex]
\hline
\hline
& & & & & & \\
$\castat$ & $\beta$ & $L_1/a$ &$L_0/a$ & $\kappa_\mathrm{s}$ & $a\Gamma_\mathrm{stat}(L_1)$ &$a\Gamma_\mathrm{stat}(L_0)$\\
& & & & & & \\
\hline
& & & & & &\\[-2ex]
1-loop& & & & &0.2558(18)&0.22717(91)\\[-1.6ex]
 &6.2110 &12  &6 &0.134766  & &\\[-1.6ex]
tree-level& & & & &0.2558(18)&0.22587(91)\\
& & & & & &\\[-2ex]
\hline
& & & & & &\\[-2ex]
1-loop& & & & &0.2154(11)&0.19575(92)\\[-1.6ex]
 &6.4200 &16  &8 &0.135015  & &\\[-1.6ex]
tree-level& & & & &0.2154(11)&0.19497(93)\\
& & & & & &\\[-2ex]
\hline
& & & & & &\\[-2ex]
1-loop& & & & &0.1663(17)&0.15608(83)\\[-1.6ex]
 &6.7370 &24  &12 &0.134801  & &\\[-1.6ex]
tree-level& & & & &0.1662(17)&0.15564(85)\\
& & & & & &\\[-2ex]
\hline
& & & & & &\\[-2ex]
1-loop& & & & &0.1426(14)&0.13545(66)\\[-1.6ex]
 &6.9630 &32  &16 & 0.134526 & &\\[-1.6ex]
tree-level& & & & &0.1427(14)&0.13524(65)\\
& & & & & &\\[-2ex]
\hline
\hline
\end{tabular}
\mycaption{Collection of results at finite lattice spacing, for the computation of the static
step scaling functions of the meson mass.}
\label{tab:Gamma_stat_data}
\end{table}

\clearpage

\begin{table}
\centering
\footnotesize
\begin{tabular}{ccccccc}
\hline
\hline
& & & & & &  \\
$\castat$ & $\beta$ &  \multicolumn{2}{c}{$L_1/a$} &$\kappa_\mathrm{s}$ & \multicolumn{2}{c}{$X_\mathrm{SF}(L_1)$} \\
& & & & & &  \\
\hline
& & & & & &  \\[-2ex]
1-loop& & &   & &\multicolumn{2}{c}{-2.221(13)}   \\[-1.6ex]
 &6.2110 &   \multicolumn{2}{c}{12}   &0.134766 & & \\[-1.6ex]
tree-level&  & & &  &\multicolumn{2}{c}{-2.102(12)} \\
& & & & & \\[-2ex]
\hline
& & & &  & & \\[-2ex]
1-loop&  & & & &\multicolumn{2}{c}{-2.266(13)} \\[-1.6ex]
 &6.4200 &   \multicolumn{2}{c}{16}   &0.135015 & & \\[-1.6ex]
tree-level&  & & & &\multicolumn{2}{c}{-2.159(12)} \\
& & & & & \\[-2ex]
\hline
& & & & &  & \\[-2ex]
1-loop& &  & & &\multicolumn{2}{c}{-2.279(28)} \\[-1.6ex]
 &6.7370 &   \multicolumn{2}{c}{24}  &0.134801 & & \\[-1.6ex]
tree-level&  & & & &\multicolumn{2}{c}{-2.192(27)} \\
& & & & &  &\\[-2ex]
\hline
& & & & &  & \\[-2ex]
1-loop& &  & & &\multicolumn{2}{c}{-2.344(35)}  \\[-1.6ex]
 &6.9630 &  \multicolumn{2}{c}{32} &0.134526 & & \\[-1.6ex]
tree-level& & & & &\multicolumn{2}{c}{-2.266(34)} \\
& & & & & &\\[-2ex]
\hline
\hline
& & & & & & \\
$\castat$ & $\beta$ &  \multicolumn{2}{c}{$L_1/a$} &$\kappa_\mathrm{crit}$ & $\Xi^{-1}(L_1)$ & $\Xi(L_1)^{(0)}$ \\
& & & & & & \\
\hline
& & & & & & \\[-2ex]
1-loop& & & & &-0.4934(24) & \\[-1.6ex]
 &6.2110 &  \multicolumn{2}{c}{12} &0.135625 & & \\[-1.6ex]
tree-level& & & & &-0.5164(25) &-1.6019540566018 \\
& & & & & &\\[-2ex]
\hline
& & & & & & \\[-2ex]
1-loop& & & & &-0.4787(28) & \\[-1.6ex]
 &6.4200 &  \multicolumn{2}{c}{16} &0.135616 & & \\[-1.6ex]
tree-level& & & & &-0.4987(29)  &-1.6027594410020 \\
& & & & & &\\[-2ex]
\hline
& & & & & & \\[-2ex]
1-loop& & & & &-0.4772(33) & \\[-1.6ex]
 &6.7370 &  \multicolumn{2}{c}{24} &0.135235 & & \\[-1.6ex]
tree-level&  & & & &-0.4939(34) &-1.6033361949926 \\
& & & & & &\\[-2ex]
\hline
& & & & & & \\[-2ex]
1-loop& & & & &-0.4712(30) & \\[-1.6ex]
 &6.9630 &  \multicolumn{2}{c}{32} &0.134832 & & \\[-1.6ex]
tree-level&  & & & &-0.4860(31) &-1.6035384024722 \\
& & & & & &\\[-2ex]

\hline
\hline
& & & & &  & \\
$\castat$ & $\beta$ & $L_1/a$ &  $L_0/a$ &$\kappa_\mathrm{s}$ & $X_\mathrm{SF}(L_1)$  &$X_\mathrm{SF}(L_0)$  \\
& & & & & &  \\
\hline
& & & & & & \\[-2ex]
1-loop& & & & &-2.221(13)  & -1.8054(34)\\[-1.6ex]
 &6.2110 &12  & 6&0.134766  & & \\[-1.6ex]
tree-level&  & & & &-2.102(12)  & -1.7170(32) \\
& & & & & &\\[-2ex]
\hline
& & & & & & \\[-2ex]
1-loop& & & & &-2.266(13)  &-1.8372(46) \\[-1.6ex]
 &6.4200 & 16& 8&0.135015  & & \\[-1.6ex]
tree-level&  & & & &-2.159(12)  &-1.7540(44)  \\
& & & & & &\\[-2ex]
\hline
& & & & & & \\[-2ex]
1-loop& & & & &-2.279(28)  &-1.8806(63) \\[-1.6ex]
 &6.7370 & 24&12 &0.134801  & & \\[-1.6ex]
tree-level&  & & & &-2.192(27)  &-1.8122(60)  \\
& & & & & &\\[-2ex]
\hline
& & & & & & \\[-2ex]
1-loop& & & & &-2.344(35)  & -1.8986(65)\\[-1.6ex]
 &6.9630 & 32&16 &0.134526  & & \\[-1.6ex]
tree-level&  & & & &-2.266(34)   &-1.8370(63)  \\
& & & & & &\\[-2ex]
\hline
\hline
\end{tabular}
\mycaption{Collection of results at finite lattice spacing for the static pseudoscalar decay constant. All
numbers are obtained with the 1-loop expression of $\bastat$.}
\label{tab:Y_stat_data}
\end{table}

\addtocounter{table}{-1}
\clearpage
\begin{table}[ht]
\centering
\footnotesize
\begin{tabular}{ccccccc}
\hline
\hline
& & & & &  & \\
$\castat$ & $\beta$ & \multicolumn{2}{c}{$L_0/a$}  &$\kappa_\mathrm{s}$ & \multicolumn{2}{c}{$X_\mathrm{SF}(L_0)$} \\
& & & & & &  \\
\hline
& & & & &  & \\[-2ex]
1-loop& &  & & &\multicolumn{2}{c}{-1.8372(46)}  \\[-1.6ex]
 &6.4200 &   \multicolumn{2}{c}{8} &0.135015  & & \\[-1.6ex]
tree-level& & & & &\multicolumn{2}{c}{-1.7540(44)} \\
& & & & & &\\[-2ex]
\hline
& & & & &  & \\[-2ex]
1-loop& &  & & &\multicolumn{2}{c}{-1.8806(63)}  \\[-1.6ex]
 &6.7370 &   \multicolumn{2}{c}{12} &0.134801  & & \\[-1.6ex]
tree-level& & & & &\multicolumn{2}{c}{-1.8122(60)} \\
& & & & & &\\[-2ex]
\hline
& & & & &  & \\[-2ex]
1-loop& &  & & &\multicolumn{2}{c}{-1.8986(65)}  \\[-1.6ex]
 &6.9630 &   \multicolumn{2}{c}{16} & 0.134526 & & \\[-1.6ex]
tree-level& & & & &\multicolumn{2}{c}{-1.8370(63)} \\
& & & & & &\\[-2ex]
\hline
& & & & &  & \\[-2ex]
1-loop& &  & & &\multicolumn{2}{c}{-1.9182(102)}  \\[-1.6ex]
 &7.3000 &   \multicolumn{2}{c}{24} &0.134041  & & \\[-1.6ex]
tree-level& & & & &\multicolumn{2}{c}{-1.8660(99)} \\
& & & & & &\\[-2ex]
\hline
\hline
& & & & & & \\
$\castat$ & $\beta$ &   \multicolumn{2}{c}{$L_0/a$}  &$\kappa_\mathrm{crit}$ & $\Xi^{-1}(L_0)$ & $\Xi^{(0)}(L_0)$ \\
& & & & & & \\
\hline
& & & & & & \\[-2ex]
1-loop& & & & &-0.5467(13) & \\[-1.6ex]
 &6.4200 &    \multicolumn{2}{c}{8} &0.135616  & & \\[-1.6ex]
tree-level& & & & &-0.5686(13)  & -1.5996643156321 \\
& & & & & &\\[-2ex]
\hline
& & & & & & \\[-2ex]
1-loop& & & & &-0.53272(64) & \\[-1.6ex]
 &6.7370 &    \multicolumn{2}{c}{12} &0.135235  & & \\[-1.6ex]
tree-level& & & & &-0.55109(66)  & -1.6019540566018 \\
& & & & & &\\[-2ex]
\hline
& & & & & & \\[-2ex]
1-loop& & & & &-0.5246(13) & \\[-1.6ex]
 &6.9630 &    \multicolumn{2}{c}{16} &0.134832  & & \\[-1.6ex]
tree-level& & & & &-0.5407(14)  &-1.6027594410020  \\
& & & & & &\\[-2ex]
\hline
& & & & & & \\[-2ex]
1-loop& & & & &-0.5182(13) & \\[-1.6ex]
 &7.3000 &    \multicolumn{2}{c}{24} &0.134235  & & \\[-1.6ex]
tree-level& & & & &-0.5319(13)  &-1.6033361949926  \\
& & & & & &\\[-2ex]
\hline
\hline
\end{tabular}
\mycaption{(continued)}
\end{table}

An accurate data analysis reveals that the quantities
\be 
\sigma_\mathrm{f}^\mathrm{stat}(L_2)\,,\quad\sigma_\mathrm{f}^\mathrm{stat}(L_1)\,\quad \mbox{and} \quad Y_\mathrm{SF}(L_0,1/L_0)
\ee
are slightly correlated, because they share the same data for $X_\mathrm{SF}(L_1)$
and $X_\mathrm{SF}(L_0)$. However, the error on the final result for the decay constant
is dominated by the uncertainty stemming from the large volume simulations.
Furthermore we observe that, if we approximate $\fBs$, computed through \eq{fBs_final},
to simply have a linear dependence on the static decay constant, the relations
\begin{align}
\fBs&\propto Y_\mathrm{SF}(L_0,1/L_0)\cdot \sigma_\mathrm{f}^\mathrm{stat}(L_1) \cdot \sigma_\mathrm{f}^\mathrm{stat}(L_2)\\
&\propto X_\mathrm{SF}(L_0)\cdot \frac{X_\mathrm{SF}(L_1)}{X_\mathrm{SF}(L_0)}\cdot \frac{1}{X_\mathrm{SF}(L_1)}\,,
\end{align}
show that this correlation is expected to further reduce the 
uncertainty quoted in (\ref{eq:fBs_Joi_result}). We neglect it,
because it is small in comparison with the large volume
uncertainties.



\clearpage

\begin{figure}[b]
\centering
\begin{minipage}[b]{0.5\textwidth}
\centering \includegraphics[width=6.5cm,height=5.3cm]{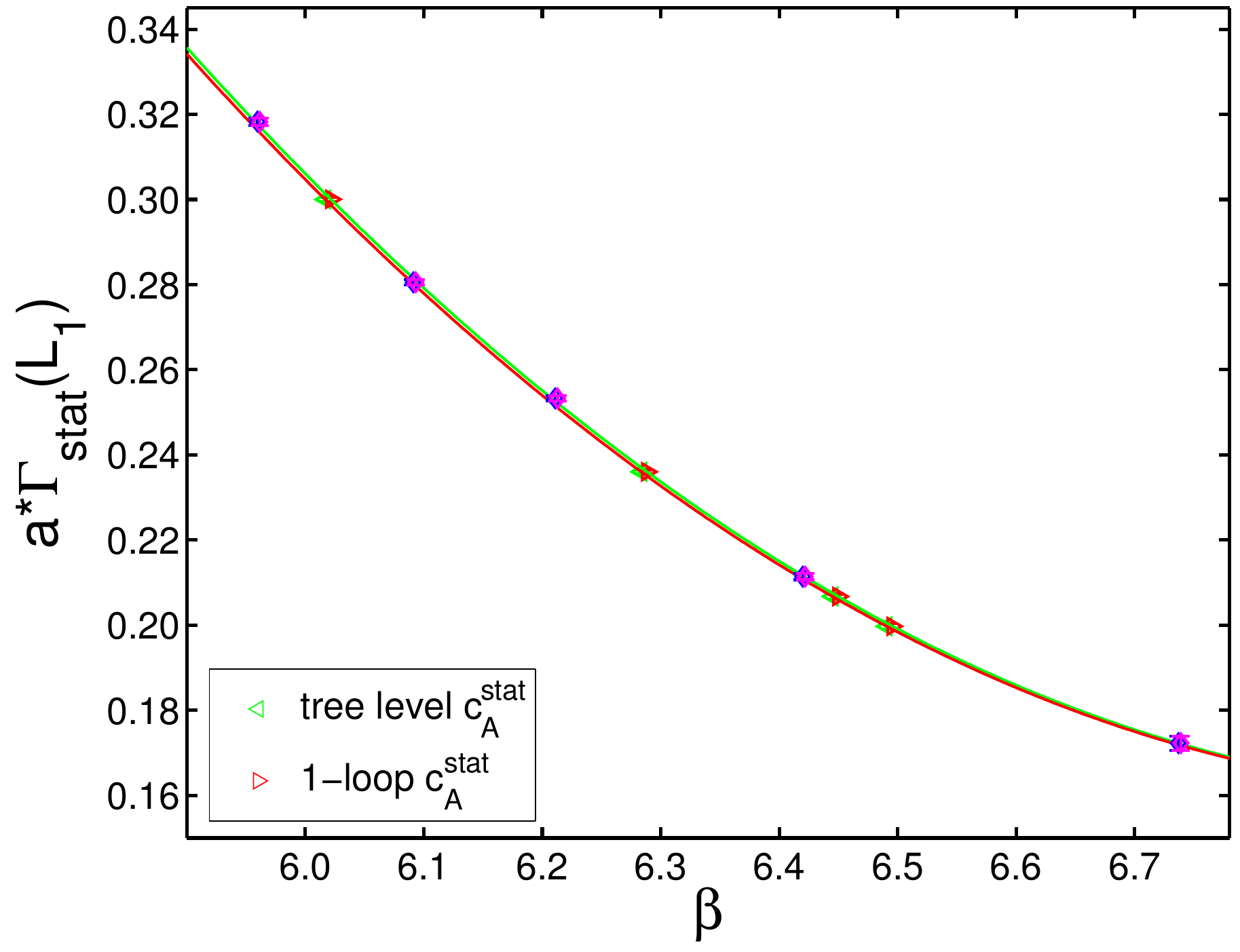} 
\end{minipage}%
\begin{minipage}[b]{0.5\textwidth}
\includegraphics[width=6.5cm,height=5.3cm]{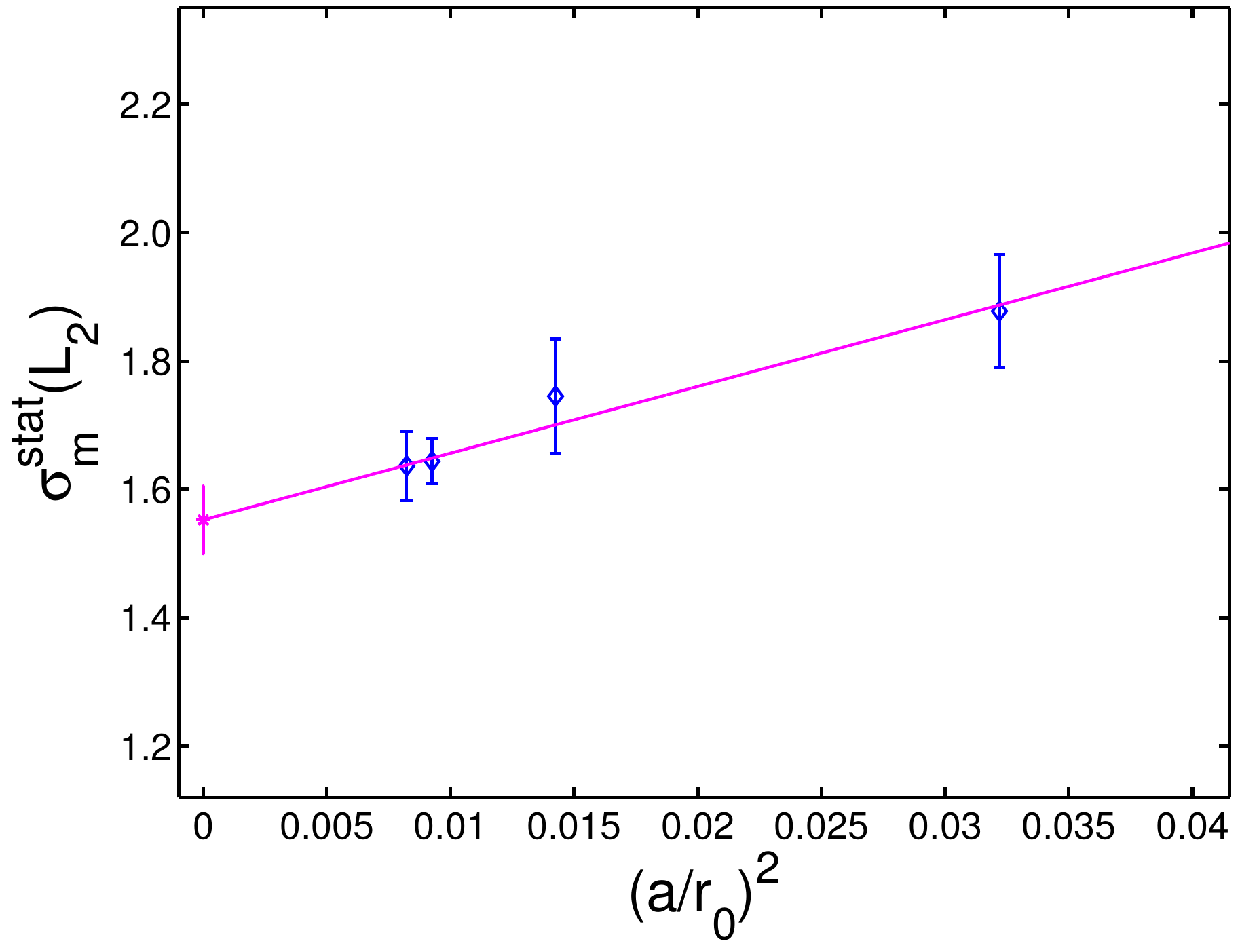}
\end{minipage}
\vspace{-0.5cm}
\mycaption{On the left: $\beta$-dependence of $a\Gamma_\mathrm{stat}(L_1)$. The red curve shows the quadratic fit of
the data with $\castat$ at 1-loop (blue points), while the green one fits the data with $\castat$
at tree-level (magenta points). The two curves almost overlap.
On the right: continuum extrapolation to get $\sigma_\mathrm{m}^\mathrm{stat}(L_2)$.}\label{fig:CL_S2_M}
\end{figure}

\vspace{1cm}
\begin{figure}[t]
\centering
\begin{minipage}[t]{0.5\textwidth}
\centering \includegraphics[width=6.5cm,height=5.3cm]{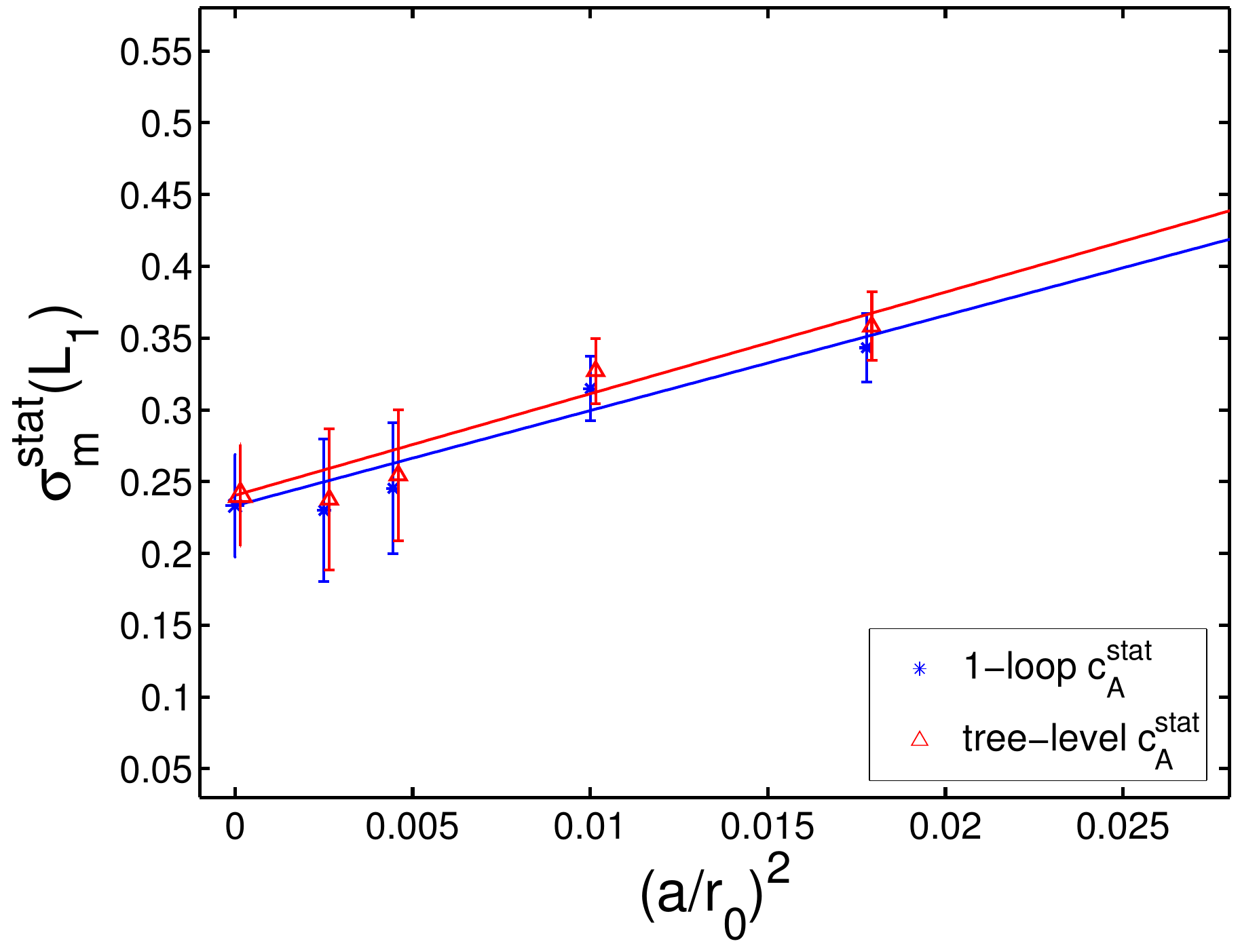} 
\end{minipage}%
\hfill\begin{minipage}[t]{0.5\textwidth}
\includegraphics[width=6.5cm,height=5.3cm]{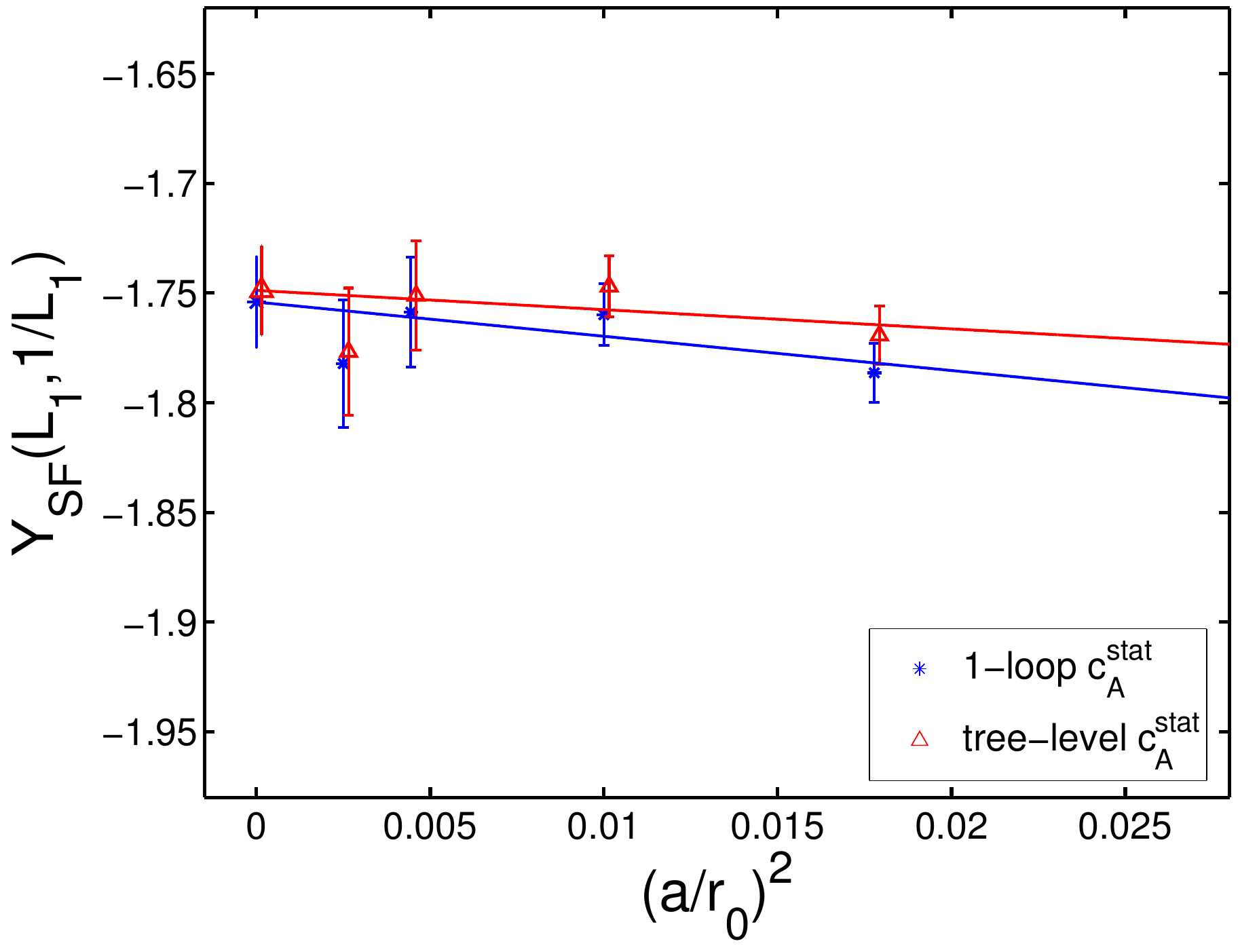}
\end{minipage}
\vspace{-0.5cm}

\vspace{1cm}
\centering
\begin{minipage}[t]{0.5\textwidth}
\centering \includegraphics[width=6.5cm,height=5.3cm]{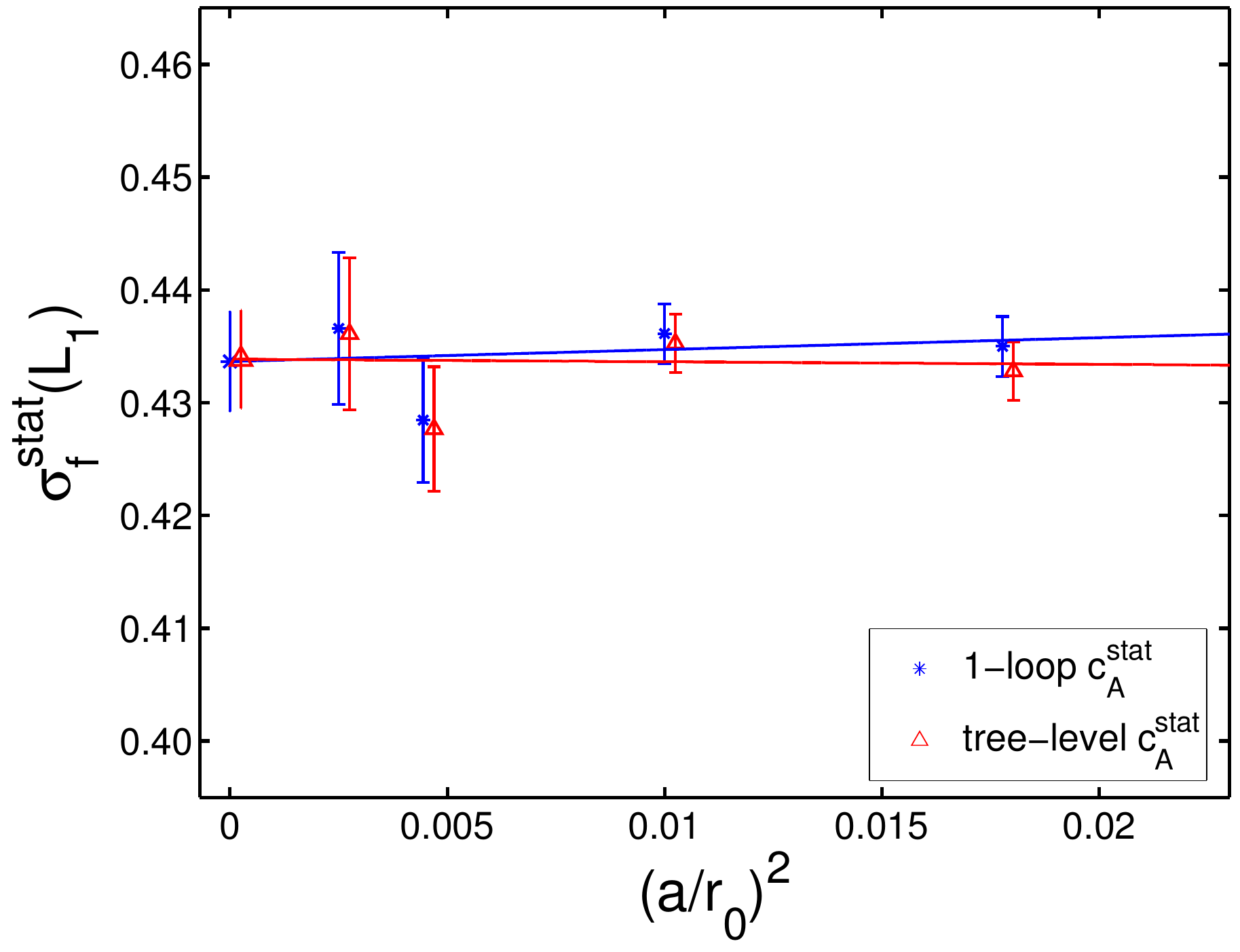} 
\end{minipage}%
\begin{minipage}[t]{0.5\textwidth}
\includegraphics[width=6.5cm,height=5.3cm]{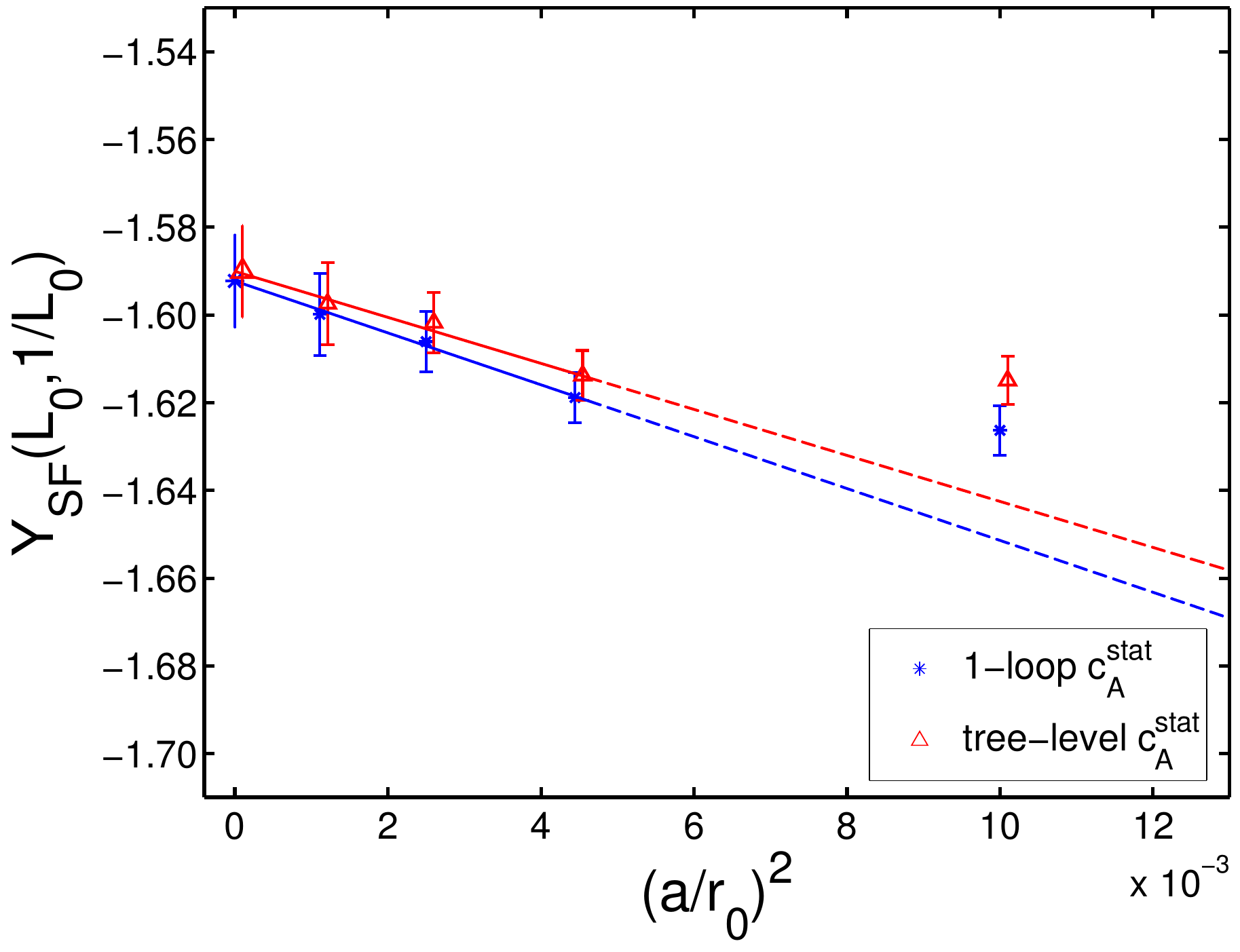}
\end{minipage}
\vspace{-0.5cm}
\mycaption{Various HQET continuum limit extrapolations for the Step Scaling Method.}\label{fig:CL_stat}
\end{figure}

\clearpage

\section{Relativistic QCD data}\label{asect:CL_QCD}

The data at finite heavy quark mass have been produced by the Tor Vergata
group, and published in \cite{romeII:mb,romeII:fb}. 
In these references one can find all simulation parameters, which have been determined
with the method explained in \sect{sect:SSM_simu_para}. 
In this section we report the new data analysis,
collecting the results of the continuum limit
extrapolations.

The latter are performed according to the method outlined in \sect{sect:cont_lim}.
Here, the way followed to define the Line of Constant Physics (cf.~\sect{sect:Oa_impr})
has to be clarified. Let us assume that we are at fixed finite volume $L$ and lattice spacing,
and consider an observable $\mathcal{O}$. The quark mass definition is firstly only one, 
say \eq{eq:M_deg}. 
The observable is computed for several 
quark-antiquark couples. Therefore, there exist a bijective correspondence
between the computed observable $\mathcal{O}$ and the couples $(LM_\mathrm{q_2}, LM_\mathrm{q_1})$.
Since we are interested in the heavy-strange meson properties,
the observable is interpolated in such a way that 
one of the two quark masses matches the physical strange quark mass;
then the remaining quark mass is matched to a set of some conveniently chosen 
values of $LM_\mathrm{h}$. This two-step interpolation procedure is repeated for
all discretizations belonging to the volume $L$. We are now able to perform the continuum
limit extrapolations of the observable $\mathcal{O}$, where the couples $(LM_\mathrm{s}, LM_\mathrm{h})$ are
kept fixed in all discretizations.
By repeating the whole procedure for several quark mass definitions, whose details
are given in \sect{sect:quark_masses_SSM}, we have the setup described in \sect{sect:cont_lim}.
Therefore, at fixed lattice spacing one has several determinations of $\mathcal{O}$, differing by being
associated with different quark mass definitions. These determinations share
the same gauge configurations, and are statistically correlated.
Each of these definitions follows a different Line of Constant Physics, but universality imposes
that they are indistinguishable in the continuum limit. Furthermore, 
one can infer 
from the $\Oa$-improvement, that, at non-vanishing but small lattice spacing,
these definitions differ by $\Oasq$-terms.  

\begin{figure}[t]
\centering
\begin{minipage}[t]{0.5\textwidth}
\centering \includegraphics[width=6.5cm,height=4.8cm]{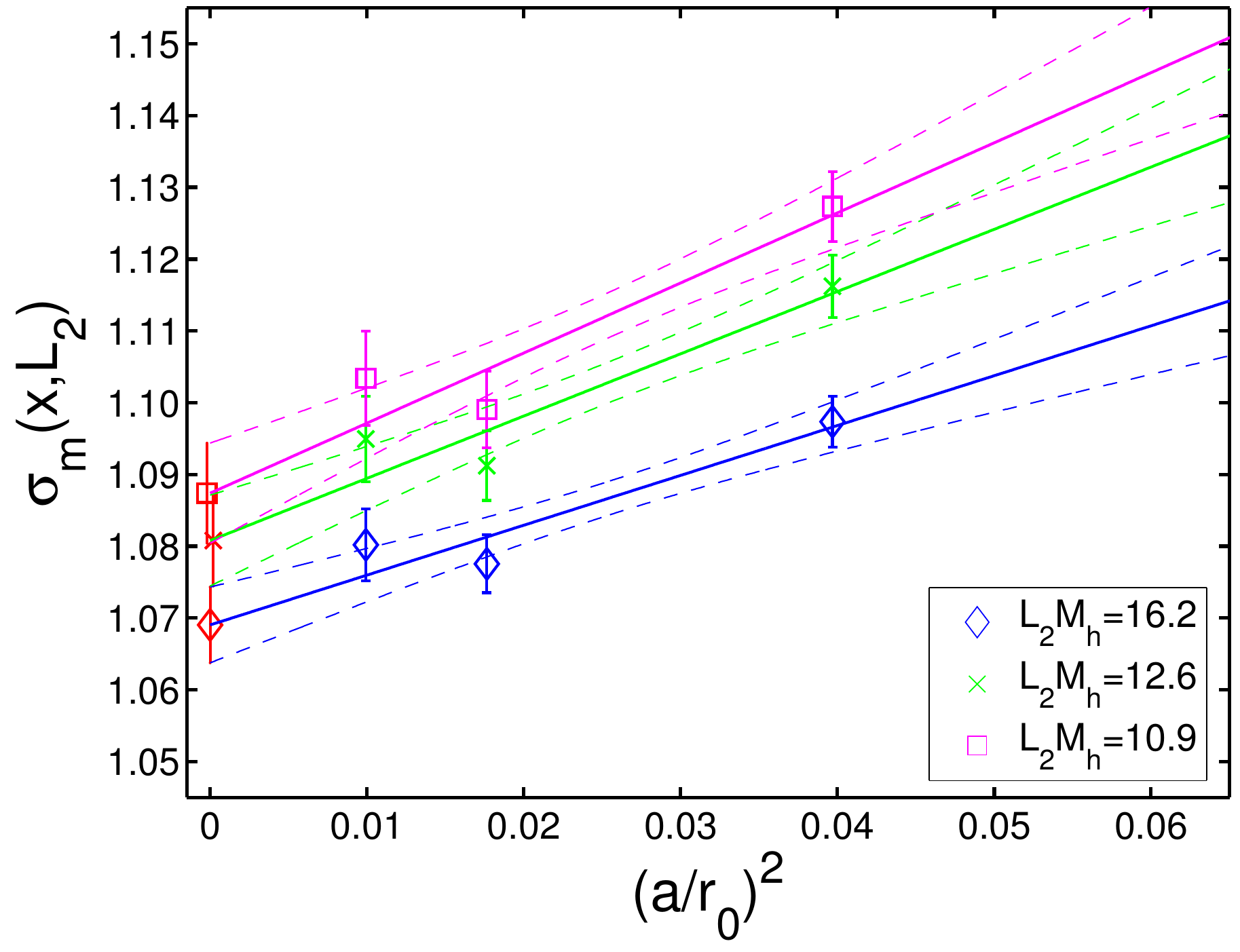} 
\end{minipage}%
\begin{minipage}[t]{0.5\textwidth}
\includegraphics[width=6.5cm,height=4.8cm]{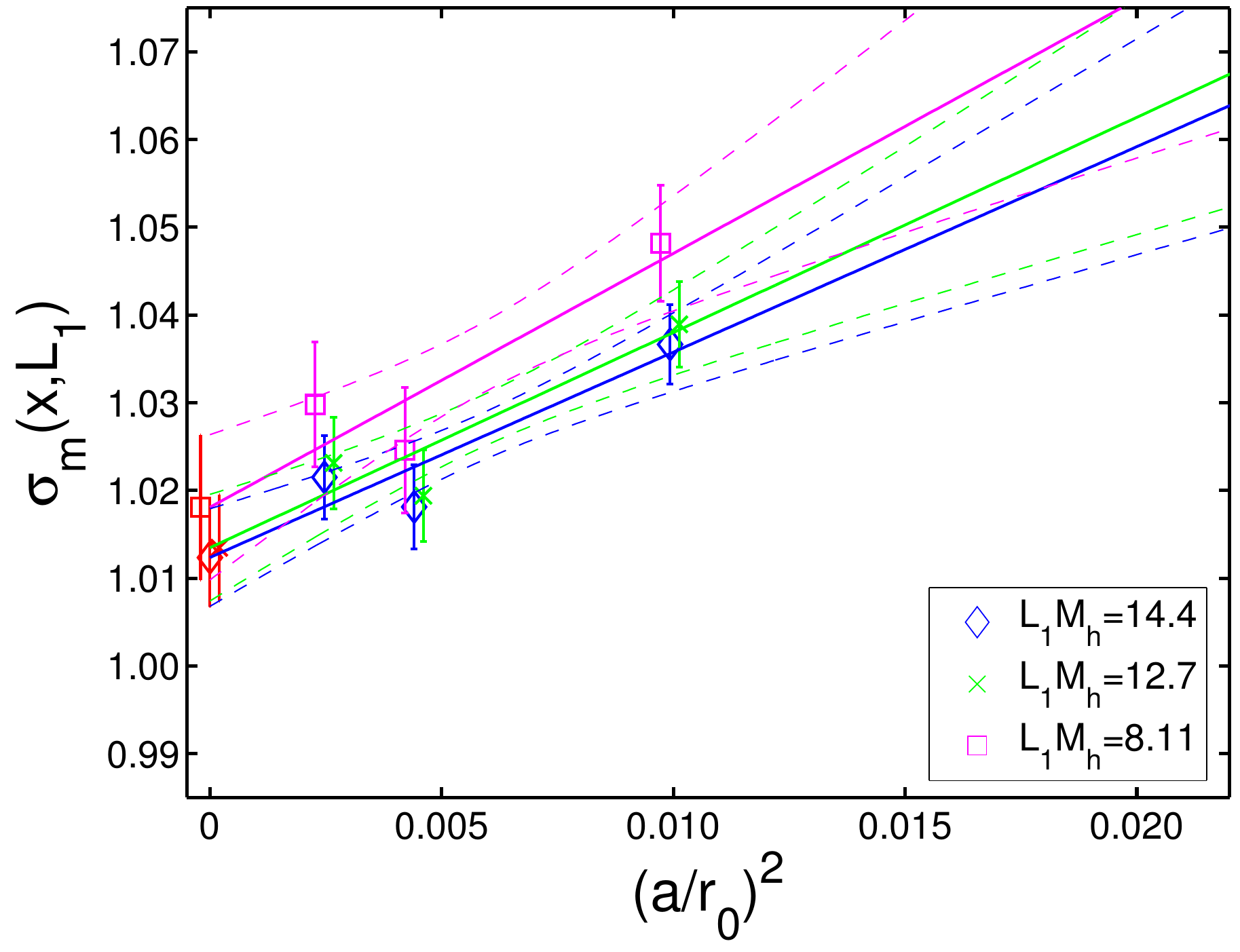}
\end{minipage}
\vspace{-0.5cm}

\centering
\begin{minipage}[t]{0.5\textwidth}
\centering \includegraphics[width=6.5cm,height=4.8cm]{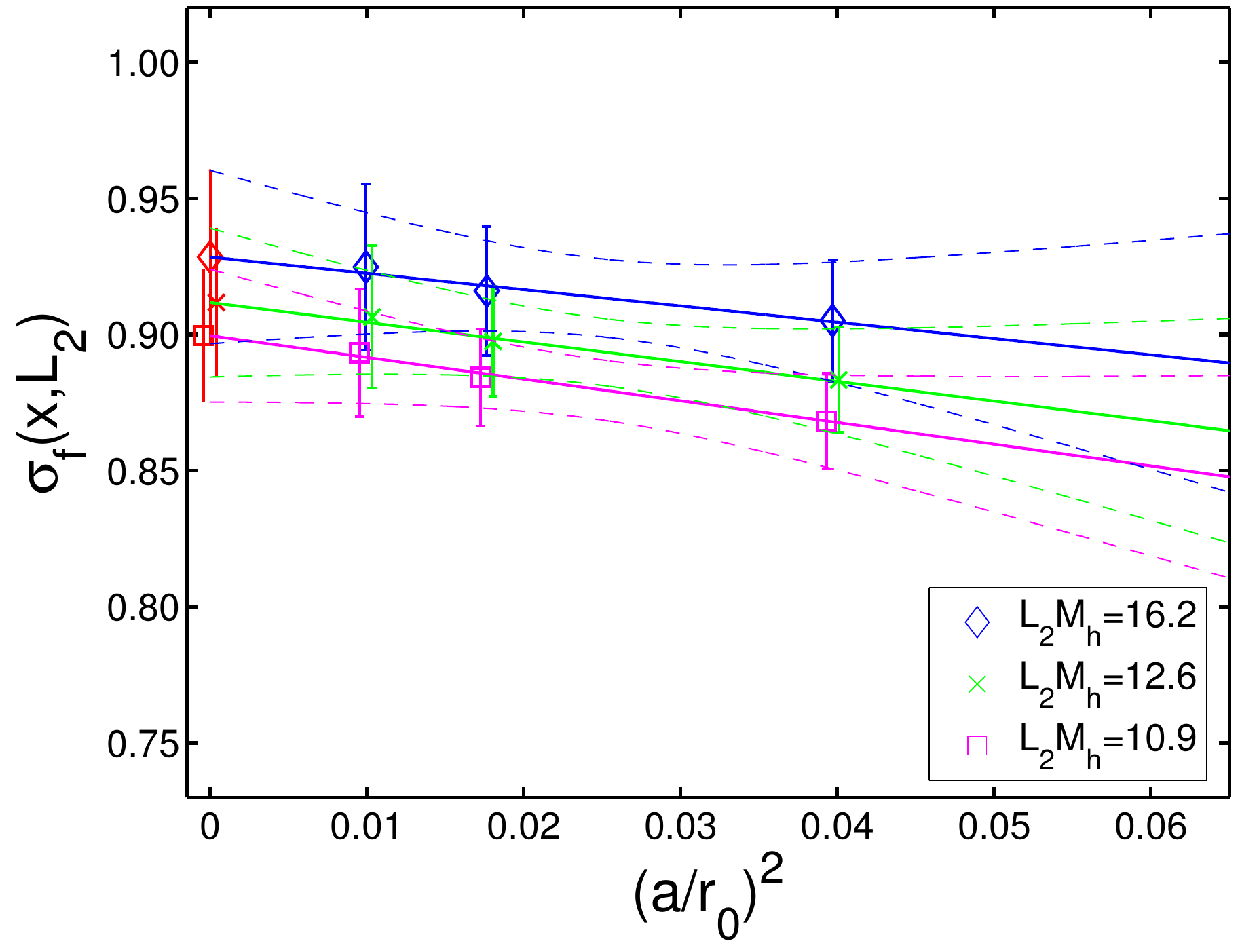} 
\end{minipage}%
\begin{minipage}[t]{0.5\textwidth}
\includegraphics[width=6.5cm,height=4.8cm]{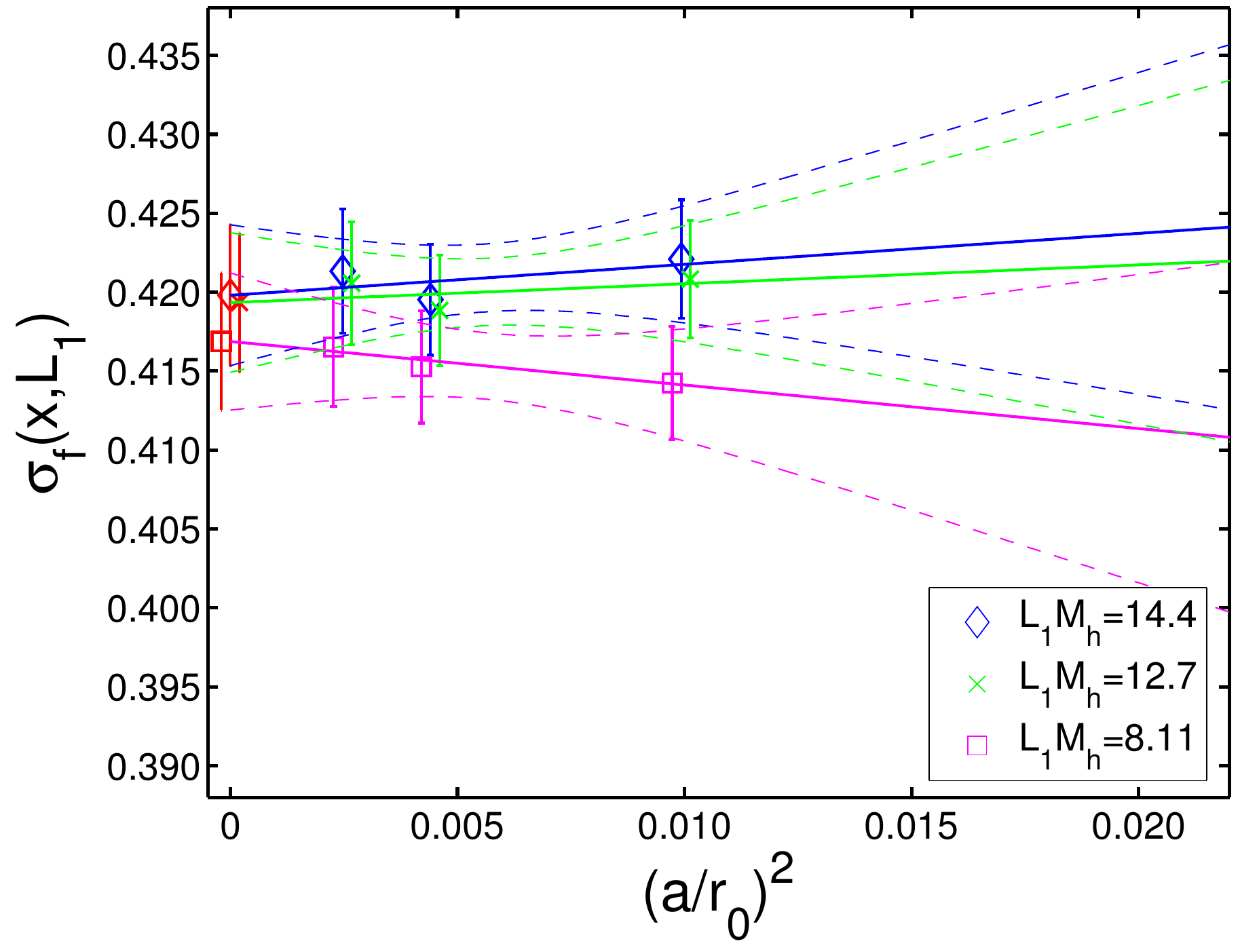}
\end{minipage}
\vspace{-0.5cm}

\centering
\begin{minipage}[t]{0.5\textwidth}
\centering \includegraphics[width=6.5cm,height=4.8cm]{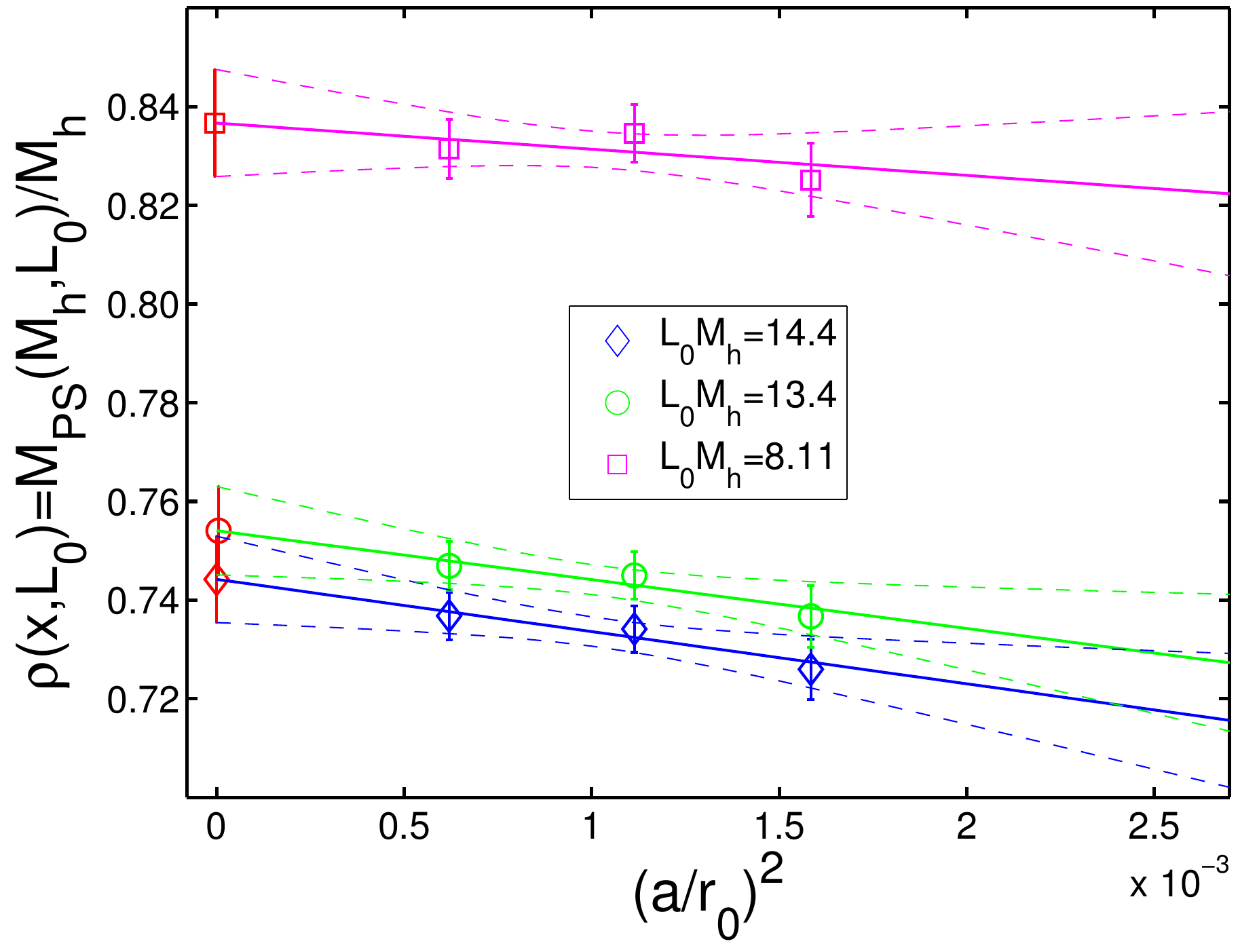} 
\end{minipage}%
\begin{minipage}[t]{0.5\textwidth}
\includegraphics[width=6.5cm,height=4.8cm]{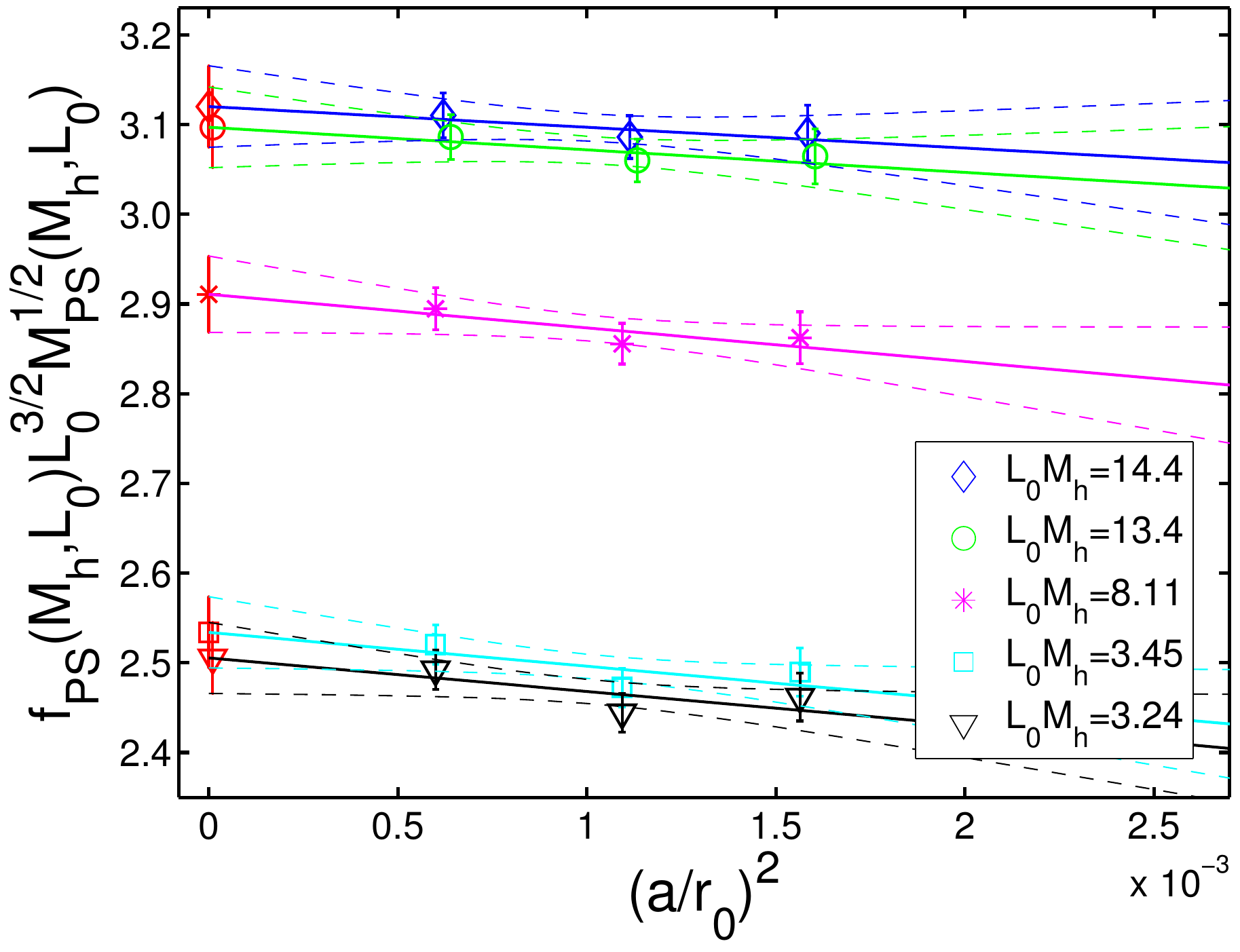}
\end{minipage}
\vspace{-0.5cm}

\centering
\begin{minipage}[t]{0.5\textwidth}
\centering \includegraphics[width=6.5cm,height=4.8cm]{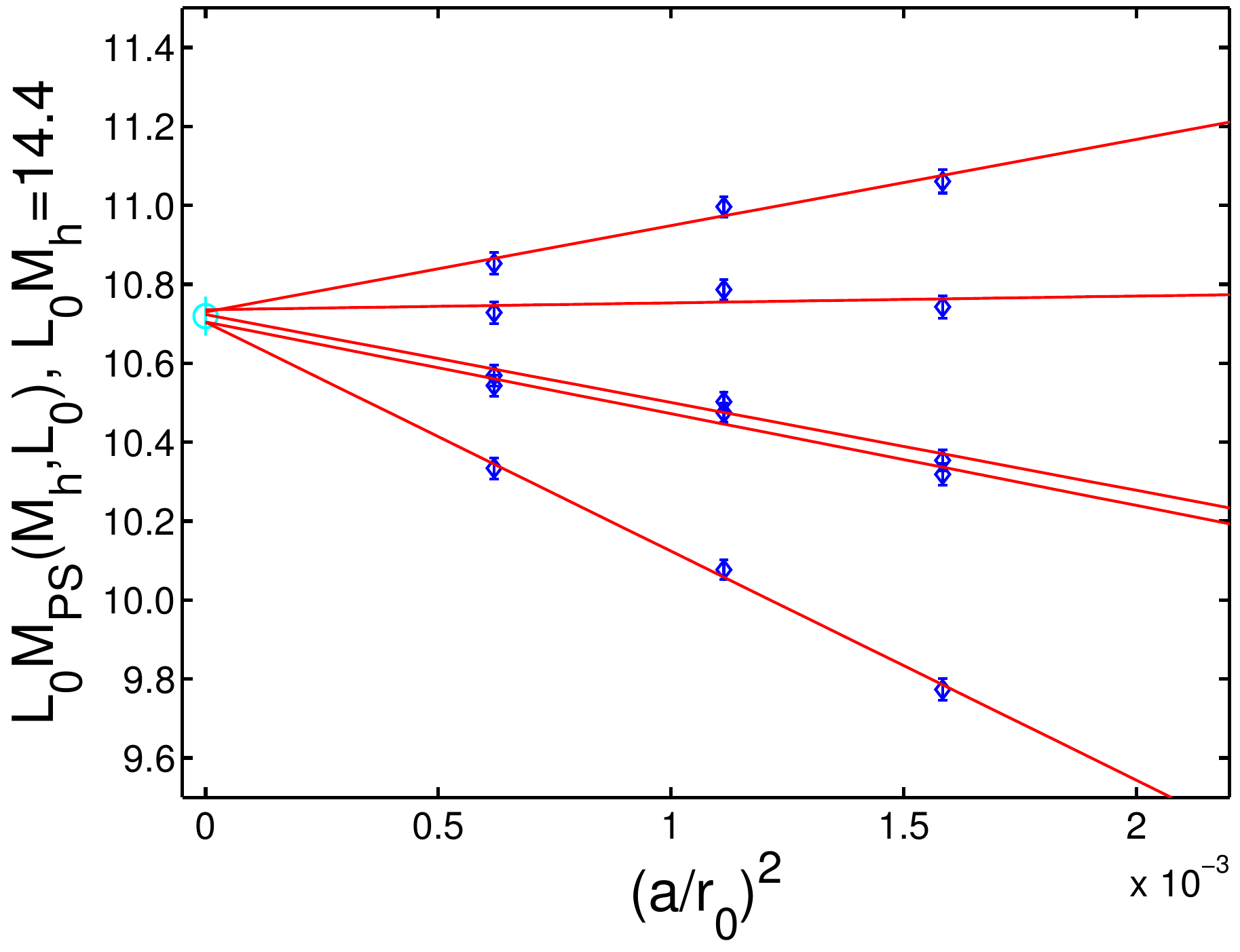} 
\end{minipage}%
\begin{minipage}[t]{0.5\textwidth}
\includegraphics[width=6.5cm,height=4.8cm]{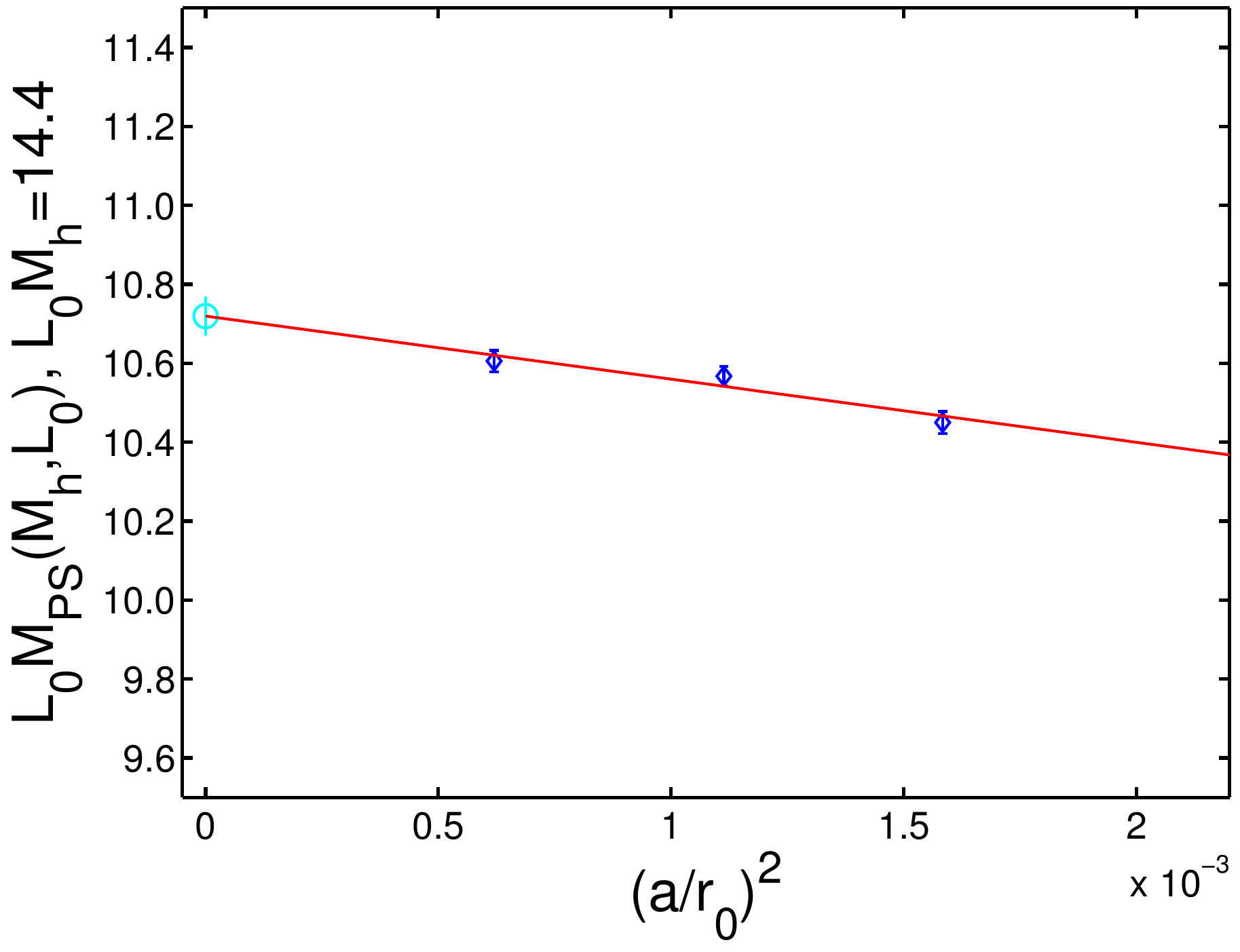}
\end{minipage}
\vspace{-0.5cm}
\mycaption{Continuum limit extrapolations for the relativistic QCD data used in the Step Scaling Method.}\label{fig:CL_rel_QCD}
\end{figure}

\tab{tab:res_sigmaPS} summarizes all continuum limit results,
and the extrapolations are shown in \fig{fig:CL_rel_QCD}. The plot on the lower
right corner shows the continuum limit (cyan point) for $L_0M_\mathrm{PS}(L_0)$, performed 
according to the averaging method of \sect{sect:cont_lim}. On the lower left corner
each definition is separately fitted. The cyan point of the r.h.s.~plot is pasted
here to show the very good agreement between each single definition and the averaged one.

In all continuum limit extrapolations the uncertainty on the improvement
coefficients $\ca$ and $\ba$ is found to be negligible.

\clearpage

\begin{table}[c]
\centering
\footnotesize
\begin{tabular}{cccc}
\hline
\hline
& & & \\
$L_2M_\mathrm{h}$ &$L_2M_\mathrm{PS}(L_2)$& $\sigma_\mathrm{m}(L_2)$ & $\sigma_\mathrm{f}(L_2)$\\
& & & \\
\hline
& & & \\[-1.6ex]
16.2&17.221(44) &1.0690(53) & 0.929(32)\\[-1.6ex]
& & & \\
\hline
& & & \\[-1.6ex]
12.6&14.918(43) &1.0808(63) & 0.912(27)\\[-1.6ex]
& & & \\
\hline
& & & \\[-1.6ex]
10.9&13.894(43) &1.0874(70) & 0.900(24)\\[-1.6ex]
& & & \\
\hline
\hline
& & & \\
$L_1M_\mathrm{h}$ &$L_1M_\mathrm{PS}(L_1)$& $\sigma_\mathrm{m}(L_1)$ & $\sigma_\mathrm{f}(L_1)$\\
& & & \\
\hline
& & & \\[-1.6ex]
14.4&12.438(52) &1.0123(56) &0.4198(45) \\[-1.6ex]
& & & \\
\hline
& & & \\[-1.6ex]
12.7&11.319(52) &1.0135(61) &0.4193(45) \\[-1.6ex]
& & & \\
\hline
& & & \\[-1.6ex]
8.11&8.307(52) &1.0181(83) &0.4169(43) \\[-1.6ex]
& & & \\
\hline
\hline
& & & \\
$L_0M_\mathrm{h}$ &$L_0M_\mathrm{PS}(L_0)$& $\rho(L_0)$ & $f_\mathrm{PS}(L_0)\sqrt{L_0^3M_\mathrm{PS}(L_0)}$\\
& & & \\
\hline
& & & \\[-1.6ex]
14.4 &10.719(47)  &0.7442(88)  & 3.120(45) \\[-1.6ex]
& & & \\
\hline
& & & \\[-1.6ex]
13.4 &10.097(47)  &0.7541(90)  & 3.097(45) \\[-1.6ex]
& & & \\
\hline
& & & \\[-1.6ex]
8.11 &6.793(46)  &0.8367(108)  & 2.911(43) \\[-1.6ex]
& & & \\
\hline
& & & \\[-1.6ex]
3.45 &3.613(44)  & - &2.534(40)  \\[-1.6ex]
& & & \\
\hline
& & & \\[-1.6ex]
3.24 &3.466(44)  & - & 2.505(40) \\[-1.6ex]
& & & \\
\hline
\hline
\end{tabular}
\mycaption{Continuum limit results for the relativistic QCD data used in the Step Scaling Method.}
\label{tab:res_sigmaPS}
\end{table} 

\clearpage

\clearpage



\chapter{Correlated errors in both coordinates}\label{app:error_xy}

This appendix contains a short digression 
about error analysis. A few formulae are derived
to deal with the case of a two dimensional fit
with correlated variables on the $y$- and $x$-axes.

Let us consider the case where we want to fit a certain 
quantity $y$ vs. $x$, and the uncertainties
$\Delta y$ and $\Delta x$, respectively on $y$ and $x$, are 
correlated. To simplify the problem let us assume that 
the fit is of the kind $y=a+bx$, but
the following considerations hold more generally.
Having the set of data $\{y_i,x_i,i=1,\ldots, N\}$ with 
correlated 
errors $\{\Delta y_i,\Delta x_i,i=1,\ldots, N\}$, 
following \cite{NumRec},
it is straightforward to write down the 
appropriate $\chi^2$ for this case,
\be\label{eq:merit_function}
\chi^2(a,b)=\sum_{i=1}^{N}\frac{(y_i-a-bx_i)^2}{\sigma_i^2}\,.
\ee
The variance appearing in the denominator of \eq{eq:merit_function} can be understood
both as the variance in the direction of the smallest $\chi^2$ between each data point and the line
with slope $b$, as well as the variance of the linear combination $z_i=y_i-a-bx_i$ of the variables $y_i$
and $x_i$,
\begin{align}
\sigma_i^2=\mathrm{Var}(z_i)&= \left(\frac{\partial z_i}{\partial y_i}\right)^2\mathrm{Var}(y_i)
+\left(\frac{\partial z_i}{\partial x_i}\right)^2\mathrm{Var}(x_i)\nonumber\\
& \nonumber\\[-1.3ex]
&  +2\frac{\partial z_i}{\partial y_i}\frac{\partial z_i}{\partial x_i}\mathrm{Cov}(y_i;x_i)\label{eq:variance1_numrec}\\
&  \nonumber\\[-1.3ex]
&= \mathrm{Var}(y_i)+b^2\mathrm{Var}(x_i)-2b\mathrm{Cov}(y_i;x_i)\,.\label{eq:variance2_numrec}
\end{align}
The expressions (\ref{eq:variance1_numrec}) and (\ref{eq:variance2_numrec}) can be generalized to
the case of $y_i$ depending on the set of 
variables $\{\omega_{i,k},k=1,\ldots,N\}$, and $x_i$ depending on $\{\omega_{i,l},l=N+1,\ldots,M\}$,
with the $\omega$'s correlated with each other.
In that case $z_i=z_i(\omega_{i,s})$ with $s=1,\ldots,M$ and one can write a compact expression for $\mathrm{Var}(z_i)$, which reads
\be\label{eq:general1}
\mathrm{Var}(z_i)=\sum_{s,s'=1}^{M}\frac{\partial z_i}{\partial \omega_{i,s}}\mathrm{Cov}(\omega_{i,s};\omega_{i,s'})
\frac{\partial z_i}{\partial \omega_{i,s'}}\,.
\ee
Let us take as example a linear fit of $y(\omega_1,\omega_2)$ vs. $x(\omega_2,\omega_3)$,
with the couples $\{\omega_1,\omega_2\}$ and $\{\omega_2,\omega_3\}$ statistically uncorrelated. 
It is clear that the uncertainty on
the $y_i$'s is correlated with the one on the $x_i$'s, because they all depend on $\omega_2$.
In this case \eq{eq:general1} reduces to
\begin{align}
\mathrm{Var}(z_i)&= \left(\frac{\partial y_i}{\partial \omega_{i,1}}\Delta \omega_{i,1}\right)^2+\left(\frac{\partial y_i}{\partial \omega_{i,2}}\Delta \omega_{i,2}\right)^2
+b^2\left(\frac{\partial x_i}{\partial \omega_{i,2}}\Delta \omega_{i,2}\right)^2\nonumber\\
&  +b^2\left(\frac{\partial x_i}{\partial \omega_{i,3}}\Delta \omega_{i,3}\right)^2-2b\frac{\partial y_i}{\partial \omega_{i,2}}
\frac{\partial x_i}{\partial \omega_{i,2}}(\Delta \omega_{i,2})^2\,.\label{eq:special_123}
\end{align}
Coming back to the general case (\ref{eq:general1}), we want to minimize \eq{eq:merit_function}
with respect to $a$ and $b$. Unfortunately,
the occurrence of $b$ in the denominator of \eq{eq:merit_function} makes the resulting
equation for the slope $\partial \chi^2/\partial b=0$ nonlinear. In most cases it is sufficient to approximate
the slope $b$ by performing the fit without the uncertainties coming from the $x_i$'s, then one computes
the $\sigma_i$'s according to (\ref{eq:general1}) and repeats the fit. The procedure has to be iterated
with the new $b$ until successive estimates of $b$ agree within the desired tolerance. 

\chapter{Fit of the correlated relativistic QCD data}\label{app:fit_QCD_corr}

This appendix is devoted to the explanation
of the method used to take into account
the correlations between the relativistic QCD data.
The latter are characterized by a non-vanishing
correlation between the observables belonging to the same lattice
spacing, but differing by being associated with different definitions
of the quark mass.
In addition, observables computed on the same lattice but for
different masses share the same set of gauge configurations, and are
indeed correlated.

\section{Performing the continuum limits}\label{sect:cont_lim}

Let us consider an observable $\mathcal{O}(m_\mathrm{h})$ depending
on the heavy quark mass $m_\mathrm{h}$. In our case the light quark
is matched to be the strange one in all cases, and the corresponding dependence
is omitted in the following to simplify the notation. The same holds
for the volume dependence; it plays no role for the present discussion.

To compute the continuum limit of the quantity $\mathcal{O}(m_k)$, 
with fixed $m_k$, the setup of the data can be
summarized by introducing the notation
\begin{align}
x_i&= \left(\frac{a}{L}\right)_i^2\,,\label{eq:discretizz}\\
&  \nonumber\\
y_{ij}(m_k)&= \mathcal{O}(\left(\frac{a}{L}\right)_i^2,\mbox{mass def}=j,\mbox{heavy mass}=m_k).\label{eq:points_nz_a}
\end{align}
The $x_i$ variable refers to the different discretizations, while $y_{ij}(m_k)$
refers
to the observable $\mathcal{O}$ with discretization $i$, quark mass definition $j$ and heavy quark mass $m_k$.
The several $y_{ij}(m_k)$'s with the same $i$ are correlated, because they are obtained from the same set of gauge
configurations.

One possibility to perform the continuum limit is the following procedure. First of all, at fixed lattice
spacing, the observables are averaged over the different mass definitions according to
\begin{align}
\bar{y}_i(m_k)&= \sum_jy_{ij}(m_k)/N,,\;\;\;N_i=\sum_j1=N,\;\forall i\,,\label{eq:mittelwert_method4} \\
&  \nonumber\\
\mbox{remarking that}&\quad  \frac{\partial \bar{y}_i(m_k)}{\partial y_{ij}(m_k)}=1/N,\;\forall j\,,\label{eq:dyl_dyls}\\
&  \nonumber\\
(\Delta \bar{y}_i(m_k))^2&= \frac{1}{N^2}\sum_{l,m}\mathrm{Cov}(y_{il}(m_k), y_{im}(m_k))\,,\label{eq:properr_method4}
\end{align}
taking into account the correlations as shown in \eq{eq:properr_method4}. Remembering
that the averaged quantities $\bar{y}_i(m_k)$ at different lattice spacing $i$ are statistically uncorrelated,
we can perform the continuum limit in the usual way, i.e.~by minimizing the $\chi^2$:
\be\label{eq:chi_sq_method4}
\chi^2 = \sum_{i}\left(\frac{\bar{y}_{i}(m_k)-a_0-a_1x_i}{\Delta \bar{y}_i(m_k)}\right)^2\,.
\ee
The result of the continuum limit is finally $a_0=\bar{y}(m_k)$.
For reasons which will be clear in the following section, it is useful to compute and store
\be\label{eq:dy_dyl}
\frac{\partial \bar{y}(m_k)}{\partial \bar{y}_l(m_k)}=\frac{1}{\Delta_{T} \cdot
(\Delta \bar{y}_l(m_k))^2}\left(T_{xx}-x_lT_x\right)\,,
\ee
where
\begin{align}
\Delta_{T}&= TT_{xx}-T_{x}^2,\;\;\;\;\hspace{0.95cm}T=\sum_i\frac{1}{(\Delta \bar{y}_i(m_k))^2}\\
T_{xx}&= \sum_i\frac{x_i^2}{(\Delta \bar{y}_i(m_k))^2},\;\;\;\;T_{x}=\sum_i\frac{x_i}{(\Delta \bar{y}_i(m_k))^2}\,.
\end{align}

\section{Fitting the mass dependence}\label{sect:mass_dependence}

The procedure described in the previous section is repeated for several heavy quark masses, and
it allows to arrange the data for a fit $\bar{y}(m)$ vs. $z=1/(Lm)$ with $L$
fixed.
Eventually the points $z_k$ can be chosen to be $1/(LM(m_k))$, where $M(m_k)$ is the meson mass associated with $m_k$.
The choice of the definition of $z_k$ is irrelevant for the present discussion,
because here we consider
as source of uncertainty only the statistical error on $\bar{y}(m_k)$, while the one associated with $z_k$ and other systematics
are neglected.

The first step for the fit $\bar{y}(m)$ vs. $z=1/(Lm)$ can be performed by dealing
with the $\bar{y}(m_k)$'s as independent quantities.
The merit function $\chi^2$
\be\label{eq:merit_funct}
\chi^2 = \sum_{k}\left(\frac{\bar{y}(m_k)-f(z_k,{c})}{\Delta \bar{y}(m_k)}\right)^2\,,
\ee
is minimized with respect to the parameters ${c}$.
Taking as example the case of a linear fit $f(z,\{c\})=c_0+c_1z$. The parameters $\{c\}$ are given by
\be
c_0=\frac{S_{xx}S_y-S_xS_{xy}}{\Delta}\,,\quad c_1=\frac{SS_{xy}-S_xS_{y}}{\Delta}\,,
\ee
where
\begin{align}
S&= \sum_k\frac{1}{(\Delta \bar{y}(m_k))^2}\,,\quad S_x=\sum_k\frac{z_k}{(\Delta
\bar{y}(m_k))^2}\,,\quad
S_y=\sum_k\frac{\bar{y}(m_k)}{(\Delta \bar{y}(m_k))^2}\nonumber\\
S_{xx}&= \sum_k\frac{z_k^2}{(\Delta \bar{y}(m_k))^2}\,,\quad S_{xy}=\sum_k\frac{z_k\bar{y}(m_k)}{(\Delta \bar{y}(m_k))^2}\,,
\quad\Delta=SS_{xx}-S_{x}^2\,.\nonumber
\end{align}
The estimate of the uncertainty on the parameters $\{c\}$ and their correlation takes into account the
correlation between the data as follows
\begin{align}
(\Delta c_0)^2&= \sum_{k,k'}\frac{\partial c_0}{\partial \bar{y}(m_k)}\mathrm{Cov}(\bar{y}(m_k),\bar{y}(m_{k'}))
\frac{\partial c_0}{\partial \bar{y}(m_{k'})}\,,\\
&  \nonumber\\
(\Delta c_1)^2&= \sum_{k,k'}\frac{\partial c_1}{\partial \bar{y}(m_k)}\mathrm{Cov}(\bar{y}(m_k),\bar{y}(m_{k'}))
\frac{\partial c_1}{\partial \bar{y}(m_{k'})}\,,\\
&  \nonumber\\
\mathrm{Cov}(c_0,c_1)&= \sum_{k,k'}\frac{\partial c_0}{\partial \bar{y}(m_k)}\mathrm{Cov}(\bar{y}(m_k),\bar{y}(m_{k'}))
\frac{\partial c_1}{\partial \bar{y}(m_{k'})}\,,
\end{align}
where one uses
\begin{align}
&  \hspace{0.875cm}\frac{\partial c_0}{\partial \bar{y}(m_k)}=\frac{1}{\Delta \cdot (\Delta \bar{y}(m_k))^2}
\left(S_{xx}-z_kS_x\right)\,,\\
&  \nonumber\\
&  \hspace{0.875cm}\frac{\partial c_1}{\partial \bar{y}(m_k)}=\frac{1}{\Delta \cdot (\Delta \bar{y}(m_k))^2}
\left(z_kS-S_x\right)\,,\\
&  \nonumber\\
&  \hspace{-1.2cm}\mathrm{Cov}(\bar{y}(m_k),\bar{y}(m_{k'}))=\sum_{l,m}\frac{\partial \bar{y}(m_k)}{\partial \bar{y}_l(m_k)}\mathrm{Cov}(\bar{y}_l(m_k),\bar{y}_m(m_{k'}))
\frac{\partial \bar{y}(m_{k'})}{\partial \bar{y}_m(m_{k'})}\,.\label{eq:cov_lev3}
\end{align}
The sums in \eq{eq:cov_lev3} can be simplified, because $\mathrm{Cov}(\bar{y}_l(m_k),\bar{y}_m(m_{k'}))=0$ for
$l\neq m$. Thus
\be\label{eq:cov_lev3_s}
\mathrm{Cov}(\bar{y}(m_k),\bar{y}(m_{k'}))=\sum_{l}\frac{\partial \bar{y}(m_k)}{\partial \bar{y}_l(m_k)}\mathrm{Cov}
(\bar{y}_l(m_k),\bar{y}_l(m_{k'}))\frac{\partial \bar{y}(m_{k'})}{\partial \bar{y}_l(m_{k'})}\,.
\ee
The term $\partial \bar{y}(m_k)/\partial \bar{y}_l(m_k)$ is taken from \eq{eq:dy_dyl}, while for the covariance
matrix we write
\be
\mathrm{Cov}(\bar{y}_l(m_k),\bar{y}_l(m_{k'}))=\sum_{s,t}\frac{\partial \bar{y}_l(m_k)}{\partial y_{ls}(m_k)}
\mathrm{Cov}(y_{ls}(m_k),y_{lt}(m_{k'}))
\frac{\partial \bar{y}_l(m_{k'})}{\partial y_{lt}(m_{k'})}\,,
\ee
where $\partial \bar{y}_l(m_k)/\partial y_{ls}(m_k)$ is taken from (\ref{eq:dyl_dyls}), and the 
covariance matrix of the r.h.s.~is directly computed by jackknife (cf.~\eq{eq:cov_jack}).
This procedure can be straightforwardly generalized to the other fitting functions $f(z,\{c\})$.
For more complicated fitting functions the procedure may become quite knotty.
In that case, it may be convenient to numerically approximate the partial derivatives 
of the coefficients $\{c\}$ with respect to $\bar{y}(m_k)$.

\begin{figure}[h]
\vspace{1cm}
\centering
\begin{minipage}[b]{0.5\textwidth}
\centering \includegraphics[width=6.3cm,height=4.8cm]{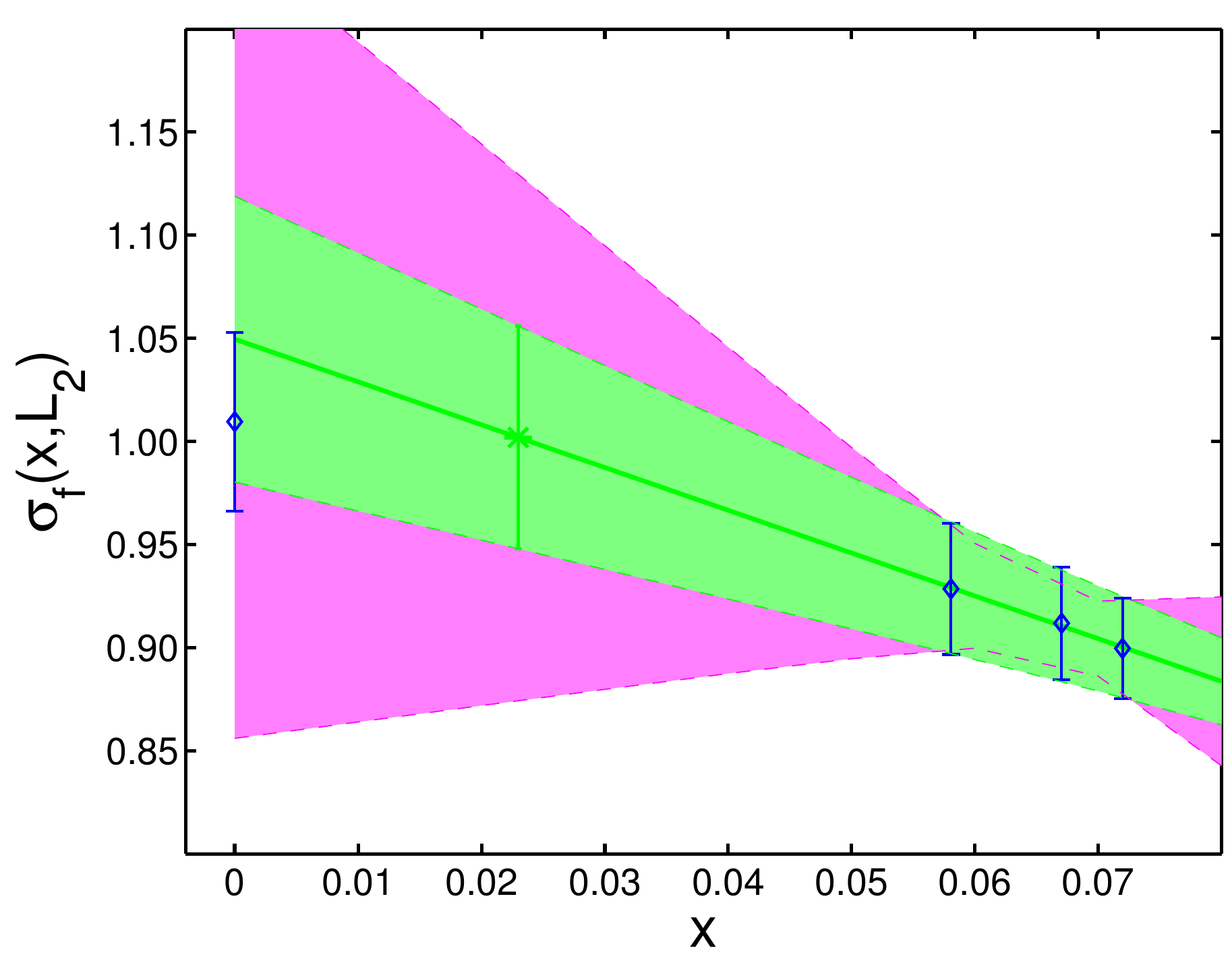} 
\end{minipage}%
\begin{minipage}[b]{0.5\textwidth}
\includegraphics[width=6.3cm,height=4.8cm]{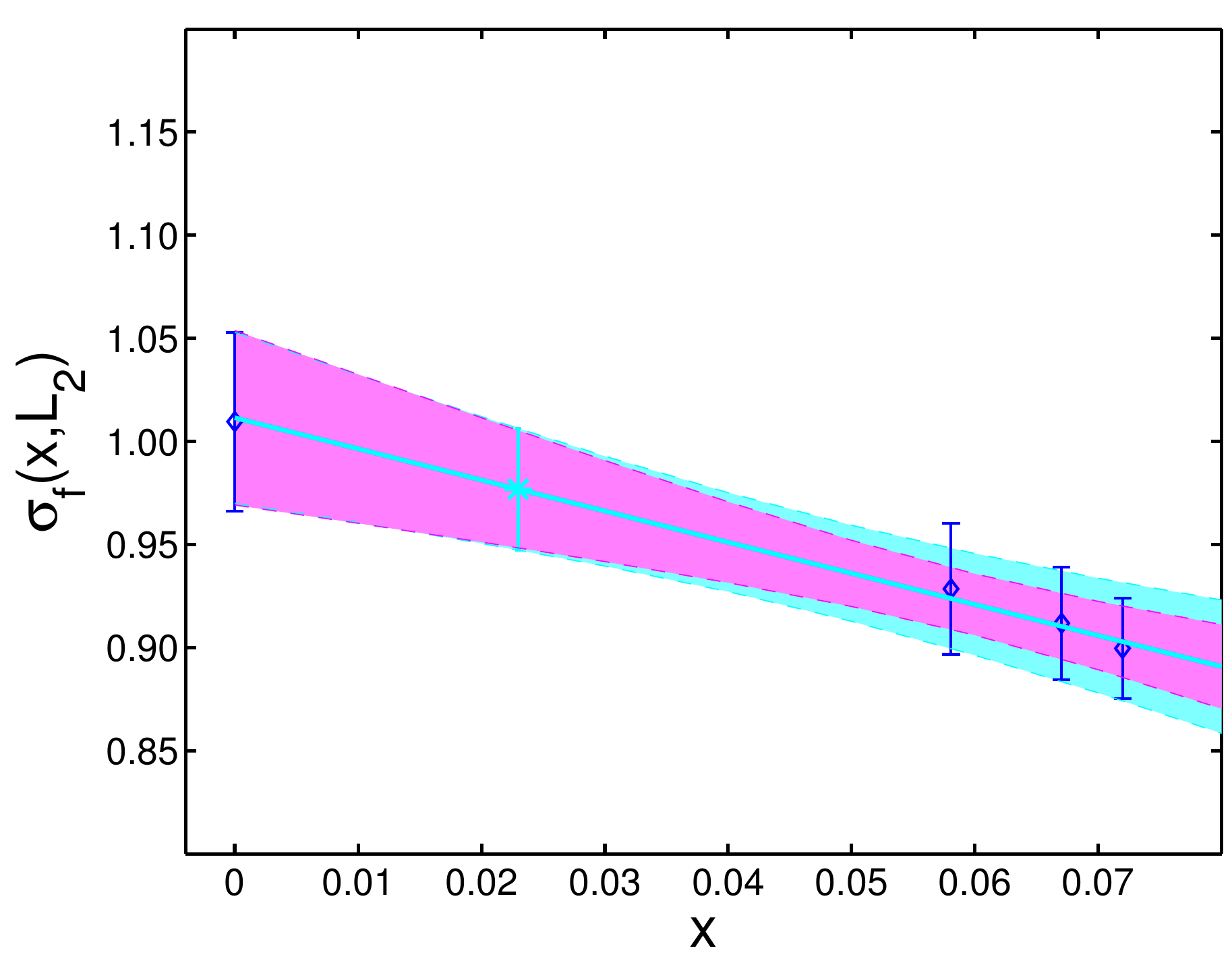}
\end{minipage}
\vspace{-0.5cm}
\mycaption{Importance of data correlation for the step scaling function
$\sigma_\mathrm{f}(L_2)$.}\label{fig:corr_DC_S2}
\end{figure}

The importance of the correlation between the data can be better understood
by looking at \fig{fig:corr_DC_S2}.
The blue data points (diamonds) and the interpolated green and cyan points (asterisks) 
are the same appearing in \fig{fig:DC_S2}. For the plot on the left only the data at finite
heavy quark mass have been considered in the fits. The magenta area (the bigger one) covers one standard deviation
for the fitting line if the correlation between the data associated with different heavy quark masses
is not included, while the green area is computed taking the correlation into account.
Similarly for the plot on the right. Here the static data point is included. The cyan area (the bigger one)
has been computed considering the correlation, while the magenta one ignores it.

\chapter{Estimate of the uncertainty related to the scale setting}\label{app:err_r0}

In this appendix we explain how the statistical 
uncertainty on $r_0/a$ propagates
into the quantities of interest in the work 
combining HQET and relativistic QCD in the
Step Scaling Method.

The quantity $r_0/a$ has been computed in \cite{pot:r0_SU3,pot:intermed,Guagnelli:2002ia} for
the range of couplings covering all our SSM simulations with a statistical accuracy
between 0.5\% and 2.0\%. Here we show that for our observables and our precision 
this uncertainty is negligible.
The analysis is carried out for the static observables, but it remains valid
for the corresponding observables at finite heavy quark mass too, because the
uncertainty on $r_0/a$ can be seen as an uncertainty on the volume, and the light
degrees of freedom,
which are dealt with by HQET and QCD in the same way,
are much more sensitive than the heavy quark to volume changes.
Bearing in mind that in our computations the light quark is fixed to be the strange
one, we consider an observable $\mathcal{O}(L/a,r_0/a)$ and observe that the uncertainty
$\Delta(r_0/a)$ propagates as
\be\label{eq:general_uncertainty}
\Delta(\mathcal{O}(r_0/a,L/a))=\frac{\partial \mathcal{O}(r_0/a,L/a)}{\partial (r_0/a)}\Delta(r_0/a)\,.
\ee
The arguments of $\mathcal{O}$ differs from the ones appearing
in the fourth chapter. There the propagation of the uncertainty 
on $r_0/a$ is not discussed, and the explicit dependence on it is omitted
to lighten the notation.
The statistical error on $r_0/a$ can be seen as an uncertainty on the volume, thus
suggesting the following strategy for the computation of (\ref{eq:general_uncertainty}).
One approximates the derivative in \eq{eq:general_uncertainty} by computing
the observable $O$ with simulation parameters $(\beta,L/a,\kappa_\mathrm{s},\kappa_\mathrm{crit})$, and 
then repeats the computation
with parameters $(\beta,L'/a,k_\mathrm{s},k_\mathrm{c})$. The change $L/a\to L'/a$ at
fixed $\beta$ clearly
plays the role of varying the volume in physical units. 
In order to optimally approximate the derivative in \eq{eq:general_uncertainty},
the values of $L/a$ and $L'/a$ should be as near as possible. The values of the
hopping parameters can be left identical for
the two computations, because they are protected by changes of the volume by the axial Ward identity.

\section{Static decay constant}\label{app:err_r0_dc}

We borrow the notation from the fourth chapter, and consider the quantities
\begin{align}
\YRGI(L/r_0)&= I^\mathrm{stat}(r_0/L)\cdot\lim_{a/L\to 0}\,\frac{\Xi^{(0)}(L/a)}{\Xi(r_0/a,L/a)}X_\mathrm{SF}(r_0/a,L/a)\,,
\label{eq:YRGI_CL1}\\
&  \nonumber\\[-1.7ex]
&= I^\mathrm{stat}(r_0/L)\cdot\lim_{a/L\to 0}Y_\mathrm{SF}(r_0/a,L/a)\,.\label{eq:YRGI_CL2}
\end{align}
at fixed $L/r_0$. The aforementioned strategy is applied to $Y_\mathrm{SF}$ on the small volume with $L/r_0=0.8$, giving
the results in the following 
table for 
\be
(\beta,\kappa_\mathrm{s},\kappa_\mathrm{crit})=(6.4200,0.135015,0.135616)\,.\nonumber
\ee 
{\hspace{1cm}
\footnotesize{
\begin{tabular}[h]{ccccc}
\hline
\hline
 & & & & \\[-0.3ex]
$L/a$ & $X_\mathrm{SF}(r_0/a,L/a)$ & $[\Xi(r_0/a,L/a)]^{-1}$ & $\Xi^{(0)}(L/a)$ & $Y_\mathrm{SF}(r_0/a,L/a)$ \\[-0.3ex]
 & & & & \\
\hline
        & & & & \\[-1ex]
    8    &-1.8372(46) &-0.54673(130) &-1.5996643156321 &-1.6068(55) \\[-1ex]
        & & & & \\
\hline
        & & & & \\[-1ex]
    10    &-1.9048(37) &-0.52842(37) &-1.6011462370857 &-1.6116(33) \\[-1ex]
        & & & & \\
\hline
\hline
& & & &\\[-1ex]
\end{tabular}}}
\normalsize
\newline 
where the values of $\Xi^{(0)}$ are taken from \cite{zastat:pap3}. By observing that the ratio 
\be
Y_\mathrm{SF}(r_0/a,L/a=8)/Y_\mathrm{SF}(r_0/a,L/a=10)=1.003(4)\,,\nonumber
\ee 
one concludes that even increasing
the volume by 25\%, no appreciable change is observed in $Y_\mathrm{SF}$, and the uncertainty on $r_0/a$
can thus be considered negligible. The factor $I^\mathrm{stat}$ is already extrapolated to the continuum,
and this discussion does not affect it.
It is finally straightforward to extend the validity of this result to the step
scaling functions of the decay constant, i.e. $Y_\mathrm{RGI}(2L/r_0)/(2^{3/2}Y_\mathrm{RGI}(L/r_0))$.

\begin{figure}[t]
\centering
\includegraphics[scale=0.6]{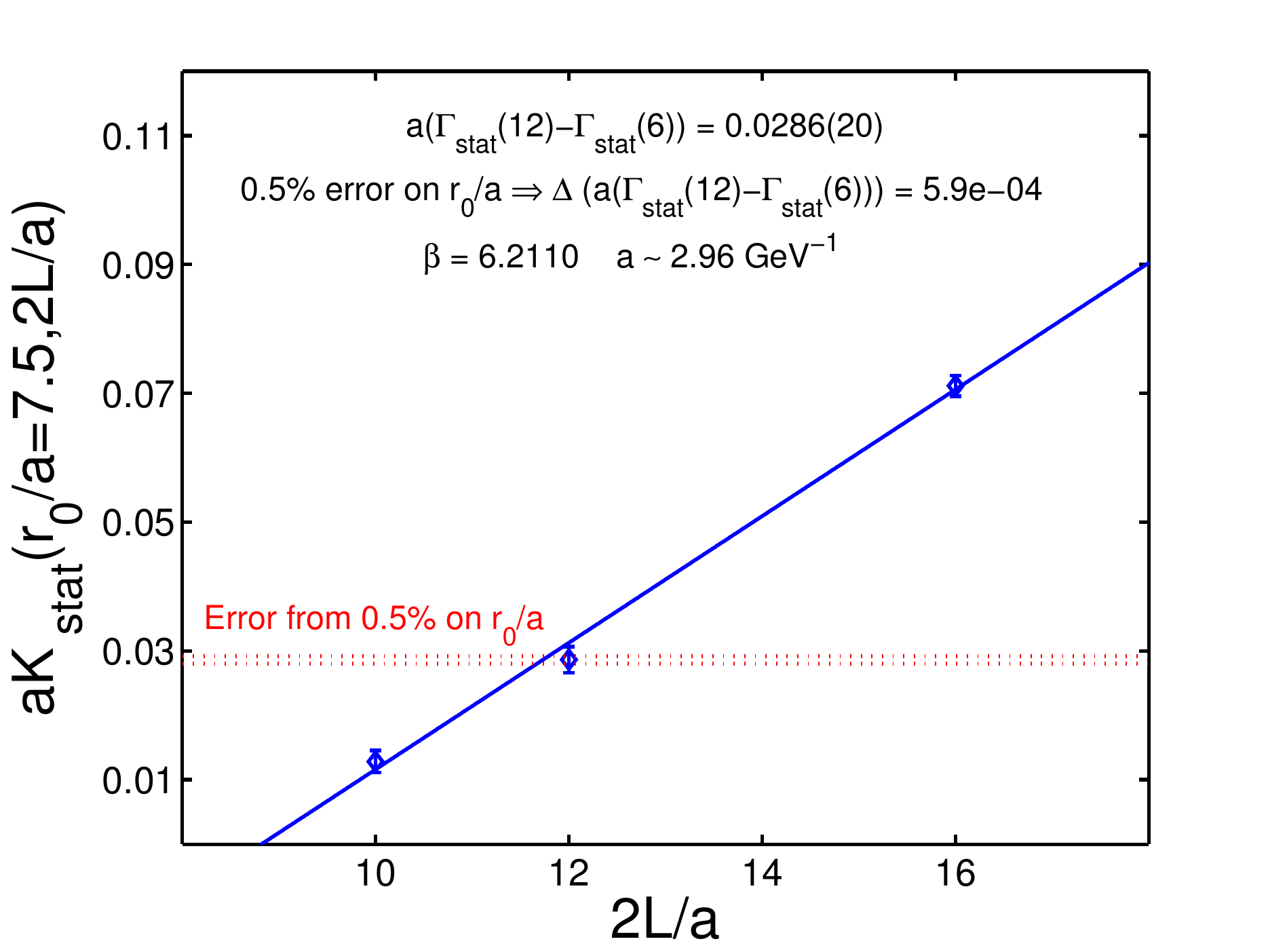} 
\vspace{-0.5cm}
\mycaption{Estimate on the propagation of the uncertainty on $r_0/a$ for
the static meson step scaling function.}\label{fig:sigmam_l_error}
\end{figure}


\section{Static meson step scaling function}\label{app:err_r0_m}

With the notation borrowed from the fourth chapter 
\begin{align}
\Gamma_\mathrm{stat}(r_0/a,L/a)&= \frac{1}{2a}\ln\left[\frac{\fastatimpr(r_0/a,L/a,x_0-a)}
{\fastatimpr(r_0/a,L/a,x_0+a)}\right]_{x_0=L}\,,\\
 &  \nonumber\\[-1ex]
\Sigma_\mathrm{m,stat}(r_0/a,2L/a)&= 2L[\Gamma_\mathrm{stat}(r_0/a,2L/a)-\Gamma_\mathrm{stat}(r_0/a,L/a)]\,,
\end{align}
we write the propagation of the uncertainty on $r_0/a$ as
\begin{align}
\frac{\Delta \Sigma_\mathrm{m,stat}(r_0/a,2L/a)}{\Delta(r_0/a)}&= \frac{2L}{a}\cdot 
\frac{\partial (aK_\mathrm{stat}(r_0/a,2L/a))}{\partial (r_0/a)}\,,\\
&  \nonumber\\[-1ex]
K_\mathrm{stat}(r_0/a,2L/a)&= \Gamma_\mathrm{stat}(r_0/a,2L/a)-\Gamma_\mathrm{stat}(r_0/a,L/a)\,.
\end{align}

\noindent On a volume with $L/r_0=0.8$ and with simulation parameters 
\be
(\beta,\kappa_\mathrm{s})=(6.2110,0.134766)\,,\nonumber
\ee
we follow the strategy used for the static
decay constant, and get the results in \tab{tab:Kstat}.

\begin{table}[hb]
\centering
\footnotesize
\begin{tabular}[h]{cccc}
\hline
\hline
   & & &\\[-1ex]
$L/a$ :& 5 & 6 & 8\\[-1ex]
   & & &\\
\hline
   & & &\\[-1ex]
$aK_\mathrm{stat}(r_0/a,2L/a)$ :&  0.0128(17) & 0.0286(20)& 0.0711(16)\\[-1ex]
   & & &\\
\hline
\hline
\end{tabular}
\mycaption{Results of the test for $aK_\mathrm{stat}$.}
\label{tab:Kstat}
\end{table}

These results are graphically represented in \fig{fig:sigmam_l_error}. From the slope
of the linear fitting curve we estimate that, for the point with $L/a=6$,
the uncertainty introduced by $r_0/a$ is more than three times smaller
than the statistical error from the Monte Carlo simulation, and it can thus be neglected.
The Monte Carlo uncertainties computed here are of the same order (in physical units)
of the ones appearing in \tab{tab:Gamma_stat_data}. It justifies the validity
of our estimate for all computed data.

\chapter{Perturbation theory results}\label{app:PT_results}

In this appendix we report the details 
on the results relative to the perturbative
computations described in chapter \ref{chap:Zspin}. 
All numbers are given with 15 decimal digits, although
the last two or three of them may be insignificant.
The expectation values of the one-loop diagrams are
normalized by the corresponding tree-level values.
We indicate with the shorthand $\mathcal{P}_3$ the Polyakov 
loop (\ref{eq:Polyakov_loop_def}) in the $z$-direction, and with 
$\mathcal{P}_{3,\mathrm{ins}}$ the loop with the insertion of the clover leaf operator.
We use lattice units throughout the appendix.

\section{Tree-level}\label{asect:PT_tree_level}

With the boundary condition defined in \app{app:Not_Conv}, the tree-level
of the Polyakov loop with and without insertion of the 
clover leaf operator is given by
\begin{align}
&  L^2\langle \tr(\mathcal{P}_3(x)E_1(x)) \rangle_{g_{\hspace{-0.005cm}0}^{\phantom{()}}\hspace{-0.1cm}=0}=
L^2\sum_{m=1}^3\exp\left\{\frac{i}{L}\left[x_0\angleprime_m+(L-x_0)\angle_m\right]\right\}\nonumber\\
&  \hspace{5cm}\times\sin\left\{\frac{1}{L^2}(\angleprime_m-\angle_m)\right\}\,,\\
&  \langle \tr(\mathcal{P}_3(x)) \rangle_{g_{\hspace{-0.005cm}0}^{\phantom{()}}\hspace{-0.1cm}=0}=
\sum_{m=1}^3\exp\left\{\frac{i}{L}\left[x_0\angleprime_m+(L-x_0)\angle_m\right]\right\}\,.
\end{align}
With the choice of the angles $\angleprime$ and $\angle$ defining the ``point A'' and $x_0=L/2$, we get
\be\label{eq:Zspin_TL_expr}
Z^{(0)}=\left.L^2\frac{ \langle \tr(\mathcal{P}_3(x)E_1(x)) \rangle}{\langle \tr(\mathcal{P}_3(x)) \rangle }\right|_{g_{\hspace{-0.005cm}0}^{\phantom{()}}\hspace{-0.1cm}=0}=\frac{\pi}{6}\cdot\frac{1+\sqrt{3}}{2-\sqrt{3}}+\mathrm{O}\left((1/L)^4\right)\,.
\ee
The raw data are presented in \Tab{tab:TL_PT}.

\begin{table}[b]
\centering
\footnotesize
\begin{tabular}{ccc}
\hline
\hline
& & \\
$L$ &$Z^{(0)}$ \\
& & \\
\hline
   &   \\[-1ex]
4  &    5.328760954837700 \\[0.4ex]                                       
6  &    5.336729884747326 \\[0.4ex]                                       
8  &    5.338071744079632 \\[0.4ex]                                       
10  &   5.338438481936987 \\[0.4ex]                                      
12  &   5.338570217499896 \\[0.4ex]                                      
14  &   5.338626689965802 \\[0.4ex]                                       
16  &   5.338654098887551 \\[0.4ex]                                       
18  &   5.338668685775827 \\[0.4ex]                                       
20  &   5.338677021409115 \\[0.4ex]                                      
22  &   5.338682062421030 \\[0.4ex]                                       
24  &   5.338685255099240 \\[0.4ex]                                       
26  &   5.338687356279386 \\[0.4ex]                                       
28  &   5.338688784675057 \\[0.4ex]                                       
30  &   5.338689783019587 \\[0.4ex]                                       
32  &   5.338690497745269 \\[0.4ex]                                       
34  &   5.338691020276737 \\[0.4ex]                                       
36  &   5.338691409429812 \\[0.4ex]                                       
38  &   5.338691704052065 \\[0.4ex]                                       
40  &   5.338691930408358 \\[0.4ex]                                       
42  &   5.338692106630467 \\[0.4ex]                                       
44  &   5.338692245472193 \\[0.4ex]                                       
46  &   5.338692356057547 \\[0.4ex]                                      
48  &   5.338692445014839 \\[0.4ex]
   &   \\[-2ex]
\hline
\hline
\end{tabular}
\mycaption{Raw data for $Z^{(0)}$ defined in \eq{eq:Zspin_TL_expr}.}
\label{tab:TL_PT}
\end{table}

\clearpage

\section{One-loop}\label{asect:PT_one_loop}
$ $\\
\begin{table}[hc]
\centering
\footnotesize
\begin{tabular}{cccc}
\hline
\hline
& & &\\[-1ex]
\multicolumn{4}{c}{Observable $\mathcal{P}_3^{\phantom{()}}$}\\[-1ex]
& & &\\
\hline
& & \\
$L$ &$\langle \tr\{W^{(2a)}_{\vec\ell}\}\rangle_0^{\phantom{()}}$ &
$\langle \tr\{W^{(2b)}_{\vec\ell}\}\rangle_0^{\phantom{()}}$ & 
$\langle \tr\{W^{(1)}_{\vec\ell}S_\mathrm{tot}^{(1)}\}\rangle_{0,\mathrm{g}}^{\phantom{()}}$ \\
& & \\
\hline
   &   \\[-1ex]
4  &   -0.205624873077021 & -0.413925832829039  & -0.052711610791353\\[0.4ex]
6  &   -0.335983123704290 & -0.619721517834779  & -0.047296572492624\\[0.4ex]
8  &   -0.466256052696041 & -0.826190695109537  & -0.045862904701804\\[0.4ex]
10  &  -0.596569618044621 & -1.032755647453710  & -0.045306225863750\\[0.4ex]
12  &  -0.726911000985525 & -1.239339863635439  & -0.045026523495318\\[0.4ex]
14  &  -0.857269180261336 & -1.445927930087319  & -0.044864334845029\\[0.4ex]
16  &  -0.987637860416361 & -1.652515987946074  & -0.044761438013021\\[0.4ex]
18  &  -1.118013423667861 & -1.859103059892027  & -0.044691916905935\\[0.4ex]
20  &  -1.248393696882175 & -2.065688990758283  & -0.044642684612822\\[0.4ex]
22  &  -1.378777313656793 & -2.272273864188386  & -0.044606518969622\\[0.4ex]
24  &  -1.509163378918473 & -2.478857821155862  & -0.044579158321204\\[0.4ex]
26  &  -1.639551285119290 & -2.685441002278846  & -0.044557951823340\\[0.4ex]
28  &  -1.769940607019591 & -2.892023531121173  & -0.044541178560616\\[0.4ex]
30  &  -1.900331038967060 & -3.098605511382531  & -0.044527680992760\\[0.4ex]
32  &  -2.030722356124227 & -3.305187028572897  & -0.044516656863397\\[0.4ex]
34  &  -2.161114389713687 & -3.511768152828316  & -0.044507535716189\\[0.4ex]
36  &  -2.291507010754891 & -3.718348941705426  & -0.044499902806717\\[0.4ex]
38  &  -2.421900119103333 & -3.924929442602486  & -0.044493450675978\\[0.4ex]
40  &  -2.552293635894514 & -4.131509694747048  & -0.044487947452020\\[0.4ex]
42  &  -2.682687498227030 & -4.338089730788150  & -0.044483215574556\\[0.4ex]
44  &  -2.813081655352524 & -4.544669578061099  & -0.044479117195471\\[0.4ex]
46  &  -2.943476065899737 & -4.751249259588366  & -0.044475543966035\\[0.4ex]
48  &  -3.073870695820916 & -4.957828794872859  & -0.044472409769887\\[0.4ex]
   &   \\[-2ex]
\hline
\hline
\end{tabular}
\mycaption{Expectation values for the one-loop diagrams of $\mathcal{P}_3$. The tadpole contributions
in the fourth column are restricted to the ghost and gluon cases.}
\label{tab:P3_PT_gauge}
\end{table}

\begin{table}
\centering
\footnotesize
\begin{tabular}{cccc}
\hline
\hline
& & &\\[-1ex]
\multicolumn{4}{c}{Observable $\mathcal{P}_{3,\mathrm{ins}}^{\phantom{()}}$}\\[-1ex]
& & &\\
\hline
& & &\\
$L$ &$\langle \tr\{W^{(2a)}_{\vec\ell}\}\rangle_0^{\phantom{()}}$ &
$\langle \tr\{W^{(2b)}_{\vec\ell}\}\rangle_0^{\phantom{()}}$ & 
$\langle \tr\{W^{(1)}_{\vec\ell}S_\mathrm{tot}^{(1)}\}\rangle_{0,\mathrm{g}}^{\phantom{()}}$\\
& & &\\
\hline
   &   \\[-1ex]
4  &   -0.168913284554384 &  -0.828842066442129  & -0.103736065134993\\[0.4ex]
6  &   -0.328214103185254 &  -1.033387450326697  & -0.095754952883402\\[0.4ex]
8  &   -0.473097009376418 &  -1.239118347533455  & -0.089957016052714\\[0.4ex]
10  &  -0.613350760169861 &  -1.445417860912832  & -0.086224254118902\\[0.4ex]
12  &  -0.751325152405088 &  -1.651913219737156  & -0.083638868463676\\[0.4ex]
14  &  -0.887922338616729 &  -1.858481585753028  & -0.081746166996522\\[0.4ex]
16  &  -1.023587362197832 &  -2.065078524083960  & -0.080302220743496\\[0.4ex]
18  &  -1.158574780275292 &  -2.271686295222065  & -0.079165010168635\\[0.4ex]
20  &  -1.293045010271327 &  -2.478297336786256  & -0.078246482909905\\[0.4ex]
22  &  -1.427106264182078 &  -2.684908327463667  & -0.077489242249884\\[0.4ex]
24  &  -1.560835300983253 &  -2.891517830373068  & -0.076854314324669\\[0.4ex]
26  &  -1.694288710683514 &  -3.098125281857394  & -0.076314322909446\\[0.4ex]
28  &  -1.827509507624724 &  -3.304730531080386  & -0.075849482178954\\[0.4ex]
30  &  -1.960531207496043 &  -3.511333621025847  & -0.075445140515130\\[0.4ex]
32  &  -2.093380466272667 &  -3.717934681183583  & -0.075090218683396\\[0.4ex]
34  &  -2.226078853805829 &  -3.924533874158024  & -0.074776184725976\\[0.4ex]
36  &  -2.358644083951775 &  -4.131131369165570  & -0.074496362329055\\[0.4ex]
38  &  -2.491090890892763 &  -4.337727329248135  & -0.074245452725766\\[0.4ex]
40  &  -2.623431667925845 &  -4.544321905574580  & -0.074019197063928\\[0.4ex]
42  &  -2.755676942497500 &  -4.750915235397367  & -0.073814133373627\\[0.4ex]
44  &  -2.887835735723776 &  -4.957507441885920  & -0.073627418607532\\[0.4ex]
46  &  -3.019915838719822 &  -5.164098634849609  & -0.073456696294221\\[0.4ex]
48  &  -3.151924027957572 &  -5.370688911857737  & -0.073299996770239\\[0.4ex]
   &   \\[-2ex]
\hline
\hline
\end{tabular}
\mycaption{Expectation values for the one-loop diagrams of $\mathcal{P}_{3,\mathrm{ins}}$. The tadpole contributions
in the fourth column are restricted to the ghost and gluon cases.}
\label{tab:P3ins_PT_gauge}
\end{table}

\begin{table}
\centering
\footnotesize
\begin{tabular}{ccc}
\hline
\hline
& & \\[-1ex]
& Observable $\mathcal{P}_{3}^{\phantom{()}}$& Observable $\mathcal{P}_{3,\mathrm{ins}}^{\phantom{()}}$\\[-1ex]
& & \\
\hline
& & \\
$L$ & $\langle \tr\{W^{(1)}_{\vec\ell}S_\mathrm{tot}^{(1)}\}\rangle_{0,\mathrm{q}}^{\phantom{()}}$& 
$\langle \tr\{W^{(1)}_{\vec\ell}S_\mathrm{tot}^{(1)}\}\rangle_{0,\mathrm{q}}^{\phantom{()}}$\\
& & \\
\hline
   &   \\[-1ex]
4   &  0.102777840715519& 0.040970131432917\\[0.4ex]
6   &  0.067072789721730& 0.029695487491120\\[0.4ex]
8   &  0.054503845697270& 0.024694965023325\\[0.4ex]
10  &  0.049999831469804& 0.022511928275922\\[0.4ex]
12  &  0.048081890180969& 0.021373558175997\\[0.4ex]
14  &  0.047098380274648& 0.020672378807829\\[0.4ex]
16  &  0.046518689745521& 0.020190464451924\\[0.4ex]
18  &  0.046143744791383& 0.019835881834839\\[0.4ex]
20  &  0.045885452353958& 0.019562935855346\\[0.4ex]
22  &  0.045699248493680& 0.019345912965656\\[0.4ex]
24  &  0.045560279263445& 0.019169041088309\\[0.4ex]
26  &  0.045453664962997& 0.019022036326953\\[0.4ex]
28  &  0.045370006920341& 0.018897879688656\\[0.4ex]
30  &  0.045303112072779& 0.018791605608642\\[0.4ex]
32  &  0.045248756321208& 0.018699597263028\\[0.4ex]
34  &  0.045203974156447& 0.018619155375739\\[0.4ex]
36  &  0.045166631835496& 0.018548223386947\\[0.4ex]
38  &  0.045135161130017& 0.018485206249965\\[0.4ex]
40  &  0.045108387852467& 0.018428847480900\\[0.4ex]
42  &  0.045085418331307& 0.018378143643499\\[0.4ex]
44  &  0.045065562413796& 0.018332283573234\\[0.4ex]
46  &  0.045048280116459& 0.018290604352713\\[0.4ex]
48  &  0.045033143949942& 0.018252558876089\\
\hline
  & &\\[-1.5ex]
\multicolumn{3}{c}{$N_\mathrm{f}=2$,\hspace{0.5cm} $\theta=-\pi/3$,\hspace{0.5cm} $m_0=0$} \\
  & &  \\[-1.5ex]
\hline
\hline
\end{tabular}
\mycaption{Expectation values for the one-loop quark tadpole contributions.}
\label{tab:P_PT_quark}
\end{table}

\begin{table}
\centering
\footnotesize
\begin{tabular}{ccc}
\hline
\hline
& & \\[-1ex]
 \multicolumn{3}{c}{Observable $\mathcal{P}_{3,\mathrm{ins}}^{\phantom{()}}$}\\[-1ex]
& & \\
\hline
& & \\
$L$ &$\langle \tr\{W^{(1)}_{\vec\ell}\delta S_\mathrm{tot,b}^{(1)}\}\rangle_0^{\phantom{()}}$ & 
$\langle \tr\{W^{(1)}_{\vec\ell}\delta S_\mathrm{tot,b}^{(1)}\}\rangle_0^{\phantom{()}}$\\
& & \\[-1ex]
&$N_\mathrm{f}=0$ &$N_\mathrm{f}=2$ \\
& & \\
\hline
& & \\[-1ex]
4  &  0.044525081527306&  0.012686646544393\\[0.4ex]
6  &  0.029669953296613&  0.008453936467964\\[0.4ex]
8  &  0.022250779285418&  0.006339972043808\\[0.4ex]
10 &  0.017800255298336&  0.005071872742815\\[0.4ex]
12 &  0.014833435923651&  0.004226529231324\\[0.4ex]
14 &  0.012714333177513&  0.003622727809534\\[0.4ex]
16 &  0.011125024343901&  0.003169881936371\\[0.4ex]
18 &  0.009888902397903&  0.002817670515825\\[0.4ex]
20 &  0.008900007976896&  0.002535902272878\\[0.4ex]
22 &  0.008090914043919&  0.002305365047637\\[0.4ex]
24 &  0.007416669872386&  0.002113250913414\\[0.4ex]
26 &  0.006846155994550&  0.001950692919840\\[0.4ex]
28 &  0.006357144340312&  0.001811357565460\\[0.4ex]
30 &  0.005933334383781&  0.001690600299307\\[0.4ex]
32 &  0.005562500760730&  0.001584937716757\\[0.4ex]
34 &  0.005235294679453&  0.001491706042430\\[0.4ex]
36 &  0.004944444866595&  0.001408833453618\\[0.4ex]
38 &  0.004684210848468&  0.001334684302318\\[0.4ex]
40 &  0.004450000249276&  0.001267950071027\\[0.4ex]
42 &  0.004238095433409&  0.001207571484223\\[0.4ex]
44 &  0.004045454700235&  0.001152681862284\\[0.4ex]
46 &  0.003869565341325&  0.001102565252704\\[0.4ex]
48 &  0.003708333433511&  0.001056625028544\\[0.4ex]
& & \\[-2ex]
\hline
\hline
\end{tabular}
\mycaption{$\Oa$-improvement counterterms for $\mathcal{P}_{3,\mathrm{ins}}$ at one-loop order.
The simple Polyakov loop $\mathcal{P}_3$ does not require any counterterm.}
\label{tab:P_PT_impr}
\end{table}






\chapter*{Acknowledgements}

Here I would like to thank all persons whose contribution
has been indispensable for a successful completion of this thesis.

\bi
\item I would like to thank Rainer Sommer for proposing me the subject of
this thesis, and for supporting me all the while with inestimable suggestions and 
encouragements. His help has been for me invaluable.
\item Many thanks to Ulli Wolff for fruitful discussions, especially on
perturbation theory and its implementation.
\item I thank Nazario Tantalo for providing the relativistic QCD data and
for useful discussions about the combination of them with HQET.
\item I am grateful to Michele Della Morte, Nicolas Garron, Jochen Heitger, {Bj\"orn} Leder and Harvey Meyer 
for an enjoyable and fruitful collaboration.
\item Thanks to Hubert Simma for introducing me to the TAO programming language.
I thank DESY for allocating computer time on the APEmille computers at DESY Zeuthen
and the APE group for its help, in particular Dirk Pleiter for the prompt assistance.
\item I thank all collaborators of DESY Zeuthen and the Humboldt University in Berlin
      with whom I spent the three years of my PhD studies.
\item Special thanks to Griseldis. She is part of my life.
\item Finally I would like to thank my family, which I have always felt near me.
\ei


\chapter*{Lebenslauf}

\begin{tabular}{ll}

Name: & \dcauthorname~\dcauthorsurname \\[5ex]

07.1999 & Abitur am ``Liceo Scientifico Statale B.~Touschek''\\
        & in Grottaferrata, Rom (Italien) \\[2ex]

10.1999--12.2003 & Studium an der ``Universit\`a di Roma II Tor Vergata''\\
                 & in der Fachrichtung Physik\\[2ex]

01.2004--03.2004 & Wissenschaftlicher Mitarbeiter an\\
                 & der ``Universit\`a di Roma II Tor Vergata''\\[2ex]

seit 04.2004     & Wissenschaftlicher Mitarbeiter beim \\
                 & Deutschen Elektronen-Synchrotron DESY Zeuthen  

\end{tabular}

\selectlanguage{ngerman}
\chapter*{Selbst"andigkeitserkl"arung}

\noindent Hiermit erkl\"are ich, die vorliegende Arbeit selbst\"andig ohne fremde Hilfe verfa{\ss}t und nur die angegebene Literatur und Hilfsmittel verwendet zu haben.\\

\vspace{5cm}
\noindent\dcauthorname~\dcauthorsurname \\  
\dcdatesubmitted \\

%


\end{document}